%% file: thesis2e.tex
\documentclass[12pt]{book}
\usepackage{epsf,psfig}

\epsfverbosetrue
\begin{document}

\frontmatter

\include{titlepg}

\newpage

\tableofcontents

\include{preface}

\include{note}

\mainmatter

\include{introd}

\include{hubbge}

\include{hilbert}

\include{handgap}

\include{elepp}

\include{corre}
\include{conclusi}

\end{document}

%% file: titlepg.tex
\begin{titlepage}
\begin{center}
\vspace*{1cm}
\large
Nuno Miguel Machado Reis Peres\\
\vspace{20pt}
\Large {\sc The Many-Electron Problem\\in\\Novel 
Low-Dimensional Materials} \\
\vspace{20pt}
\medskip
University of \'Evora\\
\vspace{1cm}
\begin{figure}[htbp]
\begin{center}
\leavevmode
\hbox{%
\psfig{figure=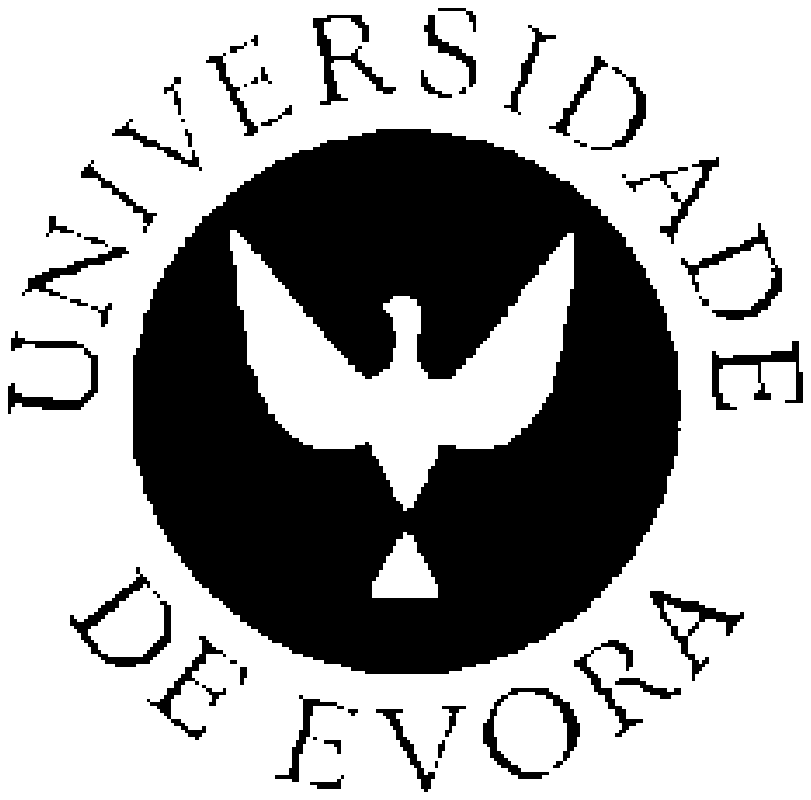,width=2.0in,angle=0}}
\end{center}
\end{figure}
\vspace{2cm}
A dissertation submitted for the degree of Doctor\\
\smallskip
of Philosophy in Physics (Solid State)\\
at the University of \'Evora.\\
\vspace{30pt}
\medskip
\'Evora -- 1998\\
\end{center}
\end{titlepage}


%% file: preface.tex
\pagestyle{headings}
\chapter*{Preface}
\addcontentsline{toc}{section}{Preface}
\pagestyle{myheadings}
\markboth{Preface}{Preface}

The work contained in this Thesis has mostly been done at the University of \'{E}vora, Portugal. Some  developments took place at the University of Urbana-Champaign, U. S. A..  I had Prof. Jos\'e Manuel Pereira Carmelo as supervisor, and with him I did most of my research work. To him I want to express my gratitude for suggesting me the topic of research developed in this Thesis. With Prof. Jos\'e
Carmelo I have learned valuable tools in theoretical physics. 

I also want to express my gratitude to Prof. Ant\'{o}nio Luciano Leite Veideira. He has been a true friend and with him I have learned many things beyond physics. 

I thank Prof. Dionys Baeriswyl for is interest in this work and for many stimulating discussions on related subjects, that took place on the later stage of my research work.

I want to acknowledge the librarian of the Faculty of Sciences of the University of Porto, Mrs. Rosalina Neves, for her efficiency in sending me by surface mail all the articles I requested.

Part of the material in this Thesis is published in the following papers:
\vskip .3cm
J. M. P. Carmelo and N. M. R. Peres, {\it Ground States of Integrable Quantum Liquids},
Phys. Rev. B {\bf 51}, 7481 (1995).
\vskip .3cm
J. M. P. Carmelo and N. M. R. Peres, {\it Symmetries and Pseudoparticles Transformations of 1D Non-Abelian Integrable Quantum Liquids}, J. Low Temp. Phys. {\bf  99}, 571 (1995).
\vskip .3cm
J. M. P. Carmelo and N. M. R. Peres, {\it Topological Ground-State Excitations and Symmetry in the 
Many-Electron One-Dimensional Problem}, Nucl. Phys. B {\bf  458}, 579 (1996). 
\vskip .3cm
N. M. R. Peres, J. M. P. Carmelo, D. K. Campbell, and A. W. Sandvik,
{\it Pseudoparticle Description of  the 1D Hubbard Model 
Electronic Transport Properties}, Z. Phys. B {\bf  103}, 217 (1997).
\vskip .3cm
J. M. P. Carmelo and N. M. R. Peres, {\it Complete Pseudohole and 
Heavy-Pseudoparticle Operator Representation for the 
Hubbard Chain}, Phys. Rev. B {\bf  56}, 3717 (1997).
\vskip .3cm
N. M. R. Peres, P. D. Sacramento, and J. M. P. Carmelo, {\it Charge and Spin Currents of the 1D Hubbard Model at Finite Energy}, submitted for publication in Phys. Rev. B (1998).
\vskip .3cm
 
The project associated with the research developed in this Thesis was partially financed by PRODEP 5.2, Ref. N.193.007.

%% file: note.tex
\chapter*{}
\addcontentsline{toc}{section}{Note}
\pagestyle{myheadings}
\markboth{Note}{Note}

\begin{flushright}
{\bf To Augusta,\\
Filipa\\
and\\
Augusto Luis.}\\
\vspace{5cm}     
Uno puede escribir un art\'{\i}culo o\\
un cuento porque se lo encomiendan (...);\\
a veces se le propone haste el tema,\\
y nada de ello me parece grave si uno\\
logra hacer suyo el proyeto y se\\
divierte escribi\'{e}ndolo.\\
Es m\'{a}s, s\'{o}lo concibo escribir\\
algo si me diverto, y s\'{o}lo puedo\\
divertirme si me intereso.\\ 
\vspace{1.5cm}
Javier Mar\'{\i}as,\\
in {\it Cuando fui mortal},\\
Alfaguara, Madrid, 1996.
\end{flushright}       

%% file: introd.tex
\pagestyle{headings}
\setcounter{chapter}{0}
\chapter{Introduction}
\label{introdu}

This Thesis deals with the one-dimensional (1D) Hubbard model
\cite{1Hubbard63,1Gutzwiller63,1Kanamori63}, which describes 
single-band
interacting (correlated) electrons in a (1D) lattice.
According to Lieb \cite{1Lieb95}:

\vspace{.5cm}

`` The Hubbard model is to the problem of electron correlations as the Ising
model is to the problem of spin-spin interactions; it is the simplest
possible model displaying many ``real world" features."

\vspace{.5cm}

The model depends on the
parameter $t$, the hopping integral, characterizing 
the kinetic energy of the electrons 
when they hop between two adjacent
lattice sites, and on the parameter $U$, 
the on-site Coulomb integral, characterizing the Coulomb
interaction energy when two electrons occupy the same lattice
site (Chapter \ref{hm}).
 
Strictly one-dimensional systems are not found in real
solid-state materials, however there are quasi-one-dimensional
materials in nature. Theoretically, a quasi-one-dimensional solid
-- the definition of which is quite broad \cite{1Little94}-- may
be represented by a three-dimensional
array of one-dimensional chains, such that
the hopping integrals
between the chains are much smaller than the hopping integral
along them \cite{1Bourbonnais91}. 
The correspondence with real materials -- for which $t$ and $U$
can be estimated from infrared spectroscopy data \cite{1Jacobsen86a,1Jacobsen86b,1Mila95}-- is then made by means
of quantum chemistry 
calculations \cite{1Parr50}, where the hopping integrals 
and the Coulomb on-site energy
are computed by means of the atomic (molecular)
wave functions of the atoms (molecules) of the solid under 
consideration. 

Obviously, approximating real materials by a simple one-dimensional
model is not enough in general to make a quantitative description 
of the solid properties. In spite of this, a description
based on the parameters $t$ and $U$ contains many
``real world" features
\cite{1Jacobsen86a,1Jacobsen86b}. On the other hand, effects like electron-phonon
coupling and more general Coulomb interactions, for example,
are needed to account for a more quantitative description of the
physical properties of real quasi-one-dimensional materials
\cite{1Su80,1Mazumdar83,1Baeriswyl85,1Campbell90,
1Pedron94,1Penc94,1Julien96}.

The systems described by the 1D Hubbard model 
(and its generalizations) have
narrow energy bands, and, therefore, the tight-binding approximation
for independent electrons can be applied \cite{1Ashcroft76}.
\section{Materials}
\pagestyle{myheadings}
\markboth{1 Introduction}{1.1 Materials}

There are several classes of quasi-one-dimensional materials 
in nature which can 
can be described by 1D electron models, 
the benzene molecule (C$_6$H$_6$) being
one of the simplest systems to which the Hubbard model can be
applied. [The molecular
$\pi$ orbitals of the six carbon atoms can be thought of as a lattice
with six sites populated by six electrons. The Coulomb interaction
have been computed by Parr {\it et al.} 
\cite{1Parr50} and are
$U=16.93$ eV, for the on-site Coulomb integral, and $V=9.027$ eV,
for the next-nearest-neighbour Coulomb integral (Chapter
\ref{hm}).]

In what follows we describe very briefly some of the most typical 
quasi-one-dimensional materials. 

\vspace{.5cm}
{\bf 1.)Inorganic Materials} 
\vspace{.5cm}

{\bf 1.a) Potassium
cyanoplatinates}: a typical example is 
K$_2$Pt(CN)$_4$Br$_{0.3}$.3H$_2$O which was synthesized 
during the 19th century. The chains are built by the
superposition of the {\it d} orbitals of the Pt atoms. 
The ratio of the conductivity along the chains 
($\sigma_{\parallel}$) over the
conductivity 
perpendicular to the chains ($\sigma_{\perp}$) is 
$\sigma_{\parallel}/\sigma_{\perp}\sim 10^5$, and the 
electronic density $n$ -- the ratio of the number of electrons $N$ 
over the number of lattice sites $N_a$ -- is 0.85
\cite{1Gruner94}. The values of 
$\sigma_{\parallel}$ and  $\sigma_{\perp}$ are a measure for
the atomic (or molecular) orbitals overlap along and 
perpendicular to the chain direction, 
respectively \cite{1Wooten72}.
Therefore,  $\sigma_{\parallel}/\sigma_{\perp}$ provides
a measure of the relative values of the hopping integrals.

\vspace{.5cm}
{\bf 1.b) Transition 
metalchalcogenides}: MX$_3$ and (MX$_4$)$_n$Y, 
with the metals M=Nb,Ta, the chalcogenides X=S,Se, 
and the halides Y=I,Br,Cl. The basic units of the chains
in MX$_3$ are MX$_6$ triangular prisms stacked
on top of each other, and in (MX$_4$)$_2$Y
are M$_2$X$_4$ octaedra.
Typical examples are
NbSe$_3$, TaS$_3$, (TaSe$_4$)$_2$I, and
(NbSe$_4$)$_2$I, with $\sigma_{\parallel}/\sigma_{\perp}\sim
(10-10^3)$ and $n \sim 1/4$ \cite{1Gruner94}. 

\vspace{.5cm}
{\bf 1.c) Transition metal bronzes}: A$_{0,3}$MoO$_3$, with
A=K,Rb,Tl. These materials are called blue bronzes due to
their metallic blue brightness. The chains are 
formed by the MoO$_6$
octaedra, and, in these materials,
$\sigma_{\parallel}/\sigma_{\perp}$ $\sim$
(10-10$^3$) and $n \sim 3/4$ \cite{1Gruner94}. 

\vspace{.5cm}
{\bf 1.d) Copper oxide compounds}: typical examples are the
1D antiferromagnetic Sr$_2$CuO$_3$ and SrCuO$_2$
materials \cite{1Motoyama96}. These systems can be modeled by 
the 1D Hubbard Model with $U>>t$ and $n=1$
(Chapter \ref{hm}). 
The Heisenberg exchange interaction $J=4t^2/U$ 
for these two compounds is 
$J=(2200 \pm 200)$ K and $J=(2100 \pm 200)$ K, respectively. 
The interaction between chains, $J'$,
obeys the ratio $J/J' \sim 10^5$ for Sr$_2$CuO$_3$
and $J/\vert J'\vert \sim$ ($5$ to $10$) for SrCuO$_2$. 
Intimately related to the 
spin-chain compounds are the spin-ladder compounds. 
These latter systems can form two, three, or more leg ladders.
Typical examples are SrCu$_2$O$_3$ (a two-leg spin 1/2 systems) and
Sr$_2$Cu$_2$O$_5$ (a three-leg spin 1/2 ladder)
\cite{1Dagotto96}. An 
interesting result is that the even-leg spin 1/2 ladders present
a gap in their excitation spectra, while the odd-leg spin
1/2 ladders do not. Upon doping (this means $n=1/2-\delta$, 
where $\delta$ is usually small when compared to 1/2)
these ladder materials present
both metallic and superconducting phases 
\cite{1Dagotto96,1Maekawa96}.

\vspace{.5cm}

{\bf 2.) Organic Linear Chain Compounds}

\vspace{.5cm}

{\bf 2.a) Polymers}: the typical example is polyacetylene. It
consists of a long chain of (CH)$_x$ units with alternating 
single and double bonds between the carbon atoms 
\cite{1Su80,1Baeriswyl85,1Baeriswyl86}.
The values of the transfer and
Coulomb integrals are taken to be $4t \sim 10$ eV and $U \sim 6.5$ eV,
respectively, and $n=1$ \cite{1Su80,1Baeriswyl85,1Baeriswyl86}.

\vspace{.5cm}

{\bf 2.b) Stacks of organic molecules}: these materials are made
of planar organic molecules stacked together. The organic 
molecules have $\pi$ molecular orbitals perpendicular to
the plane of the molecule. These orbitals overlap forming a 1D
chain. A typical example is the incommensurate 
(the average number of carriers per molecule can not be expressed by 
a simple rational number) tetrathiafulvalena-
tetracyanoquinodimethane -- TTF-TCNQ -- with $4t \sim 0.61$ eV,
$U \sim 1.1$ eV, and $n \sim 0.55$ 
\cite {1Jacobsen86a,1Jacobsen86b}. Other examples
are the salts M$_2$X, with
M=TMTSF,TMTTF (where TMTSF denotes tetramethyltetraselenafulvaleno and
TMTTF denotes tetramethyltetrathiafulvaleno) and 
X=Br,PF$_6$,ClO$_4$. These salts are quarter filling systems
(quarter-filled hole band: $n=1/4$), with the $t$ and $U$ parameters being of the order
$t \sim 0.25$ eV and $U \sim 1$ eV
($t_b$ and $t_c$, the interchain couplings, are of the order of $25$ meV and $1$ meV, respectively)
\cite{1Schwartz98}. These materials can also present a 
superconducting phase \cite{1Jerome82}.
\vspace{.5cm}

There are still other realizations of quasi-one-dimensional 
systems like in heterostructures
(an example is the AlGaAl-GaAs-AlGaAl material) and in nanotubes 
(an example is the B$_{1-\delta}$C$_{2+\delta}$N material) 
\cite{1Chang92,1Daniel93,1Angel97}.

Many of the materials we have briefly described are not
metals at zero temperature. Usually they are metals at room
temperature and undergo some type of phase transition
when the temperature is lowered. Depending
on the material and on the physical values of certain external 
parameters (like doping, external fields, pressure and cooling rate) they may, for example, undergo
structural, metal-insulator, spin density wave, or
superconducting phase transitions
\cite{1Jerome82,1Gruner94,1Rice79,1Conwell85,1Greene86,1Kotliar96}. 
 
When one performs
zero-temperature calculations and applies the results to
finite-temperature data, no more than a qualitative agreement
is to be expected. 
Nevertheless, the calculation of, for example, the 
zero-temperature 
optical conductivity has for some quasi-one-dimensional
materials been used with some success in describing the
optical conductivity of these materials in the (finite temperature)
metallic phase \cite{1Jacobsen86a,1Jacobsen86b,1Julien96}. However, due to its quasi-one-dimensional character, these materials can only be approximate by an one-dimensional 
model at moderate temperatures. At low temperatures 3D effects start to play a role. These effects may drive  phase transitions in these materials, as those indicated above.

\section{Theoretical results}
\pagestyle{myheadings}
\markboth{1 Introduction}{1.2 Theoretical results}

As the Hubbard chain is the simplest possible model for the study of
interacting electrons in a lattice, the theoretical study of the
model is by it self of interest \cite{1Lieb95}.
In this section we summarize very briefly some of
the previous work on the 1D Hubbard model. This includes
its exact solution, as well
as recent new results, some of which are presented in this Thesis.
The 1D Hubbard model is exactly solvable by the
Bethe-ansatz technique, first introduced by Bethe 
\cite{1Bethe31} in his solution of the 1D spin 1/2
Heisenberg chain. Roughly speaking, the model has charge 
and spin excitations
-- usually called holon and spinon
excitations -- which are decoupled
from each other. This feature is usualy
taken as a signature of 1D electronic correlated systems. The experimental
observation of the holon and spinon spectra has recently
been achieved by photoemission spectroscopy by Kim {\it et al.}
\cite{1Shen96} in the 1D SrCuO$_2$ material (but not in the 2D related 
Sr$_2$CuO$_2$Cl$_2$ cuprate). Moreover, 
these authors find that the Hubbard model in the limit
$U\gg 4t$ (the parameters used are $t=0.60$ eV and $U=7.2$ eV)
``agrees strikingly" with the data.

Using the same method as Yang for a continuum multicomponent model \cite{1Yang67},
Lieb and Wu \cite{1Lieb68}
computed the ground-state energy at half filling
($n=1$) and zero magnetization ($m=0$). They
also found that for $n=1$ the charge excitations have a gap --
the Mott-Hubbard gap $\Delta_{MH}$--, for any value of the on-site
Coulomb integral $U$. 
Thus the model is an insulator at half filling and zero temperature. Experimental
evidence of 1D Mott-Hubbard insulating behavior
has been found in the quasi-one-dimensional HMTSF-TCNQF$_4$,
HMTTF-TCNQF$_4$, and DBTSF-TCNQF$_4$ organic compounds
\cite{1Jacobsen86b,1Torrance80} (all these three compounds
have $n=1$, and HMTSF, HMTTF, and TCNQF$_4$
denote hexamethylenetetraselenafulfalene, 
hexamethylenetetrathifulfalene, tetrafluoro-tetracyano-$p$-
quinodimethane, respectively).

Shiba \cite{1Shiba72} performed some numerical work for densities lower 
than half filling and computed the ground state energy and
the magnetic susceptibility. In the limit
$U\gg 4t$, it has been possible to compute 
analytically 
the ground-state energy for $n \leq 1$ \cite{1Baeriswyl88}. 
The thermodynamics of the Hubbard chain
has been described by Takahashi \cite{1Takahashi72}, who obtained
the full solution of the Lieb and Wu equations \cite{1Lieb68}.

The spectra 
and the momenta of
the low-lying excitations were determined by Ovchinnikov
\cite{1Ovchinnikov70} for the half-filled band case
and by Coll for any filling \cite{1Coll74}. 
Ovchinnikov \cite{1Ovchinnikov70} also remarked 
that for 
$U\gg 4t$ the spin spectrum is the same as for
the anti-ferromagnetic Heisenberg chain
\cite{1Faddeev81}. This fact is an 
indication of the factorization of the
Bethe-ansatz wave function into a free-spinless fermion
Slater-determinant and into the
anti-ferromagnetic Heisenberg wave function
\cite{1Bethe31}, as discussed later
by Ogata and Shiba \cite{1Ogata90}. 

In the presence of a magnetic field, the charge and spin 
low-lying excitations transform into 
more exotic $c$ and
$s$ excitations, and their energy spectra 
and momenta, as well as the low-energy thermodynamics
and static properties have been 
analytically computed  
\cite{1Carm90,1Carm91,1Carm92}
for electronic
and magnetization densities such that $n\leq 1$ and $m\leq n$,
respectively. The 
study of excitations of topological
character in a finite-size chain has been made by
Carmelo and Peres \cite{1Nuno95,1Nuno96}.
In addition to the $c$ and $s$ modes, the
1D Hubbard model has finite-energy charge ($c,\gamma$) 
and spin ($s,\gamma$) 
excitations, which have been considered
by Woynarovich \cite{1Woynarovich82a,1Woynarovich82b}. 
These charge and spin
excitations can be seen as bound states of $c$ and $s$ 
pseudoparticles. Their operator generators,
excitation spectra, and momenta have been computed by
Carmelo and Peres \cite{1Nuno97a}.

Due to its involved form, the Bethe-ansatz wave function is not
ideally suitable for the computation of operator mean values 
and correlation functions. These latter quantities can be
obtained in some limits by combining the Bethe-ansatz solution
with conformal-field theory
\cite{1Frahm90,1Frahm91}. 
These calculations require a simple representation 
for the eigenstates of the model. In
particular, the possibility of finding 
a set of operators (in second quantization)
which diagonalize the Hamiltonian is
a major step in that direction. An example of this general 
idea is  the bosonization technique
\cite{1Haldane81,1Schulz94}, where the g-ology Hamiltonians
\cite{1Solyom79} can be diagonalized by boson operators. 
The low-energy correlation functions can  then be
computed by expressing both the fermionic fields and the 
eigenstates of these models in terms of the boson operators. 

A central problem in this model is the study 
of the transport properties: quantities like
the charge and spin currents and the corresponding conductivity
spectra are of importance
\cite{1Maldague77,1Loh88,1Fye91,1Horsch93}. The low-frequency
properties of the 1D Hubbard model have been studied both
by means of bosonization techniques \cite{1Schulz94} 
and by means of the pseudoparticle representation
\cite{1Carm92a,1Carm92b}. However, the study of the finite-energy transport and spectral properties requires an operator representation
for the Bethe-ansatz solution at all energy scales.

Following previous  
results for the low-energy Hilbert sub-space
\cite{1Carm94a,1Carm94b}, Carmelo and
Peres \cite{1Nuno97a} suceeded in writing the Hamiltonian, 
the momentum operator, and the generators for all 
$4^{N_a}$ eigenstates of the Hubbard model in a
suitable second quantized operational basis. This basis
is defined by the set of operators 
$a_{q,\alpha,\beta}^\dag$ and 
$b^\dag_{q,\alpha,\gamma}$, which we call pseudohole
and heavy-pseudoparticle operators, respectively
(Chapter \ref{harmonic}). 
In this basis the eigenstates of the 1D Hubbard model
have a simple Slater-determinant form. In its diagonalized form the model
is written as an infinite sum of terms
involving only forward scattering interactions among the
pseudoparticles at all energy scales.

The use of the pseudohole and
heavy-pseudoparticle operational representation allows the evaluation of
the charge and spin current mean 
values, as well
as of the charge and spin transport masses of the pseudoparticle
carriers \cite{1Nuno97b,1Nuno97c}.
Moreover, combining this operational representation
with a new generalized conformal-field theory \cite{1Carm97a},
 the optical conductivity of the 1D Hubbard model around some
finite energy values can be computed \cite{1Carm97b}. The set of these
energy values, $\omega=\omega_0$, correspond to the new conformal
critical theories of $(\omega - \omega_0)$ low energy.
 
The present Thesis is organized as follows. In Chapter \ref{hm},
a derivation of the Hubbard model is given and some of its
well known properties are discussed. In Chapter \ref{harmonic},
a detailed presentation of the operational representation 
of the model in terms of pseudoholes and heavy-pseudoparticles
is given. The general expressions for the low-lying and finite-energy 
excitation spectra are provided in Chapter \ref{PseudoPT}. In Chapter \ref{finitesize},
the topological excitations 
in a finite size system are discussed. In Chapter \ref{transport},
the charge and spin transport currents and the charge 
and spin transport masses are evaluated. Qualitative results on the optical conductivity of the Hubbard
model are discussed and some Monte Carlo
calculations are presented. Finally, in Chapter \ref{concl}, we summarize
and discuss future research directions.


%% file: hubbge.tex
\pagestyle{headings}
\setcounter{chapter}{1}
\chapter{The Hubbard Model}
\label{hm}
\section{The Hubbard model}
\label{hubbardmodel}
\pagestyle{myheadings}
\markboth{2 The Hubbard Model}{2.1 The Hubbard model}

The theoretical study of electron correlations
in solids  requires the use of a suitable Hamiltonian such that 
it includes, in some way, the Coulomb interaction among the 
relevant electrons of the solid (those of the outer shells). 
If one is interested in both the thermodynamics and the 
transport properties, one must include the necessary external fields to the 
Hamiltonian. Response to an 
electric field requires the introduction
of a vector potential $\vec{A}$ (taken to be small).
In first quantization, the general Hamiltonian including Coulomb
interactions reads

\begin{equation}
        \hat {H} = \sum_i \left[\frac {\hbar^2}{2m}(-i\vec{\nabla}_i-
        \frac{e\vec{A}_i}{c\hbar})^2 +V(\vec{r}_i)\right]
        + \frac 1{2} \sum_{i,j}
        \frac {e^2}{\vline \, \vec{r}_i- \vec{r}_j \vline}  
        \,,
\label{hamilt1}
\end{equation}
where $m$ is the electron mass, $e$ is the electron charge $(e<0)$,
and $V(\vec{r})$ is the lattice potential, which will have
the suitable symmetry (the effects of disorder will not be considered).
The treatment of this Hamiltoninan is best accomplished in second quantization.
We write the electron fields $\Psi_{\sigma,l}^{\dag}(\vec{r})$ 
($\sigma$ and $l$ are the spin and band labels, respectively)
in terms of Wannier functions which perserve the lattice symmetry 
(for practical calculations,
one uses atomic wave functions). These fields
are localized at the lattice sites
$\vec{R}_i$ and are of the form
$\Psi_{\sigma,l}^{\dag}(\vec{r})=
\sum_i\phi(\vec{r}-\vec{R}_i)
e^{-i\frac{e}{c\hbar}\int_{\vec{R}_i}^{\vec{r}}\vec{A}(\vec{x})
\cdot d\vec{x}}c_{i,\sigma,l}^{\dag}$,
where $c_{i,\sigma,l}^{\dag}$ creates one electron of spin $\sigma$
at the band $l$ and lattice site $\vec{R}_i$. 
The operators $c_{i,\sigma,l}^{\dag}$
and $c_{j,\sigma',l'}$ obey the usual anticommuting fermionic relations 
and the exponential term in the field definition is required to
mantain gauge invariance. As a first approximation we consider
only one band and neglect all the band structure of the solid. We then
omit the band index in what follows.
In the Wannier representation, the
Hamiltonian (\ref{hamilt1}) reads

\begin{equation}
    \hat{H}=-\sum_{i,j,\sigma} t_{i,j}e^{-i\frac{e}{c\hbar}
    \vec{A}\cdot(\vec{R}_i-\vec{R}_j)}
    c_{i,\sigma}^{\dag}c_{j,\sigma}
     + \sum_{i,j,l,m}\sum_{\sigma,\sigma'}U_{i,j,l,m}
    c_{i,\sigma}^{\dag}c_{j,\sigma'}^{\dag}
    c_{l,\sigma'}c_{m,\sigma}\,,
\label{hamilt2}
\end{equation}
where the hopping integral $t_{i,j}$ and the Coulomb integral
$U_{i,j,l,m}$ are given, respectively, by

\begin{equation}
   t_{i,j}=-\int d\vec{r}\phi^{\ast}(\vec{r}-\vec{R}_i)
            \left[\frac {-\hbar^2}{2m}\vec{\nabla}^2 +V(\vec{r})\right]
            \phi(\vec{r}-\vec{R}_j)\,, 
\label{hopping}
\end{equation}
and 

\begin{equation}
   U_{i,j,l,m}=\int\int d\vec{r} d\vec{r}\,'
   \phi^{\ast}(\vec{r}-\vec{R}_i)
   \phi^{\ast}(\vec{r'}-\vec{R}_j)
   \frac {e^2}{\vline\, \vec{r}- \vec{r}\,' \,\vline}
   \phi(\vec{r'}-\vec{R}_l)
   \phi(\vec{r}-\vec{R}_m)
   \,.
\label{coulomb}
\end{equation}
The hopping integral (\ref{hopping}) gives the 
amplitude for an electron to jump from the atomic
orbital centered at the lattice site $\vec{R}_i$ to the atomic
orbital centered at the lattice site $\vec{R}_j$. The Coulomb
integral accounts for the electronic correlations among the
electrons at the different atomic orbitals. Hubbard estimated
\cite{2Hubbard63}
that the most important contribution to the Coulomb integral
comes from the term where two electrons are at the same 
atomic orbital, that is $U_{i,i,i,i}\equiv U$
(the order of magnitude of the bare $U$ is roughly of
10-20 eV). The second most important contribuition to the interaction
energy, $U_{i,j,i,j}$, comes from the interaction of two electrons 
at nearest neighbor sites (the order of magnitude is roughly 2
to 3 eV). If the electrons belong to narrow free-electron bands,
the most important contribution to the kinetic energy comes
from hopping between nearest neighbour sites; this is the usual
tight-binding approach. In view of this, the simplest Hamiltonian
we can think of for describing correlated electrons in a 
regular solid is the Hubbard model. It consists in
considering, out of all the hopping  and Coulomb integrals, only
the nearest neighbour hopping $t$ and the on-site Coulomb
repulsion $U$. The Hubbard Hamiltonian then reads
\cite{2Hubbard63,2Gutzwiller63,2Kanamori63}

\begin{equation}
    \hat{H}=-t\sum_{<i,j>,\sigma}^{N_a} e^{-i\frac{e}{c\hbar}
    \vec{A}\cdot(\vec{R}_i-\vec{R}_j)}
    c_{i,\sigma}^{\dag}c_{j,\sigma}
     + U\sum_{i}^{N_a}\hat{n}_{i,\uparrow}\hat{n}_{i,\downarrow}\,,
\label{hamilt3}
\end{equation}
where $<i,j>$ means summing over nearest neighbor sites 
$i$ and $j$ only, $N_a$ is the number of lattice sites, and 
$\hat{n}_{i,\sigma}=c_{i,\sigma}^{\dag}c_{i,\sigma}$ is the
number operator at site $i$ for $\sigma$ electrons. 
For the spin label, $\sigma$, we have $\sigma=\uparrow,\downarrow$
when used has an index and $\sigma=\pm 1$ otherwise.
This Hamiltonian
has only two parameters, $t$ and $U$, and thus the only energy scale
is $U/t$. This ratio is present in all physical quantities.
In what follows
we consider $c_{N_a+1\sigma}=c_{1\sigma}$, units such that $\hbar=1$,
the lattice spacing $a=1$, and the electron charge $e=-1$.
\section{The SO(4) symmetry of the Hubbard model}
\label{so4symmetry}
\pagestyle{myheadings}
\markboth{2 The Hubbard Model}{2.2 The SO(4) symmetry of ...}

The SO(4) symmetry of the Hubbard model
can be traced back to its
spin-up/spin-down and particle-hole symmetries 
\cite{2Nowak81,2Sutherland85,2Yang89,2Yang90}. Let us define the 
number and magnetization operators as
$\hat{N}=\sum_{i,\sigma}\hat{n}_{i,\sigma}$ and 
$\hat{S}^s_z = -{1\over 2}\sum_{i,\sigma}\sigma\hat{n}_{i,\sigma }$, 
respectively. ($N_\sigma$ is the number of $\sigma$ electrons and
$N=N_\sigma+N_{-\sigma}$.) If we perform the
spin-up/spin-down transformation, $c_{i,\sigma}^{\dag}\rightarrow
c_{i,-\sigma}^{\dag}$, the local number operator for 
$\sigma$ electrons transforms as
$\hat{n}_{i,\sigma }
\rightarrow \hat{n}_{i,-\sigma }$. This implies that
the number operator, $\hat N$, and the
Hamiltonian (\ref{hamilt3})
remain unchanged, but that the magnetization operator transforms as
$\hat{S}^s_z \rightarrow -\hat{S}^s_z$. That is,
under this transformation we go from 
the sector where
$N_\sigma > N_{-\sigma}$ to the sector where $N_\sigma < N_{-\sigma}$. 
This can be accomplished by the ladder operators

\begin{equation}
\hat{S}^s_- = \sum_{j}c^{\dagger }_{j\uparrow}c_{j\downarrow} 
\, , \hspace{1cm}
\hat{S}^s_+ = \sum_{j} c^{\dagger }_{j\downarrow}
c_{j\uparrow} \,. 
\end{equation}

On the other hand, if we perform the particle-hole transformation $c_{i,\sigma}\rightarrow
(-1)^{i}c_{i,-\sigma}^{\dag}$, the local number $\sigma$ 
operator transforms as $\hat{n}_{i,\sigma }
\rightarrow (1-\hat{n}_{i,-\sigma }$). This transformation implies that
the magnetization operator and
the kinetic energy ($\vec{A}=0$) of the Hamiltonian (\ref{hamilt3}) remain
unchanged, whereas the number operator and the interaction term of the
Hamiltonian (\ref{hamilt3}) transform as 

\begin{equation}
      \hat{N} \rightarrow 
      2N_a-\hat{N}\,,
\end{equation}
and
\begin{equation}
       U\sum_{i}\hat{n}_{i,\uparrow}\hat{n}_{i,\downarrow}
       \rightarrow U\sum_{i}\hat{n}_{i,\uparrow}\hat{n}_{i,\downarrow}
       +UN_a-UN\,, 
\end{equation}       
respectively. Therefore, the interaction
term can be cast in an invariant form if we write it as 
$U\sum_{i}(\hat{n}_{i,\uparrow}-1/2)(\hat{n}_{i,\downarrow}-1/2)$. 
This transformation then takes us from the sector where $N < N_a$ to the
sector where $N > N_a$. At half-filling the number operator
remains invariant. This
change in the electronic
density sector without changing the
magnetization density can be accomplished by means of suitable operators. The simplest way for generating these operators is to perform the transformation  
$c_{i,\uparrow}\rightarrow
(-1)^{i}c_{i,\uparrow}^{\dag}$
in the $\hat{S}^s_-,\hat{S}^s_+,$ and $\hat{S}^s_z$ operators. This leads
to the following set of operators

\begin{equation}
       \hat{S}^c_z = -{1\over 2}[N_a - \sum_{\sigma}
       \hat{N}_{\sigma }] \, , \hspace{1cm}
       \hat{S}^c_- = \sum_{j} (-1)^j 
       c_{j\uparrow}c_{j\downarrow} \, , \hspace{1cm}
       \hat{S}^c_+=\sum_{j} (-1)^j c^{\dagger }_{j\downarrow}
       c^{\dagger }_{j\uparrow} \,.
\end{equation}

Both the set of generators $\{ \hat{S}^s_-,\hat{S}^s_+,\hat{S}^s_z\}$ and 
$\{ \hat{S}^c_-,\hat{S}^c_+,\hat{S}^c_z\}$ 
form two independent SU(2) algebras. The first set takes 
into account the spin
part and the second set the charge part. If we write the Hamiltonian 
(\ref{hamilt3}) as 

\begin{equation}
       \hat{H}_{SO(4)} = -t\sum_{\langle j,i \rangle ,\sigma}
       c_{j\sigma}^{\dag }c_{i,\sigma}  +
       U\sum_{i}(\hat{n}_{i,\uparrow}-1/2)(\hat{n}_{i,\downarrow}-1/2)\,,
       \label{hamiltso4}
\end{equation}
the Hubbard model becomes SO(4) invariant \cite{2Nowak81,2Yang89,2Yang90}. Note that the $SO(4)$ symmetry is not
fully equivalent to the $SU(2)\otimes SU(2)$ symmetry \cite{2Nowak81,2Yang89,2Yang90}.
All the eigenstates $\vert \phi \rangle$
of the Hubbard model can be choosen as 
simultaneous eigenstates of $\hat{H},\hat{S}^c_z,\hat{S}^s_z,
(\hat{S}^s)^2, (\hat{S}^c)^2,$ and the lattice translation operator.
The commutators 

\begin{equation}
      [\hat{S}_+^{c},\hat N]=-2\hat{S}_+^{c}\,,\hspace{1cm}
      [\hat{S}_+^{\alpha},\hat{S}_z^{\alpha}]=-	\hat{S}_+^{\alpha}\,,\hspace{1cm}
      \alpha=c,s\,,
\end{equation} 
and
\begin{equation}
      [\hat{S}_+^{c},\hat{P}]=-\pi\hat{S}_+^{c}\,,\hspace{1cm} 
      (\hat{P}=\sum_{k,\sigma}
      kc^{\dag}_{k,\sigma}c_{k,\sigma})
\end{equation}
where $\hat{P}$ is the momentum operator,
imply that if a state $\vert \phi \rangle$ has $N$ electrons,
magnetization $(N_{\downarrow}-N_{\uparrow})/2$, and momentum $P$, then
the state $\hat{S}_+^{c}\vert \phi \rangle$ has $N+2$ electrons and
momentum $P+\pi$, and the state $\hat{S}_+^{s}\vert \phi \rangle$
has magnetization $(N_{\downarrow}-N_{\uparrow})/2+1$
and momentum $P$. These results
introduce a enormous simplification in the description of the Hilbert space in any dimension. 

The inclusion of terms that break the SO(4) symmetry in a trivial
way, like the chemical-potential term and the Zeeman-coupling term, 
do not change 
the simplicity of the SO(4) description.
The inclusion in the Hamiltonian (\ref{hamiltso4}) 
of such terms reduces, in general, the symmetry of the model. The
resulting Hamiltonian reads

\begin{equation}
      \hat{H} = \hat{H}_{SO(4)} + 2\mu\hat{\eta }_z + 
       2\mu_0 H\hat{S}_z\,.
\label{hamiltonian}
\end{equation}

For finite values of both the magnetic field and the chemical potential
the symmetry of the quantum problem is reduced to 
$U(1)\otimes U(1)$, with $\hat{S^c}_z$ and $\hat{S^s}_z$ 
commuting with $\hat{H}$. 
The values of $\mu$ and $H$ 
determine the different symmetries of the Hamiltonian
(\ref{hamiltonian}). When $\mu\neq 0$ and $H\neq 0$ the symmetry is 
$U(1)\otimes U(1)$, for $\mu= 0$ and $H\neq 0$ it is $SU(2)\otimes U(1)$, when $\mu\neq 0$ 
and $H= 0$ it is $U(1)\otimes SU(2)$, and at $\mu= 0$ 
and $H= 0$ the Hamiltonian symmetry is SO(4). 

There are four $U(1)\otimes U(1)$ sectors 
of parameter space corresponding to $S^c_z< 0$ and $S^s_z< 0$, 
$S^c_z< 0$ and $S^s_z> 0$, $S^c_z> 0$ and $S^s_z< 0$, and 
$S^c_z> 0$ and $S^s_z> 0$. We follow the works
\cite{2Nuno95,2Nuno96}
and call these sectors $(l_c,l_s)$, where

\begin{equation}
      l_{\alpha} = {S^{\alpha }_z\over |S^{\alpha }_z|} \, .
\label{sector}
\end{equation}
The sectors $(-1,-1)$, $(-1,+1)$, $(+1,-1)$, and $(+1,+1)$
refer to electronic densities and spin densities $0<n<1$ and  
$0<m<n$, $0<n<1$ and $-n<m<0$, $1<n<2$ and $0<m<(2-n)$,
and $1<n<2$ and $-(2-n)<m<0$, respectively.
  
There are two $(l_s)$ sectors of $SU(2)\otimes U(1)$ Hamiltonian 
symmetry [and two $(l_c)$ sectors of $U(1)\otimes SU(2)$ Hamiltonian 
symmetry] which correspond to $S^s_z< 0$ and $S^s_z> 0$ for
$l_s=-1$ and $l_s=+1$, respectively, (and to $S^c_z< 0$ and $S^c_z> 0$ 
for $l_c=-1$ and $l_c=+1$, respectively). There is one SO(4) 
sector of parameter space [which is constituted only
by the $S^c_z= 0$ (and $\mu =0$) and $S^s_z= 0$ canonical
ensemble].
\section{Basic features of the Hubbard model}
\label{basic}
\pagestyle{myheadings}
\markboth{2 The Hubbard Model}{2.3 Basic features of the ...}

The Hubbard model (\ref{hamilt3}) [or its SO(4) invariant version
(\ref{hamiltso4})] depends on two parameters, namely 
the hopping integral $t$ 
and the Coulomb integral $U$. The model shows very different physical behavior 
depending on the relative strength of these two parameters and on the band filling $n$. In the case of
$U=0$, the Hubbard model reduces to the simple tight-binding model

\begin{equation}
       \hat{H}_0=-t\sum_{<i,j>,\sigma} e^{-\frac{i}{c}
       \vec{A}\cdot(\vec{R}_i-\vec{R}_j)}
       c_{i,\sigma}^{\dag}c_{j,\sigma}\,,
\label{tight}
\end{equation}
and a simple exact solution follows from \cite{2Krantz91}
the introduction
of momentum-space operators $c_{\vec {k},\sigma}^{\dag}$ 
and $c_{\vec {k},\sigma}$.
These operators are related to the Wannier representation by
$c_{i,\sigma}^{\dag}=N_a^{-1/2}
\sum_ie^{i\vec{k}\cdot \vec{R}_i}c_{\vec {k},\sigma}^{\dag}$. The form of the band then
depends on the type of lattice considered. For the one-dimensional case,
with one atom per unit cell,
the diagonalized Hamiltonian is

\begin{equation}
      \hat{H}_0=-2t\sum_{k,\sigma}\cos\left(k-\frac{A}{c}\right)
      c_{k,\sigma}^{\dag}c_{k,\sigma}\,.
\label{tightk}
\end{equation}
It is a simple matter to compute both the 
exact partition function 
and the conductivity for the Hamiltonian (\ref{tightk}). 
This model is a metal for
all electronic densities, exception made to $N/N_a=2$, 
where it has two electrons per lattice site, and for $N/N_a=1$, when the system
is fully polarized.

Another limit where the model (\ref{hamilt3}) has an exact solution 
is the atomic limit \cite{2Hubbard63}. In this limit
the hopping integral is zero and the electrons
are localized at the lattice sites. Since the electrons have 
zero kinetic energy, the model represents an insulator for all values of $U$,
possesing only two (flat) bands, 
one located at the chemical potencial $\mu$ and
another at the energy $\mu+U$. 
This can be put in a more formal way by means of the Green's function
formalism. A very simple calculation of the
one-particle Green's function at finite temperature (suitable
to deal with the massive degeneracy of the model eigenstates in
this limit), leads to

\begin{equation}
      {\cal G}^{atomic}_{\sigma_i,\sigma_j}(\omega)=
      \delta_{i,j}\delta_{\sigma_i,\sigma_j}\left[
      \frac {1-\langle n_{\sigma} \rangle}{i\omega +\mu}+ 
      \frac {\langle n_{\sigma} \rangle}{i\omega +\mu-U} \right]\,,
\label{green}
\end{equation}
where $i\omega$ is the usual finite-temperature Matsubara frequency
and $\langle n_{\sigma} \rangle$ is the electronic occupation number
at finite temperature for the single site Hamiltonian, 
$\hat{h}_i=\sum_{\sigma}(U/2
\hat{n}_{i,\sigma}\hat{n}_{j,-\sigma}-\mu \hat{n}_{i,\sigma})$.

The question now is whether the system undergoes or not a 
phase transition when the parameters varie from the tigth-binding 
$U/t=0$ limit to the atomic $U/t\rightarrow \infty$ limit, for some finite value of $U/t$.
In the strong-coupling regime $U\gg t$,
the states with two electrons in the same site have a much larger energy
-- of order $U$ -- relatively to those states with single-site occupancy
only.
Thus, for the half-filled band we expect to have an insulator
in the large $U$ limit, and we show below, within perturbation theory,
that this is indeed the case. 

Let us then divide the Hilbert
space of the Hubbard model into three subspaces: 
the subspace of single-occupied-site states $\{\vert S\rangle  \}$, 
the subspace of one double-occupied site states $\{\vert D_1\rangle  \}$, and
the subspace of more than one double-occupied-site
states $\{\vert D_n\rangle  \}$. Since $U\gg t$, 
the kinetic term of the Hamiltonian
(\ref{hamilt3}) can be treated as a perturbation. 
Note that due to the massive
degeneracy in each of the three considered subspaces,
degenerate perturbation theory must be applied. Moreover, in order to
extract 
information on the nature of the ground state of the model,
we compute the average energy relative to the states belonging the 
subspace $\{\vert S\rangle  \}$. 

Let us consider the system away from half filling.
All states belonging to the subspace $\{\vert S\rangle  \}$ have zero
Coulomb energy. Then, up to second order in perturbation theory,
the energy of a given state belonging to the subspace $\{\vert S\rangle  \}$
is given by

\begin{equation}
      E_{\vert S\rangle}=\langle S \vert \hat{T}\vert S\rangle-\frac 1{U}
      \sum_{\vert D_1\rangle} \langle S \vert \hat{T} \vert D_1\rangle
      \langle D_1 \vert \hat{T} \vert S\rangle\,,
\label{enersub}
\end{equation}
where $\hat{T}= -t\sum_{<i,j>,\sigma} 
c_{i,\sigma}^{\dag}c_{j,\sigma}$. The states belonging to the subspace
$\{\vert D_n\rangle  \}$ have not been considered because the
operator $\hat{T}$ does not connect that subspace with
$\{\vert S\rangle  \}$.  The operator
$\hat{P}_{D_1}=\sum_{\vert D_1\rangle} \vert D_1\rangle\langle D_1 \vert$ 
can be regarded
as the projection operator for the subspace $\{\vert D_1\rangle  \}$.
Thus, up to this order in perturbation theory, the
operator $\hat{P}_{D_1}$ can be replaced in Eq.\,(\ref{enersub}) 
by $\sum_{i}\hat{n}_
{i,\uparrow}\hat{n}_{i,\downarrow}$. It should be emphasized
that this substitution is valid only if these calculations are restricted
to the subspaces $\{\vert S\rangle  \}$ and
$\{\vert D_1\rangle  \}$. From Eq. (\ref{enersub}), one then arrives to
an effective Hamiltonian valid
for the subspace $\{\vert S\rangle  \}$ that reads

\begin{eqnarray}
      \hat{H}^{eff.}_{n<1}&=&\hat{P}_s \left[ \hat{T} - \frac {t^2}{U}
      \sum_{\sigma,\sigma'}\sum_{i,m}
      c_{m,\sigma}^{\dag}c_{i,\sigma}
      \hat{n}_{i,\uparrow}\hat{n}_{i,\downarrow}
      c_{i,\sigma'}^{\dag}c_{m,\sigma'}\right.\nonumber\\
      &-&\left. \frac {t^2}{U}
      \sum_{\sigma,\sigma'}\sum_{l\neq i,m}
      c_{l,\sigma}^{\dag}c_{i,\sigma}
      \hat{n}_{i,\uparrow}\hat{n}_{i,\downarrow}
      c_{i,\sigma'}^{\dag}c_{m,\sigma'}\right]\hat{P}_s\,,
\label{tjmodel}
\end{eqnarray}
where $\hat{P}_s=\sum_{\vert S\rangle} \vert S\rangle\langle S \vert$
is the projector onto the subspace $\{\vert S\rangle  \}$. The third
term in the rhs of Eq. (\ref{tjmodel}) is usually neglected
and the remaining two terms are called the $t-J$
model \cite{2Assa94}. 
At half filling, the first and the third terms in the rhs of Eq.
(\ref{tjmodel}) vanish and we are left with another effective Hamiltonian

\begin{equation}
       \hat{H}^{eff.}_{n=1}=\frac J{2}\sum_{i,j} [
       \vec{S}_i\cdot\vec{S}_j-\frac {\hat{n}_i\hat{n}_j}{4}]\,,
\label{heismodel}
\end{equation}
with $J=4t^2/U$. This is 
the Heisenberg anti-ferromagnetic model. 
The second term in the rhs
of Eq.\,(\ref{tjmodel}) was rewritten by using the 
electronic representation for the
spin operators defined by $\hat{S}_z^s$, $\hat{S}_+^s$, and $\hat{S}_-^s$.

The above results show (assuming that the perturbative treatment holds) that for strong electronic correlations 
($U\gg t$) 
the system is an antiferromagnetic insulator at half filling. 
Due to the restrictions
imposed on electronic movements by the reduced dimensionality,
for one-spatial dimension
the Hubbard model is an insulator at half-filling for any positive
value of $U$ \cite{2Lieb68}. 
In one dimension the system is not a simple Fermi liquid. The most difficult problem is the study of the intermediate-$U$ case and of how the model properties change as the
ratio $U/t$ goes from small to large values. The study of this regime
requires non-perturbative methods. In the following we consider the case of the 1D lattice where the exact Bethe-ansatz solution is available.
\section{Basic ideas on the Bethe ansatz}
\pagestyle{myheadings}
\markboth{2 The Hubbard Model}{2.4 Basic ideas on ...}

From here on, we will be dealing with the one-dimensional
Hubbard model. Following Yang's solution \cite{2Yang67} 
of the one-dimensional
many-body problem with a delta-function interaction,
Lieb and Wu \cite{2Lieb68} computed the ground state
of the Hubbard model.
Later, Takahashi \cite{2Takahashi72}
gave the complete solution of the Lieb and Wu 
Bethe-ansatz equations. The basic ingredients of the Bethe
ansatz are:  
the simple topology of the 1D space,
the continuity condition at the same point of the lattice, the boundary conditions, and Pauli's principle. 
However, the implementation 
of these simple concepts in order to achieve the solution of the general 
eigenvalue problem is a task of considerable complexity 
\cite{2Baeriswyl91}. In this section, the Bethe ansatz
technique is illustrated by the solution of the two-electron problem
in a chain of $N_a$ sites.

The eigenstates $\vert \Psi \rangle$
of the two-electron problem can be written as a superposition 
of Wannier states 
$c_{x_2,\sigma_2}^{\dag}c_{x_1,\sigma_1}^{\dag}\vert 0 \rangle$

\begin{equation}
      \vert \Psi \rangle=\sum_{x_1,\sigma_1}\sum_{x_2,\sigma_2}
      \psi(x_1,\sigma_1;x_2,\sigma_2)
      c_{x_2,\sigma_2}^{\dag}c_{x_1,\sigma_1}^{\dag}\vert 0 \rangle\,,
\label{psiwave}
\end{equation}
where $\vert 0 \rangle$ is the electronic vacuum and
$x_i$ stands for the position of the electron $i$.  
Obviously, the problem consists
in determining the coefficients $\psi(x_1,\sigma_1;x_2,\sigma_2)$. 
The starting point is the eigenvalue equation 
$\hat{H}\vert \Psi \rangle=E\vert \Psi \rangle$ which imposes the following
relation among the coefficients $\psi(x_1,\sigma_1;x_2,\sigma_2)$

\begin{eqnarray}
      &-&t[\psi(x_1+1,\sigma_1;x_2,\sigma_2)
      +\psi(x_1-1,\sigma_1;x_2,\sigma_2)+\nonumber\\
      &+& \psi(x_1,\sigma_1;x_2+1,\sigma_2)+\psi(x_1,\sigma_1;x_2+1,\sigma_2)]+
      \nonumber\\
      &+&U\delta{x_1,x_2}\delta{\sigma_1,-\sigma_2}
      \psi(x_1,\sigma_1;x_2,\sigma_2)=E\psi(x_1,\sigma_1;x_2,\sigma_2)\,.
\label{eigenvalue}
\end{eqnarray}
The spatial dimensionality of the system allows  the
definition of two regions, I and II, for which $x_1\leq x_2$
and $x_1\geq x_2$, repectively. The electronic antisymmetric character 
under the interchange of positions 
implies that 

\begin{equation}
      \psi_I(x_1,\sigma_1;x_2,\sigma_2)=
      -\psi_{II}(x_2,\sigma_2;x_1,\sigma_1)\,.
\label{pauli}
\end{equation}
The next step
is to choose a simple representation for the anti-symmetric property
(\ref{pauli}). We introduce the Bethe ansatz

\begin{equation}
       \psi_I(x_1,\sigma_1;x_2,\sigma_2)=A(p_1,\sigma_1;p_2,\sigma_2)
       e^{i(p_1x_1+p_2x_2)}-A(p_2,\sigma_1;p_1,\sigma_2)
       e^{i(p_2x_1+p_1x_2)}
\label{psiI}       
\end{equation}
and

\begin{equation}
       \psi_{II}(x_2\sigma_2,x_1,\sigma_1)=A(p_2,\sigma_2;p_1,\sigma_1)
       e^{i(p_1x_1+p_2x_2)}-A(p_1,\sigma_2;p_2,\sigma_1)
       e^{i(p_2x_1+p_1x_2)}\,.
\label{psiII}       
\end{equation}
It is clear that the interchange  
$(x_1,\sigma_1)\leftrightarrow (x_2,\sigma_2)$ in 
$\psi_I(x_1,\sigma_1;x_2\sigma_2)$ or $\psi_{II}(x_2\sigma_2,x_1,\sigma_1)$
agrees with the antisymmetry condition (\ref{pauli}). If $x_1\neq x_2$,
introduction of
both $\psi_I(x_1,\sigma_1;x_2\sigma_2)$ and 
$\psi_{II}(x_2\sigma_2,x_1,\sigma_1)$ in Eq. (\ref{eigenvalue})
leads to $E=-2t\cos(p_1)-2t\cos(p_2)$ (we are considering the lattice
spacing $a$ equal to 1).

At the same lattice point, $x_1=x_2=x$, we must have

\begin{equation}
      \psi_I(x,\sigma_1;x,\sigma_2)=
      \psi_{II}(x,\sigma_2;x,\sigma_1)\,,
\label{continuity}
\end{equation}
which is the continuity equation. Combining Eq.
(\ref{continuity}) with the eigenvalue equation for the two electrons
at the same site and imposing the same functional
form, $E=-2t\cos(p_1)-2t\cos(p_2)$, for the energy eigenvalue, 
after some algebra the following relation among three of the amplitudes
$A(p_j,\sigma_i;p_m,\sigma_n)$ (with $j,i,m,n=1,2$) is obtained

\begin{equation}
       \pmatrix{A(p_2,\sigma_1;p_1,\sigma_2)\cr
       A(p_2,\sigma_2;p_1,\sigma_1)}=\frac 1{k_{12}+ic}
       \pmatrix{ic&k_{12}\cr k_{12}& ic}
       \pmatrix{A(p_1,\sigma_1;p_2,\sigma_2)\cr
       A(p_1,\sigma_2;p_2,\sigma_1)}\,.
\label{amplitudes}
\end{equation}
Here $k_{12}\equiv \sin(p_1)-\sin(p_2)$ and $c \equiv U/(2t)$.
Solution of the problem in closed form involves the imposition of
periodic boundary conditions (other boundary conditions are also possible). Since we have two electrons, we must impose
periodic boundary conditions on both $x_1$ and $x_2$ 
(some care is needed in taking into account the regions I and II). These
are

\begin{eqnarray}
      {\rm Region\,I: }&& x_1<x_2 \rightarrow x_1+N_a>x_2,\nonumber\\
      &&\psi_I(x_1,\sigma_1;x_2\sigma_2)=
      \psi_{II}(x_1+N_a,\sigma_1;x_2,\sigma_2)\,
\label{bund1}
\end{eqnarray}
and
\begin{eqnarray}
      {\rm Region\,II: }&& x_1>x_2 \rightarrow x_2+N_a>x_1,\nonumber\\
      &&\psi_{II}(x_1,\sigma_1;x_2\sigma_2)=
      \psi_I(x_1,\sigma_1;x_2+N_a,\sigma_2)\,.
\label{bund2}
\end{eqnarray}
Combinig Eqs. (\ref{psiI}) and (\ref{psiII}) with Eqs. (\ref{bund1})
and (\ref{bund2}), we arrive to the two following relations

\begin{equation}
       e^{ip_1 N_a}\pmatrix{A(p_2,\sigma_1;p_1,\sigma_2)\cr
       A(p_2,\sigma_2;p_1,\sigma_1)}=
       \pmatrix{A(p_1,\sigma_2;p_2,\sigma_1)\cr
       A(p_1,\sigma_1;p_2,\sigma_2)}\,,
\label{per1}
\end{equation}
and
\begin{equation}
 e^{ip_2 N_a}\pmatrix{A(p_1,\sigma_1;p_2,\sigma_2)\cr
       A(p_1,\sigma_2;p_2,\sigma_1)}=
       \pmatrix{A(p_2,\sigma_2;p_1,\sigma_1)\cr
       A(p_2,\sigma_1;p_1,\sigma_2)}\,.
\label{per2}
\end{equation}
By combining Eqs. (\ref{per1}) and (\ref{amplitudes}), 
we end up with the simple
eigenvalue problem

\begin{equation}
      e^{-ip_1 N_a}\vert A \rangle=
       \hat{R}
       \vert A \rangle\,,
\label{eigfi}
\end{equation}
and equivalently for the eigenvalue $e^{-ip_2 N_a}$. The state 
$\vert A \rangle$ and the operator $\hat{R}$ are written as

\begin{equation}
      \vert A \rangle= \pmatrix{A(p_1,\sigma_1;p_2,\sigma_2)\cr
       A(p_1,\sigma_2;p_2,\sigma_1)}\,, \hspace{1cm} \hat{R}=
       \frac 1{k_{12}+ic}
       \pmatrix{0&1\cr 1&0\cr}
       \pmatrix{ic&k_{12}\cr k_{12}& ic}.
\end{equation}
The solution to this simple problem can be cast in the form

\begin{equation}
       e^{ip_jN_a}=\frac{\sin(p_j)-\Lambda+ic/2}{\sin(p_j)-\Lambda-ic/2}
       \hspace{1cm} j=1,2\,,
\label{b1a}       
\end{equation}
and

\begin{equation}
       \prod_{i=1}^{2}\frac{\sin(p_i)-\Lambda+ic/2}
       {\sin(p_i)-\Lambda-ic/2}=1\,,
\label{b3}
\end{equation}
where $\Lambda =\sin (p_1)+\sin (p_2)$. These are the Bethe-ansatz 
equations for one up-spin and one down-spin electron. 

Finally, by considering a magnetic flux 
$\phi$ through the ring, 
the general Bethe-ansatz equations become \cite{2Lieb68,2Shastry90}

\begin{equation}
        e^{ik_j N_a}=e^{i\phi}\prod_{\delta=1}^{N_\downarrow}
        \frac{\sin(k_j)-\Lambda_\delta+iU/4}{\sin(k_j)-
        \Lambda_\delta-iU/4}\, ,
        \hspace{1cm} (j=1,\ldots,N)
\label{inter1}
\end{equation}
and
\begin{equation}
        \prod_{j=1}^{N}
        \frac{\sin(k_j)-\Lambda_\delta+iU/4}
        {\sin(k_j)-\Lambda_\delta-iU/4}
        =-\prod_{\beta=1}^{N_\downarrow}
        \frac{\Lambda_\beta-\Lambda_\delta+iU/2}
        {\Lambda_\beta-\Lambda_\delta-iU/2}\, ,
        \hspace{1cm} (\delta=1,\ldots,N_\downarrow)\,.
\label{inter2}
\end{equation}

%% file: hilbert.tex
\pagestyle{headings}
\setcounter{chapter}{2}
\chapter{Algebraic Representation for the Hilbert Space of the $1$D Hubbard Model}
\label{harmonic}

\section{Introduction}
\pagestyle{myheadings}
\markboth{3 Algebraic Representation for ...}{3.1 Introduction}

In this chapter we present the algebraic solution for the 1D Hubbard 
model, which shows some basic similarities with the algebraic
solution of the isotropic harmonic oscillator. In order to clarify the
basic ideas underlying  the problem,
we begin by presenting in the following the main results for the 
algebraic solution of the harmonic oscillator. Afterwards, we discuss
the general 
similarities and differences between the two cases. 

The isotropic harmonic oscillator has SO(3) symmetry. The SO(3) group
can be defined by the following algebra envolving the angular momentum
operators

\begin{equation}
      [L_+,L_z]=-L_+\,,\hspace{1cm}[L_-,L_z]=L_-\,,\hspace{1cm}
      [L_+,L_-]=2L_z\,,
\label{comml}
\end{equation}
where $L=(L_x,L_y,L_z)$ and $L_{\pm}=L_x \pm iL_y$. Moreover,
$[L^2,L]=0$ and $[H,L]=0$, where $H$ represents the isotropic harmonic oscillator Hamiltonian. The commutators (\ref{comml})
imply that $L^2$, $L_z$, and $H$ can be simultaneously diagonalized. In second quantization, $H$, $L_z$, and $L^2$ can be written as ($\hbar =1$)

\begin{equation}
	H=\frac {\omega}{2}\left(3+a^{\dag}_{0}a_{0}+
        a^{\dag}_{\frac 1{2}}a_{\frac 1{2}}+
        a^{\dag}_{-\frac 1{2}}a_{-\frac 1{2}}\right)\,,
\end{equation}

\begin{equation}
      L_z=\frac 1{2}(a^{\dag}_{\frac 1{2}}a_{\frac 1{2}}
      -a^{\dag}_{-\frac 1{2}}a_{-\frac 1{2}})\,,
\end{equation}

\begin{equation}
       L^2=L_-L_+ +L_z^2+L_z\,,
\end{equation}
with

\begin{equation}
      L_-=a^{\dag}_{-\frac 1{2}}a_{\frac 1{2}}\,, \hspace{1cm}
      L_+=a^{\dag}_{\frac 1{2}}a_{-\frac 1{2}}\,.
\end{equation}
The operators $a^{\dag}_0$, $a^{\dag}_{\frac 1{2}}$, and
$a^{\dag}_{-\frac 1{2}}$ are bosonic operators and
the angular momentum operators are written in the
Schwinger representation \cite{Sakurai}. 
The eigenstates $\vert n_{\frac 1{2}},n_0,n_{-\frac 1{2}}\rangle$ 
of $H$, $L_z$,
and $L^2$, corresponding to a given
$n=n_{\frac 1{2}}+n_0+n_{-\frac 1{2}}$, a given
$\ell=n-2n_0$, and given 
$m$ (the eigenvalue of $L_z$), read

\begin{equation}
      \vert n_{\frac 1{2}},n_0,n_{-\frac 1{2}}\rangle= 
      \frac{
      L_+^{(n-2n_{0}+m)}(a^{\dag}_{-\frac 1{2}})
      ^{2(n-2n_{0})}
      (a^{\dag}_{0})^{4n_{0}}}
      {\sqrt{(4n_{0})!}\sqrt{(2n-4n_{0})!}
      \sqrt{(n-2n_{0}+m)!}}\
      \vert 0,0,0 \rangle\,.
\label{harmstates}
\end{equation}

For a given value of $n$, the quantum number
$n_0$ assumes the values $0,1,2,\ldots,n/2$, 
for $n$ even, and $0,1,2,\ldots,(n-1)/2$, for $n$ odd
[the number $n_0$ is used to generate, for example, all the LWS ($m=-\ell$, see below) for a given $n$].
The states of the form $\vert n_{\frac 1{2}},n_0,0\rangle$ are called
highest-weight states (HWS's), 
since $L_+\vert n_{\frac 1{2}},n_0,0\rangle=0$,
and the states $\vert 0,n_0,n_{-\frac 1{2}}\rangle$ are called
lowest-weight states (LWS's), 
since $L_-\vert 0,n_0,n_{-\frac 1{2}}\rangle=0$.
The states such that 
$L_{\pm}\vert n_{\frac 1{2}},n_0,n_{-\frac 1{2}}\rangle \neq 0$ are
called non-LWS's or non-HWS's. To go from the LWS of a given tower, 
characterized by the angular momentum number $\ell=n-2n_{0}$, to the
respective HWS (or vice-versa), one applies the rising operator $L_+$ 
(or the lowering operator $L_-$). For a given $\ell$, the
number $m$ can have the values $m=-\ell,-\ell+1,\ldots,\ell-1,\ell$.
All the states in the $\ell$
tower differ only in the $L_z$ eigenvalue, and all the states with a given
$n$ have degeneracy $(n^2+3n+2)/2$.

From the above analysis, we see that the isotropic harmonic oscillator is
characterized by three types of particles: the $a^{\dag}_{0}$ bosons, which
do not contribute to the angular momentum operators, and the 
$a^{\dag}_{\frac 1{2}}$
and $a^{\dag}_{-\frac 1{2}}$ bosons. 
The latter two particles describe the towers of
angular momentum. Note that the HWS's (LWS's) 
have only $a^{\dag}_{\frac 1{2}}$
($a^{\dag}_{-\frac 1{2}}$) bosons. The non-LWS's or non-HWS's 
have both types of particles. The ladder operators $L_{\pm}$ act upon the 
$\pm \frac 1{2}$ bosons interchanging the $\pm \frac 1{2}$ quantum numbers.

A similar algebra occurs in the 1D  Hubbard model. 
Obviously, the algebraic description of this model is much more
involved, for we are dealing with a many-body system
with discrete translational symmetry. 
The Bethe-ansatz solution of the Hubbard model refers
usually to the symmetry
sector $(-1,-1)$ \cite{Yang67,Lieb68,Takahashi72} 
(Sec. \ref{so4symmetry}). 
In this sector
it has been proved by Essler {\it et al.} 
\cite{Essler92a,Essler92b,Essler92c}
that the Bethe-ansatz eigenstates $\vert \psi \rangle$
are LWS's of the SO(4) algebra, that is 
$\hat{S}^{\alpha}_-\vert \psi \rangle=0$. We can then construct the
towers of spin and eta-spin by acting onto the states $\vert \psi \rangle$
with the ladder operators $\hat{S}^{\alpha}_+$ and reach the respective
HWS's. This procedure generates
all the $4^{N_a}$ eigenstates of the Hubbard chain, which span the whole
Hilbert space associated with all the
nine symmetry sectors of the model.

In the algebraic solution of the Hubbard chain 
the 
eigenstates of the model are described
by the set of operators $a_{q,\alpha,\beta}^{\dag}$ and
$b_{q,\alpha,\gamma}^{\dag}$ which refer to the quantum objects
that we call pseudoholes and 
heavy-pseudoparticles, respectively
\cite{Nuno95,Nuno96,Nuno97a}. These operators obey the
usual anticommutation relations. 
The quantum numbers that label the operators 
$a_{q,\alpha,\beta}^{\dag}$ and  $b_{q,\alpha,\gamma}^{\dag}$ are the
following:
a) the $q$ is the pseudomomentum and is to be chosen
out of the available pseudo-Brillouin zone values
(to be defined later); b) there are two colors of 
pseudohole and heavy-pseudoparticle,
which are called $\alpha=c$ and $\alpha=s$; c) these different types
of heavy pseudoparticles can populate different $\gamma$ bands
($\gamma=1,2,...,\infty$). In the
case of the $\alpha,\beta$ pseudoholes, the band index number is $\gamma=0$ (this was omitted for convenience of notation); d) the $\beta=\pm 1/2$ 
is related to the SO(4) algebra.

In the 1D Hubbard model the Bethe-ansatz
eigenstates can be classified as LWS's I and LWS's II
\cite{Essler92a,Essler92b,Essler92c}. 
The former have neither
heavy-pseudoparticle nor $\alpha,\beta=1/2$ pseudohole occupancy
and all the $\alpha,\beta$ pseudoholes which populate the $\alpha,0$ band have $\beta=-1/2$. On the other hand,
the
latter eigenstates have both heavy-pseudoparticles 
and $\beta=\pm 1/2$ pseudoholes. The SO(4) generators involve only the 
$a^{\dag}_{q,\alpha,\beta}$ pseudohole operators, 
as the SO(3) generators,
in the isotropic oscillator, involve only the
$a^{\dag}_{\mp \frac 1{2}}$ boson operators. 
The two colors $\alpha=c$ and $\alpha=s$ are directly
related to the charge and spin degrees of freedom, respectively.
The generators of the SO(4) algebra are given, in
the generalized Schwinger representation, by
\cite{Nuno96,Nuno97a}

\begin{equation}
      \hat{S}^{\alpha }_z =\frac 1{2} \sum_{q}
      (a^{\dag }_{q,\alpha,
      {1\over 2}}a_{q,\alpha,{1\over 2}} 
      -a^{\dag }_{q,\alpha,
      -\frac 1{2}}a_{q,\alpha,-\frac 1{2}}) 
      \,, \hspace{1cm}
      {\hat{S}^{\alpha }}_{\pm} =\sum_{q}a^{\dag }_{q,\alpha,
      \pm{1\over 2}}a_{q,\alpha,\mp{1\over 2}} \, .
\label{offdiagonal}
\end{equation}
The spin and eta-spin ``angular" quantum numbers $S^{\alpha}$ 
are given by

\begin{equation}
S^{\alpha} = {1\over 2}(N^h_{\alpha}
-\sum_{\gamma =1}^{\infty}2\gamma N_{\alpha,\gamma})\,,
\label{angulars}
\end{equation}
where  $N_{\alpha}^h=\sum_{\beta}N^h_{\alpha,\beta}$,
 $N^h_{\alpha,\beta}$ is the number of $\alpha,\beta$ pseudoholes,  
and $N_{\alpha,\gamma}$ is the number of heavy $\alpha,\gamma$
pseudoparticles.
As in the SO(3) algebra, the
ladder operators interchange only the $\beta=\pm 1/2$ quantum number. 
Thus, as in
the case of the harmonic oscillator, all the pseudoholes have
$\beta=-1/2$ in the LWS's I 
and $\beta=+1/2$ in the HWS's I. The
non-LWS's I (and non-HWS's I) have both $\beta=+1/2$ and $\beta=-1/2$
pseudohole occupancy.

The number of discrete pseudomomentum 
values in each ${\alpha,\gamma}$ band ($\alpha,\gamma$-band 
Fock-space dimension) is given by the Bethe-ansatz 
solution
\cite{Takahashi72} 
and reads

\begin{eqnarray}
	&&d_{c,0}=N_a\\
      &&d_{\alpha,\gamma}=
      N^h_{\alpha}+N_{\alpha,\gamma}-\sum_{\gamma'=1}^{\infty}
      [\gamma + \gamma' - |\gamma - \gamma'|]N_{\alpha,\gamma'} 
	\,, \hspace{1cm} \alpha,\gamma \neq c,0 \, . 
\label{bzlim}
\end{eqnarray}
In the summations and products over $q$ the pseudomomentum 
$q$ takes on values 

\begin{equation}
      q_j={2\pi\over {N_a}}I_j^{\alpha,\gamma} \, ,
\label{qpseudo}      
\end{equation}
where in contrast to the usual momentum, $I_j^{\alpha,\gamma}$ are 
consecutive integers or half-odd integers for ${\bar{N}}_{\alpha,\gamma}$ 
odd and even, respectively
\cite{Lieb68,Takahashi72}. Here 

\begin{equation}
      {\bar{N}}_{c,0} = \sum_{\alpha}{N^h_{\alpha}\over 2} -
      \sum_{\gamma =1}^{\infty}N_{c,\gamma} = {N_a\over 2} -
      N_{s,0} - \sum_{\alpha ,\gamma =1}^{\infty}N_{\alpha,\gamma} \, ,
\label{bound1}
\end{equation}
and 

\begin{equation}
      {\bar{N}}_{\alpha,\gamma}=
      d_{\alpha,\gamma} \, , 
\label{bound2}      
\end{equation}
for values of $\alpha ,\gamma $ other than $c, 0$.
It follows that for each $\alpha,\gamma$ 
band, $q^{(-)}_{\alpha,\gamma}\leq q\leq q^{(+)}_{\alpha,\gamma}$, 
with the limits of the pseudo-Brillouin zones given by

\begin{equation}
      q^{(\pm)}_{c,0}=\pm\pi[1-{1\over N_a}] \, ,
\label{bza}      
\end{equation}
for ${\bar{N}}_{c,0}$ even and 

\begin{equation}
      q^{(+)}_{c,0}=\pi \, , \hspace{1cm}
      q^{(-)}_{c,0}=-\pi[1-{2\over N_a}] \, ,
\label{bzb}      
\end{equation}
or 

\begin{equation}
       q^{(+)}_{c,0}=\pi[1-{2\over N_a}] \, , \hspace{1cm}
       q^{(-)}_{c,0}=-\pi \, , 
\label{bzc}       
\end{equation}
for ${\bar{N}}_{c,0}$ odd and simply given by 

\begin{equation}
      q^{(\pm)}_{\alpha,\gamma}=\pm{\pi\over N_a}
      [d_{\alpha,\gamma}-1] \, ,
\label{bzd}      
\end{equation}
for all the remaining ${\alpha,\gamma}$ bands. 

In the LWS's II (or HWS's II) the situation is 
somewhat more complex, and
has no analogon in the case of the harmonic oscillator. The origin of these 
states is in the lack of continuous translation symmetry.
In these states, 
which can be interpreted as bound states of $\alpha,0$ pseudoparticles, we have both
$\beta =1/2$ and $\beta =-1/2$ pseudohole occupancy. The 
general second quantized representation for all Hubbard-chain eigenstates is \cite{Nuno97a}

\begin{equation}
      |\psi;\{N^h_{\alpha ,\beta}\},\{N_{\alpha ,\gamma}\}\rangle = 
      {1\over \sqrt{C}}\prod_{\alpha} A_{\alpha} \prod_{\gamma =1}^{\infty}\left[ 
      [\hat{S}^{\alpha }_+]^{N^h_{\alpha ,{1\over 2}}}
      \prod_{q,q'}a^{\dag }_{q,\alpha ,-{1\over 2}}
      b^{\dag }_{q',\alpha ,\gamma}\right] 
      |V\rangle \,,
\label{bethestates}
\end{equation}
where the symbols $\{N^h_{\alpha ,\beta}\}$ and $\{N_{\alpha ,\gamma}\}$ 
are abbreviations for the sets $\{N^h_{c,{1\over 2}},
N^h_{c,-{1\over 2}},$ $N^h_{s,{1\over 2}},N^h_{s,-{1\over 2}}\}$ and
$\{N_{c ,1},...,N_{c ,\infty}, N_{s ,1},..., N_{s,\infty}\}$, 
respectively, $C=\prod_{\alpha}(N^h_{\alpha }!/
N^h_{\alpha ,-{1\over 2}}!)$, $A_{\alpha}=
\prod_{\beta}\Theta (N^h_{\alpha,\beta}-
\sum_{\gamma =1}^{\infty}\gamma N_{\alpha ,\gamma})$,
and the ladder operators $S_+^{\alpha}$ are defined in Eq. (\ref{offdiagonal}). 
Note that the states (\ref{bethestates}) have a very simple form 
and present a remarkable similarity with the states (\ref{harmstates}).

As in the harmonic oscillator, the Hubbard model
Hamiltonian (and other operators) can be written in terms of the 
operators $a_{q,\alpha,\beta}^{\dag}$ and
$b_{q,\alpha,\gamma}^{\dag}$ and the energy of the LWS's and
HWS's is independent of the $\beta$ quantum number. 
In general, the non-LWS's
(non-HWS's) and the LWS's II (HWS's II) have an energy gap 
relatively to the corresponding ground state (which 
is always a LWS's I or HWS's I). 
In the remaining sections, we present the details of our
algebraic solution for the 1D Hubbard model. 
\section{Pseudoholes: the $\beta$ quantum number in LWS's I/HWS's I}
\pagestyle{myheadings}
\markboth{3 Algebraic Representation for ...}{3.2 Pseudoholes: the $\beta$ ...}

As was referred above, the description of the non-LWS's, not included in the
Bethe-ansatz solution
requires the introduction of the $\beta=\pm 1/2$ pseudohole 
numbers. We start the analysis with the study of the LWS's I. The Bethe-ansatz 
solution in the (-1,-1) sector (Sec. \ref{so4symmetry})
gives for the numbers of $\alpha,0$ pseudoparticles,
$N_{\alpha,0}$,
and for the total numbers of permitted orbitals, $d_{\alpha,0}$ (number
of states in the respective pseudo-Brillouin zone),
the following values \cite{Nuno95,Nuno96}

\begin{equation}
      N_{c,0}=N\,, \hspace{1cm} N_{s,0}=\frac 1{2}
	\left ( N_{c,0}-N_{\uparrow}+
      N_{\downarrow}\right ) \,,
\end{equation} 
\begin{equation}
      d_{c,0}=N_a\,, \hspace{1cm} d_{s,0}=N_{c,0}-N_{s,0}\,.
\end{equation}
The use of the $SO(4)$ symmetry
(Sec.\,\ref{so4symmetry}) allows one to obtain these numbers in the 
remaining three U(1)$\otimes$U(1) symmetry sectors. The number 
of pseudoholes in the $\alpha,0$ band is given by
$N_{\alpha}^h=d_{\alpha,0}-N_{\alpha,0}$. The corresponding results
for the LWS's I/HWS's I
in the four U(1)$\otimes$U(1)  symmetry sectors are listed in Table 
\ref{pseudou1}. The numbers $N$ and $N_{\sigma}$ in Table 
\ref{pseudou1} are those of the respective $(l_c,l_s)$
sector, and are related
to those in different sectors by the particle/hole and spin-up/spin-down symmetries discussed in Sec. \ref{so4symmetry}. 

Defining $N_e$ as the electron number in a given eta-spin LWS, the
respective eigenvalue of $\hat{S}_z^c$ is $S^c_z=-S^c=-\frac 1{2}(N_a-N_e)$
in terms of electrons, or $S^c_z=-S^c=
-\frac 1{2}N^h_{c,-\frac 1{2}}$ (see Table \ref{pseudou1}), in terms of 
pseudoholes. On the other hand, the corresponding HWS has $2N_a-N_e$ 
electrons and $S^c_z=S^c=\frac 1{2}(N_a-N_e)$, 
or,  equivalently, $S^c_z=S^c=
\frac 1{2}N^h_{c,\frac 1{2}}$ (see Table \ref{pseudou1}). Obviously, 
equivalent results hold for the spin towers.
\begin{table}[ht]
\begin{center}
\begin{tabular}{ccccc}
\hline
&$(-1,-1)$ & $(-1,1)$ & $(1,-1)$ &$(1,1)$\\ 
\hline
\hline
 $N_{c,0}$& $N$ & $N$ & $2N_a-N$ 
& $2N_a-N$\\ 
$d_{c,0}$ & $N_a$ & $N_a$ 
& $N_a$ & $N_a$\\ 
$N_{c,-\frac 1{2}}^h$& $N_a-N$ & $N_a-N$ 
&$0$ & $0$\\ 
$N_{c,\frac 1{2}}^h$& $0$ & $0$ 
&$N-N_a$ & $N-N_a$\\ 
$N_{s,0}$& $N_{\downarrow}$ & $N_{\uparrow}$ & $N_a-N_{\uparrow}$ 
& $N_a-N_{\downarrow}$\\ 
$d_{s,0}$ & $N_{\uparrow}$ & $N_{\downarrow}$ 
& $N_a-N_{\downarrow}$ & $N_a-N_{\uparrow}$\\ 
$N_{s,-\frac 1{2}}^h$& $N_{\uparrow}-N_{\downarrow}$ & $0$ 
&$N_{\uparrow}-N_{\downarrow}$ & $0$\\
$N_{s,\frac 1{2}}^h$& $0$ & $N_{\downarrow}-N_{\uparrow}$ 
&$0$ & $N_{\downarrow}-N_{\uparrow}$\\  
\hline
\end{tabular}
\end{center}
\caption{Values for the total number of pseudoparticles $N_{\alpha,0}$,
total number of accessible orbitals $d_{\alpha,0}$, and total number of pseudoholes 
$N^h_{\alpha,\beta=\pm \frac 1{2}}$ for LWS's I/HWS's I in
the four sectors of symmetry U(1)$\otimes$U(1). The
numbers $N$ and $N_{\sigma}$ are those of the respective $(l_c,l_s)$ 
sector.}
\label{pseudou1}
\end{table}
It follows from the results of Table (\ref{pseudou1}) that the number 
of pseudoholes in any $(l_c,l_s)$ sector 
is equal to $2\vert S_z^{\alpha} \vert$ ---the eigenvalues
of $\hat {S}_z^{\alpha}$---, defined in Sec. \ref{so4symmetry} in terms
of electronic operators and in Eq. (\ref{offdiagonal}) in terms of
pseudohole operators. Then, in going from a LWS, 
belonging to a
given $l_{\alpha}$ sector, to the respective $S_{\alpha}$ HWS,
we have to apply the ladder operators 
$S_+^{\alpha}$ a $N_{\alpha}^h$ number of times, with $N_c^h=
N_a-N_e$ for a eta-spin tower and $N_s^h=N_{\uparrow}-N_{\downarrow}$
for a spin tower
(the numbers $N_{\uparrow}$ and $N_{\downarrow}$ are those for the respective
LWS).
Hence, as in the isotropic harmonic oscillator,
we have in the LWS's I occupancies of $\beta=-1/2$
pseudoholes only. On the other hand, the
HWS's I show only $\beta=1/2$ 
pseudoholes occupancy. In between 
we have non-LWS's or non-HWS's whose occupancies refer
to mixtures of both 
$\beta=-1/2$ and $\beta=+1/2$ pseudoholes. The number of pseudoholes is such
that it gives the correct values for the $\hat{S}_z^{\alpha}$
eigenvalues and the correct number of states for a given $S^{\alpha}$
tower.

\section{Heavy-pseudoparticles: the $\beta$ quantum number in LWS's II/HWS's II}
\pagestyle{myheadings}
\markboth{3 Algebraic Representation for ...}
{3.3 Heavy-pseudoparticles: the $\beta$ ...}

In the LWS's II/HWS's II there is both pseudohole 
and heavy-pseudoparticle occupancy. In the $(-1,-1)$ sector, the Bethe-ansatz solution 
gives the following results for the numbers of $\alpha,0$ pseudoparticles
and accessible $d_{\alpha,0}$ orbitals 
\cite{Takahashi72,Nuno97a}

\begin{equation}
      N_{c,0}=N-2\sum_{\gamma=1}^{\infty}\gamma N_{c,\gamma}\,, \hspace{1cm}
      N_{s,0}=N_{\downarrow}-\sum_{\gamma=1}^{\infty}\gamma N_{c,\gamma}
      -\sum_{\gamma=1}^{\infty}(\gamma +1)N_{s,\gamma}\,,      
\end{equation}
and

\begin{equation}
      d_{c,0}=N_a\,, \hspace{1cm} d_{s,0}=N-2N_{\downarrow}+N_{s,0}+2\sum_{\gamma=1}^{\infty}
\gamma N_{s,\gamma}\,,
\end{equation}
respectively. From the above two equations, it follows that the number of 
$\alpha,0$ pseudoholes 
populating the LWS's II is given by

\begin{equation}
      N_{c}^h=N_a-N+2\sum_{\gamma=1}^{\infty}\gamma N_{c,\gamma}\,,
      \hspace{1cm} N_{s}^h=N-2N_{\downarrow}+
      2\sum_{\gamma=1}^{\infty}\gamma N_{s,\gamma}\,.
\label{nholeslwsii}
\end{equation}
Equation (\ref{nholeslwsii}) shows
that the number of holes $N_{\alpha}^h$ in the LWS's II is larger 
than the number of
holes in the LWS's I by the addicional term
$2\sum_{\gamma=1}^{\infty}\gamma N_{\alpha,\gamma}$. 
This requires, if we want Eq. (\ref{offdiagonal})
for $\hat{S}_z^{\alpha}$ to give the correct eigenvalues, that half of the
pseudoholes in excess, relatively to the respective LWS I, have 
$\beta=+1/2$. That is, in any LWS/HWS, the numbers 
of $\beta=\pm 1/2$
pseudoholes must be given by

\begin{equation}
       N^h_{c,\beta}={N^h_c\over 2}-\beta [N_a-N] \, ,  \hspace{1cm}
       N^h_{s,\beta}={N^h_s\over 2}-\beta [N_{\uparrow}-
       N_{\downarrow}] \, , 
\label{holebeta}
\end{equation}
the total number of $\alpha,0$ pseudoholes being 
$N^h_{\alpha} = \sum_{\beta}N^h_{\alpha,\beta}$.                

Equation (\ref{holebeta}) is a generalization of the pseudohole
expressions introduced in the context of LWS's I/HWS's I \cite{Nuno96}.
It is convenient to introduce the numbers $N^z_{\alpha}$ such that

\begin{equation}
      N^z_{\alpha} = S^{\alpha }-\vert S_z^{\alpha} \vert =
     {1\over 2}[\sum_{\beta}(1-2\beta l_{\alpha})\hat{N}^h_{\alpha ,\beta}
     -\sum_{\gamma =1}^{\infty}2\gamma N_{\alpha,\gamma}]\, .
\label{nza}      
\end{equation}
This number indicates whether a given eigenstate is an LWS (or HWS).
The knowledge of all the $N^h_{\alpha,\beta}$
pseudohole and $N_{\alpha,\gamma}$ heavy-pseudoparticle
($\gamma >0$) numbers provides the knowledge of all pseudoparticle
numbers $N_{\alpha,\gamma}$ ($\gamma =0,1,2,3...$) and
$N^z_{\alpha }$ numbers, the inverse being also true. 
While the set of $N_{\alpha,\gamma}$ numbers
are directly given by the BA solution \cite{Takahashi72}, 
the $\beta $-dependent pseudohole numbers $N^h_{\alpha,\beta}$ 
and the numbers $N^z_{\alpha }$ are not considered by that solution.

It follows from Eq. (\ref{holebeta}) that the electron numbers 
are exclusive functions of the pseudohole numbers (this had to be so, as required from the form of the $SO(4)$ generators) and read 

\begin{equation}
      N_{\uparrow} = {N_a\over 2} + \sum_{\beta}\beta[N^h_{c,\beta}
      - N^h_{s,\beta}] \, , \hspace{1cm}
      N_{\downarrow} = {N_a\over 2} + \sum_{\beta}\beta[N^h_{c,\beta}
      + N^h_{s,\beta}] \, . 
\label{nupndown}
\end{equation}

The pseudomomentum-number operators

\begin{equation}
       \hat{N}^h_{\alpha,\beta}(q)=a^{\dag }_{q,\alpha 
       ,\beta}a_{q,\alpha ,\beta} \, , \hspace{1cm}
       \hat{N}_{\alpha ,\gamma}(q)=
       b^{\dag }_{q,\alpha ,\gamma}b_{q,\alpha ,\gamma} \, , 
	 \hspace{1cm} \gamma =1,2,...\,, 
\label{nqop}       
\end{equation}
and 

\begin{equation}
      \hat{N}_{\alpha ,0}(q)\equiv 1
      -\sum_{\beta}\hat{N}^h_{\alpha,\beta}(q)\,,
	\hspace{1cm} (\gamma=0) \, , 
\label{nalphazero}
\end{equation}
play a central role in the generalized theory 
(see Sec. \ref{perturbation}). 
The operator (\ref{nalphazero}) has the following 
alternative representation
in terms of $\alpha,0$ pseudoparticle operators

\begin{equation}
      \hat{N}_{\alpha ,0}(q)=
      b^{\dag }_{q,\alpha ,0}b_{q,\alpha ,0} \, . 
\end{equation}
The number operators can be expressed in terms of
the pseudomomentum distributions as follows

\begin{equation}
       \hat{N}^h_{\alpha ,\beta}=\sum_{q}\hat{N}^h_{\alpha ,\beta}(q) 
       \, , \hspace{1cm}
       \hat{N}_{\alpha,\gamma}=\sum_{q}\hat{N}_{\alpha 
       ,\gamma}(q) \, . 
\end{equation}

We emphasize that while all Hamiltonian eigenstates are
also eigenstates of the operator $\hat{N}^h_{\alpha,\beta}$
because these numbers are good quantum numbers, this does 
not hold for the operator $\hat{N}^h_{\alpha,\beta}(q)$;
obviously, only the states I are eigenstates of the 
latter operator. This fact follows directy from the
structure of the state (\ref{bethestates}).
On the other hand, all Hamiltonian eigenstates are
also eigenstates of the pseudoparticle operators 
$\hat{N}_{\alpha ,\gamma}(q)$.

The generalization of the LWS's BA total momentum expression
\cite{Essler92b,Essler92c,Takahashi72}, which we denote by $P_{BA}$, 
to all Hamiltonian eigenstates, requires the addition of an 
extra term associated with $\eta$ pairing 
\cite{Yang89,Yang90}. This is a direct implication 
of the commutators introduced
in Sec. (\ref{so4symmetry}) and leads to

\begin{equation}
       P = P_{BA} + \pi[S^c + S^c_z] \,.
\end{equation}
When this expression provides values of momentum such that
$|P|>\pi$, the total momentum is defined as the corresponding 
value at the first Brillouin zone.  
In operator form and for sub-canonical ensembles such that 
$S^{\alpha }_z\neq 0$, and with $l_{\alpha }$ given by Eq. (\ref{sector}),
we have that $N^h_{\alpha ,-\frac {l_{\alpha}}{2}}<
N^h_{\alpha ,{l_{\alpha}\over 2}}$,
and the present representation leads to the
following simple expressions for the momentum operator 
$\hat{P}$ and its eigenvalue $P$

\begin{eqnarray}
       \hat{P} & = & \sum_{q,\alpha}
       \sum_{\gamma =0}^{\infty} qC_{\alpha,\gamma}
       \hat{N}_{\alpha,\gamma}(q) +
       \pi [\hat{N}^h_{c 
       ,{-l_c\over 2}}+\sum_{\gamma =1}\hat{N}_{c,\gamma}] \, ;
\nonumber \\
      P & = & \sum_{q,\alpha}
      \sum_{\gamma =0}^{\infty} qC_{\alpha,\gamma}
      N_{\alpha,\gamma}(q) +
      \pi\sum_{\gamma =1}\left[(1+\gamma)N_{c,\gamma} + N^z_c\right]\,,
\label{pmomentum}
\end{eqnarray}
respectively, where $C_{\alpha,0}=1$ and 
$C_{c,\gamma}=-1$ for $\gamma >0$. We note that the momentum term, 
$\pi\sum_{\gamma =1}\left[(1+\gamma)N_{c,\gamma} + N^z_c\right]$,
is always a multiple of $\pm\pi$.

The pseudoparticle perturbation theory introduced in the works
\cite{Carm90,Carm91,Carm92a,Carm92b,Carm92c,Carm93a}
and developed in a suitable operator basis 
\cite{Nuno95,Nuno96,Carm93b,Carm94a,Carm94b} refers to the Hilbert 
subspace spanned by the Hamiltonian eigenstates I. At finite 
values of the magnetic field and chemical potential and at 
constant electronic numbers the low-energy excitations I are 
described by pseudoparticle-pseudohole processes relative to
the canonical-ensemble GS. The latter state, as well 
as all excited states I with the same electron numbers, are simple 
Slater determinants of pseudoparticle $\alpha,0$ levels 
\cite{Nuno95,Nuno96,Carm93b,Carm94a,Carm94b}. That is, in 
Eq. (\ref{bethestates}) either $N_{\alpha,\frac 1{2}}=0$ for LWS's or
$N_{\alpha,-\frac 1{2}}=0$ for HWS's.
\section{LWS's and non-LWS's: the criterium}
\pagestyle{myheadings}
\markboth{3 Algebraic Representation for ...}{3.4 LWS's and non.LWS's ...}

A $S^{\alpha}$ LWS is such that 

\begin{equation}
       N^h_{\alpha ,{1\over 2}}=\sum_{\gamma 
       =1}^{\infty}\gamma N_{\alpha ,\gamma} \, ,
\label{nholea}
\end{equation}
whereas for a $S^{\alpha}$ HWS we have that

\begin{equation}
      N^h_{\alpha ,-{1\over 2}}=\sum_{\gamma =1}^
      {\infty}\gamma N_{\alpha,\gamma} \, , 
\label{nholeb}      
\end{equation}
with $N^h_{\alpha ,-{1\over 2}}\geq N^h_{\alpha 
,{1\over 2}}$ for a LWS and $N^h_{\alpha ,{1\over 2}}\geq N^h_{\alpha 
,-{1\over 2}}$ for a HWS. 

For the particular case of states I, we have that
$N_{\alpha ,\gamma}=0$ for all $\gamma =1,2,...$ branches in 
the former equations. Therefore, the states I are
characterized by zero occupancies of the 
$\alpha ,\gamma$ heavy-pseudoparticle
bands (with $\gamma >0$). It then
follows from the general Eqs. (\ref{nholea}) and (\ref{nholeb})
that a $S^{\alpha}$ LWS I is such that 

\begin{equation}
      N^h_{\alpha ,{1\over 2}} = 0 \, ,
\label{nh120}      
\end{equation}
whereas for a $S^{\alpha}$ HWS I, we have that

\begin{equation}
      N^h_{\alpha ,-{1\over 2}}= 0 \, , 
\label{nh210}      
\end{equation}
with $N^h_{\alpha ,-{1\over 2}}\geq 0$ for a LWS I and 
$N^h_{\alpha ,{1\over 2}}\geq 0$ for a HWS I. This
leads to the simple Slater-determinant form (\ref{bethestates})
found
for the states I 
\cite{Nuno96}.
These 
results also confirm that in the case of states
I the $\alpha ,\beta$ pseudoholes are such that
$\beta ={l_{\alpha }\over 2}$ but that this equality
does not hold true in the general case. This is
because in the case of the states I the Slater
determinant (\ref{bethestates}) either involves $\alpha, +{1\over 2}$
or $\alpha, -{1\over 2}$ pseudoholes only, whereas in the case
of the non-LWS's and non-HWS's or in the case of
states II (i.e. LWS's or HWS's containing $\alpha ,\gamma$ heavy 
pseudoparticles) the state (\ref{bethestates}) has both $\alpha, +{1\over 2}$
and $\alpha, -{1\over 2}$ pseudohole occupancies. 

In the case of the non-LWS's and non-HWS's, 
all or some of the off-diagonal generators
of the rhs of Eq. (\ref{bethestates}) create new states from a reference
BA LWS. The $S^{\alpha}$ non-LWS's and non-HWS's outside 
the BA are such that 

\begin{equation}
      N^h_{\alpha}-\sum_{\gamma =1}^{\infty}2\gamma 
      N_{\alpha ,\gamma}>|N^h_{\alpha ,{1\over 2}}-
      N^h_{\alpha ,-{1\over 2}}| \, .
\label{condition}      
\end{equation}

In contrast, in the case of the LWS's II, the $\beta $ 
flips generated by these operators create a number $\gamma $ of 
($\alpha ,{+1\over 2}$) pseudoholes for each $\alpha ,\gamma$ heavy 
pseudoparticle. Following Eqs. (\ref{offdiagonal}), (\ref{angulars}),
and (\ref{nza}) this is required for the LWS 
condition $S^{\alpha }=-S^{\alpha }_z$. Therefore, in the 
latter, cases there is no relation between the quantity
$l_{\alpha }$ defined in Eq. (\ref{sector}) and the pseudohole quantum 
number $\beta$.

\section{The generalized ground state concept}
\label{generalized}
\markboth{3 Algebraic Representation for ...}{3.5 The generalized ...}

The possible pseudohole and heavy-pseudoparticle numbers 
are constrained by Eqs. (\ref{offdiagonal}),
(\ref{angulars}), and (\ref{nza}) 
and together with the
number of discrete pseudomomentum values in each band
leads to $4^{N_a}$ possible orthonormal Hamiltonian 
eigenstates of form (\ref{bethestates}), in agreement with the
results of
Essler {\it et al.} \cite{Essler92a,Essler92b,Essler92c}. 

The conservation of the pseudohole and heavy-pseudoparticle
numbers permits dividing the Hilbert space into
subspaces spanned by the set of Hamiltonian 
eigenstates (\ref{bethestates}) with the same $\{N^h_{\alpha ,\beta}\}$
and $\{N_{\alpha ,\gamma}\}$ numbers, which
correspond to the same sub-canonical ensemble.
Obviously, a canonical ensemble (with constant electron 
numbers $N_{\sigma}$ and thus with constant values of $N^h_{\alpha 
,{1\over 2}}-N^h_{\alpha ,-{1\over 2}}$ for both $\alpha=c,s$) 
is usually realized by several $\{N^h_{\alpha ,\beta}\}$,
$\{N_{\alpha ,\gamma}\}$ sub-canonical ensembles. 
Let us introduce the generalized ground state (GGS) as the 
Hamiltonian eigenstate(s) (\ref{bethestates}) of lowest energy in each
Hilbert subspace. This type of eigenstate
is very useful for the construction of
the pseudoparticle perturbation theory of Sec. \ref{perturbation}.
They are of the form

\begin{eqnarray}
      |GGS;\{N^h_{\alpha ,\beta}\},\{N_{\alpha ,\gamma}\}\rangle &=& 
      {1\over \sqrt{C}}\prod_{\alpha} A_{\alpha} \prod_{\gamma =1}^{\infty}
      \left[[\hat{S}^{\alpha }_+]^{N^h_{\alpha ,{1\over 2}}}
      \prod_{q=q_{\alpha,0}^{(-)}}^{{\bar{q}}_{F\alpha,0}^{(-)}} 
      \prod_{q={\bar{q}}_{F\alpha,0}^{(+)}}^{q_{\alpha,0}^{(+)}}\right . 
      \nonumber\\
      &&\left . \prod_{q'=q_{c,\gamma}^{(-)}}^{q_{Fc,\gamma}^{(-)}} 
      \prod_{q'=q_{Fc,\gamma}^{(+)}}^{q_{c,\gamma}^{(+)}} 
      \prod_{q''=q_{Fs ,\gamma}^{(-)}}^{q_{Fs,\gamma}^{(+)}}
      a^{\dag }_{q,\alpha ,-{1\over 2}}
      b^{\dag }_{q'',c 
      ,\gamma}
      b^{\dag }_{q',s 
      ,\gamma}\right]|V\rangle \, ,
\label{ggs}
\end{eqnarray}
and when (i) $\alpha =c$, $\gamma >0$, and $N_{c,\gamma}$ is even or 
(ii) $\alpha =s$ or $\alpha =c$ and $\gamma =0$ and 
$N_{\alpha,\gamma}$ is odd (even) and $I_j^{\alpha,\gamma}$ are 
integers (half-odd integers) the pseudo-Fermi points are symmetric 
and given by 

\begin{equation}
	q_{F\alpha,\gamma}^{(\pm)}=\pm [q_{F\alpha,\gamma}
	- C_{\alpha ,\gamma}{\pi\over {N_a}}] \, , 
\end{equation}
where 

\begin{eqnarray}
	q_{Fc,\gamma} & = & {\pi[d_{c,\gamma}-N_{c,\gamma}]\over N_a} \, ;
	\hspace{1cm} \gamma >0 \, , \nonumber \\
	q_{Fc,0} & = & {\pi N_{c,0}\over N_a} \, , \nonumber \\ 
	q_{Fs,\gamma} & = & {\pi N_{s,\gamma}\over N_a} \, .
\end{eqnarray}
On the other hand, when (i) $\alpha =c$, $\gamma >0$, and 
$N_{c,\gamma}$ is odd or (ii) $\alpha =s$ or $\alpha =c$ and
$\gamma =0$ and $N_{\alpha,\gamma}$ is odd (even) and 
$I_j^{\alpha,\gamma}$ are half-odd integers (integers) we have that 

\begin{equation}
	q_{F\alpha,\gamma}^{(+)}=q_{F\alpha,\gamma} \, , \hspace{1cm}
	q_{F\alpha,\gamma}^{(-)} =-[q_{F\alpha,\gamma}-
	C_{\alpha ,\gamma}{2\pi\over N_a}] \, , 
\end{equation}
or 

\begin{equation}
	q_{F\alpha,\gamma}^{(+)}=q_{F\alpha,\gamma}-
	C_{\alpha ,\gamma}{2\pi\over {N_a}} 
	\, , \hspace{1cm}
	q_{F\alpha,\gamma}^{(-)}= -q_{F\alpha,\gamma} \, . 
\end{equation}

The GS associated with a given canonical ensemble is always a state 
I \cite{Nuno96}, which is
a particular case of the general GGS expression 
(\ref{ggs}).
It is useful to denote the GS's by $|GS;S^c_z,S^s_z\rangle$,
where $S^c_z$ and $S^s_z$ are the eigenvalues of the
corresponding canonical ensemble.
The GS expression associated with the $(l_c,l_s)$ 
sector of Hamiltonian symmetry $U(1)\otimes U(1)$ reads 

\begin{equation}
      |GS;S^c_z,S^s_z\rangle = 
      \prod_{q=q_{c,0}^{(-)}}^{{\bar{q}}_{Fc,0}^{(-)}} 
      \prod_{q={\bar{q}}_{Fc,0}^{(+)}}^{q_{c,0}^{(+)}} 
      a^{\dag }_{q,c,{l_c\over 2}} 
      \prod_{q=q_{s,0}^{(-)}}^{{\bar{q}}_{Fs,0}^{(-)}} 
      \prod_{q={\bar{q}}_{Fs,0}^{(+)}}^{q_{s,0}^{(+)}} 
      a^{\dag }_{q,s,{l_s\over 2}} 
      |V \rangle \, .
\label{gsop}      
\end{equation}

In all sectors of Hamiltonian symmetry there are states I. In the
particular case of the SO(4) zero-chemical potential
and zero-magnetic field canonical ensemble there is only one state 
I, the pseudohole and heavy-pseudoparticle vacuum, 
$|V \rangle$, which is nothing but the SO(4) GS
\cite{Nuno96}. The same applies to the sectors 
of Hamiltonian symmetry $SU(2)\otimes U(1)$ and $U(1)\otimes SU(2)$, 
the GS being always a state I. (In addition, in these
sectors there is a large number of excited states I.)

In the case of the two $(l_s)$ $SU(2)\otimes U(1)$ 
sectors, the GS is both a LWS and HWS of the $\eta$-spin algebra. 
Therefore, it is empty of $c$ pseudoholes and reads

\begin{equation}
      |GS;0,S^s_z\rangle = 
      \prod_{q=q_{s,0}^{(-)}}^{{\bar{q}}_{Fs,0}^{(-)}} 
      \prod_{q={\bar{q}}_{Fs,0}^{(+)}}^{q_{s,0}^{(+)}} 
      a^{\dag }_{q,s,{l_s\over 2}} 
      |V\rangle \, .
\end{equation}

In the case of the $(l_c)$ $U(1)\otimes SU(2)$ sector 
the GS is both a LWS and a HWS of the spin algebra
and is empty of $s$ pseudoholes. It reads

\begin{equation}
       |GS;S^c_z,0\rangle = 
       \prod_{q=q_{c,0}^{(-)}}^{{\bar{q}}_{Fc,0}^{(-)}} 
       \prod_{q={\bar{q}}_{Fc,0}^{(+)}}^{q_{c,0}^{(+)}} 
       a^{\dag }_{q,c,{l_c\over 2}} |V\rangle \, .
\label{gs3}
\end{equation}

Finally, the $S^c=S^c_z=0$ (and $\mu =0$) and $S^s=S^s_z=0$ SO(4) 
ground state is, at the same time, a LWS and HWS of both the $\eta$-spin 
and spin algebras 
\cite{Nuno95,Nuno96}. Therefore, it is empty of both $c$ and $s$
pseudoholes and it is the vacuum of the pseudohole and
heavy-pseudoparticle theory. All the remaining eigenstates can be 
described by particular cases of the general expression
(\ref{bethestates}), being the only difference the values of the momenta 
$q$ in the product of operators $a^{\dag}_{\alpha,q,\beta}$
and $b^{\dag}_{q,\alpha,\gamma}$.

%% file: handgap.tex
\pagestyle{headings}
\setcounter{chapter}{3}
\chapter{Pseudoparticle Perturbation Theory}
\label{PseudoPT}

\section{Pseudoparticle Hamiltonian}
\label{pphamilt}
\pagestyle{myheadings}
\markboth{4  Pseudoparticle Perturbation Theory}{4.1 Pseudoparticle Hamiltonian}

Using the operational representation for the Hilbert space of the one-dimensional Hubbard model, we develop
a generalized perturbation theory whose small parameter is the density of excited pseudoparticles. A similar theory
for the LWS-I has been developed previously in the literature \cite{0Carm94b}.

For simplicity, let us denote the general Hamiltonian
eigenstates $|\psi;\{N^h_{\alpha ,\beta}\},\{N_{\alpha 
,\gamma}\}\rangle$ by $|\psi\rangle$ (and the GGS's 
(\ref{ggs}) by $|GGS\rangle$). These states are
also eigenstates of the $\alpha,\gamma$ 
pseudomomentum-distribution operators (\ref{nqop}) such that

\begin{equation}
      \hat{N}_{\alpha,\gamma}(q)
      |\psi\rangle  = N_{\alpha,\gamma}(q)|\psi\rangle \, ,
\label{qdistrib}      
\end{equation}
where $N_{\alpha,\gamma}(q)$ represents the 
eigenvalue of the operator (\ref{nqop}), which is given by $1$ and $0$ 
for occupied and empty values of $q$, respectively. 

It follows from the form of the state-generator of the rhs of
Eq. (\ref{bethestates}) that all $2S^{\alpha }+1$ Hamiltonian eigenstates 
constructed from the same $S^{\alpha }=-S^{\alpha }_z$ LWS by applying 
onto it $1,2,...,2S^{\alpha }$ number of times the off-diagonal operator 
${\hat{S}}_+^{\alpha }$ (\ref{offdiagonal}) 
have the same $N_{\alpha,\gamma}(q)$
pseudomomentum distribution.

As in the case of the spatial wave functions, the expression of 
the energy in terms of the quantum numbers $I_j^{\alpha ,\gamma}$
of Eq. (\ref{qpseudo}) involves the BA rapidities \cite{0Lieb68,0Takahashi72}. 
These are complex functions
of the numbers $I_j^{\alpha ,\gamma}$. Following our pseudomomentum
definition, Eq. (\ref{qpseudo}), in our description they are functions of
the pseudomomentum $q$. The real part of the BA rapidities, which we 
denote by $R_{\alpha,\gamma}(q)$, are in the present 
basis eigenvalues of corresponding rapidity operators
$\hat{R}_{\alpha,\gamma}(q)$, and obey the eigenvalue equation

\begin{equation}
       \hat{R}_{\alpha,\gamma}(q)|\psi\rangle = 
       R_{\alpha,\gamma}(q)|\psi\rangle \, ,
\label{ralpha}       
\end{equation}
where for the particular case of $R_{c,0}(q)$ we also consider
the associate rapidity $K(q)$, which is defined by the following 
equation

\begin{equation}
      R_{c,0}(q) = {\sin K(q)\over u} \, .
\end{equation}
The rapidity $K(q)$ is the 
eigenvalue of the rapidity operator $\hat{K}(q)$ such that

\begin{equation}
       \hat{K}(q)|\psi\rangle = K(q)|\psi\rangle \, .
\label{kalpha}       
\end{equation}
The eigenvalues $K(q)$, $R_{c,\gamma}(q)$, and
$R_{s,\gamma}(q)$ obey the following non-linear algebraic
equations \cite{0Takahashi72,0Nuno97a} 

\begin{eqnarray}
      K(q) & = & q - {2\over N_a}\sum_{\gamma =0}\sum_{q'}N_{s,\gamma}(q')
      \tan^{-1}\Bigl({R_{c,0}(q)-R_{s,\gamma}(q')\over 1+\gamma}\Bigr)
      \nonumber \\
      & - & {2\over N_a}\sum_{\gamma =1}\sum_{q'}N_{c,\gamma}(q')
      \tan^{-1}\Bigl({R_{c,0}(q)-R_{c,\gamma}(q')
      \over \gamma}\Bigr) \, ,
\label{tak1}      
\end{eqnarray}

\begin{eqnarray}
      &&2Re \sin^{-1}\Bigl(u(R_{c,\gamma}(q)+i\gamma)\Bigr) 
      =  q + {2\over N_a}\sum_{q'}N_{c,0}(q')
      \tan^{-1}\Bigl({R_{c,\gamma}(q)-R_{c,0}(q')\over \gamma}\Bigr)
      \nonumber \\
      & + & {1\over N_a}\sum_{\gamma' =1}\sum_{q'}N_{c,\gamma'}(q')
      \Theta_{\gamma,\gamma'}\Bigl(R_{c,\gamma}
      (q)-R_{c,\gamma'}(q')\Bigr) \, ,
\label{tak2}      
\end{eqnarray}
and 

\begin{eqnarray}
      0 & = & q - {2\over N_a}\sum_{q'}N_{c,0}(q')
      \tan^{-1}\Bigl({R_{s,\gamma}(q)-R_{c,0}(q')\over 1+\gamma}\Bigr)
      \nonumber \\
      & + & {1\over N_a}\sum_{\gamma' =0}\sum_{q'}N_{s,\gamma'}(q')
      \Theta_{1+\gamma, 1+\gamma'}\Bigl(R_{s,\gamma}(q)-
      R_{s,\gamma'}(q')\Bigr) \, ,
\label{tak3}      
\end{eqnarray}
where

\begin{eqnarray}
       \Theta_{\gamma,\gamma'}\Bigl(x\Bigl) & = &
       \Theta_{\gamma',\gamma}\Bigl(x\Bigl) =
       \delta_{\gamma ,\gamma'}\{2\tan^{-1}\Bigl({x\over 2\gamma}\Bigl)
       + \sum_{l=1}^{\gamma -1}4\tan^{-1}\Bigl({x\over 2l}\Bigl)\}
       \nonumber \\
       & + & (1-\delta_{\gamma ,\gamma'})\{
       2\tan^{-1}\Bigl({x\over |\gamma-\gamma'|}\Bigl)
       + 2\tan^{-1}\Bigl({x\over \gamma+\gamma'}\Bigl)\nonumber \\
       & + & \sum_{l=1}^{{\gamma+\gamma'-|\gamma-\gamma'|\over 2}
       -1}4\tan^{-1}\Bigl({x\over |\gamma-\gamma'|+2l}\Bigl)\} \, .
\label{ftheta1}       
\end{eqnarray}
The limits of the pseudo-Brillouin zones, 
$q_{\alpha,\gamma}^{(\pm)}$, associated with the pseudomomentum 
summations are given in Eqs. (\ref{bza})-(\ref{bzd}). 
Although the integral Eqs. (\ref{tak1}-\ref{tak3}) are coupled, note
that each of these equations defines the rapidity
associated with one of the $\alpha,\gamma $ bands  
in terms of other rapidities. 

The key point is that the eigenvalue $N_{\alpha,\gamma}(q)$
is common to the whole tower
of $2S^{\alpha }+1$ Hamiltonian eigenstates constructed
from the same $S^{\alpha }$ LWS. Also the associate
rapidity eigenvalue $R_{\alpha,\gamma}(q)$ of Eq. (\ref{ralpha})
is common to these $2S^{\alpha }+1$ Hamiltonian eigenstates.
This follows from the fact that rapidity solutions
$R_{\alpha,\gamma}(q)$ of Eqs. (\ref{tak1}), (\ref{tak2}), and (\ref{tak3}) are functionals
of the $N_{\alpha,\gamma}(q)$ distributions and $\beta $
independent. We thus conclude that the LWS rapidities extracted from
the BA provide full information on the non-LWS's
rapidity operators.

The rapidity eigenvalues of Eqs. $(\ref{ralpha})$ and $(\ref{kalpha})$ 
are independent of the 
pseudohole numbers $N^h_{\alpha,\beta}$ and only involve the 
pseudoparticle distributions $N_{\alpha,\gamma}(q)$. The
same holds true for the associate rapidity, pseudohole,
and pseudoparticle operators. Since the SO(4) Hamiltonian
$(2)$ and other physical quantities turn out to be exclusive functionals
of the rapidity operators, in studies involving such quantities
it is often more convenient to use the $\alpha,\gamma$ pseudoparticle 
representation (with $\gamma =0,1,2,3,...$) than the $\alpha,\beta$
pseudohole and $\alpha,\gamma$ heavy-pseudoparticle representation
(with $\gamma =1,2,3,...$).

By direct insertion in each of Eqs. (\ref{tak1})-(\ref{tak3}) 
of the corresponding 
$q_{\alpha,\gamma}^{(\pm)}$ pseudomomenta we find that the functions 
$R_{\alpha,\gamma}(q)$ have {\it for all} eigenstates the following 
boundary values at the limits of the pseudo-Brillouin zones

\begin{equation}
      K(q^{(\pm )}_{c,0}) = \pm\pi \, , \hspace{1cm}
      R_{c,0}(q^{(\pm )}_{c,0}) = 0 \, ,
\end{equation}
for $\alpha =c$ and $\gamma =0$ and

\begin{equation}
      R_{\alpha,\gamma}(q^{(\pm )}_{\alpha,\gamma}) = \pm\infty \, ,
\end{equation}
for the remaining choices of the quantum numbers $\alpha,\gamma$.

It is useful to consider the rapidity functions
$K^{(0)}(q)$ and $R_{\alpha,\gamma}^{(0)}(q)$ which
are the GGS eigenvalues such that

\begin{equation}
      \hat{K}(q)|GGS\rangle =
      K^{(0)}(q)|GGS\rangle \, ,
\label{kalpha0}
\end{equation}
and

\begin{equation}
      \hat{R}_{\alpha,\gamma}(q)|GGS\rangle =
      R^{(0)}_{\alpha,\gamma}(q)|GGS\rangle \, .
\label{ralpha0}
\end{equation}
These are defined by Eqs. (\ref{tak1}-\ref{tak3}) with the pseudomomentum
distribution given by its GGS value, which we denote
by $N^{(0)}_{\alpha,\gamma}(q)$ and is such that 

\begin{equation}
      \hat{N}_{\alpha,\gamma}(q)|GGS\rangle =
      N^{(0)}_{\alpha,\gamma}(q)|GGS\rangle \, .
\end{equation}

Following the GGS expression (\ref{ggs}), $N^{(0)}_{\alpha,0}(q)$ and
$N^{(0)}_{s,\gamma}(q)$ 
have the simple free particle/hole Fermi-like form

\begin{eqnarray}
     N^{(0)}_{\alpha,\gamma}(q) & = & \Theta\Bigl(
     C_{\alpha,\gamma}(q_{F\alpha,\gamma}^{(+)}
      - q) \Bigl) \, , \hspace{1cm} 0<q<q_{\alpha,\gamma}^{(+)}
     \nonumber \\
     & = & \Theta\Bigl(C_{\alpha,\gamma}
     (q - q_{F\alpha,\gamma}^{(-)})\Bigl) \, , 
     \hspace{1cm} q_{\alpha,\gamma}^{(-)}<q<0 \,.
\label{dist1}     
\end{eqnarray}
For the particular case of a GS, these distributions read

\begin{eqnarray}
      N^{(0)}_{\alpha,0}(q) & = & \Theta\Bigl(q_{F\alpha,0}^{(+)}
      - q \Bigl) \, , \hspace{1cm} 0<q<q_{\alpha,0}^{(+)}
      \nonumber \\
      & = & \Theta\Bigl(q - q_{F\alpha,0}^{(-)}\Bigl) \, , 
      \hspace{1cm} q_{\alpha,0}^{(-)}<q<0 \, ,
      \nonumber \\
      N^{(0)}_{\alpha,\gamma}(q) & = & 0 \, , \hspace{2cm} \gamma > 0  \, .
\label{gsdist}      
\end{eqnarray}

In Appendix \ref{normalorder}
we study the rapidity solutions of Eqs.
(\ref{tak1})-(\ref{tak3}) both for GGS's and for Hamiltonian eigenstates differing
from a GGS by a small density of excited pseudoparticles.
In equations (\ref{rhoc}) and (\ref{rhos}) the parameters $Q^{(\pm)}$ and 
$r^{(\pm)}_{\alpha,\gamma }$ are defined combining the
two equations

\begin{equation}
      Q^{(\pm)} = K^{(0)}(q^{(\pm )}_{Fc,0}) \, , \hspace{1cm}
      r^{(\pm)}_{\alpha,\gamma} = R^{(0)}_{\alpha,\gamma}(q^{(\pm 
      )}_{F\alpha,\gamma}) \, ,
\label{raplimit}
\end{equation}
with equations (\ref{intrhoc}) and (\ref{intrhos}). 
For later use, it is convenient 
to introduce the associate parameters

\begin{equation}
      Q = K^{(0)}(q_{Fc,0}) \, , \hspace{1cm}
      r_{\alpha,\gamma} = R^{(0)}_{\alpha,\gamma}(q_{F\alpha,\gamma}) \, .
\label{limits}      
\end{equation}

In the thermodynamic limit, we 
can take the pseudomomentum continuum limit $q_j\rightarrow q$ and 
the real part of the rapidities become functions of $q$, which we have 
called here $R_{\alpha,\gamma} (q)$. In that limit, the BA system of 
algebraic equations are replaced by the system of infinite coupled 
integral equations presented in Appendix \ref{normalorder}. 

We have mentioned that the BA solution
is most naturally expressed in the pseudoparticle basis.
One consequence of this is the simple expression for 
the Hamiltonian (\ref{hamiltonian}) in that basis. 
From the SO(4) symmetry we know that the SO(4) energy is 
the same for all eigenstates of a given SO(4) tower. 
Thus, it suffices to know the SO(4) energy for the tower's LWS, which
is given by the Bethe ansatz. In the pseudoparticle basis, where
the non-LWS's/HWS's are characterized by the $\beta$ quantum number,
the SO(4) Hamiltonian is written, as it should be,
in terms of the pseudoparticle numbers
 
\begin{eqnarray}
       \hat{H}_{SO(4)} & = & -2t\sum_{q}\hat{N}_{c,0}(q)\cos [\hat{K}(q)]
       + 2t\sum_{q,\gamma =1}\hat{N}_{c,\gamma}(q)
       \sum_{j=\pm 1}\sqrt{1-[u(\hat{R}_{c,\gamma}(q)+ji\gamma)]^2}
        \nonumber \\
        & + & U[{N_a\over 4} - {\hat{N}_{c,0}\over 2}
        - \sum_{\gamma =1}\gamma \hat{N}_{c,\gamma}] \, .
\label{hamiltba}               
\end{eqnarray}

Contrary to the SO(4) term, the chemical potencial and the Zeeman term
are $\beta$ sensitive, because the respective terms in the Hamiltonian
(\ref{hamiltonian}) do not commute with the ladder operators of the 
SO(4) algebra. Thus, the inclusion in the Hamiltoninan (\ref{hamiltba})
of a chemical potential and a Zeeman terms must be written in terms of 
pseudoholes, and reads

\begin{equation}   
      \hat{H} = \hat{H}_{SO(4)} + 2\mu\sum_{\beta}\beta\hat{N}^h_{c,\beta} 
      + 2\mu_0 H\sum_{\beta}\beta\hat{N}^h_{s,\beta} \, .
\label{hamiltbabeta}
\end{equation}
This is the pseudoparticle expression of the Hamiltonian 
(\ref{hamiltonian}) 
at all energy scales. The Hamiltonian (\ref{hamiltbabeta}) then implies
that all energy eigenvalues of a given tower read
$E=E_{SO(4)}+2\mu (S^c+S^c_z)+2\mu_0 H (S^s+S^s_z)$.
Despite its simple appearance, the Hamiltonian (\ref{hamiltba}) 
describes a many-pseudoparticle problem, the reason being that the 
expression for the rapidity operator in terms of 
the operators $\hat{N}_{\alpha,\gamma}(q)$ contains many-pseudoparticle
interacting terms.

Unsurprisingly, it is difficult to solve the BA operator Eqs.\, 
(\ref{tak1}), (\ref{tak2}), and (\ref{tak3}) directly 
and to obtain the explicit expression for the rapidity 
operators in terms of the pseudomomentum distribution operators 
(\ref{nqop}). In contrast, it is easier to calculate their normal-ordered 
expression in terms of the normal-ordered operators. 
The rapidity operators $\hat{R}_{\alpha,\gamma}(q)$
contain all information about the many-pseudoparticle
interactions of the quantum-liquid Hamiltonian. There are two 
fundamental properties which imply the central role that the  
rapidity operators of Eqs. (\ref{kalpha0}) and  (\ref{ralpha0}) 
have in the present quantum problem:

(a) As we find in Sec. \ref{perturbation}, 
each of the normal-ordered rapidity 
operators $:\hat{R}_{\alpha}(q):$ (relative to the suitable
GGS or GS) can be written exclusively in terms of the 
pseudomomentum distribution operators (\ref{nqop}).

(b) The normal-ordered version of the SO(4) Hamiltonian (\ref{hamiltba}) can 
be written, exclusively, in terms of the pseudomomentum 
distribution operators (\ref{nqop}), but all the corresponding 
many-pseudoparticle interaction terms can be written in terms 
of the rapidity operators $:\hat{R}_{\alpha,\gamma}(q): $.
It follows that the rapidity operators commute with the Hamiltonian. 

In the ensuing section, we introduce the pseudoparticle perturbation 
theory which leads to the normal-ordered expressions for the 
Hamiltonian and rapidity operators. Despite the non-interacting 
form of the Hamiltonian eigenstates (\ref{bethestates}), the normal-ordered 
Hamiltonian includes pseudoparticle interaction terms and is, therefore, 
a many-pseudoparticle operator. However, we find in Sec. \ref{perturbation}
that
these pseudoparticle interactions have a pure forward-scattering, 
zero-momentum transfer, character. 

\section{Pseudoparticle perturbation theory}
\label{perturbation}
\pagestyle{myheadings}
\markboth{4  Pseudoparticle Perturbation Theory}{4.2 Pseudoparticle perturbation theory}

We will be mostly interested in GS - GGS transitions 
followed by pseudoparticle - pseudohole excitations
involving a small density of $\alpha,\gamma$ pseudoparticles 
relative to both the GS and GGS distributions.  Let us then 
introduce the normal-ordered pseudomomentum distribution
operator

\begin{equation}
      :\hat{N}_{\alpha,\gamma}(q): =
      \hat{N}_{\alpha,\gamma}(q) - N^{(0)}_{\alpha,\gamma}(q) \, .
\label{normaln}      
\end{equation}
When we choose $N^{(0)}_{\alpha,\gamma}(q)$ to be 
the GS pseudomomentum distribution (\ref{gsdist}) we have

\begin{equation}
      :\hat{N}_{\alpha,\gamma}(q): =
      \hat{N}_{\alpha,\gamma}(q) \, , \hspace{2cm} \gamma > 0 \, .
\end{equation}
The normal-ordered distribution (\ref{normaln})
 obeys the eigenvalue equation

\begin{equation}
      :\hat{N}_{\alpha,\gamma}(q):|\psi\rangle =
      \delta N_{\alpha,\gamma}(q)|\psi\rangle \, .
\label{eigenn}      
\end{equation}
We also consider the normal-ordered rapidity operator

\begin{equation}
      :\hat{R}_{\alpha,\gamma}(q): =
      \hat{R}_{\alpha,\gamma}(q) - R^{(0)}_{\alpha,\gamma}(q) \, ,
\end{equation}
where $R_{\alpha,\gamma}^0(q)$ is the GGS eigenvalue
of Eq. (\ref{ralpha0}).

In the pseudoparticle basis, the normal-ordered rapidity 
operators $:\hat{R}_{\alpha,\gamma}(q):$ contain
an infinite number of terms, as we shall demonstrate below. The first of
these terms is linear in the pseudomomentum distribution
operator $:\hat{N}_{\alpha,\gamma}(q):$, 
(\ref{normaln}), whereas the remaining terms 
consist of products of two, three,.....,infinite, 
of these operators. The number of $:\hat{N}_{\alpha,\gamma}(q):$ 
operators which appears in these products equals the order of the
scattering in the corresponding rapidity term.

A remarkable property is that in the pseudoparticle 
basis the seemingly ``non-perturbative'' quantum liquids 
become perturbative: while the two-electron
forward scattering amplitudes and vertices diverge, the
two-pseudoparticle $f$ functions (given by Eq. (\ref{ffunction}) 
below) are finite. 

By ``perturbative'' we also mean here the 
following: at each point of parameter space (canonical
ensemble), the excited eigenstates are of form (\ref{bethestates}) 
and correspond to quantum-number configurations involving a 
density of excited pseudoholes and heavy pseudoparticles relative to the GS 
configuration (\ref{gsop}) and (\ref{gsdist}). We introduce the following
density 

\begin{equation}
       n_{ex}' = \sum_{\alpha,\beta}n_{ex}^{\alpha,\beta} +
       \sum_{\alpha,\gamma =1}n_{ex}^{\alpha,\gamma} \, ,
\label{nexprime}       
\end{equation}
which is kept small. Here

\begin{equation}
      n_{ex}^{\alpha,\beta} = {1\over {N_a}}\sum_{q}
      [1 - N_{\alpha,\beta}^{h,(0)}(q)]\delta N_{\alpha, \beta}(q) \, , 
\label{nexgama}      
\end{equation}
and

\begin{equation}
      n_{ex}^{\alpha,\gamma} = {1\over {N_a}}\sum_{q}
      N_{\alpha,\gamma}(q) \, ,  
\label{nexalphagamma}      
\end{equation}
define the densities of excited $\alpha,\beta$ pseudoholes
and $\alpha,\gamma$ heavy pseudoparticles, respectively, 
associated with the Hamiltonian eigenstate $|\psi\rangle$.

In Eq. (\ref{nexgama}), $N_{\alpha,\beta}^{h,(0)}(q)$ is the GS pseudohole
distribution which reads

\begin{eqnarray}
      N_{\alpha,-{1\over 2}}^{h,(0)}(q) & = & 1 - N_{\alpha,0}^{(0)}(q)
      \, ,\nonumber \\
      N_{\alpha,+{1\over 2}}^{h,(0)}(q) & = & 0 \, ,
\end{eqnarray}
if the GS is a $S^{\alpha }$ LWS and 

\begin{eqnarray}
      N _{\alpha,-{1\over 2}}^{h,(0)}(q) & = & 0
      \, ,\nonumber \\
      N_{\alpha,+{1\over 2}}^{h,(0)}(q) & = & 1 - N_{\alpha,0}^{(0)}(q) \, ,
\end{eqnarray}
if the GS is a $S^{\alpha }$ HWS, where $N_{\alpha,0}^{(0)}(q)$
is the $\alpha,0$ pseudoparticle GS distribution (\ref{gsdist}).
Our perturbation theory refers to general GS transitions 
to Hamiltonian eigenstates $|\psi\rangle$ involving
(a) a GS - GGS transition and (b) a Landau-liquid 
pseudoparticle-pseudohole excitation around the GGS. Therefore,
the initial state is a GS and we have assumed the GS distribution
(\ref{gsdist}) for the $\alpha,\gamma$ heavy pseudoparticles. Following Eq.
(\ref{gsdist}), we thus have that $N_{\alpha,\gamma}(q) =
\delta N_{\alpha,\gamma}(q)$ in the rhs of Eq. (\ref{nexalphagamma}).  

The expectation values of the SO(4) Hamiltonian 
(\ref{hamiltba}) in the final 
states $|\psi\rangle$ are functions of the density 
of excited pseudoparticles only. This density is given by

\begin{equation}
      n_{ex} = \sum_{\alpha,\gamma =0}n_{ex}^{\alpha,\gamma} \, ,
\label{nextotal}      
\end{equation}
where $n_{ex}^{\alpha,\gamma}$ is given by

\begin{equation}
      n_{ex}^{\alpha,0} = {1\over {N_a}}\sum_{q}
      [1 - N_{\alpha,0}^{(0)}(q)]\delta N_{\alpha,0}(q) \, , 
\label{nexalpha0}      
\end{equation}
for $\gamma =0$ and by (\ref{nexalphagamma}) otherwise. 
The density (\ref{nextotal}) 
is small provided the density (\ref{nexprime}) is also small.
This then implies that all the densities $n_{ex}^{\alpha,\gamma}$ are 
small and that we can expand the expectation values in these densities. 
The perturbative character of the quantum liquid rests on the 
fact that the evaluation of the expectation values up to the
$n^{th}$ order in the densities (\ref{nextotal}) requires considering only the 
corresponding operator terms of scattering orders less than or
equal to $n$. This follows from the linearity of the density of 
excited $\alpha,\gamma$ pseudoparticles, which are the elementary 
``particles'' of the quantum liquid, in 

\begin{equation}
      \delta N_{\alpha,\gamma}(q)=\langle\psi|:\hat{N}_{\alpha,\gamma}(q):
      |\psi\rangle \, ,
\label{devenorm}
\end{equation}
and from the form of (\ref{nexalphagamma}) and (\ref{nexalpha0}). 
The perturbative character 
of the quantum liquid implies, for example, that, to second order 
in the density of excited pseudoparticles, the energy involves only one-
and two-pseudoparticle Hamiltonian terms, 
as in the case of the quasiparticle terms of a Fermi-liquid 
energy functional \cite{0Pines,0Baym}. 
\subsection{Pseudoparticle bands and $f$-functions}

We will develop our perturbation
theory considering the two cases where (i) the
operator normal-ordering is relative to the initial GS
and (ii) that ordering refers to the GGS associated
with a GS - GGS transition. In the following we consider the 
general case of normal ordering relative to a GGS. Later on 
we will specify when that ordering refers either to a GS
(which is a particular case of a GGS) or to a (non-GS)
GGS.

The expressions for the rapidity operators 
$:{\hat{R}}_{\alpha, \gamma}(q):$ corresponds to expanding the
expressions of the operators $:{\hat{R}}_{\alpha, \gamma}(q):$ in terms
of increasing pseudoparticle scattering order. It is convenient 
to define these expressions in terms of the operators 
$:{\hat{X}}_{\alpha, \gamma}(q):$. These are related to the rapidity 
operators as follows

\begin{equation}
      :{\hat{R}}_{\alpha, \gamma}(q): = 
      R_{\alpha, \gamma}^0(q + :{\hat{X}}_{\alpha, \gamma}(q):) 
      - R_{\alpha, \gamma}^0(q) \, ,
\label{rapexp}      
\end{equation}
where $R_{\alpha,\gamma}^{(0)}(q)$ is the suitable  ${\hat{R}}_{\alpha,\gamma}(q)$
GGS eigenvalue of Eq. (\ref{ralpha0}).

The operators $:\hat{X}_{\alpha,\gamma}(q):$ contain the same
information as the rapidity operators, and involve, exclusively, the 
two-pseudoparticle phase shifts $\Phi_{\alpha,\gamma;\alpha '
,\gamma'}(q,q ')$ that we define below. In Appendix \ref{normalorder} 
we introduce
(\ref{rapexp}) in the BA equations (\ref{tak1}-\ref{tak3}) and expand in the scattering order. 
This leads to

\begin{equation}
       :\hat{X}_{\alpha,\gamma}(q): = \sum_{j=1}^{\infty}
       \hat{X}_{\alpha,\gamma}^{(j)}(q) \, ,
\label{rapinfinity}       
\end{equation}
where $j$ gives the scattering order of the operator
term $\hat{X}_{\alpha,\gamma}^{(j)}(q)$. For the
first-order term we find

\begin{equation}
       \hat{X}_{\alpha,\gamma}^{(1)}(q) = {2\pi\over {N_a}}
       \sum_{q',\alpha',\gamma'}
       \Phi_{\alpha,\gamma;\alpha ',\gamma'}(q,q ')
       :\hat{N}_{\alpha',\gamma'}(q'):\, ,
\label{rapodr1}       
\end{equation}
where the phase-shift expressions are given below.
While the expressions for the phase shifts 
$\Phi_{\alpha,\gamma;\alpha ',\gamma'}(q,q ')$ are specific to each model
because they involve the spectral parameters, the {\it form} of 
the operator term $\hat{X}_{\alpha,\gamma}^{(1)}(q)$ (\ref{rapodr1}) is 
universal and refers to all the solvable electronic multicomponent 
quantum liquids. 

The phase shifts $\tilde{\Phi}_{\alpha,\gamma,\alpha '\gamma'}$ are 
given by

\begin{equation}
      \tilde{\Phi}_{c,0;c,0}(k,k') = 
      \bar{\Phi }_{c,0;c,0}\left({\sin k\over u},
      {\sin k'\over u}\right) \, ,
\end{equation}

\begin{equation}
      \tilde{\Phi}_{c,0;\alpha ,\gamma}(k,r') = 
      \bar{\Phi }_{c,0;\alpha ,\gamma}\left({\sin k\over u},
      r'\right) \, ,
\end{equation}

\begin{equation}
       \tilde{\Phi}_{\alpha ,\gamma;c,0}(r,k') = 
       \bar{\Phi }_{\alpha ,\gamma;c,0}\left(r,
       {\sin k'\over u}\right) \, ,
\end{equation}

\begin{equation}
       \tilde{\Phi}_{\alpha,\gamma;\alpha,\gamma'}\left(r,r'\right) = 
       \bar{\Phi }_{\alpha,\gamma;\alpha,\gamma'}\left(r,r'\right) \, ,
\end{equation}
where the phase shifts $\bar{\Phi }_{\alpha,\gamma;\alpha ',\gamma'}$ are
defined by the integral equations (\ref{f1})-(\ref{fs38}) of 
Appendix \ref{normalorder}.

The two-pseudparticle phase shifts can be defined in terms of
the phase shifts $\bar{\Phi }_{\alpha,\gamma;\alpha',\gamma'}$ as 
follows

\begin{equation}
       \Phi_{\alpha,\gamma;\alpha',\gamma'}(q,q') = 
       \bar{\Phi }_{\alpha,\gamma;\alpha',\gamma'}
       \left(R^{(0)}_{\alpha,\gamma}(q),
       R^{(0)}_{\alpha,\gamma}(q')\right) \, .
\label{phaseshift}       
\end{equation}
The quantity $\Phi_{\alpha,\gamma;\alpha',\gamma'}(q,q')$ 
represents the shift in the phase of the $\alpha ',\gamma '$ 
pseudoparticle of pseudomomentum $q'$ due to a zero-momentum 
forward-scattering collision with the $\alpha ,\gamma$ pseudoparticle
of pseudomomentum $q$.

In Appendix \ref{normhamitexp}, 
we use the Hamiltonian expression (\ref{hamiltba}) in terms of 
the rapidity operators to derive the expression 
for the normal-ordered Hamiltonian. We find that
in normal order relative to the suitable GGS 
(or GS), the Hamiltonian ${\hat{H}}_{SO(4)}$, Eq. (\ref{hamiltba}), can 
be written as

\begin{equation}
       :{\hat{H}}_{SO(4)}: = \sum_{j=1}^{\infty}\hat{H}^{(j)} \, . 
\label{hamiltinf}
\end{equation}
For example, the first and second pseudoparticle-scattering order 
terms read
 
\begin{equation}
      \hat{H}^{(1)}=\sum_{q,\alpha,\gamma}
      \epsilon^0_{\alpha ,\gamma}(q):\hat{N}_{\alpha ,\gamma}(q): \, ,
\label{hamitord1}      
\end{equation}
and 

\begin{equation}
      \hat{H}^{(2)}={1\over {N_a}}\sum_{q,\alpha ,\gamma} 
      \sum_{q',\alpha',\gamma'}{1\over 2}f_{\alpha,\gamma;
      \alpha',\gamma'}(q,q') :\hat{N}_{\alpha ,\gamma}(q):
     :\hat{N}_{\alpha' ,\gamma'}(q'): \, , 
\label{hamiltord2}     
\end{equation}
respectively. All the remaining higher-order operator terms of 
expressions (\ref{rapinfinity})  and 
(\ref{hamiltinf}) can be obtained by combining 
the rapidity equations (\ref{tak1}), (\ref{tak2}), and (\ref{tak3}) and the Hamiltonian expression 
(\ref{hamiltba}).

In Appendix \ref{normhamitexp},  it is shown that the pseudoparticle bands 
$\epsilon^0_{\alpha ,\gamma}(q)$ can be expressed
in terms of the phase shifts (\ref{phaseshift}), with the result

\begin{eqnarray}
       \epsilon_{c,0}^0(q) & = & -{U\over 2} -2t\cos K^{(0)}(q) +
       2t\int_{Q^{(-)}}^{Q^{(+)}}dk\widetilde{\Phi }_{c,0;c,0}
       \left(k,K^{(0)}(q)\right)\sin k\, ,
\label{c0qband}
\end{eqnarray}

\begin{eqnarray}
      \epsilon_{c,\gamma}^0(q) & = & -\gamma U 
      + 4t Re \sqrt{1 - u^2[R^{(0)}_{c,\gamma}(q) + i\gamma]^2}\nonumber\\
      &+& 2t\int_{Q^{(-)}}^{Q^{(+)}}dk\widetilde{\Phi }_{c,0;c,\gamma}
      \left(k,R^{(0)}_{c,\gamma}(q)\right)\sin k\, ,
\label{cgqband}
\end{eqnarray}
and

\begin{eqnarray}
      \epsilon_{s,\gamma}^0(q) & = & 2t\int_{Q^{(-)}}^{Q^{(+)}}dk
      \widetilde{\Phi }_{c,0;s,\gamma}
      \left(k,R^{(0)}_{s,\gamma}(q)\right)\sin k\, .
\label{sgqband}
\end{eqnarray}
At the 
pseudo-Brillouin zones
the band expressions have the following  values 

\begin{equation}      
\epsilon_{\alpha,\gamma}^0(q^{(\pm)}_{alpha,\gamma}) = 0 \,, 
\end{equation}
and the associate group velocities, $v_{\alpha ,\gamma}(q)$,
are given by

\begin{equation}
      v_{\alpha ,\gamma}(q) = 
      {d\epsilon^0_{\alpha ,\gamma}(q) \over {dq}} \, . 
\label{velq}      
\end{equation}
The two-pseudoparticle forward-scattering phase shifts 
$\Phi_{\alpha,\gamma;\alpha',\gamma'}(q,q')$ defined by 
Eq. (\ref{phaseshift}) and ``light'' velocities 

\begin{equation}
      v_{\alpha ,\gamma}\equiv v_{\alpha ,\gamma}(q_{F\alpha ,\gamma}) \, ,
\label{velfermi}       
\end{equation}
play an important role in the physical quantities when the
conformal regime is approached \cite{0Carm92b,0Frahm90,0Frahm91}.

All $\hat{X}_{\alpha,\gamma}^{(j)}(q)$ 
terms of the rhs of Eq. (\ref{rapinfinity}) 
are such that both the $f$ functions on the rhs of Eq. (\ref{hamiltord2}) 
and all the remaining higher order coefficients associated with the 
operators $\hat{H}^{(j)}$ of order $j>1$ have universal
forms in terms of the two-pseudoparticle phase shifts and 
pseudomomentum derivatives of the bands and coefficients 
of order $<j$. This follows from the fact that the $S$-matrix for 
$j$-pseudoparticle scattering factorizes into two-pseudoparticle 
scattering matrices, as in the case of the usual BA
$S$-matrix \cite{0Essler92c}. For
example, we show in Appendix 
\ref{normhamitexp}  that although the second-order term 
$\hat{H}^{(2)}$ of Eq. (\ref{hamiltord2}) 
involves an integral over the second-order
function $\hat{X}_{\alpha,\gamma}^{(2)}(q)$ (see Eq. (\ref{energy2ord}) of
Appendix \ref{normhamitexp} ), this function is such that $\hat{H}^{(2)}$
can be written exclusively in terms of the
first-order functions (\ref{rapodr1})
  
\begin{eqnarray}
       \hat{H}^{(2)} &=& \sum_{q,\alpha,\gamma}v_{\alpha,\gamma} (q)
	\hat{X}_{\alpha,\gamma}^{(1)}(q) :\hat{N}_{\alpha,\gamma}(q):
	\nonumber \\
	&+& {N_a\over {2\pi}}\sum_{\alpha,\gamma}\theta (N_{\alpha,\gamma})
	{|v_{\alpha,\gamma}|\over 2}
	\sum_{j=\pm 1}[\hat{X}_{\alpha,\gamma}^{(1)}(jq_{F\alpha,\gamma})]^2 \,,
\label{hamiltx2}       
\end{eqnarray} 
where $|v_{\alpha,\gamma}|=C_{\alpha,\gamma}v_{\alpha,\gamma}$,
$\theta (x)=1$ for $x>0$, and $\theta (x)=0$ for $x\leq 0$
and the ``Landau'' $f$ functions, $f_{\alpha,\gamma;\gamma'
,\alpha'}(q,q')$, are found in that Appendix to have universal 
form

\begin{eqnarray}
      &&f_{\alpha,\gamma;\alpha',\gamma'}(q,q')  =  2\pi 
	v_{\alpha,\gamma}(q)\Phi_{\alpha,\gamma;\alpha',\gamma'}(q,q')+
	2\pi v_{\alpha',\gamma'}(q')
	\Phi_{\alpha',\gamma';\alpha,\gamma}(q',q) \nonumber \\
	& + & 2\pi\sum_{j=\pm 1} 
	\sum_{\alpha''}\sum_{\gamma''=0}^{\infty}
	\theta (N_{\alpha'',\gamma''})C_{\alpha'',\gamma''}v_{\alpha'',\gamma''}
	\Phi_{\alpha'',\gamma'';\alpha,\gamma}(jq_{F\alpha''
	,\gamma''},q)\nonumber \\
	&&\Phi_{\alpha'',\gamma'';\alpha',\gamma'}
	(jq_{F\alpha'',\gamma''},q') \, ,
\label{ffunction}      
\end{eqnarray}
where the pseudoparticle group velocities are given by 
Eqs. (\ref{velq}) and (\ref{velfermi}). 

We can also write the expression of $\hat{H}^{(j)}$ for higher 
scattering orders $j>2$. The main feature here is that for all
sub-canonical ensembles, energy scales, and pseudoparticle 
scattering orders, the Landau-liquid terms of that Hamiltonian 
involve only zero-momentum forward-scattering.
\subsection{Energy-gap equations}

Let us consider that the normal-ordering of the
Hamiltonian (\ref{hamiltinf}) above refers to the initial GS.
In this case, and 
in contrast to the low-energy Landau theory 
\cite{0Carm94b,0Carm92b,0Nuno96,0Carm91,0Carm92c},
this Hamiltonian describes finite-energy transitions
involving a small density of excited pseudoparticles.

Since the relevant finite-energy excitations involve a 
topological GS - GGS transition (to be discussed in Chapter \ref{finitesize}),
it is convenient to define the normal
ordering relative to the suitable final GGS of energy 

\begin{equation}
       \omega_0 = E_{GGS}-E_{GS} \, . 
\label{oomega}       
\end{equation}
In this case the Hamiltonian acts directly on the 
Hilbert subspace of the final GGS where all states have
the same choices concerning the integer or half-odd integer
character of the numbers $I_j^{\alpha ,\gamma}$ of Eq. (\ref{qpseudo}). 

Normal-ordering relative to the GGS implies
separating the problem (\ref{hamiltinf}) into (a) a 
finite-energy term associated with the GS - GGS transitions
and (b) a low $(\omega -\omega_0)$ energy theory associated
with the Landau-liquid pseudoparticle-pseudohole processes
around the final GGS. In most cases the excitation energy
(\ref{oomega}) associated with the GS - GGS transitions is finite. 
The finite energies are the energy gaps of the states 
II and (or) non-LWS's and non-HWS's relative to the 
initial GS.

At finite values of $S^c_z$ and $S^s_z$ the low-energy Hilbert 
space is entirely spanned by states I \cite{0Carm94b,0Nuno96,0Carm94a}.
We note that, for these states, fixing the electron numbers fixes 
the pseudohole numbers \cite{0Nuno96} and the $\alpha,\gamma$ 
bands are empty for $\gamma >0$. Therefore, if we fix the 
electronic 
numbers we have that at energy scales smaller 
than the gaps (\ref{oomega}) for the states II and non-LWS's and non-HWS's 
only Landau-liquid excitations (b) are allowed.
This justifies the Landau-liquid character of the problem at
low energies \cite{0Carm94b,0Carm92b,0Nuno96,0Carm92a,0Carm93b}.
Obviously, if we change the electron numbers, there occur
at low energy GS - GS transitions which are particular
cases of the above GS - GGS transitions (a). Understanding
these GS - GS transitions permits one to express the 
electron operators in terms of the pseudohole operators
and topological momentum 
shifts (Chapter \ref{finitesize}).

Let us evaluate the general $\omega_0$ expression for GS - GGS 
transitions such that the final sub-canonical ensemble 
is characterized by vanishing or by small values of 
$N^h_{\alpha ,{-{l_{\alpha}\over 2}}}/N^h_{\alpha }$. Here, 
$l_{\alpha }=\pm 1$ is defined by Eq. (\ref{sector}). 
We also assume that both the initial GS and final GGS 
correspond to canonical ensembles such that $S^{\alpha}_z\neq 0$
and belonging the same sector of parameter space, i.e. with the
same $l_{\alpha }$ numbers. We note that since GS's are
always $S^{\alpha}$ LWS's I or HWS's I there are neither 
$\alpha ,{-{l_{\alpha}\over 2}}$ pseudoholes [see Eqs. (\ref{nh120}) 
and (\ref{nh210})] nor $\alpha,\gamma$ heavy pseudoparticles ($\gamma >0)$ 
in the initial GS, and that $N^z_{\alpha }=0$ [see Eq. (\ref{nza})]
for that state. From Eqs. 
(\ref{hamiltinf})-(\ref{hamiltord2}), we find for the GS - GGS gap

\begin{equation}
      \omega_0 = \sum_{\alpha } \epsilon^0_{\alpha ,0}(q_{F\alpha,0})
      \Delta N_{\alpha ,0} + \sum_{\gamma =1}
      \epsilon^0_{s,\gamma }(0)N_{s ,\gamma} 
      +2\mu \Delta S_z^c+2\mu_0 H \Delta S_z^s\,.
\label{ggap}      
\end{equation}
Inserting in Eqs. 
(\ref{holebeta}) both the pseudohole expressions (\ref{nholeslwsii}) and 
the expressions $S^c_z=-{1\over 2}[N_a-N]$ and
$S^s_z=-{1\over 2}[N_{\uparrow}-N_{\downarrow}]$, we find,
after use of Eq. (\ref{nza}), the following
general expressions for $\Delta N_{c,0}$ and $\Delta N_{s,0}$

\begin{eqnarray}
      \Delta N_{c,0} &=& - 2l_c\Delta S^c_z - 
      2N^z_c - \sum_{\gamma =1}2\gamma 
      N_{c,\gamma } = \nonumber\\
      &-&[N^h_{c ,{-{l_{c}\over 2}}}
      - \Delta N^h_{c ,{l_{c}\over 2}}] 
      - 2N^z_c - \sum_{\gamma =1}2\gamma 
N_{c,\gamma }  \, ,
\label{delc0}
\end{eqnarray}
and

\begin{eqnarray}
       \Delta N_{s,0}  =  &-& \sum_{\alpha }
       [l_{\alpha}\Delta S^{\alpha}_z + N^z_{\alpha}]
       - \sum_{\gamma =1}\gamma N_{c,\gamma }  
       - \sum_{\gamma =1}(1+\gamma) N_{s,\gamma } \nonumber \\
        = &-& {1\over 2}\sum_{\alpha }[N^h_{\alpha ,{-{l_{\alpha}\over 2}}}
       - \Delta N^h_{\alpha ,{l_{\alpha}\over 2}} 
       + 2N^z_{\alpha}] \nonumber\\
       &-& \sum_{\gamma =1}\gamma N_{c,\gamma } 
       - \sum_{\gamma =1}(1+\gamma) N_{s,\gamma } \, .
\label{delso}       
\end{eqnarray}
Using these and Eq. (\ref{nza}) in the rhs of 
Eq. (\ref{ggap}) leads to the gap 
expression (\ref{bgap1}) of Appendix \ref{normhamitexp}.

On the other hand, generalization of the results contained in the work
\cite{0Carm91}
for the chemical-potential and 
magnetic-field expressions to 
all sectors of parameter space leads to

\begin{equation}
      |\mu | = - \epsilon^0_{c,0}(q_{Fc,0})
      - {\epsilon^0_{s,0}(q_{Fs,0})\over 2}  
      \, , \hspace{1cm}
      2\mu_0 |H| = - \epsilon^0_{s,0}(q_{Fs,0}) \, .
\end{equation}

From the use of Eqs. (\ref{delc0}) and (\ref{delso}) in the gap equation 
(\ref{ggap}), we can rewrite Eq. (\ref{bgap1}) as follows

\begin{equation}
       \omega_0 = 2|\mu|N^h_{c,-{l_c\over 2}} 
       + 2\mu_0|H|[N^h_{s,-{l_s\over 2}}
       + \sum_{\gamma =1}^{\infty}N_{s,\gamma}] 
       + \sum_{\gamma =1}\epsilon^0_{s,\gamma }(0)
       N_{s,\gamma} \, ,
\label{andgap}
\end{equation}
which is the general expression for the gap for GS - GGS
transitions. It follows from Eqs. (\ref{nza}) and
(\ref{offdiagonal}) that the
gap expression (\ref{andgap}) can be rewritten as

\begin{equation}
      \omega_0 = 2|\mu|\Bigl(\sum_{\gamma =1}\gamma  
      N_{c,\gamma} + N^z_c\Bigl)
      + 2\mu_0|H|\Bigl(\sum_{\gamma =1}(1+\gamma ) 
       N_{s,\gamma} + N^z_s\Bigl)
      + \sum_{\gamma =1}\epsilon^0_{s,\gamma }(0)
      N_{s,\gamma} \, .
\label{finalgap}      
\end{equation}

For GS - GGS transitions to pure LWS's or HWS's 
II, the use of Eq. (\ref{nza}) leads to

\begin{equation}
      N^z_{\alpha } = 0 \, ,
\end{equation}
for both $\alpha =c,s$. It follows that for such transitions
the gap is given by expression (\ref{finalgap}) with $N^z_c = 0$ and 
$N^z_s = 0$. Therefore, the single ($\gamma >0$) $\alpha,\gamma $ 
heavy-pseudoparticle gap is 
$2\gamma |\mu|$ and 
$2(1+\gamma)\mu_0|H|+\epsilon^0_{s,\gamma }(0)$ 
for $c$ and $s$, respectively.

Given a final sub-canonical ensemble we can write 

\begin{equation}
      :{\hat{H}}: = {\hat{H}}_{0} + {\hat{H}}_{Landau} \, , 
\label{ham1}
\end{equation}
where ${\hat{H}}_{0}$ has eigenvalue $\omega_0$ 
and corresponds to the GS - GGS transition (a) and 
${\hat{H}}_{Landau}$ is normal-ordered relative
to the GGS and is of the form

\begin{equation}
      {\hat{H}}_{Landau} = \hat{H}^{(1)}_L + \hat{H}^{(2)}_L \, , 
\label{ham2}
\end{equation}

where

\begin{equation}
       \hat{H}^{(1)}_L=\sum_{q,\alpha,\gamma}
       \epsilon_{\alpha ,\gamma}(q):\hat{N}_{\alpha ,\gamma}(q): \, ,
\label{ham3}       
\end{equation}

\begin{equation}
       \epsilon_{\alpha ,0}(q) = \epsilon^{(0)}_{\alpha ,0}(q) 
       - \epsilon^{(0)}_{\alpha ,0}(q_{F\alpha,0}) 
       \, , \hspace{1cm}
       \epsilon_{\alpha ,\gamma}(q) = \epsilon^{(0)}_{\alpha ,\gamma}(q) 
       - \delta_{\alpha,s}\epsilon^{(0)}_{\alpha ,\gamma}(0) \, ,
\end{equation}
and $\hat{H}^{(2)}$ is given in Eq. (\ref{hamiltord2}). 
The Hamiltonian (\ref{ham2})
describes the Landau-liquid excitations (b) relative to the GGS. 
Here (\ref{ham2}) are the Landau-liquid Hamiltonian terms 
which are relevant at low energy $(\omega-\omega_0)$. 
Therefore, the corresponding second-order Hamiltonian is 
suitable to study the physics at low positive energy above 
the gap, {\it i.e.}, small $(\omega -\omega_0)$. 

In general, the different final GGS's of the Hilbert subspace 
where a given initial GS is transformed upon excitations
involving a small density of pseudoparticles have different 
energies (\ref{andgap})-(\ref{finalgap}). 
This implies that the study of the 
quantum-liquid physics at energy-scale $\omega_0$ involves, in 
general, transitions to one sub-cannonical ensemble only. However, 
if two or several possible final GGS's had the same energy gap $\omega_0$,
the physics would involve the Hilbert subspace spanned by all
Hamiltonian eigenstates associated with the corresponding 
different sub-canonical ensembles. 

For sub-canonical ensembles such that $S^{\alpha}_z\neq 0$ and 
characterized by small values of $N^h_{\alpha ,{-{l_{\alpha}\over 
2}}}/N^h_{\alpha }$ (where $l_{\alpha }$ is given in Eq. (\ref{sector})), 
the gap expression (\ref{andgap}) leads directly to 

\begin{equation}
{\hat{H}}_{0} = 2|\mu|{\hat{N}}^h_{c,-{l_c\over 2}}+ 
2\mu_0|H|[{\hat{N}}^h_{s,-{l_s\over 2}}
+\sum_{\gamma =1}^{\infty}{\hat{N}}_{s,\gamma}] 
+ \sum_{\gamma =1}\epsilon^{(0)}_{s,\gamma }(0)
{\hat{N}}_{s,\gamma} \, .
\end{equation}

There is a remarkable similarity between the general Hamiltonian 
(\ref{ham1}) and its version in the Hilbert subspace spanned by the states I 
\cite{0Carm94b,0Nuno96,0Carm94a}. Moreover, in the case that 
the GGS is the GS it self, the gap (\ref{andgap})-(\ref{finalgap}) 
vanishes and the Hamiltonian 
(\ref{ham1}) reduces to the above low-energy Hamiltonian. In this case it 
refers to Landau-liquid $\alpha ,0$ pseudoparticle-pseudohole 
excitations around the GS.
\begin{appendix}
\input{apnoro}

\input{apnohe}

\end{appendix}


%% file: apnoro.tex
\chapter{Normal-ordered rapidity operator expressions}
\label{normalorder}
\pagestyle{myheadings}
\markboth{ Normal-ordered rapidity ...}{Appendix A of Chapter 4}

Following the discussion of Sec. \ref{perturbation}, 
the perturbative character of 
the system implies the equivalence between expanding in the 
pseudoparticle scattering order and/or in the pseudomomentum 
deviations (\ref{devenorm}). In this Appendix we give a short 
description of the calculation of the normal-ordered operator 
expansion for the pseudoparticle rapidities (\ref{rapinfinity}). 
We focus our 
study on the evaluation of the functions $X_{\alpha,\gamma}^{(1)}(q)$ 
and $X_{\alpha,\gamma}^{(2)}(q)$ and associate functions
$Q_{\alpha,\gamma}^{(1)}(q)$ and $Q_{\alpha,\gamma}^{(2)}(q)$ 
(eigenvalues of the operators $\hat{X}_{\alpha}^{(1)}(q)$ and 
$\hat{X}_{\alpha}^{(2)}(q)$ and $\hat{Q}_{\alpha}^{(1)}(q)$ and 
$\hat{Q}_{\alpha}^{(2)}(q)$, respectively).

We evaluate here the first-order and second-order
terms of the eigenvalues of the operator (\ref{rapinfinity}). 
Equation (\ref{eigenn}) then 
allows the straightforward calculation of the corresponding 
operator expressions. All the quantities are given in the
limit established by Eq. (\ref{nexprime}).

In the thermodynamic limit, Eqs. (\ref{tak1})-(\ref{tak3}) lead to
the following equations 
 
\begin{equation}
       K(q) = q - {1\over \pi}
       \int_{q^{(-)}_{s,0}}^{q^{(+)}_{s,0}}dq'
       N_{s,0}(q')\tan^{-1}(R_{c,0}(q)-R_{s,0}(q'))
\label{integral1}
\end{equation}

\begin{eqnarray}
      q
      & = & 2 Re \sin^{-1}\Bigl(u(R_{c,\gamma}(q)+i\gamma)\Bigr) \nonumber\\
      &-& {1\over\pi}\int_{q^{(-)}_{c,0}}^{q^{(+)}_{c,0}}dq'
      N_{c,0}(q')\tan^{-1}\Bigl({R_{c,\gamma}(q)-R_{c,0}(q')\over 
      \gamma}\Bigr)
\label{integral2}
\end{eqnarray}
and 

\begin{eqnarray}
       0 & = & q - {1\over\pi}
       \int_{q^{(-)}_{c,0}}^{q^{(+)}_{c,0}}dq'
       N_{c,0}(q')\tan^{-1}\Bigl({R_{s,\gamma}(q)-R_{c,0}(q')
       \over 1+\gamma}\Bigr)\nonumber \\
       & + & {1\over 2\pi}
\int_{q^{(-)}_{s,0}}^{q^{(+)}_{s,0}}dq'
N_{s,0}(q')\Theta_{1+\gamma, 1}
\Bigl(R_{s,\gamma}(q)-R_{s,0}(q')\Bigr) \, .
\label{integral3}
\end{eqnarray}

In what follows, it is useful to define the GGS rapidities $K^{(0)}(q)$ and 
$R_{\alpha,\gamma}^{(0)}(q)$ in terms of their inverse functions 
$2\pi\rho_{\alpha,\gamma}(r)$ as 

\begin{equation}
      q = \int_{0}^{K^{(0)}(q)}
      dk 2\pi\rho_{c,0}(k) \, , 
\label{intrhoc}      
\end{equation}

\begin{equation}
      q = \int_{0}^{R^{(0)}_{\alpha,\gamma}(q)}
      dr 2\pi\rho_{\alpha,\gamma}(r) \, . 
\label{intrhos}      
\end{equation}

We start by considering the GGS eigenstate rapidities of 
Eqs. (\ref{kalpha0}) and (\ref{ralpha0}). 
If we insert in Eqs. (\ref{tak1})-(\ref{tak3}) the 
GGS distributions (\ref{dist1}), 
after some algebra we find that the 
functions of the rhs of Eqs. (\ref{intrhoc})
and (\ref{intrhos}) defining the inverse
of these GGS rapidities are solutions of the 
following integral 
equations

\begin{equation}
      2\pi\rho_{c,0}(k)  =  1
      + {\cos k\over u} 
      \int_{r^{(-)}_{s,0}}^{r^{(+)}_{s,0}}dr 
      {2\pi\rho_{s,0}(r)\over \pi
      \left[1+(r - {\sin k\over u})^2\right]}
	\, ,
\label{rhoco}
\end{equation}

\begin{equation}
      2\pi\rho_{c,\gamma}(r)  =  2 Re \Bigl({u\over
      \sqrt{1 - u^2[r + i\gamma]^2}}\Bigl)
      - \int_{Q^{(-)}}^{Q^{(+)}}dk {2\pi\rho_{c,0}(k)\over
     \pi\gamma\left[1+({{\sin k\over u} 
     -r\over\gamma})^2\right]}\, ,
\label{rhoc}
\end{equation}

and

\begin{equation}
      2\pi\rho_{s,\gamma}(r)  =  
      \int_{Q^{(-)}}^{Q^{(+)}}dk {2\pi\rho_{c,0}(k)\over
      \pi(1 +\gamma)\left[1+({{\sin k\over u} 
      -r\over 1+\gamma})^2\right]} 
-\int_{r_{s,0}^{(-)}}^{r_{s,0}^{(+)}}\frac{dr\,'}{2\pi}
\Theta^{[1]}_{1,1+\gamma}(r-r\,')2\pi\rho_{s,0}(r\,')\, ,
\label{rhos}
\end{equation}
where

\begin{eqnarray}
\Theta^{[1]}_{\gamma,\gamma'}\Bigl(x\Bigl) & = &
\Theta^{[1]}_{\gamma',\gamma}\Bigl(x\Bigl) =
{d\Theta_{\gamma,\gamma'}\Bigl(x\Bigl)\over dx} \nonumber \\
& = & \delta_{\gamma ,\gamma'}\{{1\over \gamma[1+({x\over 
2\gamma})^2]}+ \sum_{l=1}^{\gamma -1}{2\over l[1+({x\over 2l})^2]}\}
\nonumber \\ 
& + & (1-\delta_{\gamma ,\gamma'})\{
{2\over |\gamma-\gamma'|[1+({x\over |\gamma-\gamma'|})^2]}
+ {2\over (\gamma+\gamma')[1+({x\over \gamma+\gamma'})^2]}\nonumber \\
& + & \sum_{l=1}^{{\gamma+\gamma'-|\gamma-\gamma'|\over 2}
-1}{4\over (|\gamma-\gamma'|+2l)[1+({x\over |\gamma-\gamma'|+2l})^2]}\} \, ,
\label{derivativeteta}
\end{eqnarray}
is the derivative of the function (\ref{ftheta1})
 and the parameters 
$Q^{(\pm)}$ and $r^{(\pm)}_{\alpha,\gamma }$ are defined 
combining Eqs. (\ref{intrhoc}),
(\ref{intrhos}), and (\ref{raplimit}).

Let us now consider small deviations from a GGS or GS.
The eigenvalue form of Eq. (\ref{rapexp}) is  

\begin{equation}
      \delta K(q) = K^{(0)}(q + 
      \delta X_{c,0}(q)) - K^{(0)}(q) \, , 
\label{delkq}
\end{equation}
for $\alpha =c$ and $\gamma =0$ and

\begin{equation}
      \delta R_{\alpha,\gamma}(q) = R_{\alpha,\gamma}^0(q + 
      \delta X_{\alpha,\gamma}(q)) - R_{\alpha,\gamma}^0(q) \, ,
\label{delrq}
\end{equation}
for all remaining values of the quantum numbers $\alpha $
and $\gamma $. Here $\delta K(q)$, $\delta R_{\alpha,\gamma}(q)$, and 
$\delta X_{\alpha,\gamma}(q)$ are the eigenvalues of the operators 
$:\hat{K}(q):$, $:\hat{R}_{\alpha,\gamma}(q):$, and 
$:\hat{X}_{\alpha,\gamma}(q):$, respectively. From Eq. (\ref{rapinfinity})
$\delta X_{\alpha,\gamma}(q)$ can be written as

\begin{equation}
      \delta X_{\alpha,\gamma}(q) = X_{\alpha,\gamma}^{(1)}(q) + 
      X_{\alpha,\gamma}^{(2)}(q)
      + ... \, , 
\label{delxq}
\end{equation}
where $X_{\alpha,\gamma}^{(i)}(q)$ is the eigenvalue of the operator 
$\hat{X}_{\alpha,\gamma}^{(i)}(q)$. Expanding the 
$\delta R_{\alpha,\gamma}(q)$ expressions 
(\ref{delkq}) and (\ref{delrq}) we find

\begin{equation}
     K(q) = \sum_{i=0}^{\infty} K^{(i)}(q) \, ,
\label{kqinf}
\end{equation}
and

\begin{equation}
      R_{\alpha,\gamma}(q) = \sum_{i=0}^{\infty} 
      R_{\alpha,\gamma}^{(i)}(q) \, ,
\label{rqinf}
\end{equation}
respectively, [and $\delta K(q) = \sum_{i=1}^{\infty} K^{(i)}(q)$ and
$\delta R_{\alpha,\gamma}(q)=\sum_{i=1}^{\infty}
R_{\alpha,\gamma}^{(i)}(q)$] where the zero-order GGS
functions $K^{(0)}(q)$ and $R_{\alpha,\gamma}^{(0)}(q)$ are 
defined by Eqs. Eqs. (\ref{intrhoc})
and (\ref{intrhos}). From the resulting equations we can
obtain all derivatives of the GGS functions $K^{(0)}(q)$ and
$R_{\alpha,\gamma}^{(0)}(q)$ with respect to $q$. The terms of the 
rhs of Eqs. (\ref{kqinf}) and (\ref{rqinf}) 
involve these derivatives. For instance,
the first-order and second-order terms read

\begin{equation}
     K^{(1)}(q) =  {d K^{(0)}(q)\over {dq}} 
     X_{c,0}^{(1)}(q) \, ,
\label{k1q}
\end{equation}

\begin{equation}
      R_{\alpha,\gamma}^{(1)}(q) =  {d R_{\alpha,\gamma}^{(0)}(q)\over {dq}} 
      X_{\alpha,\gamma}^{(1)}(q) \, ,
\label{r1q}
\end{equation}
and

\begin{equation}
      K^{(2)}(q) =  {d K^{(0)}(q)\over {dq}} 
      X_{c,0}^{(2)}(q) + 
      {1\over 2}{d^2 K^{(0)}(q)\over 
      {dq^2}}[X_{c,0}^{(1)}(q)]^2 \, ,
\label{k2q}
\end{equation}

\begin{equation}
       R_{\alpha,\gamma}^{(2)}(q) =  {d R_{\alpha,\gamma}^{(0)}(q)\over {dq}} 
       X_{\alpha,\gamma}^{(2)}(q) + 
       {1\over 2}{d^2 R_{\alpha,\gamma}^{(0)}(q)\over 
       {dq^2}}[X_{\alpha,\gamma}^{(1)}(q)]^2 \, ,
\label{r2q}
\end{equation}
respectively, and involve the first and second derivatives.

From Eqs. (\ref{integral1})-(\ref{integral3}) 
[with $N_{\alpha,\gamma}(q')$ given
by the GGS distribution $(75)$] we find that
the first derivatives $d K_{\alpha}^{(0)}(q)\over 
{dq}$ and $d R_{\alpha,\gamma}^{(0)}(q)\over 
{dq}$ can be expressed in terms of the functions 
(\ref{rhoco})-(\ref{rhos}) as follows 

\begin{equation}
       {d K^{(0)}(q)\over {dq}} ={1\over 2\pi\rho_{c,0}(K^{(0)}(q))}
       \, ,
\label{dkq0}
\end{equation}
and

\begin{equation}
      {d R^{(0)}_{\alpha,\gamma}(q)\over {dq}} =
      {1\over 2\pi\rho_{c,\gamma}(R^{(0)}_{\alpha,\gamma}(q))}
      \, .
\label{drq0}
\end{equation}
The second derivatives $d^2 K^{(0)}(q)\over {dq^2}$ and
$d^2 R_{\alpha,\gamma}^{(0)}(q)\over {dq^2}$ then read

\begin{equation}
       {d^2 K^{(0)}(q)\over {dq^2}} =
       - {1\over [2\pi\rho_{c,0}(K^{(0)}(q))]^3}
       2\pi{d \rho_{c,0}(k) \over {dk}}|_{k=K^{(0)}(q)} \, ,
\label{d2kq0}
\end{equation}
and

\begin{equation}
      {d^2 R_{\alpha,\gamma}^{(0)}(q)\over {dq^2}} 
      =-{1\over [2\pi\rho_{c,\gamma}(R^{(0)}_{\alpha,\gamma}(q))]^3}
      2\pi{d \rho_{\alpha,\gamma}(r) \over 
      {dr}}|_{r=R_{\alpha,\gamma}^{(0)}(q)} \, .
\label{d2rq0}
\end{equation}

By introducing both the distributions

\begin{equation}
     N_{\alpha,\gamma}(q) = N_{\alpha,\gamma}^{(0)}(q) + 
     \delta N_{\alpha,\gamma}(q) \, ,
\label{distribalga}
\end{equation} 
and the first-order and second-order functions 
(\ref{k1q})-(\ref{d2rq0}) 
into Eqs. (\ref{integral1})-(\ref{integral3}), 
we find after expanding to second order that for
$j=1$ and $j=2$ the functions $X^{(j)}_{\alpha,\gamma}(q)$ can 
be written as 

\begin{equation}
      X_{\alpha,\gamma}^{(j)}(q) = Q_{\alpha,\gamma}^{(j)}(q) + 
      Y_{\alpha,\gamma}^{(j)}(q) \, ,
\label{xopq}
\end{equation}
where $Y_{\alpha,\gamma}^{(j)}(q)$ is even in $q$

\begin{equation}
      Y_{\alpha,\gamma}^{(j)}(q) = Y_{\alpha,\gamma}^{(j)}(-q) \, ,
\label{yopq}
\end{equation}
and does not contribute to the physical quantities to second
scattering order (and to second order in the density of excited
pseudoparticles). Up to this order only 
$\hat{Q}_{\alpha,\gamma}^{(j)}(q)$ contributes. 

To derive this result we have expanded the expression
for the even functions $Y^{(j)}_{\alpha,\gamma}(q)$
for $j=1$ and $j=2$ to second order in the density of excited 
pseudoparticles. Following the perturbative character of the quantum
liquid in the pseudoparticle basis, the obtained expression 
is exact up to $j=2$ pseudoparticle scattering order.
Moreover, we find that the even function $Y^{(1)}_{\alpha,\gamma}(q)$ 
does not contribute to the physical quantities up
to that order and can, therefore, be omitted.
 
Let us introduce the functions $\bar{Q}_{\alpha,\gamma}^{(1)}(r)$
such that

\begin{equation}
      Q_{\alpha,\gamma}^{(1)}(q) = 
      \bar{Q}_{\alpha,\gamma}^{(1)}(R^{(0)}_{\alpha,\gamma}(q)) \, ,
\label{qbarq}
\end{equation}
for all values of $\alpha$ and $\gamma$, with
$R^{(0)}_{c,0}(q) = {\sin K^{(0)}(q)\over u}$.
It is also useful to define the function
$\widetilde{Q}^{(1)}(k)$ such that

\begin{equation}
      \widetilde{Q}^{(1)}(k) =
      {\bar{Q}}_{c,0}^{(1)}({\sin k\over u}) \, .
\label{qtilq}
\end{equation}

By introducing both the distributions (\ref{distribalga})
and the 
first-order functions defined in Eq. (\ref{qbarq}) into 
Eqs. (\ref{integral1})-(\ref{integral3}), we find after expanding to first order that
the functions $\bar{Q}_{\alpha,\gamma}^{(1)}(r)$ are 
defined by the following system of coupled integral
equations

\begin{eqnarray}
      {\bar{Q}}_{c,0}^{(1)}(r) & = &
      - \sum_{\gamma =0}\int_{q_{s,\gamma}^{(-)}}^{q_{s,\gamma}^{(+)}}dq 
      \delta N_{s,\gamma}(q){1\over \pi}
      \tan^{-1}\Bigl({r - R_{s,\gamma}^{(0)}(q)\over 1+\gamma}\Bigl)
\nonumber \\ 
      & - & \sum_{\gamma =1}\int_{q_{c,\gamma}^{(-)}}^{q_{c,\gamma}^{(+)}}dq 
      \delta N_{c,\gamma}(q){1\over \pi}
      \tan^{-1}\Bigl({r - R_{c,\gamma}^{(0)}(q)\over \gamma}\Bigl)
\nonumber\\      
      &+& \int_{r^{(-)}_{s,0}}^{r^{(+)}_{s,0}}
      dr'{\bar{Q}_{s,0}^{(1)}(r')\over 
      \pi\left[1 + (r -r')^2\right]} \, ,
\label{qbarc0r}
\end{eqnarray}

\begin{eqnarray}
      {\bar{Q}}_{c,\gamma}^{(1)}(r) & = &
      \int_{q_{c,0}^{(-)}}^{q_{c,0}^{(+)}}dq
      \delta N_{c,0}(q){1\over \pi}
      \tan^{-1}\Bigl({r - R_{c,0}^{(0)}(q)\over \gamma}\Bigl)
\nonumber \\
      & + & \sum_{\gamma' =1}\int_{q_{c,\gamma'}^{(-)}}^{q_{c,\gamma'}^{(+)}}dq 
      \delta N_{c,\gamma'}(q){1\over 2\pi}\Theta_{\gamma,\gamma'}
      \Bigl(r - R_{c,\gamma'}^{(0)}(q)\Bigl)
\nonumber\\      
&-& \int_{r^{(-)}_{c,0}}^{r^{(+)}_{c,0}}
      dr'{\bar{Q}_{c,0}^{(1)}(r')\over 
      \pi\gamma\left[1 + ({r -r'\over\gamma })^2\right]} \, ,
\label{qbarcgr}
\end{eqnarray}

and

\begin{eqnarray}
       {\bar{Q}}_{s,\gamma}^{(1)}(r) & = &
       - \int_{q_{c,0}^{(-)}}^{q_{c,0}^{(+)}}dq
       \delta N_{c,0}(q){1\over \pi}
       \tan^{-1}\Bigl({r - R_{c,0}^{(0)}(q)\over 1+\gamma}\Bigl)
\nonumber \\
       & + & \sum_{\gamma' =0}\int_{q_{s,\gamma'}
       ^{(-)}}^{q_{s,\gamma'}^{(+)}}dq 
       \delta N_{s,\gamma'}(q){1\over 2\pi}\Theta_{1+\gamma,1+\gamma'}
       \Bigl(r - R_{s,\gamma'}^{(0)}(q)\Bigl)
\nonumber \\       
       &+& \int_{r^{(-)}_{c,0}}^{r^{(+)}_{c,0}}
       dr'{\bar{Q}_{c,0}^{(1)}(r')\over 
       \pi(1+\gamma)\left[1 + ({r -r'\over 1+\gamma })^2\right]} 
\nonumber \\
       & - & \int_{r^{(-)}_{s,0}}^{r^{(+)}_{s,0}}
       dr'\bar{Q}_{s,0}^{(1)}(r'){1\over 
       2\pi}\Theta^{[1]}_{1+\gamma,1}
       \Bigl(r - r'\Bigl) \, .
\label{qbarsgr}
\end{eqnarray}

(The use of Eq. (\ref{qbarc0r}) in Eqs. 
(\ref{qbarcgr}) and (\ref{qbarsgr}) allows the 
expression of both ${\bar{Q}}_{c,\gamma}^{(1)}(r)$
and ${\bar{Q}}_{s,\gamma}^{(1)}(r)$ in terms of
free terms and integrals involving $\bar{Q}_{s,0}^{(1)}(r)$.)
Combining Eqs. (\ref{qbarc0r})-(\ref{qbarsgr})
with Eqs. (\ref{qbarq}) and (\ref{qtilq}) leads to 
Eq. (\ref{rapodr1}) with the phase shifts defined below. Note that at 
first order we can either consider the function 
$\hat{X}_{\alpha,\gamma}^{(1)}(q)$ or the associate function 
$\hat{Q}_{\alpha,\gamma}^{(1)}(q)$ of Eq. (\ref{xopq}). 
Both functions are of the form (\ref{rapodr1}). However,
the general expression for the phase shifts associated
with the function $\hat{X}_{\alpha,\gamma}^{(1)}(q)$
of Eq. (\ref{rapodr1}) have extra terms. These arise from the
function $\hat{Y}_{\alpha,\gamma}^{(1)}(q)$ of the rhs
of Eq. (\ref{xopq}). If we expand the physical quantities involving 
the phase shifts to first (and second) order in the density 
of excited pseudoparticles these extra terms lead to vanishing 
contributions. For instance, expression (\ref{rapodr1}) is identical if 
we use in it either choice for the phase shift expression.
For simplicity, we omit here the phase-shift extra terms 
associated with the function $\hat{Y}_{\alpha,\gamma}^{(1)}(q)$.
We find that the phase shifts associated with the function
$\hat{Q}_{\alpha,\gamma}^{(1)}(q)$ are defined by the following 
coupled integral equations

\begin{equation}
\bar{\Phi }_{c,0;c,0}\left(r,r'\right) = 
{1\over{\pi}}\int_{r_{s,0}^{(-)}}^{r_{s,0}^{(+)}}
dr''{\bar{\Phi }_{s,0;c,0}\left(r'',r'\right) 
\over {1+(r-r'')^2}} \, ,
\label{f1}
\end{equation}

\begin{equation}
\bar{\Phi }_{c,0;c,\gamma}\left(r,r'\right) = 
-{1\over{\pi}}\tan^{-1}({r-r'\over \gamma}) +
{1\over{\pi}}\int_{r_{s,0}^{(-)}}^{r_{s,0}^{(+)}}
dr''{\bar{\Phi }_{s,0;c,\gamma}\left(r'',r'\right) 
\over {1+(r-r'')^2}} \, ,
\end{equation}

\begin{equation}
\bar{\Phi }_{c,0;s,\gamma}\left(r,r'\right) = 
-{1\over{\pi}}\tan^{-1}({r-r'\over 1+\gamma}) + {1\over{\pi}}
\int_{r_{s,0}^{(-)}}^{r_{s,0}^{(+)}}
dr''{\bar{\Phi }_{s,0;s,\gamma}\left(r'',r'\right) 
\over {1+(r-r'')^2}} \, ,
\end{equation}

\begin{equation}
{\bar{\Phi }}_{c,\gamma;c,0}\left(r,r'\right) = 
{1\over{\pi}}\tan^{-1}({r-r'\over {\gamma}}) -
{1\over{\pi}}\int_{r_{c,0}^{(-)}}^{r_{c,0}^{(+)}}
dr''{{\bar{\Phi }}_{c,0;c,0}\left(r'',r'\right) 
\over {\gamma[1+({r-r''\over {\gamma}})^2]}} \, ,
\end{equation}

\begin{equation}
\bar{\Phi }_{c,\gamma;c,\gamma'}\left(r,r'\right) = 
{1\over{2\pi}}\Theta_{\gamma,\gamma'}\Bigl(r-r'\Bigr)
- {1\over{\pi}}\int_{r_{c,0}^{(-)}}^{r_{c,0}^{(+)}}
dr''{\bar{\Phi }_{c,0;c,\gamma'}\left(r'',r'\right) 
\over {\gamma[1+({r-r''\over\gamma})^2]}} \, ,
\end{equation}

\begin{equation}
\bar{\Phi }_{c,\gamma;s,\gamma'}\left(r,r'\right) = 
- {1\over{\pi}}\int_{r_{c,0}^{(-)}}^{r_{c,0}^{(+)}}
dr''{\bar{\Phi }_{c,0;s,\gamma'}\left(r'',r'\right) 
\over {\gamma[1+({r-r''\over\gamma})^2]}} \, ,
\end{equation}

\begin{eqnarray}
{\bar{\Phi }}_{s,\gamma;c,0}\left(r,r'\right) & = &
-{1\over{\pi}}\tan^{-1}({r-r'\over {1+\gamma}}) + 
{1\over{\pi}}\int_{r_{c,0}^{(-)}}^{r_{c,0}^{(+)}}
dr''{{\bar{\Phi }}_{c,0;c;0}\left(r'',r'\right) 
\over {(1+\gamma)[1+({r-r''\over {1+\gamma}})^2]}} 
\nonumber \\
& - & \int_{r_{s,0}^{(-)}}^{r_{s,0}^{(+)}}
dr''{\bar{\Phi }}_{s,0;c,0}\left(r'',r'\right)
{\Theta^{[1]}_{1+\gamma,1}\Bigl(r-r''\Bigr)\over{2\pi}} \, ,
\label{fs36}
\end{eqnarray}

\begin{eqnarray}
{\bar{\Phi }}_{s,\gamma;c,\gamma'}\left(r,r'\right) & = &
{1\over{\pi}}\int_{r_{c,0}^{(-)}}^{r_{c,0}^{(+)}}
dr''{{\bar{\Phi }}_{c,0;c;\gamma'}\left(r'',r'\right) 
\over {(1+\gamma)[1+({r-r''\over {1+\gamma}})^2]}} 
\nonumber \\
& - & \int_{r_{s,0}^{(-)}}^{r_{s,0}^{(+)}}
dr''{\bar{\Phi }}_{s,0;c,\gamma'}\left(r'',r'\right)
{\Theta^{[1]}_{1+\gamma,1}\Bigl(r-r''\Bigr)\over {2\pi}} \, ,
\end{eqnarray}

\begin{eqnarray}
{\bar{\Phi }}_{s,\gamma;s,\gamma'}\left(r,r'\right) & = &
{\Theta_{1+\gamma,1+\gamma'}\Bigl(r-r'\Bigr)\over{2\pi}} +
{1\over{\pi}}\int_{r_{c,0}^{(-)}}^{r_{c,0}^{(+)}}
dr''{{\bar{\Phi }}_{c,0;s;\gamma'}\left(r'',r'\right) 
\over {(1+\gamma)[1+({r-r''\over {1+\gamma}})^2]}} 
\nonumber \\
& - & \int_{r_{s,0}^{(-)}}^{r_{s,0}^{(+)}}
dr''{\bar{\Phi }}_{s,0;s,\gamma'}\left(r'',r'\right)
{\Theta^{[1]}_{1+\gamma,1}\Bigl(r-r''\Bigr)\over{2\pi}} \, .
\label{fs38}
\end{eqnarray}

For $\gamma =0$ Eqs. (\ref{fs36}) and (\ref{fs38}) can be rewritten as

\begin{equation}
\bar{\Phi }_{s,0;c,0}\left(r,r'\right) = 
-{1\over{\pi}}\tan^{-1}(r-r') + 
{1\over{\pi}}\int_{r_{c,0}^{(-)}}^{r_{c,0}^{(+)}}
dy''G(y,y''){\bar{\Phi }}_{s,0;c,0}\left(y'',x'\right) \, ,
\end{equation}
and

\begin{eqnarray}
\bar{\Phi }_{s,0;s,0}\left(r,r'\right) & =
& {1\over{\pi}}\tan^{-1}({r-r'\over{2}}) -
{1\over{\pi^2}}\int_{r_{c,0}^{(-)}}^{r_{c,0}^{(+)}} 
dr''{\tan^{-1} 
(r''-r')\over{1+(r-r'')^2}}
\nonumber \\
& + & \int_{r_{s,0}^{(-)}}^{r_{s,0}^{(+)}}
dr''G(r,r''){\bar{\Phi }}_{s,0;s,0}\left(r'',r'\right) \, ,
\end{eqnarray}
and the kernel $G(r,r')$ reads \cite{0Carm92b}

\begin{equation}
G(r,r') = - {1\over{2\pi}}\left[{1\over{1+((r-r')/2)^2}}\right]
\left[1 - {1\over 2}
\left(t(r)+t(r')+{{l(r)-l(r')}\over{r-r'}}\right)\right] \, ,
\end{equation}
with

\begin{equation}
t(r) = {1\over{\pi}}\left[\tan^{-1}(r + r_{c,0}^{(+)}) 
- \tan^{-1}(r + r_{c,0}^{(-)})\right]\, ,
\end{equation}
and

\begin{equation}
l(r) = {1\over{\pi}}\left[
\ln (1+(r + r_{c,0}^{(+)})^2) -  
\ln (1+(r + r_{c,0}^{(-)})^2)\right] \, .
\end{equation}

In order to evaluate the second-order functions
$Q_{\alpha,\gamma}^{(2)}(q)$ of the rhs of Eq. (\ref{xopq})
for $j=2$ 
we introduce the functions (\ref{k1q})-(\ref{r2q})  
and distributions (\ref{distribalga}) in Eqs. 
(\ref{integral1})-(\ref{integral3}). 
Expanding to second order we find after some algebra
 
\begin{equation}
      Q_{\alpha,\gamma}^{(2)}(q) = Q_{\alpha,\gamma}^{(2,*)}(q) + 
      {1\over 2} {d\over {dq}}[[Q_{\alpha,\gamma}^{(1)}(q)]^2] \, , 
\label{qbarast1}
\end{equation}
where

\begin{equation}
      Q_{\alpha,\gamma}^{(2,*)}(q) = 
      \widetilde{Q}_{\alpha,\gamma}^{(2,*)}(R_{\alpha,\gamma}^{(0)}(q)) \, .
\label{qbarast2}
\end{equation}
It is also useful to define the function
$\widetilde{Q}^{(2,*)}(k)$ such that

\begin{equation}
      \widetilde{Q}^{(2,*)}(k) =
      {\bar{Q}}_{c,0}^{(2,*)}({\sin k\over u}) \, .
\label{qbarast0}      
\end{equation}
The functions $\bar{Q}_{\alpha,\gamma}^{(2,*)}(r)$ are 
defined by the following system of coupled integral equations

\begin{eqnarray}
\bar{Q}_{c,0}^{(2,*)}(r) & = & \sum_{\gamma =0}
\int_{q_{s,\gamma}^{(-)}}^{q_{s,\gamma}^{(+)}}dq \delta N_{s,\gamma}(q)
{Q_{s,\gamma}^{(1)}(q)\over 
2\pi\rho_{s,\gamma}\Bigl(R^{(0)}_{s,\gamma}(q)\Bigl)}
{1\over \pi (1+\gamma)\left[1 + ({r - R^{(0)}_{s,\gamma}(q)
\over 1+\gamma})^2\right]}
\nonumber\\
& + & \sum_{\gamma =1}\int_{q_{c,\gamma}^{(-)}}^{q_{c,\gamma}^{(+)}}dq 
\delta N_{c,\gamma}(q){Q_{c,\gamma}^{(1)}(q)\over 
2\pi\rho_{c,\gamma}\Bigl(R^{(0)}_{c,\gamma}(q)\Bigl)}
{1\over \pi\gamma\left[1 + ({r - R^{(0)}_{c,\gamma}(q)
\over\gamma})^2\right]}
\nonumber\\
& + & \sum_{\gamma =0}{1\over 2\pi\rho_{s,\gamma}
\Bigl(r_{s,\gamma}\Bigl)}\sum_{j=\pm 1}
{[Q_{s,\gamma}^{(1)}(jq_{Fs,\gamma})]^2
\over 2\pi (1+\gamma )\left[1 + ({r-jr_{s,\gamma}
\over 1+\gamma})^2\right]}
\nonumber\\
& + & \sum_{\gamma =1}{1\over 2\pi\rho_{c,\gamma}
\Bigl(r_{c,\gamma}\Bigl)}\sum_{j=\pm 1}
{[Q_{c,\gamma}^{(1)}(jq_{Fc,\gamma})]^2
\over 2\pi\gamma\left[1 + ({r-jr_{c,\gamma}
\over\gamma})^2\right]} 
\nonumber\\
& + & \int_{r^{(-)}_{s,0}}^{r^{(+)}_{s,0}}
dr'{\bar{Q}_{s,0}^{(2,*)}(r')\over 
\pi\left[1 + (r -r')^2\right]} \, ,
\label{qintast1}
\end{eqnarray}

\begin{eqnarray}
\bar{Q}_{c,\gamma}^{(2,*)}(r) & = & 
- \int_{q_{c,0}^{(-)}}^{q_{c,0}^{(+)}}dq 
\delta N_{c,0}(q){\cos K^{(0)}(q)\over u}
{Q_{c,0}^{(1)}(q)\over 
2\pi\rho_{c,0}\Bigl(k^{(0)}(q)\Bigl)}
{1\over \pi\gamma\left[1 + ({r - R^{(0)}_{c,0}(q)
\over\gamma})^2\right]}
\nonumber\\
& - & \sum_{\gamma' =1}\int_{q_{c,\gamma'}^{(-)}}^{q_{c,\gamma'}^{(+)}}dq 
\delta N_{c,\gamma'}(q){Q_{c,\gamma'}^{(1)}(q)\over 
2\pi\rho_{c,\gamma'}\Bigl(R^{(0)}_{c,\gamma'}(q)\Bigl)}
{1\over 2\pi}\Theta^{[1]}_{\gamma ,\gamma'}\Bigl(r - 
R^{(0)}_{c,\gamma'}(q)\Bigl)
\nonumber\\
& - & {\cos Q\over u}{1\over 2\pi\rho_{c,0}
\Bigl(Q\Bigl)}\sum_{j=\pm 1}
{[Q_{c,0}^{(1)}(jq_{Fc,0})]^2
\over 2\pi\gamma\left[1 + ({r-jr_{c,0}
\over\gamma})^2\right]} 
\nonumber\\
& - & \sum_{\gamma' =1}{1\over 2\pi\rho_{c,\gamma'}
\Bigl(r_{c,\gamma'}\Bigl)}{1\over 2}\sum_{j=\pm 1}
[Q_{c,\gamma'}^{(1)}(jq_{Fc,\gamma'})]^2
{1\over 2\pi}\Theta^{[1]}_{\gamma ,\gamma'}
\Bigl(r-jr_{c,\gamma'}\Bigl)
\nonumber\\
& - & \int_{r^{(-)}_{c,0}}^{r^{(+)}_{c,0}}
dr'{\bar{Q}_{c,0}^{(2,*)}(r')\over 
\pi\gamma\left[1 + ({r -r'\over\gamma})^2\right]} \, ,
\end{eqnarray}
and

\begin{eqnarray}
\bar{Q}_{s,\gamma}^{(2,*)}(r) & = & 
\int_{q_{c,0}^{(-)}}^{q_{c,0}^{(+)}}dq 
\delta N_{c,0}(q){\cos K^{(0)}(q)\over u}
{Q_{c,0}^{(1)}(q)\over 
2\pi\rho_{c,0}\Bigl(k^{(0)}(q)\Bigl)}
{1\over \pi (1+\gamma )\left[1 + ({r - R^{(0)}_{c,0}(q)
\over 1+\gamma})^2\right]}
\nonumber\\
& - & \sum_{\gamma' =0}\int_{q_{s,\gamma'}^{(-)}}^{q_{s,\gamma'}^{(+)}}dq 
\delta N_{s,\gamma'}(q){Q_{s,\gamma'}^{(1)}(q)\over 
2\pi\rho_{s,\gamma'}\Bigl(R^{(0)}_{s,\gamma'}(q)\Bigl)}
{1\over 2\pi}\Theta^{[1]}_{1+\gamma ,1+\gamma'}\Bigl(r - 
R^{(0)}_{s,\gamma'}(q)\Bigl)
\nonumber\\
& + & {\cos Q\over u}{1\over 2\pi\rho_{c,0}
\Bigl(Q\Bigl)}\sum_{j=\pm 1}
{[Q_{c,0}^{(1)}(jq_{Fc,0})]^2
\over 2\pi (1+\gamma)\left[1 + ({r-jr_{c,0}
\over 1+\gamma})^2\right]} 
\nonumber\\
& - & \sum_{\gamma' =0}{1\over 2\pi\rho_{s,\gamma'}
\Bigl(r_{s,\gamma'}\Bigl)}{1\over 2}\sum_{j=\pm 1}
[Q_{s,\gamma'}^{(1)}(jq_{Fs,\gamma'})]^2
{1\over 2\pi}\Theta^{[1]}_{1+\gamma ,1+\gamma'}
\Bigl(r-jr_{s,\gamma'}\Bigl)
\nonumber\\
& + & \int_{r^{(-)}_{c,0}}^{r^{(+)}_{c,0}}
dr'{\bar{Q}_{c,0}^{(2,*)}(r')\over 
\pi(1+\gamma)\left[1 + ({r -r'\over 1+\gamma })^2\right]} 
\nonumber \\
& - & \int_{r^{(-)}_{s,0}}^{r^{(+)}_{s,0}}
dr'\bar{Q}_{s,0}^{(2,*)}(r'){1\over 
2\pi}\Theta^{[1]}_{1+\gamma,1}
\Bigl(r - r'\Bigl) \, .
\label{qintast3}
\end{eqnarray}

As for the first-order case, the use of either the function
$Q_{\alpha,\gamma}^{(2)}(q)$ defined by 
Eqs. (\ref{qbarast1})-(\ref{qbarast0})
or of the full function $X_{\alpha,\gamma}^{(2)}(q)$
of Eq. (\ref{xopq}) for $j=2$ leads to the same results for
the physical quantities up to second order in the
density of excited pseudoparticles.

Note that the free terms of Eqs. (\ref{qintast1})-(\ref{qintast3}) involve the 
first-order functions only. This implies that the unique 
solutions of these integral equations can be expressed in 
terms of the first-order functions. Therefore, following Eqs. 
(\ref{rapodr1}) and (\ref{xopq}) the second-order functions can
also be expressed in terms of the pseudoparticle
phase shifts.

%% file: apnohe.tex
\chapter{Normal-ordered Hamiltonian expression}
\label{normhamitexp}
\pagestyle{myheadings}
\markboth{Normal-ordered Hamiltonian ...}{Appendix B of Chapter 4}

In order to derive the first-order and second-order Hamiltonian
terms of Eqs. (\ref{hamitord1}) and (\ref{hamiltord2}), 
we again consider eigenvalues 
and deviations. The energy associated with the Hamiltonian
(\ref{hamiltba}) reads

\begin{eqnarray}
E_{SO(4)} & = & -{2tN_a\over 2\pi}
\int_{q^{(-)}_{c,0}}^{q^{(+)}_{c,0}}N_{c,0}(q)\cos K(q)\nonumber\\
&+& {4tN_a\over 2\pi}\sum_{\gamma =1}\int_{q^{(-)}_{c,
\gamma}}^{q^{(+)}_{c,\gamma}}N_{c,\gamma}(q)
Re \sqrt{1-[u(R_{c,\gamma}(q)+i\gamma)]^2}
\nonumber \\
& + & U[{N_a\over 4} - {N_{c,0}\over 2}
- \sum_{\gamma =1}\gamma N_{c,\gamma}] \, .
\label{energyso44}
\end{eqnarray}
To caculate the bands (\ref{c0qband})-(\ref{sgqband}) 
we introduce in 
this energy Eqs. (\ref{k1q}), (\ref{r1q}), and (\ref{distribalga}) 
and expand the obtained 
expression to first order in the deviations with the result

\begin{eqnarray}
\Delta E^{(1)}_{SO(4)} & = & {N_a\over 2\pi}\{
\int_{q^{(-)}_{c,0}}^{q^{(+)}_{c,0}} \delta N_{c,0}(q)[
- 2t\cos K^{(0)}(q) - {U\over 2}]
\nonumber \\
& + & \sum_{\gamma =1}\int_{q^{(-)}_{c,\gamma}}^{q^{(+)}_{c,\gamma}}
\delta N_{c,\gamma}(q)[4t
Re \sqrt{1-[u(R^{(0)}_{c,\gamma}(q)+i\gamma)]^2} -\gamma U]
\nonumber \\
& + & 2t\int_{Q^{(-)}}^{Q^{(+)}}dk
\widetilde{Q}^{(1)}(k)\sin k\nonumber\\
&-& 4t\sum_{\gamma =1}\int_{r^{(-)}_{c,\gamma}}^{r^{(+)}_{c,\gamma}}dr
\bar{Q}^{(1)}_{c,\gamma}(r)
Re\Bigl({u^2[r + i\gamma]\over \sqrt{1 - u^2[r + i\gamma]^2}}\Bigl)\} 
\, ,
\label{enerdel1}
\end{eqnarray}
where the functions $\widetilde{Q}^{(1)}(k)$ and
$\bar{Q}^{(1)}_{c,\gamma}(r)$ are defined by 
Eqs. (\ref{qtilq})-(\ref{qbarcgr}).
The use of Eqs. (\ref{xopq}), (\ref{qbarq}), (\ref{qtilq}), 
and  (\ref{rapodr1}) in (\ref{enerdel1}) leads 
after some straightforward algebra to

\begin{equation}
      \Delta E^{(1)}_{SO(4)}= \sum_{q,\alpha,\gamma}
      \epsilon^0_{\alpha ,\gamma}(q)\delta \hat{N}_{\alpha ,\gamma}(q)\, ,
\label{ener1final}
\end{equation}
with the bands given by Eqs. (\ref{c0qband})-(\ref{sgqband}) 
in terms of the
phase shifts (\ref{f1})-(\ref{fs38}).

An equivalent representation for the band energies is

\begin{equation}
      \epsilon_{c,0}^0(q) =  -{U\over 2} -2t\cos K^{(0)}(q) 
      - \int_{r^{(-)}_{s,0}}^{r^{(+)}_{s,0}}dr 2t\eta_{s,0}(r)
{1\over \pi}\tan^{-1}\left(r -R^{(0)}_{c,0}(q)\right)\, ,
\label{eneretac0}
\end{equation}

\begin{eqnarray}
      \epsilon_{c,\gamma}^0(q)  &=&  -\gamma U 
      + 4t Re \sqrt{1 - u^2[R^{(0)}_{c,\gamma}(q) + i\gamma]^2}
\nonumber\\      
      &-& \int_{Q^{(-)}}^{Q^{(+)}}dk{1\over \pi}
      \tan^{-1}\left({{\sin k\over u} -R^{(0)}_{c,\gamma}(q)
      \over\gamma}\right)2t\eta_{c,0}(k) \, ,
\label{eneretacg}      
\end{eqnarray}
and

\begin{eqnarray}
      \epsilon_{s,\gamma}^0(q) & = & 
      - 2t\int_{Q^{(-)}}^{Q^{(+)}}dk{1\over \pi}
      \tan^{-1}\left({{\sin k\over u} -R^{(0)}_{s,\gamma}(q)
      \over 1+\gamma}\right)\sin k
\nonumber \\
      & + & \int_{r^{(-)}_{s,0}}^{r^{(+)}_{s,0}}dr 2t\eta_{s,0}(r)
      \frac 1{2\pi}
      \Theta_{1,1+\gamma}\Bigl(r-R^{(0)}_{s,\gamma}(q)\Bigl)
\nonumber\\
      &-& \int_{r^{(-)}_{c,0}}^{r^{(+)}_{c,0}}dr'
      {\tan^{-1}\left({r' -R^{(0)}_{s,\gamma}(q)\over 1+\gamma}\right)
      \over \pi^2[1 + (r- r')^2]} \, ,
\label{eneretasg}
\end{eqnarray}
where the functions $2t\eta_{c,0}(k)$ and $2t\eta_{\alpha,\gamma}(r)$
are defined by the integral equations  

\begin{equation}
      2t\eta_{c,0}(k)  =  2t\sin k 
      + {\cos k\over u} \int_{r^{(-)}_{s,0}}^{r^{(+)}_{s,0}}dr 
      {2t\eta_{s,0}(r)\over 
      \pi\left[1+(r - {\sin k\over u})^2\right]}\, ,
\label{etaco}      
\end{equation}

\begin{eqnarray}
       2t\eta_{c,\gamma}(r) & = & 
       - 4t Re\Bigl({u^2[r + i\gamma]\over
       \sqrt{1 - u^2[r + i\gamma]^2}}\Bigl)
\nonumber\\       
       &+& \int_{Q^{(-)}}^{Q^{(+)}}dk {2t\eta_{c,0}(k)\over
       \pi\gamma\left[1+({{\sin k\over u} 
       -r\over\gamma})^2\right]} \, ,
\label{etacg}       
\end{eqnarray}
and

\begin{eqnarray}
       2t\eta_{s,\gamma}(r) & = & 
       2t\int_{Q^{(-)}}^{Q^{(+)}}dk{\sin k\over 
       \pi(1+\gamma)\left[1+({{\sin k\over u} - r\over 1+\gamma})^2\right]} 
\nonumber \\
       & - & \int_{r^{(-)}_{s,0}}^{r^{(+)}_{s,0}}dr' 2t\eta_{s,0}(r')
       \frac 1{2\pi}[\Theta^{(1)}_{1,1+\gamma}\Bigl(r'- r\Bigl)
\nonumber\\       
        &-& \int_{r^{(-)}_{c,0}}^{r^{(+)}_{c,0}}dr''
       {1\over\pi^2(1+\gamma)\left[1+({r - r''\over 1+\gamma})^2\right]
       \left[1 + (r'- r'')^2\right]}] \, .
\label{etasg}       
\end{eqnarray}

The gap expressions (\ref{andgap}) and  (\ref{finalgap}) 
can be rewritten in terms of pseudoparticle bands as

\begin{eqnarray}
       \omega_0 = &-& 2\epsilon^0_{c ,0}
       (q_{Fc,0}) N^h_{c ,{-l_{c}\over 2}} 
       - \epsilon^0_{s ,0}(q_{Fs,0}) 
      \sum_{\alpha } N^h_{\alpha ,{-l_{\alpha}\over 2}} 
      \nonumber \\
      &-& 
       \sum_{\gamma =1}[\epsilon^0_{s ,0}(q_{Fs,0})
       - \epsilon^0_{s ,\gamma }(0)] N_{s ,\gamma} \, ,
\label{bgap1}       
\end{eqnarray}
and

\begin{eqnarray}
      \omega_0  = &-& [2\epsilon^0_{c ,0}(q_{Fc,0}) + 
      \epsilon^0_{s ,0}(q_{Fs,0})] 
       \Bigl(\sum_{\gamma =1}\gamma N_{c,\gamma } + N^z_c \Bigl) 
       \nonumber \\
       &-& \epsilon^0_{s ,0}(q_{Fs,0})\Bigl(\sum_{\gamma =1}
      [1+\gamma ]N_{s,\gamma } + N^z_s \Bigl) 
       + \sum_{\gamma =1}\epsilon^0_{s,\gamma }(0)
       N_{s ,\gamma} \, ,
\label{bgap2}       
\end{eqnarray}
respectively. In order to derive the expression for the second-order Hamiltonian 
(\ref{hamiltord2}) and associate $f$ functions (\ref{ffunction})
we expand the energy (\ref{energyso44})
to second-order with the result 

\begin{eqnarray}
\Delta E^{(2)}_{SO(4)} & = & {N_a\over {2\pi}}\{
\int_{q_{c,0}^{(-)}}^{q_{c,0}^{(+)}} dq \delta N_{c,0}(q)
{2t\sin K^{(0)}(q)\over 2\pi\rho_{c,0}\Bigl(K^{(0)}(q)\Bigl)}  
Q_{c,0}^{(1)}(q)
\nonumber\\
& - & \sum_{\gamma =1}\int_{q^{(-)}_{c,\gamma}}^{q^{(+)}_{c,\gamma}}
dq \delta N_{c,\gamma}(q){Q^{(1)}_{c,\gamma}(q)\over
2\pi\rho_{c,\gamma}\Bigl(R^{(0)}_{c,\gamma}(q)\Bigl)}
4t Re\Bigl({u^2[R^{(0)}_{c,\gamma}(q) + i\gamma]
\over \sqrt{1 - u^2[R^{(0)}_{c,\gamma}(q) + i\gamma]^2}}\Bigl) 
\nonumber\\
& + & {2t\sin Q\over 2\pi\rho_{c,0}(Q)}{1\over 2}
\sum_{j=\pm 1}[Q_{c,0}^{(1)}(jq_{Fc,0})]^2
\nonumber\\
& - & \sum_{\gamma =1}{1\over 2}\sum_{j=\pm 1}
{[Q^{(1)}_{c,\gamma}(jq_{c,\gamma})]^2\over
2\pi\rho_{c,\gamma}(r_{c,\gamma})}
4t Re\Bigl({u^2[r_{c,\gamma} + i\gamma]
\over \sqrt{1 - u^2[r_{c,\gamma} + i\gamma]^2}}\Bigl) 
\nonumber\\
& + & \int_{Q^{(-)}}^{Q^{(+)}}dk\widetilde{Q}^{(2,*)}(k)
2t\sin k
\nonumber\\
& - & \sum_{\gamma =1}\int_{r^{(-)}_{c,\gamma}}^{r^{(+)}_{c,\gamma}}
dr\bar{Q}^{(2,*)}_{c,\gamma}(r)
4t Re\Bigl({u^2[r + i\gamma]
\over \sqrt{1 - u^2[r + i\gamma]^2}}\Bigl)\} 
\, .
\label{energy2ord}
\end{eqnarray}
Inserting the suitable functions in the rhs of 
Eq. (\ref{energy2ord}), performing 
some integrations by using symmetry properties of the kernels of
the integral equation (\ref{qbarc0r})-(\ref{qbarsgr}) 
and (\ref{qintast1})-(\ref{qintast3}), and replacing 
deviations by pseudomomentum normal-ordered operators (\ref{normaln}) we 
find after some algebra expression (\ref{hamiltx2}). 
Note that replacing in 
Eq. (\ref{hamiltx2}) the function $X^{(1)}_{\alpha,\gamma}(q)$ by the associate
function $Q^{(1)}_{\alpha,\gamma}(q)$ leads to the same result. 
By the use of Eq. (\ref{rapodr1}) expression (\ref{hamiltord2})
can be rewritten in terms 
of the $f$ functions (\ref{ffunction}) 
as given in the rhs of Eq. (\ref{hamiltord2}).

%% file: elepp.tex
\pagestyle{headings}
\setcounter{chapter}{4}
\chapter{Excitations and Finite-Size Effects}%
\label{finitesize}
\section{Introduction}
\pagestyle{myheadings}
\markboth{5  Excitations and Finite-Size ...}{5.1 Introduction}

In this Chapter we introduce the quasiparticle concept for
the 1D Hubbard Model. Exception made to Secs. \ref{fiento}
and \ref{swaves}, we deal in this Chapter with 
LWS's I/HWS's I only.
As we shall see, these quasiparticles
have not exactly the same physical meaning as in Fermi-liquid
theory \cite{3Landau56,3Landau57,3Landau59,3Pines,3Baym}. In the latter
theory, the quasiparticle concept is used to describe 
low-energy excitations at constant or variable electronic and
magnetization densities, $n$ and $m$, respectively. 
The quasiparticle in a Fermi-liquid is
related to the electronic operators trought a factor $Z$ 
--the renormalization factor. The value of $Z$ is determined by
the electronic correlations.

In our Landau-liquid description of the Hubbard chain the quasiparticles
do not describe the elementary excitations at constant $n$ and $m$. These
are described by the $\alpha$ pseudoholes  and by the  $\alpha,\gamma$ heavy-pseudoparticles, introduced in Chapter \ref{harmonic}. In contrast, 
excitations that change $n$ and $m$ are described by composite objects
of pseudoholes, that we call (in this restricted Fermi-liquid sense)
quasiparticles. Due to the non-perturbative character of the one-dimensional
many-body problem \cite{3Meden92,3Voit93,3Voit96}, 
the renormalization factor that connects the 
quasiparticle and electronic operators vanishes as the Fermi points are approached 
\cite{3Neto96}. 

The quasiparticle operator is defined as the generator of a ground-state--ground-state
transition involving the addition or removal of one electron to or from the many-body
ground state. We find that the quasiparticle momentum equals
the expected electronic momentum $k=\pm k_{F\sigma}$. It is the combined momenta (i) from the addition or removal of pseudoparticles
and (ii) from the global
pseudomomentum shifts of $\pm \pi/N_a$ which leads to this final result. This type
of pseudomomentum shifts are also present in excitations connecting the 
ground state and LWS's II (or HWS's II). 
We will illustrate the importance of these momentum shifts 
\cite{3Dias92,3Nuno96} in
the limits
$n=1$ and $U/4t^2\gg 1$, where the model is closely related to the  anti-ferromagnetic
Heisenberg chain. 


\section{Ground-state -- ground-state 
topological excitations} 
\label{gsgstoex}
\pagestyle{myheadings}
\markboth{5  Excitations and finite-size ...}{5.2 Ground-state -- ...}

The quasiparticle operator, ${\tilde{c}}^{\dag }_{k_{F\sigma },\sigma }$, 
creates one quasiparticle with spin projection $\sigma$ and 
momentum $k_{F\sigma}$. As in Fermi liquid theory, such operator is the generator of the following ground-state--ground-state transition

\begin{equation}
      {\tilde{c}}^{\dag }_{k_{F\sigma},\sigma}
      |GS; N_{\sigma}, N_{-\sigma}\rangle =
      |GS; N_{\sigma} + 1, N_{-\sigma}\rangle \, .
\label{quasipdef}      
\end{equation}
The quasiparticle operator defines a one-to-one correspondence
between the addition of one electron to the system and the creation
of one quasiparticle. The momentum of the ground states
$|GS; N_{\sigma}, N_{-\sigma}\rangle$ associated with canonical ensembles of
U(1)$\otimes$U(1) symmetryis given in Table \ref{tabp}. Since we are 
studying ground-state--ground-state transitions, 
$\alpha,\gamma>1$ heavy-pseudoparticles are not 
involved.

\begin{table}[ht]
\begin{center}
\begin{tabular}{ccccc}
\hline
&$(-1,-1)$ & $(-1,1)$ & $(1,-1)$ &$(1,1)$\\ 
\hline
\hline
(A)\,$P$ & $\pm 2k_F$ & $\pm 2k_F$ & $\pm[2\pi-2k_F]$ 
& $\pm[2\pi-2k_F]$\\ 
(B)\,$P$ & $\pm k_{F\uparrow}$ & $\pm k_{F\uparrow}$ 
& $\pm k_{F\uparrow}$ & $\pm k_{F\uparrow}$\\ 
(C)\,$P$ & $\pm k_{F\downarrow}$ & $\pm k_{F\downarrow}$ 
& $\pm k_{F\downarrow}$ & $\pm k_{F\downarrow}$\\ 
(D)\,$P$ & $0$ & $0$ & $0$ & $0$ \\
\hline
\end{tabular}
\end{center}
\caption{Values of the ground-state momentum in the four $(l,l')$ 
sectors of Hamiltonian symmetry $U(1)\otimes U(1)$. The different 
momentum values correspond to the following parities of the numbers 
$\bar{N}_{s,0}$ and $N_{s,0}$, respectively: 
(A) even, even; (B) even, odd; (C) odd, even; and (D) 
odd, odd. In the case of the $(1,\pm 1)$ sectors, if 
$k_{F\sigma}>\pi$ then $\pm k_{F\sigma}$ should be replaced
by the first-Brillouin-zone momenta $\pm [2\pi -k_{F\sigma}]$.}
\label{tabp}
\end{table}

The study of the ground-state momentum (\ref{pmomentum})
confirms that the relative momentum of ground 
states differing in the number of $\sigma $ electrons by one equals 
the $U=0$ Fermi points, ie $\Delta P=\pm k_{F\sigma }$.

It is interesting to study the expression of the $\sigma $ quasiparticle and 
quasihole operators in the 
$\alpha,\beta$ pseudohole basis for all sectors of 
Hamiltonian symmetry. Since we are discussing the problem of 
addition or removal of one particle, the boundary conditions
(\ref{bound1}) and (\ref{bound2})
play a crucial role.
When we add or remove one electron from the many-body system we
have to consider the transitions between states with integer and
half-integer quantum numbers $I_j^{\alpha,0}$. The transition between
two ground states differing in the number of electrons
by one is then associated with two different processes: 
a backflow in the Hilbert space of the $\alpha ,\beta$ 
pseudoholes with a shift of all the pseudomomenta 
by $\pm\frac{\pi}{N_a}$ and the creation and (or) annihilation 
of one pair of $c$ and $s$ pseudoholes at the pseudo-Fermi points (or
at the limit of the pseudo-Brillouin zone for the $s$ pseudohole). 

The
backflow associated with a shift of all the pseudomomenta
by $\pm\frac{\pi}{N_a}$ is described by a topological
unitary operator such that 

\begin{equation}
      V^{\pm}_{\alpha }a^{\dag }_{q,\alpha ,\beta}
      V^{\mp}_{\alpha }= 
      a^{\dag }_{q\mp {\pi\over {N_a}},\alpha ,\beta} \, .
\end{equation}
Obviously, the pseudohole vacuum $|V\rangle$
is invariant under this operator, this is 
$V^{\pm}_{\alpha }|V\rangle =|V\rangle$.
It is simple to find that $V^{\pm}_{\alpha }$ reads

\begin{equation}
      V^{\pm}_{\alpha } = \exp\left\{
      -\sum_{q,\beta}a^{\dag }_{q\pm\frac{\pi}{N_a}
      ,\alpha ,\beta}a_{q,\alpha ,\beta}\right\} \, .
\label{vop}
\end{equation}
The structure of this operator reveals that the pseudomomentum of all pseudoholes are translated by
$\pm \pi/N_a$.
Adding all pseudohole 
contributions gives a large momentum. 
This large-momentum excitation induced by the operator
(\ref{vop}) is the $\alpha $ topological momentum shift.
Note that in the present case of the Hilbert subspace spanned by
the states I of the $(l,l')$ sector only the value 
$\beta={l\over 2}$ for $\alpha =c$ and the value
$\beta={l'\over 2}$ for $\alpha =s$ contributes to the 
$\beta$ summation of Eq. (\ref{vop})

In addition to the topological momentum shift, the quasiparticle
or quasihole excitation includes creation and (or) 
annihilation of pseudoholes. The changes in the 
pseudohole and pseudoparticle numbers and the corresponding
changes in the values of $S^c$, $S^c_z$, $S^s$, and $S^s_z$ are 
given in Tables \ref{tabe1} and \ref{tabe2} for 
the ground-state -- ground-state 
transitions $(N_{\uparrow},N_{\downarrow})\rightarrow
(N_{\uparrow}\pm 1,N_{\downarrow})$ and
$(N_{\uparrow},N_{\downarrow})\rightarrow
(N_{\uparrow},N_{\downarrow}\pm 1)$, respectively.
 
We consider below the expressions for the quasiparticles 
${\tilde{c}}^{\dag }_{k_{F\uparrow},\uparrow}$ and
${\tilde{c}}^{\dag }_{k_{F\downarrow},\downarrow}$
associated with the transitions $(N_{\uparrow},
N_{\downarrow})\rightarrow (N_{\uparrow}+1,N_{\downarrow})$ and
$(N_{\uparrow},N_{\downarrow})\rightarrow
(N_{\uparrow},N_{\downarrow}+1)$, respectively, and the 
quasiholes ${\tilde{c}}_{k_{F\uparrow},\uparrow}$ and
${\tilde{c}}_{k_{F\downarrow},\downarrow}$
associated with the transitions
$(N_{\uparrow},N_{\downarrow})\rightarrow
(N_{\uparrow}-1,N_{\downarrow})$ and
$(N_{\uparrow},N_{\downarrow})\rightarrow
(N_{\uparrow},N_{\downarrow}-1)$, respectively. 

\begin{table}[ht]
\begin{center}
\begin{tabular}{ccccc}
\hline
 &$(-1,-1)$ &$(-1,1)$ &$(1,-1)$ &$(1,1)$\\ 
\hline
\hline
$\Delta N_{c}^h$ & $\mp 1$ & $\mp 1$ & $\pm 1$ & $\pm 1$\\ 
$\Delta N_{c}$ & $\pm 1$ & $\pm 1$ & $\mp 1$ & $\mp 1$\\ 
$\Delta N_{c}^*$ & $0$ & $0$ & $0$ & $0$\\ 
$\Delta N_{s}^h$ & $\pm 1$ & $\mp 1$ & $\pm 1$ & $\mp 1$\\ 
$\Delta N_{s}$ & $0$ & $\pm 1$ & $\mp 1$ & $0$\\ 
$\Delta N_{s}^*$ & $\pm 1$ & $0$ & $0$ & $\mp 1$\\ 
$\Delta S^c$ & $\mp 1/2$ & $\mp 1/2$ & $\pm 1/2$ & $\pm 1/2$\\ 
$\Delta S^c_z$ & $\pm 1/2$ & $\pm 1/2$ & $\pm 1/2$ & $\pm 1/2$\\ 
$\Delta S^s$ & $\pm 1/2$ & $\mp 1/2$ & $\pm 1/2$ & $\mp 1/2$\\ 
$\Delta S^s_z$ & $\mp 1/2$ & $\mp 1/2$ & $\mp 1/2$ & $\mp 1/2$ \\
\hline
\end{tabular}
\end{center}
\caption{Changes in the numbers of pseudoholes, pseudoparticles,
pseudoparticle orbitals, and in the values of $S^c$, $S^c_z$, $S^s$, 
and $S^s_z$ in the ground-state--ground-state transition 
$(N_{\uparrow},N_{\downarrow})\rightarrow
(N_{\uparrow}\pm 1,N_{\downarrow})$.}
\label{tabe1}
\end{table}
\begin{table}[ht]
\begin{center}
\begin{tabular}{ccccc}
\hline
 &$(-1,-1)$&$(-1,1)$ & $(1,-1)$& $(1,1)$\\ 
\hline
\hline
$\Delta N_{c}^h$ & $\mp 1$ & $\mp 1$ & $\pm 1$ & $\pm 1$\\ 
$\Delta N_{c}$ & $\pm 1$ & $\pm 1$ & $\mp 1$ & $\mp 1$\\ 
$\Delta N_{c}^*$ & $0$ & $0$ & $0$ & $0$\\ 
$\Delta N_{s}^h$ & $\mp 1$ & $\pm 1$ & $\mp 1$ & $\pm 1$\\ 
$\Delta N_{s}$ & $\pm 1$ & $0$ & $0$ & $\mp 1$\\ 
$\Delta N_{s}^*$ & $0$ & $\pm 1$ & $\mp 1$ & $0$\\ 
$\Delta S^c$ & $\mp 1/2$ & $\mp 1/2$ & $\pm 1/2$ & $\pm 1/2$\\ 
$\Delta S^c_z$ & $\pm 1/2$ & $\pm 1/2$ & $\pm 1/2$ & $\pm 1/2$\\ 
$\Delta S^s$ & $\mp 1/2$ & $\pm 1/2$ & $\mp 1/2$ & $\pm 1/2$\\ 
$\Delta S^s_z$ & $\pm 1/2$ & $\pm 1/2$ & $\pm 1/2$ & $\pm 1/2$\\ 
\hline
\end{tabular}
\end{center}
\caption{Changes in the numbers of pseudoholes, pseudoparticles,
pseudoparticle orbitals, and in the values of $S^c$, $S^c_z$, $S^s$, 
and $S^s_z$ in the ground-state -- ground-state transition 
$(N_{\uparrow},N_{\downarrow})\rightarrow
(N_{\uparrow},N_{\downarrow}\pm 1)$.}
\label{tabe2}
\end{table}
We emphasize that because the initial ground state for the above 
two quasiparticles and two quasiholes is the same, the $\sigma $ 
quasiparticle and $\sigma $ quasihole momenta differ by $\pm 
{2\pi\over {N_a}}$. Therefore, the corresponding 
quasiparticle and quasihole expressions are not related by 
an adjunt transformation. On the other hand, the operators 
$\tilde{c}^{\dag }_{\pm k_{F\sigma },\sigma }$ and
$\tilde{c}_{\pm k_{F\sigma },\sigma }$ associated with the 
transitions $(N_{\sigma},N_{-\sigma})\rightarrow
(N_{\sigma}+1,N_{-\sigma})$ and $(N_{\sigma}+1,N_{-\sigma})\rightarrow
(N_{\sigma},N_{-\sigma})$ are obviously related by
such transformation. In this case the initial (final)
ground state of the electrons (holes) is the final
(initial) ground state of the holes (electrons).  
Moreover, let us consider the set of four operators
$\tilde{c}^{\dag }_{\pm k_{F\uparrow},\uparrow }$,
$\tilde{c}_{\pm k_{F\uparrow },\uparrow }$, 
$\tilde{c}^{\dag }_{\pm k_{F\downarrow},\downarrow }$, and
$\tilde{c}_{\pm k_{F\downarrow },\downarrow }$
such that the creation operators act on the same
initial ground state $(N_{\uparrow},N_{\downarrow})$
transforming it in the ground states  $(N_{\uparrow }+1,
N_{\downarrow})$ and $(N_{\uparrow},N_{\downarrow} +1)$,
respectively, and the hole operators act on the corresponding
latter states giving rise to the original ground state.
Let us consider the reduced Hilbert subspace spanned
by these three ground states. 
If we combine that with the electron and hole expressions 
introduced below, it is easy to show 
that the corresponding quasiparticle and quasihole operators 
$\tilde{c}^{\dag }_{\pm k_{F\uparrow},\uparrow }$,
$\tilde{c}_{\pm k_{F\uparrow },\uparrow }$, 
$\tilde{c}^{\dag }_{\pm k_{F\downarrow},\downarrow }$, and
$\tilde{c}_{\pm k_{F\downarrow },\downarrow }$ obey the usual
anticommutation relations.

The two electrons and two holes refer to the same
initial ground state. The pseudo-Fermi points and pseudohole-Fermi 
points of the expressions below refer to that initial ground state.
On the other hand, $s$ pseudohole creation and annihilation
operators at the limits of the pseudo-Brillouin zones refer to the
final and initial ground states, respectively. In the case of
the $(l,l')$ sectors of Hamiltonian symmetry $U(1)\otimes U(1)$
we consider that the initial and final ground states belong
the same sector of parameter space. In the case of the $(l')$ 
sectors of Hamiltonian symmetry $SU(2)\otimes U(1)$ [or $(l)$ 
sectors of Hamiltonian symmetry $U(1)\otimes SU(2)$] we consider that 
the initial and final ground states belong to sectors of parameter
space characterized by the same value of $(l')$ [or $(l)$]. We 
present below the electron and hole expressions found for 
different initial ground states in the nine sectors of parameter 
space. It is useful to introduce the pseudohole-Fermi
point $\bar{q}^{(\pm)}_{F\alpha}$ such that $\bar{q}^{(\pm)}_{F\alpha}=q^{(\pm)}_{F\alpha}\pm 2\pi/N_a$

For initial ground states in the $(-1,-1)$ sector of Hamiltonian
symmetry $U(1)\otimes U(1)$ we find

\begin{equation}
      \tilde{c}^{\dag }_{\pm k_{F\uparrow },\uparrow } =
      a_{{\bar{q}}_{Fc}^{(\pm)},c,-{1\over 2}}V^{\pm}_{s}
      a^{\dag }_{q_{s}^{(\pm)},s,-{1\over 2}} \, ,\hspace{1cm}
      \tilde{c}^{\dag }_{\pm k_{F\downarrow },\downarrow } =
      V^{\pm}_{c}a_{{\bar{q}}_{Fc}^{(\pm)},c.-{1\over 2}}
      a_{{\bar{q}}_{Fs}^{(\pm)},s,-{1\over 2}} \, ,
\label{qp1}
\end{equation}
for the electrons and 

\begin{equation}
\tilde{c}_{\pm k_{F\uparrow },\uparrow } =
a^{\dag }_{q_{Fc}^{(\pm)},c,-{1\over 2}}V^{\mp}_{s}
a_{q_{s}^{(\pm)},s,-{1\over 2}} \, ,\hspace{1cm}
\tilde{c}_{\pm k_{F\downarrow },\downarrow } =
V^{\mp}_{c}a^{\dag }_{q_{Fc}^{(\pm)},c,-{1\over 2}}
a^{\dag }_{q_{Fs}^{(\pm)},s,-{1\over 2}} \, ,
\end{equation}
for the holes. 

For the $(-1,1)$ sector we find

\begin{equation}
\tilde{c}^{\dag }_{\pm k_{F\uparrow },\uparrow } =
V^{\pm}_{c}a_{{\bar{q}}_{Fc}^{(\pm)},c,-{1\over 2}}
a_{{\bar{q}}_{Fs}^{(\pm)},s,{1\over 2}} \, ,\hspace{1cm}
\tilde{c}^{\dag }_{\pm k_{F\downarrow },\downarrow } =
a_{{\bar{q}}_{Fc}^{(\pm)},c,-{1\over 2}}V^{\pm}_{s}
a^{\dag }_{q_{s}^{(\pm)},s,{1\over 2}} \, ,
\end{equation}
for the electrons and

\begin{equation}
\tilde{c}_{\pm k_{F\uparrow },\uparrow } =
V^{\mp}_{c}a^{\dag }_{q_{Fc}^{(\pm)},c,-{1\over 2}}
a^{\dag }_{q_{Fs}^{(\pm)},s,{1\over 2}} \, ,\hspace{1cm}
\tilde{c}_{\pm k_{F\downarrow },\downarrow } =
a^{\dag }_{q_{Fc}^{(\pm)},c,-{1\over 2}}V^{\mp}_{s}
a_{q_{s}^{(\pm)},s,{1\over 2}} \, ,
\end{equation}
for the holes.

In the $(1,-1)$ sector the result is

\begin{equation}
\tilde{c}^{\dag }_{\pm k_{F\uparrow },\uparrow } =
V^{\mp}_{c}a^{\dag }_{q_{Fc}^{(\pm)},c,{1\over 2}}
a^{\dag }_{q_{Fs}^{(\pm)},s,-{1\over 2}} \, ,\hspace{1cm}
\tilde{c}^{\dag }_{\pm k_{F\downarrow },\downarrow } =
a^{\dag }_{q_{Fc}^{(\pm)},c,{1\over 2}}V^{\mp}_{s}
a_{q_{s}^{(\pm)},s,-{1\over 2}} \, ,
\end{equation}
for the electrons and

\begin{equation}
\tilde{c}_{\pm k_{F\uparrow },\uparrow } =
V^{\pm}_{c}a_{{\bar{q}}_{Fc}^{(\pm)},c,{1\over 2}}
a_{{\bar{q}}_{Fs}^{(\pm)},s,-{1\over 2}} \, ,\hspace{1cm}
\tilde{c}_{\pm k_{F\downarrow },\downarrow } =
a_{{\bar{q}}_{Fc}^{(\pm)},c,{1\over 2}}V^{\pm}_{s}
a^{\dag }_{q_{s}^{(\pm)},s,-{1\over 2}} \, ,
\end{equation}
for the holes.

The expressions for the $(1,1)$ sector are

\begin{equation}
\tilde{c}^{\dag }_{\pm k_{F\uparrow },\uparrow } =
a^{\dag }_{q_{Fc}^{(\pm)},c,{1\over 2}}V^{\mp}_{s}
a^{\dag }_{q_{s}^{(\pm)},s,{1\over 2}} \, ,\hspace{1cm}
\tilde{c}^{\dag }_{\pm k_{F\downarrow },\downarrow } =
V^{\mp}_{c}a^{\dag }_{q_{Fc}^{(\pm)},c,{1\over 2}}
a^{\dag }_{q_{Fs}^{(\pm)},s,{1\over 2}} \, ,
\end{equation}
for the electrons and

\begin{equation}
       \tilde{c}_{\pm k_{F\uparrow },\uparrow } =
       a_{{\bar{q}}_{Fc}^{(\pm)},c,{1\over 2}}V^{\pm}_{s}
       a_{q_{s}^{(\pm)},s,{1\over 2}} \, ,\hspace{1cm}
       \tilde{c}_{\pm k_{F\downarrow },\downarrow } =
       V^{\pm}_{c}a_{{\bar{q}}_{Fc}^{(\pm)},c,{1\over 2}}
       a_{{\bar{q}}_{Fs}^{(\pm)},s,{1\over 2}} \, ,
\label{qp8}
\end{equation}
for the holes.

According to Eqs. (\ref{qp1})-(\ref{qp8}) the $\sigma $ quasiparticles and
quasiholes are many-pseudohole objects which recombine the 
colors $c$ and $s$ (charge and spin in the 
limit $m = n_{\uparrow}-n_{\downarrow} \rightarrow 0$)  
giving rise to spin projection $\uparrow $ and $\downarrow $
and have Fermi surfaces at $\pm k_{F\sigma }$. 

Similar expressions can be derived for the sectors of
parameter space where the Hamiltonian (\ref{hamiltbabeta}) has higher
symmetry. We start by considering the sectors of
Hamiltonian symmetry $SU(2)\otimes U(1)$ where

\begin{equation}
      q_{Fc}^{(+)}=-q_{Fc}^{(-)}=
      q_{c}^{(+)}=-q_{c}^{(-)}=
      \pi[1-{1\over {N_a}}] \, .
\label{fermicq}
\end{equation}
For ground states of the $l'=-1$ sector of Hamiltonian 
symmetry $SU(2)\otimes U(1)$ the electrons read

\begin{equation}
\tilde{c}^{\dag }_{\pm k_{F\uparrow },\uparrow } =
V^{\mp}_{c}a^{\dag }_{q_{Fc}^{(\pm)},c,{1\over 2}}
a^{\dag }_{q_{Fs}^{(\pm)},s,-{1\over 2}} \, ,\hspace{1cm}
\tilde{c}^{\dag }_{\pm k_{F\downarrow },\downarrow } =
a^{\dag }_{q_{Fc}^{(\pm)},c,{1\over 2}}V^{\mp}_{s}
a_{q_{s}^{(\pm)},s,-{1\over 2}} \, ,
\label{qp9}
\end{equation}
and the holes read

\begin{equation}
\tilde{c}_{\pm k_{F\uparrow },\uparrow } =
a^{\dag }_{q_{Fc}^{(\pm)},c,-{1\over 2}}V^{\mp}_{s}
a_{q_{s}^{(\pm)},s,-{1\over 2}} \, ,\hspace{1cm}
\tilde{c}_{\pm k_{F\downarrow },\downarrow } =
V^{\mp}_{c}a^{\dag }_{q_{Fc}^{(\pm)},c,-{1\over 2}}
a^{\dag }_{q_{Fs}^{(\pm)},s,-{1\over 2}} \, .
\end{equation}

For the $l'=1$ sector of Hamiltonian symmetry $SU(2)\otimes U(1)$ 
the electrons read

\begin{equation}
\tilde{c}^{\dag }_{\pm k_{F\uparrow },\uparrow } =
a^{\dag }_{q_{Fc}^{(\pm)},c,{1\over 2}}V^{\mp}_{s}
a^{\dag }_{q_{s}^{(\pm)},s,{1\over 2}} \, ,\hspace{1cm}
\tilde{c}^{\dag }_{\pm k_{F\downarrow },\downarrow } =
V^{\mp}_{c}a^{\dag }_{q_{Fc}^{(\pm)},c,{1\over 2}}
a^{\dag }_{q_{Fs}^{(\pm)},s,{1\over 2}} \, ,
\end{equation}
and the holes read

\begin{equation}
\tilde{c}_{\pm k_{F\uparrow },\uparrow } =
V^{\mp}_{c}a^{\dag }_{q_{Fc}^{(\pm)},c,-{1\over 2}}
a^{\dag }_{q_{Fs}^{(\pm)},s,{1\over 2}} \, ,\hspace{1cm}
\tilde{c}_{\pm k_{F\downarrow },\downarrow } =
V^{\mp}_{s}a^{\dag }_{q_{Fc}^{(\pm)},c,-{1\over 2}}
a_{q_{s}^{(\pm)},s,{1\over 2}} \, .
\label{qp12}
\end{equation}

In the sectors of Hamiltonian symmetry $U(1)\otimes SU(2)$ we
have that

\begin{equation}
      q_{Fs}^{(+)}=-q_{Fs}^{(-)}=
      q_{s}^{(+)}=-q_{s}^{(-)}=
      \pi [{n\over 2}-{1\over {N_a}}] \, .
\label{fermisq}
\end{equation}
In the case of the $l=-1$ sector of Hamiltonian symmetry 
$U(1)\otimes SU(2)$ the up-spin electron and hole read

\begin{equation}
\tilde{c}^{\dag }_{\pm k_{F\uparrow },\uparrow } =
V^{\pm}_{c}a_{{\bar{q}}_{Fc}^{(\pm)},c,-{1\over 2}}
a^{\dag }_{q_{s}^{(\pm)},s,{-1\over 2}} \, ,\hspace{1cm}
\tilde{c}_{\pm k_{F\uparrow },\uparrow } =
V^{\mp}_{c}a^{\dag }_{q_{Fc}^{(\pm)},c,-{1\over 2}}
a^{\dag }_{q_{s}^{(\pm)},s,{1\over 2}} \, ,
\label{qp13}
\end{equation}
and the down-spin electron and hole read

\begin{equation}
\tilde{c}^{\dag }_{\pm k_{F\downarrow },\downarrow } =
V^{\pm}_{c}a_{{\bar{q}}_{Fc}^{(\pm)},c,-{1\over 2}}
a^{\dag }_{{\bar{q}}_{s}^{(\pm)},s,{1\over 2}} \, ,\hspace{1cm}
\tilde{c}_{\pm k_{F\downarrow },\downarrow } =
V^{\mp}_{c}a^{\dag }_{q_{Fc}^{(\pm)},c,-{1\over 2}}
a^{\dag }_{q_{s}^{(\pm)},s,-{1\over 2}} \, .
\end{equation}

For the $l=1$ sector of Hamiltonian symmetry 
$U(1)\otimes SU(2)$ the up-spin electron and hole read

\begin{equation}
\tilde{c}^{\dag }_{\pm k_{F\uparrow },\uparrow } =
V^{\mp}_{c}a^{\dag }_{q_{Fc}^{(\pm)},c,{1\over 2}}
a^{\dag }_{q_{s}^{(\pm)},s,-{1\over 2}} \, , \hspace{1cm}
\tilde{c}_{\pm k_{F\uparrow },\uparrow } =
V^{\pm}_{c}a_{{\bar{q}}_{Fc}^{(\pm)},c,{1\over 2}}
a^{\dag }_{q_{s}^{(\pm)},s,{1\over 2}} \, ,
\end{equation}
and the down-spin electron and hole read

\begin{equation}
\tilde{c}^{\dag }_{\pm k_{F\downarrow },\downarrow } =
V^{\mp}_{c}a^{\dag }_{q_{Fc}^{(\pm)},c,{1\over 2}}
a^{\dag }_{q_{s}^{(\pm)},s,{1\over 2}} \, ,\hspace{1cm}
\tilde{c}_{\pm k_{F\downarrow },\downarrow } =
V^{\pm}_{c}a_{{\bar{q}}_{Fc}^{(\pm)},c,{1\over 2}}
a^{\dag }_{q_{s}^{(\pm)},s,{-1\over 2}} \, .
\label{qp16}
\end{equation}

Finally, for the $SO(4)$ initial ground state both Eq. $(49)$
and the following equation

\begin{equation}
      q_{Fs}^{(+)}=-q_{Fs}^{(-)}=
      q_{s}^{(+)}=-q_{s}^{(-)}=
      \pi [{1\over 2}-{1\over {N_a}}] \, ,
\label{fermiso4}
\end{equation}
hold true and we find for the electrons

\begin{equation}
\tilde{c}^{\dag }_{\pm k_{F\uparrow },\uparrow } =
V^{\mp}_{c}a^{\dag }_{q_{Fc}^{(\pm)},c,{1\over 2}}
a^{\dag }_{q_{Fs}^{(\pm)},s,-{1\over 2}} \, ,\hspace{1cm}
\tilde{c}^{\dag }_{\pm k_{F\downarrow },\downarrow } =
V^{\mp}_{c}a^{\dag }_{q_{Fc}^{(\pm)},c,{1\over 2}}
a^{\dag }_{q_{Fs}^{(\pm)},s,{1\over 2}} \, ,
\label{qp18}
\end{equation}
and for the holes

\begin{equation}
\tilde{c}_{\pm k_{F\uparrow },\uparrow } =
V^{\mp}_{c}a^{\dag }_{q_{Fc}^{(\pm)},c,-{1\over 2}}
a^{\dag }_{q_{Fs}^{(\pm)},s,{1\over 2}} \, ,\hspace{1cm}
\tilde{c}_{\pm k_{F\downarrow },\downarrow } =
V^{\mp}_{c}a^{\dag }_{q_{Fc}^{(\pm)},c,-{1\over 2}}
a^{\dag }_{q_{Fs}^{(\pm)},s,-{1\over 2}} \, .
\label{qp19}
\end{equation}

Equations 
(\ref{qp18})-(\ref{qp19}) reveal that removing or adding electrons from
the $SO(4)$ ground state always involves creation
of pseudoholes. Furthermore, while in the
case of the $(l,l')$ sectors the initial and
final ground states belong in general to the same sector,
in the case of the SO(4) ground state each of the four 
possible transitions associated with adding one
up-spin or one down-spin quasiparticle or quasihole leads to
four ground states belonging to a different
$(l,l')$ sector. If the initial ground state belongs
to the $(l')$ $SU(2)\otimes U(1)$ sector [or to the  
$(l)$ $U(1)\otimes SU(2)$ sector] then two
of the final ground states belong to the $(1,l')$
[or to the $(l,1)$] sector and the remaining two
ground states to the $(-1,l')$ [or to the $(l,-1)$] sector.

\section{Low-energy excitations and symmetry}
\label{loenexsy}
\pagestyle{myheadings}
\markboth{5  Excitations and Finite-Size ...}{5.3 Low-energy excitations ...}

Given a ground state with electron numbers $(N_{\sigma},
N_{-\sigma })$, we find in this section that the set of all 
pseudoholes of different type which constitute, in pairs, 
the $\uparrow $ and $\downarrow $ electrons and $\uparrow $ and 
$\downarrow $ holes associated with the ground-state transitions  
$(N_{\uparrow},N_{\downarrow})\rightarrow 
(N_{\uparrow}\pm 1,N_{\downarrow})$ and $(N_{\uparrow},
N_{\downarrow})\rightarrow (N_{\uparrow},N_{\downarrow}\pm 1)$ 
transform as the symmetry group of the Hamiltonian 
(\ref{hamiltbabeta})
in the corresponding sector of parameter space.

Equations (\ref{offdiagonal}) tell us that the values of $S^c $ and
$S^c_z$ are fully determined by the number of $c,\beta$ pseudoholes
whereas the number of $s,\beta$ pseudoholes determines the values of
$S^s$ and $S^s_z$. In addition, note that the quasiparticle and quasihole 
operators (\ref{qp1})-(\ref{qp8}), (\ref{qp9})-(\ref{qp12}), 
(\ref{qp13})-(\ref{qp16}), 
and (\ref{qp18})-(\ref{qp19})
involve always a change in the number of $c$ and $s$
pseudoholes by one. Moreover, when acting on the suitable ground state 
these operators change the values of $S^c $ and $S^c_z$ by 
$\pm 1/2$ and $\pm sgn (S^c_z)1/2$, respectively, and the values 
of $S^s$ and $S^s_z$ by $\pm 1/2$ and $\pm sgn (S^s_z)1/2$, respectively. 
(The corresponding changes in the pseudohole numbers and in the 
values of $S^c $, $S^c_z$, $S^s$, and $S^s_z$ are shown in 
Tables \ref{tabe1} 
and \ref{tabe2}.) The analysis of the changes in the pseudohole numbers 
could lead to the conclusion that the $c,\pm {1\over 2}$ pseudoholes have 
quantum numbers $(S^c =1/2;S^s=0;S^c_z=\pm 1/2;S^s_z=0)$ and that 
the $s,\mp {1\over 2}$ pseudoholes have quantum numbers 
$(S^c =0;S^s=1/2;S^c_z =0;S^s_z=\pm 1/2)$. 
If this was true the $c$ and corresponding $\beta$ pseudohole quantum 
numbers could be identified with $S^c $ and $S^c_z$ for $S^c=1/2$ and $S_z^c=\pm 1/2$, respectively, 
and the $s$ and corresponding $\beta$ pseudohole quantum numbers could be 
identified with $S^s$ and $S^s_z$ for $S^s=1/2$ and $S_z^s=\pm 1/2$, respectively. (Note that $S^{\alpha}$ and $S_z^{\alpha}$ do not refer to the whole many-electron sustem but to a single electron) However, this is not in 
general true. This holds true in the particular case of 
zero-momentum number operators. On the other hand, the above
identities are also true for finite-momentum operators
for $c,\pm {1\over 2}$ in the limit of zero chemical 
potential and for $s,\pm {1\over 2}$ in the limit of zero magnetic 
field, as we find below. 

In order to confirm that the above equivalences are not in 
general true for finite-momentum fluctuations, we consider 
the $\alpha $-pseudohole fluctuation operator \cite{3Carm94,3Carm96}

\begin{equation}
\rho_{\alpha }(k) =
- \sum_{q,\beta}\beta a_{q,\alpha ,\beta}^{\dag }a_{q+k,\alpha,\beta} 
\, , 
\end{equation}
and the $S^c_z$- (charge) and $S^s_z$- (spin) fluctuation
operators 

\begin{equation}
\rho_{S^c_z }(k) =\sum_{k',\sigma}
\left[{1\over 2}\delta_{k,o} - 
c_{k'+k,\sigma}^{\dag }c_{k',\sigma}\right] \, , 
\end{equation}
and

\begin{equation}
\rho_{S^s_z }(k) =\sum_{k',\sigma}\sigma c_{k'+k,\sigma}^{\dag }
c_{k',\sigma} \, ,  
\end{equation}
respectively. From Eqs. (\ref{offdiagonal}) 
and (\ref{nupndown}) we find

\begin{equation}
\rho_{c}(0) =\rho_{S^c_z }(0) = N_a - N_{\uparrow }
- N_{\downarrow } \, , 
\label{fluc0}
\end{equation}
and

\begin{equation}
\rho_{s}(0) = \rho_{S^s_z }(0) = N_{\uparrow }
- N_{\downarrow } \, .  
\label{flus0}
\end{equation}
We then conclude that at zero momentum the above
equivalences hold true. The electron numbers $N_{\uparrow }$ and 
$N_{\downarrow }$ are good quantum numbers of the 
many-electron system. Since the exact Hamiltonian eigenstates are 
simple Slater determinants of $\alpha ,\beta$-pseudohole levels,
the numbers of $\alpha ,\beta$ pseudoholes are thus required to be 
also good quantum numbers. They are such that Eqs. (\ref{fluc0}) 
and (\ref{flus0}) are obeyed. 

On the other hand, the conservation of electron and
pseudohole numbers does not require the finite-momentum
$c$ and $s$ fluctuations being bare finite-momentum charge and 
spin fluctuations, respectively. For simplicity, we consider 
the smallest momentum values, $k\pm {2\pi\over {N_a}}$.
We emphasize that for $k=\pm {2\pi\over {N_a}}$, acting 
the operator $\rho_{\alpha }(k)$ onto a ground state 
of general form (\ref{gsop}) generates a sinlge-pair 
$\alpha $-pseudoparticle-pseudohole excitation where the 
$\alpha ,\beta$ pseudohole at 
$q={\bar{q}}_{F\alpha }^{(\pm)}$ moves to 
$q=q_{F\alpha }^{(\pm)}$. If in $c,\beta$
the color $c$ was eta spin and 
$\beta=S^c_z $ and in $s,\beta$ the
color $s $ was spin and $\beta=S^s_z $, we should have
that $\rho_{c}(\pm {2\pi\over {N_a}})=\rho_{S^c_z }
(\pm {2\pi\over {N_a}})$ and $\rho_{s}(\pm {2\pi\over 
{N_a}})=\rho_{S^s_z }(\pm {2\pi\over {N_a}})$, respectively. 
However, the results of the works
\cite{3Carm94,3Carm96} show that this is not true for the
sectors of Hamiltonian symmetry $U(1)\otimes U(1)$. Altough the
pseudohole summations of Eqs. (\ref{offdiagonal}) and 
(\ref{nupndown}) give $S^c$, 
$S^s$, $S^c_z$, and $S^s_z$ this does not require each $c$
pseudohole having $\eta$ spin $1/2$ and spin $0$ and each $s$
pseudohole having $\eta$ spin $0$ and spin $1/2$. Also, the fact 
that the quasiparticle or quasihole of 
Eqs. $(\ref{qp1})-(\ref{qp8})$, $(50)-(\ref{qp12})$, 
$(\ref{qp13})-(58)$, and $(\ref{qp18})-(\ref{qp19})$ has 
$S^c =1/2;S^s=1/2;S^c_z=sgn (S^c_z)1/2;
S^s_z=sgn (S^s_z)1/2$ does not tell how these
values are destributed by the corresponding 
$c$ pseudohole, $s$ pseudohole, and topological momentum shifts.

The studies of the works \cite{3Carm94,3Carm96} 
reveal that for finite 
values of the chemical potential and magnetic field there is 
a $c$ and $s$ separation of the low-energy and small-momentum 
excitations but that the orthogonal modes $c$ and $s$ are not 
in general charge and spin, respectively \cite{3Carm94,3Carm96}. On 
the other hand, these studies have found that in the limit 
of zero chemical potential the finite-momentum $c$ fluctuations 
become real charge excitations and that in the limit of zero 
magnetic field the finite-momentum $s$ fluctuations become real 
spin excitations. In the latter limit the $c$ excitations
are also real charge excitations and the $c$ and $s$
low-energy separation becomes the usual charge and
spin separation \cite{3Meden92,3Voit93,3Voit96}. 

It follows that in the case of the SO(4) canonical ensemble 
the set of pseudoholes involved in the description of the two 
electron and two hole operators (\ref{qp18})-(\ref{qp19}),
which are the $c,+{1\over 2}$; $c,-{1\over 2}$; 
$s,+{1\over 2}$; and $s,-{1\over 2}$ pseudoholes 
at the pseudo-Fermi points, transform
in the $S^c =1/2$ and $S^s=1/2$ representation of the SO(4)
group. Moreover, it can be shown from the changes in the BA quantum
numbers and from the study of the pseudohole energies
that the $S^c =1/2$ and $S^s=1/2$ elementary
excitations studied Essler {\it et al.} 
\cite{3Essler94} are simple combinations 
of one of the ground-state -- ground-state transitions
generated by the operators (\ref{qp18})-(\ref{qp19}) with a single
pseudoparticle-pseudohole process relative to the final
ground state. In addition, the usual half-filling holons and 
zero-magnetization spinons can be shown to be limiting cases 
of our pseudohole excitations. For instance, the  
$(S^c =1/2;S^s=0;S^c_z=1/2;S^s_z =0)$ anti holon and
$(S^c =1/2;S^s=0;S^c_z=-1/2;S^s_z =0)$ holon excitations
\cite{3Essler94} are at lowest energy 
generated from the SO(4) ground state
by the operators $V^{-}_{c}
a^{\dag }_{q_{Fc}^{(\pm)},c,{1\over 2}}$ and 
$V^{+}_{c}a^{\dag }_{q_{Fc}^{(\pm)},c,-{1\over 2}}$, 
respectively. Also at lowest energy, the two 
$(S^c =0;S^s=1/2;S^c_z =0;S^s_z=1/2)$ and 
$(S^c =0;S^s=1/2;S^c_z =0;S^s_z=-1/2)$ spinons
\cite{3Essler94} are generated from that ground state by the operators 
$a^{\dag }_{q_{Fs}^{(\pm)},s,-{1\over 2}}$ and $a^{\dag 
}_{q_{Fs}^{(\pm)},s,{1\over 2}}$, respectively. The full spectrum
of these excitations is obtained by adding to these generators
a suitable single pseudoparticle-pseudohole-pair operator.

Since the $SO(4)$ symmetry only allows Hamiltonian
eigenstates with integer values of $S^c_z +S^s_z$,
the holon -- spinon pairs of Eqs. (\ref{qp18})-(\ref{qp19})
cannot be separated.
This also holds true in the general case, the electron
being constituted by one $c$ pseudohole, one $s$ pseudohole, 
and one many-pseudohole topological momentum shift of large
momentum, as confirmed by Eqs. (\ref{qp1})-(\ref{qp8}), 
(\ref{qp9})-(\ref{qp12}),
and (\ref{qp13})-(\ref{qp16}). Also in this case the fact that only 
Hamiltonian eigenstates with integer values of 
$S^c_z +S^s_z$ are allowed prevents these three excitations
from being separated. 

As for the SO(4) ground state, we can relate the symmetry
of the Hamiltonian (\ref{hamiltbabeta}) 
in a given canonical ensemble by 
looking at the pseudohole contents of the corresponding two 
electrons and two holes of Eqs. (\ref{qp1})-(\ref{qp8}), 
(\ref{qp9})-(\ref{qp12}), and
(\ref{qp13})-(\ref{qp16}). For instance, Eqs. 
(\ref{qp1})-(\ref{qp8}) show that 
in the $(l,l')$ sectors of Hamiltonian 
symmetry $U(1)\otimes U(1)$ the two electrons and two holes  
involve one pair of the same type of pseudoholes, namely the 
corresponding $c,{l\over 2}$ and $s,{l'\over 2}$ pseudoholes. 
Each of these transforms in the representation of the group 
$U(1)$ and, therefore, the set of two pseudoholes transforms 
in the representation of the group $U(1)\otimes U(1)$.

In the case of the $(l')$ sectors of Hamiltonian symmetry 
$SU(2)\otimes U(1)$ $c$ is $\eta$ spin and the corresponding
quantum number $\beta$ is $S^c_z$ and 
Eqs. (\ref{qp9})-(\ref{qp12}) confirm that 
the two electrons and holes involve 
either one $c,{1\over 2}$ pseudohole or one $c,-{1\over 2}$ 
pseudohole combined with one $s,{l'\over 2}$ pseudohole. 
The $c,{1\over 2}$ and $c,-{1\over 2}$ pseudoholes transform in 
the $S^c =1/2$ representation of the 
$\eta$-spin $SU(2)$ group, whereas 
the $s,{l'\over 2}$ pseudohole 
transforms in the representation of the 
$U(1)$ group. Therefore, the set of $c,{1\over 2}$; 
$c,-{1\over 2}$; and $s,{l'\over 2}$ pseudoholes transforms 
in the $S^c =1/2$ representation of the $SU(2)\otimes U(1)$ group. 

In the case of the $(l)$ sectors of Hamiltonian symmetry 
$U(1)\otimes SU(2)$ $s$ is spin and the corresponding
quantum number $\beta$ is $\beta=S^s_z$ 
and Eqs. (\ref{qp13})-(\ref{qp16})
show that the two electrons and two holes are constituted by either 
one $s,{1\over 2}$ or one $s,-{1\over 2}$ pseudohole combined 
with one $c,{l\over 2}$ pseudohole. The $s,{1\over 2}$ and 
$s,-{1\over 2}$ pseudoholes transform in the $S^s =1/2$ 
representation of the spin $SU(2)$ group and the $c,{l\over 2}$ 
pseudohole transforms in the representation of the $U(1)$ group. It
follows that the set of the $c,{l\over 2}$; $s,{1\over 2}$; 
and $s,-{1\over 2}$ pseudoholes transforms in the $S^s =1/2$ 
representation of the $U(1)\otimes SU(2)$ group.

\section{Finite-energy topological momentum-shift operators}
\label{fiento}
\pagestyle{myheadings}
\markboth{5  Excitations and finite-size ...}{5.4 Finite-energy topological ...}

The simple form of the GGS expression (\ref{ggs}), GS expressions 
(\ref{gsop})-(\ref{gs3}), and of the general-Hamiltonian eigenstates 
(\ref{bethestates}) introduced in Chapter \ref{harmonic} has a 
deep physical meaning. It reveals that in the present
basis these eigenstates of the many-electron quantum problem 
are ``non-interacting'' states of simple 
Slater-determinant form. However, that the
numbers $I_j^{\alpha ,\gamma}$ of the rhs of Eq. $(28)$ can 
be integers or half-odd integers for different Hamiltonian 
eigenstates, makes the problem much more involved than a simple 
non-interacting case. This change in the integer or half-odd
integer character of 
some of the numbers $I_j^{\alpha ,\gamma}$ of two
states, shifts {\it all} the occupied pseudomomenta. 
As in the case of last section, transition 
between a GS $(46)$ and {\it any} eigenstate 
(\ref{bethestates}) can be separated into two types of excitations: (a) a 
topological GS -- GGS transition which involves the creation or 
annihilation of pseudoholes and (or) heavy pseudoparticles as 
well as the occurrence of topological momentum shifts and (b) a {\it 
Landau-liquid} excitation associated 
with pseudoparticle - pseudohole processes relative to the GGS. 

The topological transitions (a) 
are basically superpositions of three kinds of elementary 
transitions: (i) GS - GS transitions involving changes in
the $\sigma $ electron numbers by one, (ii) single $\beta$ 
flip processes of $\alpha,\beta$ pseudoholes 
which lead to non-LWS's and non-HWS's outside the BA,
and (iii) creation of single $\alpha,\gamma$ pseudoparticles at 
constant values of $S^c$, $S^c_z$, $S^s$, and $S^s_z$. 
While the transitions (i) are gapless, the elementary
excitations (ii) and (iii) require a finite amount of 
energy. 

The generators of the excitations (i) were 
studied in the previous section and in the work  
\cite{3Nuno96} and the ones of (ii) involve
only topological momentum shifts and $\beta $ flips (which describe either 
creation of electron pairs or spin flip processes). Consider 
as an example of a transition (iii) the creation of 
one $c,\gamma $ pseudoparticle which is found  
to be the relevant process for the finite-frequency 
conductivity \cite{3Carm97}. 
(We examine here explicitly topological momentum 
shifts of occupied bands only, {\it i.e.} 
these which generate momentum.) When $\gamma >0$ 
is even (or odd) the generator reads $V^{\pm 1}_c
G_{c,\gamma}$ (or $G_{c,\gamma}V^{\pm 1}_s$), where the 
topological-momentum-shift operator was defined above and

\begin{equation}
G_{c,\gamma}=[\hat{S}^c_+]^{\gamma}
\prod_{q=q_{Fc,0}^{(-)}}^{q_{Fc,0}^{(-)}+
{2\pi\over N_a}(\gamma -1)}\prod_{q_{Fc,0}^{(+)}-
{2\pi\over N_a}(\gamma -1)}^{q_{Fc,0}^{(+)}} 
a^{\dag }_{q,c,-{1\over 2}}b^{\dag }_{q'=0,c,\gamma} \, .
\end{equation}
Therefore, such transitions involve one $c$ or $s$
topological momentum shift, the creation of a number $\gamma$ of
$c,{1\over 2}$ pseudoholes and $\gamma$ of $c,-{1\over 2}$ pseudoholes, 
and the creation of one $c,\gamma$ pseudoparticle at
$q=0$. Similar results hold for the topological 
momentum shifts associated with the creation of one $s,\gamma$ 
pseudoparticle. In next section we show the physical importance of this latter
momentum shifts.
\section{Momentum shifts at work: the example of spin
waves}
\label{swaves}
\pagestyle{myheadings}
\markboth{5  Excitations and finite-size ...}{5.5 Momentum shifts at ...}

By use of a simple example, we now show that the topological momentum shifts contribute
to the total momentum of a given excitation. We consider
1D Hubbard model at electronic and magnetization densities
equal to 1 and 0, respectively, and in the strong coupling
regime $U/4t^2\gg 1$. 

It was shown by Ogata {\it et al.} \cite{3Ogata90} that
in the strong coupling regime the Bethe-ansatz wave function
for the Hubbard model 
decouples into the product of the 1D anti-ferromagnetic
Heisenberg model wave function \cite{3Bethe31}
and a Slater determinant of free spinless fermions. At $n=1$ the spinless fermions have no dynamics
and we are left with the antiferromagnetic Heisenberg chain
with magnetic coupling given by $4t^2/U$. 
The correct description of the anti-ferromagnetic chain 
(\ref{heismodel}) excitation
spectrum at zero magnetic field was given by Faddeev {\it et al.}
\cite{3Faddeev81}. These authors 
found that the excitation spectrum is two 
parametric, which leads to a continumn of excitations. The
excitation spectrum $E(q_1,q_2)$ and the excitation momentum $\Delta P$
are given by 

\begin{equation}
      E(q_1,q_2)=\frac {J\pi}{2}[\cos(q_1)+\cos(q_2)]\,,
      \hspace{2cm} \Delta P=\pi-q_1-q_2\,.
\label{spectrumheis}
\end{equation}
Here $J$ is the antiferromagnetic coupling and $q_1$ and $q_2$
are the momentum of two spin waves (each with spin 1/2). The factor
 $\pi$ in $\Delta P$ expression refers to the ground-state momentum 
($N_{\downarrow}$ was choosen to be odd and the number
of sites $N_a$ was choosen to be even).

The same results can be obtained for the Hubbard
chain in the strong-coupling regime. Although
the spinless fermions band is full,  all the fermions change their
momenta by $\pi/N_a$, what recovers the  
excitation
momentum of the Heisenberg chain. For $U/4t^2\gg 1$ 
and electronic density $n=1$, the energy
band of the $s,0$ pseudoparticles is given by
\cite{3Carm91}

\begin{equation}
      \varepsilon_{s,0}^0 (q)=-\frac {4t^2}{U} \frac {\pi}{2}
      \cos(q)
\label{exs}
\end{equation}
and the $c,0$ energy band by

\begin{equation}
      \varepsilon_{c,0}^0 (q)=-2t\cos(q)+{\cal O}(4t^2/U)\,.
\label{exc}
\end{equation}
At zero magnetic field the ground state is the
pseudohole vacuum, that is the SO(4) ground state. 
In this case the $s,\gamma$-energy bands collapse to the point $\varepsilon_{s,\gamma}^0(q)=0$ and $q=0$. Both the $c,0$ and the
$s,0$ bands are filled in the ground state and $\varepsilon_{s,0}^0(q_{Fs,0})=\varepsilon_{s,\gamma}^0(0)=0$.
Thus, 
creation of
the $s,\gamma$ heavy-pseudoparticles costs no energy (excitations involving creation of
$c,\gamma$ heavy-pseudoparticles have an energy gap).
That is, excitations between the ground state and LWS's II.
Let us consider that the initial ground has both $N_{\downarrow}$
and $N_{\uparrow}$ odd. From Table \ref{tabp} we see that the ground
state has zero momentum. The simplest excitation one can think
of is, under these conditions, the creation of one single $s,1$ heavy
pseudoparticle. From the numbers (\ref{nupndown}) and keeping
constant $N_{\downarrow}$ and $N_{\uparrow}$, we see that the total
number of $s,0$ pseudoholes varies by 2, 
and that the number of $c,0$ pseudoparticles remains unchanged
(the corresponding number of pseudoholes is zero). ( Equation (\ref{bzlim})
confirms that the $s,1$ pseudo-Brillouin zone is reduced to the single
 $q=0$ point.)
Equations (\ref{bound1}) and (\ref{bound2}) imply that all
$c,0$ pseudoparticles (spinless fermions in this limit) change
their momenta by $\pi/N_a$.  This leads to  a finite
momentum of $\delta q=\pi$, and the energy and momentum excitation 
spectra
are
\begin{equation}
      E(q_1,q_2,q_3)=\varepsilon_{s,1}(q_3)+
      \frac {4t^2}{U} \frac {\pi}{2}
      [\cos(q_1)+\cos(q_2)]=\frac {4t^2}{U} \frac {\pi}{2}
      [\cos(q_1)+\cos(q_2)]
\label{hubexc}
\end{equation}
and

\begin{equation}
      \Delta P=\pi+q_3-q_1-q_2=\pi-q_1-q_2\,,
\label{pex}
\end{equation}
respectively.
It is obvious that Eqs. (\ref{hubexc}) and (\ref{pex})
recover Faddeev's results (\ref{spectrumheis}) for the Heisenberg chain
and that our $s,0$ pseudoparticles are
Faddev's spin-1/2 spin waves, in agreement with
the results of Sec. \ref{loenexsy}.

The same excitation spectra, (\ref{hubexc}) and (\ref{pex}), are
obtained if we take the limit $H=0^+$ and make a triplet excitation
(flip one electron spin). In this limit and for
the same parity of the electronic numbers $N_{\sigma}$ we have just
used, 
the ground state has two holes in the $s,0$ band, 
as Eq. (\ref{bzlim}) implies. The ground-state
momentum is zero. Due to the electronic spin-flip excitation, the boundary conditions
(\ref{bound1}) and (\ref{bound2}) adjust in such a way that
the final state still has two holes in the $s,0$ band. The
number of $s,0$ pseudoparticles changes by one, and a momentum shift of $\pi/N_a$ occurs in the $c,0$ band,
what leads, at half-filling, to a finite momentum of $\pi$. We
then recover the result that at zero magnetic field the triplet and the singlet excitations
are degenerated.

Without the use of the Bethe-ansatz equations, Dias and Lopes dos Santos found
equivalent results
\cite{3Dias92}. For 
$U=\infty$ they confirmed that all the spin configurations are degenerated.
This corresponds, in the pseudoparticle formalism, to 
$\varepsilon^0_{s,\gamma \geq 0}(q)=0$. Despite this massive
degeneracy, changes in the spin configurations lead to a global momentum shift
of $\pi/N_a$ in the $c,0$ pseudoparticle band. This effect was 
represented by Dias and Lopes dos Santos as a fictitious magnetic
flux through the spinless-fermion ring. This simple representation
shows that the change in the pseudomomenta is such that 
the system minimizes its energy when both the spin configuration changes
and one electron is added to or removed from the system.



%% file: corre.tex
\pagestyle{headings}
\setcounter{chapter}{5}
\chapter{Zero-Temperature Transport}
\label{transport}
\section{Introduction}
\pagestyle{myheadings}
\markboth{6  Zero-Temperature Transport}{6.1 Introduction}

The transport properties of strongly correlated electron models for 
low-dimensional conductors has been a subject of experimental and 
theoretical interest for over twenty years. Low-dimensional conductors 
show large deviations in their transport properties from the usual 
single-particle description. This suggests that electronic correlations 
might play an important role in these systems \cite{5Jacobsen1,5Jacobsen2,5Donovan,5Danilo,5Mori,5Schwartz},
even if the on-site Coulomb repulsion is small \cite{5Mori}. 

In this chapter we use the generalized pseudoparticle theory to
study the charge and spin currents of the Hubbard chain at finite
energy. For that we solve the BA equations with a twist angle for 
all electronic densities and magnetizations. We express the charge 
and the spin-diffusion currents in terms of the elementary currents 
of the charge and spin carriers. It is shown that the latter 
carriers are the $\alpha,\gamma$ pseudoparticles of the 
pseudoparticle-perturbation theory (PPT) presented in
Chapter \ref{harmonic} and introduced in the work
\cite{5CarmeloNuno97}. 
We evaluate their couplings to charge and spin and define the 
charge and spin pseudoparticle transport masses. The ratios 
of these masses over the corresponding static mass provide
important information on the role of electronic correlations
in the transport of charge and spin in the 1D quantum liquid.
Furthermore, we find that the transport of charge and spin  
can be described by means of pseudoparticle kinetic equations. 
Our results are a generalization to finite energies of
the low-energy results on transport of charge and spin
results obtained in earlier works \cite{5Carm4a,5Carm4b}. This is possible by
means of the generalized pseudoparticle representation introduced
in
Chapters \ref{harmonic} and \ref{PseudoPT}.

Recently, the pseudoparticle description of all BA Hamiltonian 
eigenstates introduced in Chapter \ref{harmonic} \cite{5CarmeloNuno97} has allowed the evaluation of 
analytical expressions for correlation functions at finite energy 
\cite{5GCFT1,5GCFT2}. From these results one can obtain 
expressions for the absorption-band edges of the frequency-dependent 
electronic conductivity, $\sigma(\omega)$ \cite{5conductivity}. 

\section{Review of fundamental concepts in pseudoparticle theory}
\label{5rev}
\pagestyle{myheadings}
\markboth{6  Zero-Temperature Transport}{6.2 Review of fundamental ...}

Although the pseudoparticle description, introduced in Chapter
\ref{harmonic}, refers to all Hamiltonian
eigenstates \cite{5CarmeloNuno97}, in this Chapter we restrict our 
study to the Hilbert subspace involved in the zero-temperature 
charge and spin frequency-dependent conductivity \cite{5conductivity}.
This is spanned by all the Hamiltonian eigenstates contained
in the states $\hat{j}^\zeta|GS\rangle$, where $|GS\rangle$ denotes
the ground state and $\hat{j}^\zeta$ are the charge ($\zeta=\rho$)
and spin ($\zeta=\sigma_z$) current operators [given by Eqs.
(\ref{5currrho}) and (\ref{5currsigma} below, respectively].
Since these current operators  
commute with the six generators of the $\eta$-spin and spin algebras 
\cite{5Yang} (Chapter \ref{hubbardmodel}), our Hilbert subspace is in the present parameter 
space spanned only by the lowest-weight states (LWS's) of these 
algebras \cite{5Nuno3} (Chapter \ref{harmonic}). 
This refers to the Hilbert subspace directly
described by the BA solution \cite{5CarmeloNuno97}. [Therefore,
following the studies and notations of Chapter \ref{harmonic}
\cite{5CarmeloNuno97}, 
we can use the $\alpha,\gamma =0$ pseudoparticles instead of the
$\alpha,\beta$ pseudoholes (with $\beta=\pm {1\over2}$) required 
for the description of the non-LWS's outside the BA solution.]

The $c,0$ and $s,0$ pseudoparticle branches have been called 
in previous low-energy studies $c$ and $s$ pseudoparticle branches, 
respectively \cite{5Carm4a,5Carm4b}. They were shown to describe the 
low-energy excitations of the Hubbard chain and to determine the 
low-energy behavior of its charge and spin transport properties 
\cite{5Carm4a,5Carm4b}. (In the limit of low energy the 
heavy-pseudoparticle branches are empty.) On the other hand, description 
of the LWS's of the model that have a finite-energy gap, $\omega_0$, 
relatively to the ground state involves the heavy pseudoparticle 
branches $c,\gamma>0$ and $s,\gamma>0$ 
(Chapter \ref{harmonic}) \cite{5CarmeloNuno97}.

A very useful concept in this theory is that of generalized ground 
state (GGS). In Chapter \ref{harmonic} it was defined as the 
Hamiltonian eigenstate(s) of lowest energy in the Hilbert subspace 
associated with a given sub-canonical ensemble. The concept of 
sub-canonical ensemble follows from the conservation laws of the 
$\alpha ,\gamma $ pseudoparticle numbers, $N_{\alpha ,\gamma}$. 
Each Hamiltonian eigenstate has constant values for these numbers 
and a sub-canonical ensemble refers to a given choice of constant 
$N_{\alpha ,\gamma}$ numbers. 

On the other hand,  
the concept of GGS  has been recently extended \cite{5GCFT1,5GCFT2}
as follows (i) filled $\alpha ,\gamma$ pseudoparticle seas with 
compact occupations around $q=0$, i.e. for 
$q^{(-)}_{F\alpha ,\gamma,+1}\leq q\leq q^{(+)}_{F\alpha ,\gamma,+1}$,
where the pseudo-Fermi points are given by $q^{(\pm)}_{F\alpha 
,\gamma,+1}=\pm {\pi N_{\alpha ,\gamma}\over N_a}+O(1/N_a)$, and (ii)
filled $\alpha ,\gamma$ pseudoparticle seas with compact occupations 
for $q^{(-)}_{\alpha ,\gamma}\leq q\leq q^{(-)}_{F\alpha ,\gamma,-1}$
and for $q^{(+)}_{F\alpha ,\gamma,-1}\leq 
q\leq q^{(-)}_{\alpha ,\gamma}$,
where the pseudo-Fermi points are given by 
$q^{(\pm)}_{F\alpha ,\gamma,-1}=\pm [q_{\alpha ,\gamma} - 
{\pi N_{\alpha ,\gamma}\over N_a}]+O(1/N_a)$. 
Recent developments on the problem \cite{5GCFT1,5GCFT2,5conductivity}, have revealed that the creation of one $\alpha,\gamma$
pseudoparticle from the ground state involves, to
leading order, a number $2\gamma$ of electrons. Since the
currents are two-electron operators, it follows that
the creation of single $\alpha,1$ pseudoparticles from
the ground state are the most important contributions to the
transport of charge ($\alpha=c$) and spin ($\alpha=s$) at finite energies. On the other
hand, since the states $\hat{j}^\zeta|GS\rangle$ which
define our Hilbert space have zero momentum
(relatively to the ground state) and the creation 
from the ground state of single $\alpha,1$ pseudoparticles of type
(ii) is a finite-momentum excitation, for simplicity
in this chapter we restrict our study to GGS's 
of type (i). Therefore, in order to simplify our notation
we denote the pseudo-Fermi points $q^{(\pm)}_{F\alpha ,\gamma,+1}$
simply by $q^{(\pm)}_{F\alpha ,\gamma}$. These are given by
$q^{(\pm)}_{F\alpha ,\gamma}=\pm q_{F\alpha ,\gamma}+O(1/N_a)$
where the pseudo-Fermi momentum (Chapter \ref{harmonic})
\cite{5CarmeloNuno97} 

\begin{equation}
q_{F\alpha ,\gamma} = {\pi N_{\alpha ,\gamma}\over N_a} \, ,
\label{5qF}
\end{equation}
appears in several expressions below. Note, however, that the generalization
of our results to GGS's of type (ii) is straightfoward. We emphasize 
that in Chapter \ref{harmonic} the definition of GGS refers to the
above choice (i) for the $c,0$ and $s,\gamma$ branches
and to the choice (ii) for the $c,\gamma$ branch with $\gamma >0$.
Therefore, in the case of the $c,\gamma >0$ pseudoparticles
our GGS choice differs from the choice of that chapter.

The ground state is a special case of a GGS where there is no 
$\alpha,\gamma >0$ heavy-pseudoparticle occupancy 
\cite{5CarmeloNuno97} and the pseudo-Fermi points (\ref{5qF})
are of the form

\begin{equation}
q_{Fc ,0} = 2k_F \, ; \hspace{1cm } q_{Fs ,0} = k_{F\downarrow} 
\, ; \hspace{1cm } q_{F\alpha,\gamma } = 0 \hspace{0.5cm} 
\gamma >0 \, .
\label{5gspfs}
\end{equation}

Since the conservation of the electron numbers imposes the following
sum rules on the numbers $N_{\alpha,\gamma}$ 
(Chapter \ref{harmonic}) \cite{5CarmeloNuno97}

\begin{equation}
N_\downarrow = \sum_{\gamma >0}\gamma N_{c,\gamma} 
+ \sum_{\gamma}(1+\gamma)N_{s,\gamma}\, ,
\label{5nbaixo}
\end{equation}
and

\begin{equation}
N = N_{c,0} + 2\sum_{\gamma>0}\gamma N{c,\gamma}\, ,
\label{5nc}
\end{equation}
the creation of heavy pseudoparticles from the ground state
at constant electron numbers requires the annihilation
of $\alpha,0$ pseudoparticles. It follows from Eqs.
(\ref{5nbaixo}) and (\ref{5nc}) that the changes 
$\Delta N_{\alpha,0}$ associated with a 
corresponding ground-state -- GGS transition read

\begin{equation}
\Delta N_{s,0} = -\sum_{\gamma >0}\gamma N_{c,\gamma} 
- \sum_{\gamma >0}(1+\gamma)N_{s,\gamma}\, ,
\label{5DNs0}
\end{equation}
and

\begin{equation}
\Delta N_{c,0} = - 2\sum_{\gamma>0}\gamma N{c,\gamma}\, .
\label{5DNc0}
\end{equation}
For instance, the creation of one $c,\gamma $ heavy 
pseudoparticle from the ground state requires the annihilation
of a number $2\gamma $ of $c,0$ pseudoparticles and of a
number $\gamma $ of $s,0$ pseudoparticles, whereas the
creation of one $s,\gamma$ pseudoparticle involves the
annihilation of a number $1+\gamma$ of $s,0$ pseudoparticles
and conserves the number of $c,0$ pseudoparticles.

\section{Charge and spin currents: solution of the BA
Equations}
\label{5baphi}
\pagestyle{myheadings}
\markboth{6  Zero-Temperature Transport}{6.3 Charge and spin currents ...}

Within linear response theory the charge and spin currents 
of the 1D Hubbard model can be computed by performing
a spin-dependent Peierls-phase substitution \cite{5Maldague,5Shastry}
in the hopping integral of Hamiltonian (\ref{hamiltso4}), 
$t \rightarrow te^{\pm i\phi_{\sigma}/N_a}$, leading to Hamiltonian
(\ref{hamilt3}).

It has been possible to solve the Hamiltonian (\ref{hamiltso4}) 
with the additional hopping phase $e^{\pm i\phi_{\sigma}/N_a}$
by means of the coordinate BA both with twisted and toroidal 
boundary conditions, both approaches giving essentially the 
same results \cite{5Shastry,5Martins}. One obtains the energy 
spectrum of the model parameterized by a set a numbers 
$\{k_j,\Lambda_\delta\}$ which are solution of the BA
interaction equations given by

\begin{equation}
e^{ik_j N_a}=e^{i\phi_{\uparrow}}\prod_{\delta=1}^{N_\downarrow}
\frac{\sin(k_j)-\Lambda_\delta+iU/4}{\sin(k_j)-\Lambda_\delta-
iU/4}\, ,
\hspace{1cm} (j=1,\ldots,N)\, ,
\label{5inter1}
\end{equation}
and
\begin{equation}
\prod_{j=1}^{N}
\frac{\sin(k_j)-\Lambda_\delta+iU/4}{\sin(k_j)-\Lambda_\delta-iU/4}
=e^{i(\phi_{\downarrow}-\phi_{\uparrow})}
\prod_{\beta=1,\neq\delta}^{N_\downarrow}
\frac{\Lambda_\beta-\Lambda_\delta+iU/2}
{\Lambda_\beta-\Lambda_\delta-iU/2}\, ,
\hspace{1cm} (\delta=1,\ldots,N_\downarrow)\, .
\label{5inter2}
\end{equation}
These equations reduce to Eqs.\,(\ref{inter1}) and (\ref{inter2}) with $\phi_{\uparrow}=\phi_{\downarrow}$.

However, previous studies of the $\phi_\sigma\neq 0$ problem 
\cite{5Shastry,5Martins} have only considered the real
BA rapidities of Eqs. (\ref{5inter1}) and (\ref{5inter2})
which refer to low energy. Here we follow the same steps as 
Takahashi \cite{5Takahashi} for the $\phi_\sigma= 0$ 
Eqs.\,(\ref{5inter1}) and (\ref{5inter2}) and consider both
real and complex rapidities. We then arrive to the 
following $\phi_\sigma\neq 0$ equations which refer to
the real part of these rapidities 

\begin{eqnarray}
k_jN_a &=&
2\pi I_j^c+\phi_\uparrow
-\sum_{\gamma }\sum_{j'=1}^{N_{s,\gamma}}
2\tan^{-1}\left(\frac{\sin(k_j)/u-R_{s,\gamma,j'}}{(\gamma+1)} \right)
\nonumber \\
&-& \sum_{\gamma >0}\sum_{j'=1}^{N_{c,\gamma}}
2\tan^{-1}\left(\frac{\sin(k_j)/u-R_{c,\gamma,j'}}{\gamma}\right)\, ,
\label{5tak1}
\end{eqnarray}

\begin{eqnarray} 
2N_a Re \, \sin^{-1}([R_{c,\gamma,j}+i\gamma]u)
&=& 2\pi I^{c,\gamma}_j
+ \gamma (\phi_\uparrow+\phi_\downarrow)\nonumber\\
&-&\sum_{j'=1}^{N_c}2\tan^{-1}\left(\frac{\sin(k_{j'})/u-R_{c,\gamma,j}}
{\gamma}\right)\nonumber\\
&+&\sum_{\gamma'>0}\sum_{j'=1}^{N_{c,\gamma'}}
\Theta_{\gamma,\gamma'}(R_{c,\gamma,j}-R_{c,\gamma',j'})\, ,
\label{5tak2}
\end{eqnarray}
and

\begin{eqnarray}
&&\sum_{j'=1}^{N_c}2\tan^{-1}\left(\frac{R_{s,\gamma,j}-\sin(k_j')/u}
{(1+\gamma)}\right) + (1+\gamma)(\phi_\uparrow-\phi_\downarrow)=
\nonumber \\
&=& 2\pi I^{s,\gamma}_j+
\sum_{\gamma'}\sum_{j'=1}^{N_{s,\gamma'}}
\Theta_{\gamma+1,\gamma'+1}(R_{s,\gamma,j}-R_{s,\gamma',j'})\, .
\label{5tak3}
\end{eqnarray}
The functions $\Theta_{\gamma,\gamma'}(x)$ [and 
$\Theta_{\gamma+1,\gamma'+1}(x)$] of Eqs.
\,(\ref{5tak1}), (\ref{5tak2}), and (\ref{5tak3}) are
defined by Eq.\,(\ref{ftheta1}). The following definitions
for the real part of the rapidities, 
$\Lambda_\alpha^{n+1}/u=R_{s,\gamma,j}$ (with $n+1=\gamma$ and 
$\alpha=j$), $\Lambda_\alpha^{' \, n}/u=R_{c,\gamma,j}$ 
(with $n=\gamma$ and $\alpha=j$), and $\gamma=1,2,\ldots,\infty$ 
for the $N_{c,\gamma}$ sums and $\gamma=0,1,2,\ldots,\infty$ for 
the $N_{s,\gamma}$ sums, allows us to recover Takahashi's 
formulae for $\phi=0$ \cite{5Takahashi}. Here and often below
we use the notation $c\equiv c,0$, which allows the $c,\gamma$
sums to run over $1,2,3,\ldots,\infty$. Whether we are using this 
notation or the previous one will be obvious from the context.

The important numbers $I_j^c$, $I^{c,\gamma}_j$, and 
$I^{s,\gamma}_j$ which appear when going from Eqs.\,(\ref{5inter1}) 
and (\ref{5inter2}) to Eqs.\,(\ref{5tak1}), (\ref{5tak2}), and 
(\ref{5tak3}) are the quantum numbers which describe
the Hamiltonian eigenstates. 
As we have seen in Chapter \ref{harmonic},
it is convenient to describe the eigenstates of the model  
in terms of pseudomomentum $\{q_j^{\alpha,\gamma}=
2\pi I_j^{\alpha,\gamma}/N_a\}$ distributions, where 
$I_j^{c,0} \equiv I_j^c$. The energy and momentum eigenvalues are given by \cite{5CarmeloNuno97}

\begin{eqnarray}
E&=&-2t\sum_{j=1}^{N_c}\cos(k_j)+
\sum_{\gamma >0}\sum_{j=1}^{N_{c,\gamma}}4t Re
\sqrt{1-u^2[R_{c,\gamma,j}-i\gamma]^2}\nonumber\\
&+&N_a(U/4-\mu)+N(\mu-U/2)-\mu_0 H(N_\uparrow-N_\downarrow),
\label{5energy}
\end{eqnarray}
and

\begin{equation}
P = \frac{2\pi}{N_a}\left[\sum_{j=1}^{N_c}(I_j^c+\frac{\phi_{\uparrow}}{N_a})+
\sum_{\gamma}\sum_{j=1}^{N_{s,\gamma}}(I_j^{s,\gamma}-(\gamma+1)\frac{\phi_{\uparrow-\downarrow}}{N_a})
-\sum_{\gamma >0}\sum_{j=1}^{N_{c,\gamma}}(I_j^{c,\gamma}+
\gamma\frac{\phi_{\uparrow+\downarrow}}{N_a})\right]
+\pi \sum_{\gamma >0} N_{c,\gamma} \, ,
\label{5momentum}
\end{equation}
respectively. Eqs.\,(\ref{5energy}) and (\ref{5momentum})
are the eigenvalues of Eqs.\,(\ref{hamiltbabeta}) and (\ref{pmomentum}). 
We emphasize that the rapidity dependence on
$\phi_{\sigma}$ is defined by Eqs. (\ref{5tak1})-(\ref{5tak3}) 
and determines the energy-functional (\ref{5energy}) dependence on 
$\phi_{\sigma}$. The corresponding $\phi_{\sigma}=0$ expressions 
recover the rapidity equations (\ref{5tak1}), (\ref{5tak2}), and (\ref{5tak3}).

In the limit of a large system ($N_a\rightarrow \infty$, $N/N_a$ fixed), and using the same methods as in Chapters \ref{harmonic} and \ref{PseudoPT},
we can develop a generalization of the low-energy 
pseudoparticle-Landau-liquid description of the Hubbard model. This generalization refers to energies just 
above the set of energies $\omega_0$ given by

\begin{equation}
	\omega_0=2\mu\sum_{\gamma >0} \gamma N_{c,\gamma}
	+ 2\mu_0H\sum_{\gamma >0}(1+\gamma) N_{s,\gamma}+
	\sum_{\alpha,\gamma>0}\epsilon^0_{\alpha,\gamma}(0)
	N_{\alpha,\gamma} \, ,
\label{5gap}
\end{equation}
where the
Hamiltonian (\ref{hamiltinf}) truncated up to second
order describes the quantum-problem at $(\omega-\omega_0)$ low energies. Note that the choice $\omega_0=0$, which refers 
to $N_{\alpha,\gamma}=0$
for $\gamma >0$, recovers the usual low-energy theory
\cite{5Carm4a,5Carm4b}. On the other hand,
when $\omega_0>0$, in addition to finite occupancy of 
the usual $c,0\equiv c$ and $s,0\equiv s$ pseudoparticle bands, 
there is finite occupancy for some of the branches of the  
$c,\gamma$ and $s,\gamma$ heavy-pseudoparticles \cite{5CarmeloNuno97}. 

In the above thermodynamic limit the rapidity real parts
$k_j=k_j(q_j)$, $R_{s,\gamma,j}=R_{s,\gamma,j}(q_j)$, and 
$R_{c,\gamma,j}=R_{c,\gamma,j}(q_j)$ proliferate on the 
real axis. As we have done in Chapters \ref{harmonic}
and \ref{PseudoPT},
Eqs. \,(\ref{5tak1}), (\ref{5tak2}), and (\ref{5tak3}) can be 
rewritten as integral equations with an explicit dependence on the 
pseudomomentum distribution functions $N_{\alpha,\gamma}(q)$. These 
are Eqs.\, (\ref{5int1}), (\ref{5int2}), and (\ref{5int3}) of
Appendix\,\ref{5normalOrder} which refer to the case
$\phi_{\sigma}\neq 0$. In that Appendix we derive ground-state 
normal-ordered expressions for the rapidities and charge
and spin currents. 

The combination of Eqs.\,(\ref{5energy}), (\ref{5int1}), (\ref{5int2}), 
and (\ref{5int3}) allows the evaluation of several interesting 
transport quantities. This includes the charge and spin currents 
and the charge and spin pseudoparticle transport masses. The 
charge and spin current 
operators $\hat{j}^\zeta$ (with $\zeta=\rho$ for charge, and 
$\zeta=\sigma_z$ for spin) are for the 1D Hubbard model
given by \cite{5Carm4a,5Carm4b}

\begin{equation}
	\hat{j}^\rho=-eit\sum_{\sigma}\sum_{j=1}^{N_a}
	(c_{j\sigma}^{\dag}c_{j+1\sigma}-
	c_{j+1\sigma}^{\dag }c_{j\sigma})\, ,
\label{5currrho}
\end{equation}
and
\begin{equation}
	\hat{j}^{\sigma_z}=-(1/2)it\sum_{\sigma}\sum_{j=1}^{N_a}
	\sigma(c_{j\sigma}^{\dag}c_{j+1\sigma}-
	c_{j+1\sigma}^{\dag}c_{j\sigma})\, .
\label{5currsigma}
\end{equation}
Importantly, the discrete nature of the model 
implies that the commutators of the Hamiltonian (\ref{hamiltso4}) 
and of the current operators $\hat{j}^\zeta$,
Eqs. (\ref{5currrho}) and (\ref{5currsigma}), are non zero. It 
follows that the BA wave function 
does not diagonalizes simultaneously the Hamiltonian 
(\ref{hamiltso4}) and the current operators (\ref{5currrho}) 
and (\ref{5currsigma}). Since the BA solution alone only provides 
the diagonal part in the Hamiltonian-eigenstate basis
of the physical operators \cite{5CarmeloNuno97},
we can only evaluate expressions for the diagonal part
of the currents which provide the mean values of the 
charge and spin currents. These refer to all LWS's and are important 
quantities for they allow us to compute the transport masses of 
the charge and spin carriers of the system. In addition, our 
formalism defines the charge and spin carriers. These are found 
to be the $c$ (i.e., $c,0$) and $c,\gamma$ pseudoparticles for charge, and 
the $c$ and $s,\gamma$ pseudoparticles for spin. 
This follows from Eqs.\,(\ref{5int1})-(\ref{5int3})
and also from the Boltzmann transport analysis of 
Sec.\,\ref{5kinetic}. 

We emphasize that combining the generalized pseudoparticle
representation of Chapter \ref{harmonic} 
\cite{5CarmeloNuno97} with a low-energy 
$(\omega -\omega_0)$ conformal-field theory \cite{5GCFT1,5GCFT2},
leads to finite-energy current -- current correlation function 
expressions which are determined by the non-diagonal terms 
(in the Hamiltonian-eigenstate basis) of the current operators.
This is a generalization of the low-energy correlation-function
studies of several authors \cite{5Carm593,5Carm597,5Frahm90,5Frahm91}. 
However, these expressions cannot be derived within the BA solution
alone. Therefore, these studies go beyond the scope of the 
present thesis and here we consider the diagonal part of the 
charge and spin current operators only.

The mean value of the current operator $\hat{j}^\zeta$
evaluated for the LWS $\vert m\rangle$ is given by

\begin{equation}
\langle m\vert \, \hat{j}^\zeta\vert m\rangle =-\left.
\frac{d(E_m/N_a)}{d(\phi/N_a)} \right\vert_{\phi=0} \, ,
\label{5mean}
\end{equation}
where $E_m$ is the energy of the Hamiltonian eigenstate 
$\vert m\rangle$ and \cite{5Shastry}

\begin{eqnarray}
\phi & = & \phi_\uparrow=\phi_\downarrow \, ,
\hspace{1cm}  \zeta=\rho \, ,\nonumber \\
\phi & = & \phi_\uparrow=-\phi_\downarrow \, ,
\hspace{1cm}  \zeta=\sigma_z \, .
\label{5phis}
\end{eqnarray}
 
In our basis the LWS's are simply obtained 
by considering all the possible occupation distributions of the 
pseudomomenta $q_j =2\pi I_j^{\alpha,\gamma}/N_a$. Therefore, it is 
convenient to describe the matrix elements 
$\langle m\vert \, \hat{j}^\zeta\vert 
m\rangle $ in terms of the pseudomomentum occupation
distributions $N_{\alpha,\gamma}(q)$. This leads
to a functional form for the current mean values. 
The computation of $\langle m\vert \, \hat{j}^\zeta\vert m\rangle$ 
involves the expansion of Eqs. (\ref{5energy}) and 
(\ref{5int1})-(\ref{5int3}) up to first order in the flux
$\phi$. Writing Eq.\,(\ref{5energy})
in the limit of $N_a\rightarrow\infty$, expanding it up to first order
in the flux $\phi$, and using Eq.\,(\ref{5mean}) we obtain

\begin{eqnarray}
&&\langle m\vert \, \hat{j}^\zeta\vert m\rangle = -2t\frac 1{2\pi}
\int_{-q_c}^{q_c}dqN_c(q)K^{\phi}(q)\sin(K(q))\nonumber\\
&+&\sum_{\gamma >0}4t\frac 1{2\pi}
\int_{-q_{c,\gamma}}^{q_{c,\gamma}}dq N_{c,\gamma}(q)
Re\, \frac{u^2[R_{c,\gamma}(q)-i\gamma]}
{\sqrt{1-u^2[R_{c,\gamma}(q)-i\gamma]^2}}
R_{c,\gamma}^{\phi}(q)\, ,
\label{5jmean}
\end{eqnarray}
where the important functions $W^{\phi}(q)$ (with $W=K, R_{s,\gamma}$, 
and $R_{c,\gamma}$) are the derivatives of the rapidity functions
defined by Eqs. (\ref{5int1}) - (\ref{5int3}) in order to the flux 
$\phi$ at $\phi=0$. They obey a set of integral equations obtained 
from differentiation of Eqs. (\ref{5int1}) - (\ref{5int3}).

It is convenient to write $\langle m\vert \, 
\hat{j}^\zeta\vert m\rangle$ in normal order relatively to 
the ground state. To achieve this goal we expand 
all the rapidities $W(q)$ and the functions $W^{\phi}(q)$ as

\begin{eqnarray}
&&W(q)=W^0(q)+W^1(q)+\ldots\, ,
\label{5exp1}\\
&&W^{\phi}(q)=W^{0,\phi}(q)+W^{1,\phi}(q)+\ldots\, ,
\label{5exp2}
\end{eqnarray}
respectively. In these equations the functions $W^0(q)$ and 
$W^{0,\phi}(q)$ are both referred to the ground state, and 
the functions $W^1(q)$ and $W^{1,\phi}(q)$ are first-order 
functionals of the deviations $\delta N_{\alpha,\gamma}(q)$.
In Appendix \ref{5normalOrder} we show that the above expansions lead
to a ground-state normal-ordered representation. To first order in
the deviations, the normal-ordered expression for the matrix element 
(\ref{5jmean}) simply reads

\begin{equation}
\langle m\vert \, \hat{j}^\zeta\vert m\rangle=\sum_{\alpha}
\sum_{\gamma}
\int_{-q_{\alpha,\gamma}}^{q_{\alpha,\gamma}}dq
\delta N_{\alpha,\gamma}(q)
j^\zeta_{\alpha,\gamma}(q)\, ,
\label{5deltaj}
\end{equation}
where the elementary-current spectrum $j_{\alpha,\gamma}^\zeta(q)$ is 
given by

\begin{equation}
	j_{\alpha,\gamma}^\zeta(q)=
	\sum_{\alpha'}\sum_{\gamma'}
	\theta(N_{\alpha',\gamma'})
	{\cal C}^{\zeta}_{\alpha',\gamma'}\left[
	v_{\alpha,\gamma}(q)\delta_{\alpha ,\alpha'}
	\delta_{\gamma ,\gamma'} +
	F^1_{\alpha,\gamma;\alpha',\gamma'}(q)\right]\, .
\label{5jq}
\end{equation}
Here

\begin{equation}
	F^1_{\alpha,\gamma;\alpha',\gamma'}(q)={1\over 2\pi}
	\sum_{j=\pm 1}jf_{\alpha,\gamma;\alpha'
	,\gamma'}(q,jq_{F\alpha',\gamma'})\, ,
\label{5Fq}
\end{equation}
and ${\cal C}^{\zeta}_{\alpha,\gamma}$ are the coupling constants 
of the pseudoparticles to charge and spin given by

\begin{equation}
{\cal C}^{\zeta}_{\alpha,\gamma} = \delta_{\alpha,c}\delta_{\gamma,0}
+ {\cal K}^{\zeta}_{\alpha,\gamma} \, ,
\label{5coupling}
\end{equation}
where

\begin{equation}
{\cal K}_{\alpha,\gamma}^{\rho}=\delta_{\alpha,c}2\gamma \, ;
\hspace{1cm}
{\cal K}_{\alpha,\gamma}^{\sigma_z}=
-\delta_{\alpha,s}2(1+\gamma) \, .
\label{5coupling2}
\end{equation}

As in a Fermi liquid \cite{5Pine,5Baym}, the expressions of the 
elementary currents (\ref{5jq}) involve the velocities 
$v_{\alpha,\gamma}(q)$ and the interactions [or $f$-functions] 
$f_{\alpha,\gamma;\alpha',\gamma'}(q,q')$. However, the 
pseudoparticle coupling contants to charge and spin, Eqs. 
(\ref{5coupling}) and (\ref{5coupling2}), are very different
from the corresponding couplings of the Fermi-liquid 
quasiparticles. We emphasize that at low energy
Eq. (\ref{5deltaj}) recovers the expression already obtained
in previous works \cite{5Carm4a,5Carm4b} which only contains the $c\equiv c,0$ and
$s\equiv s,0$ elementary currents. The coupling constants 
(\ref{5coupling})-(\ref{5coupling2})
play an important role in the description of charge and
spin transport and are a generalization for
$\gamma >0$ of the couplings introduced previously
\cite{5Carm4a,5Carm4b}. They define the $\alpha,\gamma$ pseudoparticles 
as charge and spin carriers. We emphasize that
when ${\cal C}^{\zeta}_{\alpha,\gamma}=0$ the corresponding 
$\alpha,\gamma$ pseudoparticles do not couple to $\zeta $
(i.e. charge or spin). Therefore, for $\gamma >0$ the
$c,\gamma$ and $s,\gamma$ pseudoparticles do not couple to 
spin and charge, respectively. This is related to the
charge and spin separation of one-dimensional quantum
liquids. Importantly, when 
${\cal C}^{\zeta}_{\alpha,\gamma}=0$ the $\alpha,\gamma$
pseudoparticle -- pseudohole processes do not contribute
to the $\zeta $ correlation functions.

In contrast to the general current expression
(\ref{5jmean}), expression (\ref{5deltaj}) is only
valid for Hamiltonian eigenstates which differ from
the ground-state pseudoparticle occupancy by a small
density of psudoparticles. This is because in expression
(\ref{5deltaj}) we are only considering the first-order 
deviation term.

The velocity term of the current-spectrum expression \,(\ref{5jq}) 
is what we would expect for a 
non-interacting 
gas of pseudoparticles and the 
second term takes account for the dragging effect on a single 
pseudoparticle due to its interactions with the remaining
pseudoparticles. (This is similar to the Fermi-liquid 
quasiparticle elementary currents \cite{5Pine,5Baym}.) 
We remind that Eq.\,(\ref{5jq}) is valid for 
finite energies $\omega $ just above the energy $\omega_0$ 
corresponding to the suitable set of finite $N_{\alpha,\gamma}$ 
numbers. These numbers characterize the state 
$\vert m\rangle$. Therefore, the sum over $\gamma$ is
in Eq.\,(\ref{5jq}) restricted to the $\alpha,\gamma$ bands that have 
non-zero occupancy of pseudoparticles, as is imposed by the
presence of the step function. Within the PPT, the deviation
second-order pseudoparticle energy expansion corresponds
to the deviation first-order current expansion (\ref{5deltaj})
which refers to small positive values $(\omega -\omega_0)$ 
of the excitation energy. In contrast to Fermi liquid theory, 
our PPT is valid for finite energies [just above the 
energy values $\omega_0$, Eq. (\ref{5gap})] because
(i) there is only forward scattering among the pseudoparticles 
at all energy scales and (ii) at low $(\omega-\omega_0)$ 
energy values only two-pseudoparticle forward-scattering 
interactions are relevant. (In a Fermi liquid this
is only true for $\omega_0=0$ and $\omega\rightarrow 0$
\cite{5Pine,5Baym}.) We emphasize that the current expression
(\ref{5jmean}) includes all orders of scattering and, therefore,
applies to all energies without restrictions. 

\section{Pseudoparticle static and transport masses}
\label{5masses}
\pagestyle{myheadings}
\markboth{6  Zero-Temperature Transport}{6.4 Pseudoparticle static and ...}

In previous studies \cite{5Carm4a,5Carm4b} the charge and spin transport
masses of the $c,0$ and $s,0$ pseudoparticles were defined
and were shown to play an important role in the transport
of charge and spin.
For instance, they were shown to fully determine
the charge and spin stiffnesses
\cite{5Carm4a,5Carm4b,5Shastry,5Stafford,5Fye,5Millis,5Scalapino}.
Here we generalize the mass definitions of these studies to 
$\gamma >0$ and define the charge and spin transport masses, 
$m_{\alpha,\gamma}^{\zeta}$, as

\begin{equation} 
m^{\zeta}_{\alpha,\gamma}=
\frac {q_{F\alpha,\gamma}} {{\cal C}^{\zeta}_{\alpha,\gamma}
j^\zeta_{\alpha,\gamma}}\, ,
\label{5tmass}
\end{equation}
where $j^\zeta_{\alpha,\gamma}=j^\zeta_{\alpha,\gamma}
(q_{F\alpha,\gamma})$. These masses contain important physical information. 
As in a Fermi liquid \cite{5Pine,5Baym}, the ratio

\begin{equation} 
r^{\zeta}_{\alpha,\gamma} = 
m^{\zeta}_{\alpha,\gamma}/m^{\ast}_{\alpha,\gamma} \, ,
\label{5ratio}
\end{equation}
of the transport mass, $m^{\zeta}_{\alpha,\gamma}$, over 
the static mass, $m^{\ast}_{\alpha,\gamma}$, provides a measure 
of the electronic correlation importance in transport. Similarly to the 
$\gamma =0$ case \cite{5Carm4a,5Carm4b}, the latter is in general defined as

\begin{equation} 
m^{\ast}_{\alpha,\gamma}=
\frac {q_{F\alpha,\gamma}} {v_{\alpha,\gamma}}\, .
\label{5smass}
\end{equation}
In Appendix \ref{5static} we define the mass (\ref{5smass})
in terms of suitable functions and find some limiting
expressions.
\begin{figure}[htbp]
\begin{center}
\leavevmode
\hbox{%
\psfig{figure=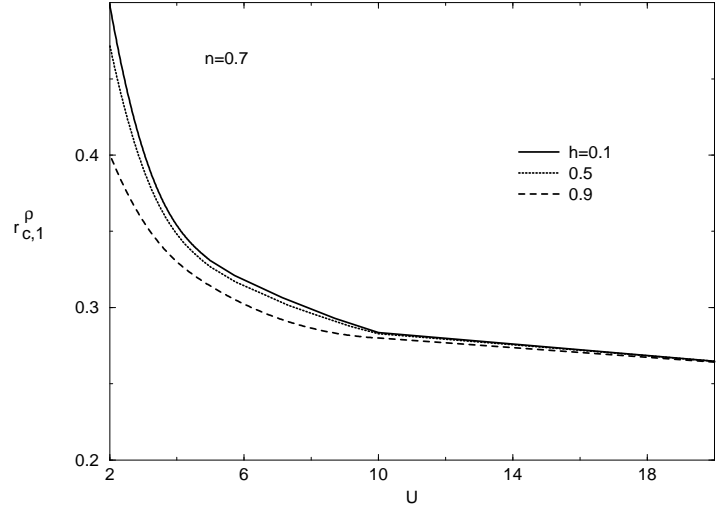,width=4.0in,angle=-90}}
\end{center}
\caption{The ratio $m_{c,1}^{\rho}/m_{c,1}^{\ast}$ as function of $U$ 
at electronic density $n=0.7$ and for values of the magnetic field 
$h=H/H_c=0.1$ (full line),
$h=0.5$ (dashed line) and $h=0.9$ (dashed-dotted line). 
For other electronic densities, the plots follow the same trends as for
$n=0.7$.}
\label{5fig:fig1}
\end{figure}

It can be shown from the transport- and static-mass
expressions that the ratio $m^{\zeta}_{\alpha,\gamma}
/m^{\ast}_{\alpha,\gamma}$ involves the Landau parameters

\begin{equation}
F^i_{\alpha,\gamma;\alpha',\gamma'}\equiv 
F^i_{\alpha,\gamma;\alpha',\gamma'}(q_{F\alpha,\gamma}) \, ;
\hspace{1cm} i = 0,1 \, ,
\label{5Fi}
\end{equation}
with $F^i_{\alpha,\gamma;\alpha',\gamma'}(q)$ given by Eq.
(\ref{5Fq}). These parameters can be written as follows

\begin{equation}
F^i_{\alpha,\gamma;\alpha',\gamma'} =
-\delta_{\alpha,\alpha'}\delta_{\gamma,\gamma'}v_{\alpha,\gamma}+
\sum_{\alpha'',\gamma''}\theta(N_{\alpha'',\gamma''}) 
v_{\alpha'',\gamma''}
[\xi^i_{\alpha'',\gamma'';\alpha,\gamma}
\xi^i_{\alpha'',\gamma'';\alpha',\gamma'}]\, ,
\end{equation}
where the quantities $\xi^i_{\alpha,\gamma;\alpha',\gamma'}$ are
given by

\begin{equation}
\xi^i_{\alpha,\gamma;\alpha',\gamma'}=
\delta_{\alpha,\alpha'}\delta_{\gamma,\gamma'}+
\sum_{l=\pm 1}l^i \Phi_{\alpha,\gamma;\alpha',\gamma'}
(q_{F\alpha,\gamma},lq_{F\alpha',\gamma'})\, .
\label{5xi}
\end{equation}
We find for the ratios $m^{\zeta}_{\alpha,\gamma}
/m^{\ast}_{\alpha,\gamma}$ the following expressions

\begin{equation}
\frac{m^{\zeta}_{\alpha,0}}{m^{\ast}_{\alpha,0}} =
\frac {v_{\alpha,0}}{{\cal C}^{\zeta}_{\alpha,\gamma}
(\sum_{\alpha',\alpha''}{\cal C}^{\zeta}_{\alpha',0}
v_{\alpha'',0}\xi^1_{\alpha'',0;\alpha,0}
\xi^1_{\alpha'',0;\alpha',0})} \, ,
\hspace{1cm}\gamma=0\, , 
\label{5ratio0}
\end{equation}
and

\begin{equation}
\frac{m^{\zeta}_{\alpha,\gamma}}{m^{\ast}_{\alpha,\gamma}} = \frac 1{
{\cal C}^{\zeta}_{\alpha,\gamma}({\cal C}^{\zeta}_{\alpha,\gamma}+
\sum_{\alpha'}{\cal C}^{\zeta}_{\alpha',0}\xi^1_{\alpha,\gamma;
\alpha',0})}\,, \hspace {1cm} \gamma > 0 \, . 
\label{5ratios}
\end{equation}
In the Table \ref{5tab} analytical limiting values for the mass ratios of form
(\ref{5ratio}) are listed. Obviously, since for $\gamma >0$ the $c,\gamma$ 
and $s,\gamma$ pseudoparticles do not couple to spin and charge,
respectively, the ratios $m_{c,\gamma}^{\sigma_z}/m^{\ast}_{c,\gamma}$ 
and $m_{s,\gamma}^\rho/m^{\ast}_{s,\gamma}$ are infinite.

\vspace{1cm}
\begin{table}[ht]
\begin{center}
\begin{tabular}{cccc}
\hline
 & $H\rightarrow H_c$ & $H\rightarrow 0$ & $n\rightarrow 1$\\
\hline
\hline
$m_{c,\gamma}^\rho/m^{\ast}_{c,\gamma}$ & $\frac 1
{2\gamma(2\gamma-\eta_{\gamma-1})}$ & 
$\frac 1
{2{\gamma}(2\gamma+\xi_{c,\gamma;c,0}^1)}$ & $\frac {1}{4{\gamma}^2}$\\ 
$m_{c,0}^\rho/m^{\ast}_{c,0}$ & $\frac {v_c}
{2t\sin(\pi n_{\uparrow})}$&
$\frac 1{(\xi_0)^2}$& $\infty$\\
$m_{s,\gamma}^{\sigma_z}/m^{\ast}_{s,\gamma}$& $\frac 1
{2(\gamma +1)(2\gamma +2-\eta_{\gamma})}$ & $\frac 1 
{2(\gamma +1)(2\gamma+2+2\xi^1_{s,\gamma;s,0})}$& 
$\frac 1 {2(\gamma +1)(2\gamma+2+2\xi^1_{s,\gamma;s,0}
-\xi^1_{s,\gamma;c,0})}$\\ 
\hline
\end{tabular}
\end{center}
\caption{Mass ratios in several limits of physical interest. 
The function $\eta_{\gamma }$ is defined as 
$\eta_{\gamma }=2/(\pi)\tan^{-1}[(\sin(n\pi))/(u[\gamma +1])]$. 
The equations for the static 
masses $m^{\ast}_{\alpha,\gamma}$ are given in Appendix 
\ref{5static}. In the case $H\rightarrow 0$, simple expressions
for the parameters $\xi_{\alpha,\gamma;\alpha',0}^1$, Eq. (\ref{5xi}), 
can be obtained from the results of Appendix \ref{5zeroH}. 
The ratios $m_{c,\gamma}^{\sigma_z}/m^{\ast}_{c,\gamma}$ 
and $m_{s,\gamma}^\rho/m^{\ast}_{s,\gamma}$ are infinite. The 
dependence on $U$ and $n$ of the parameter $\xi_0$ has been studied by
Frahm and Korepin \protect\cite{5Frahm90,5Frahm91}.}
\label{5tab}
\end{table}
\vspace{1cm}
\begin{figure}[htbp]
\begin{center}
\leavevmode
\begin{minipage}[b]{.32\linewidth}
\hbox{%
\psfig{figure=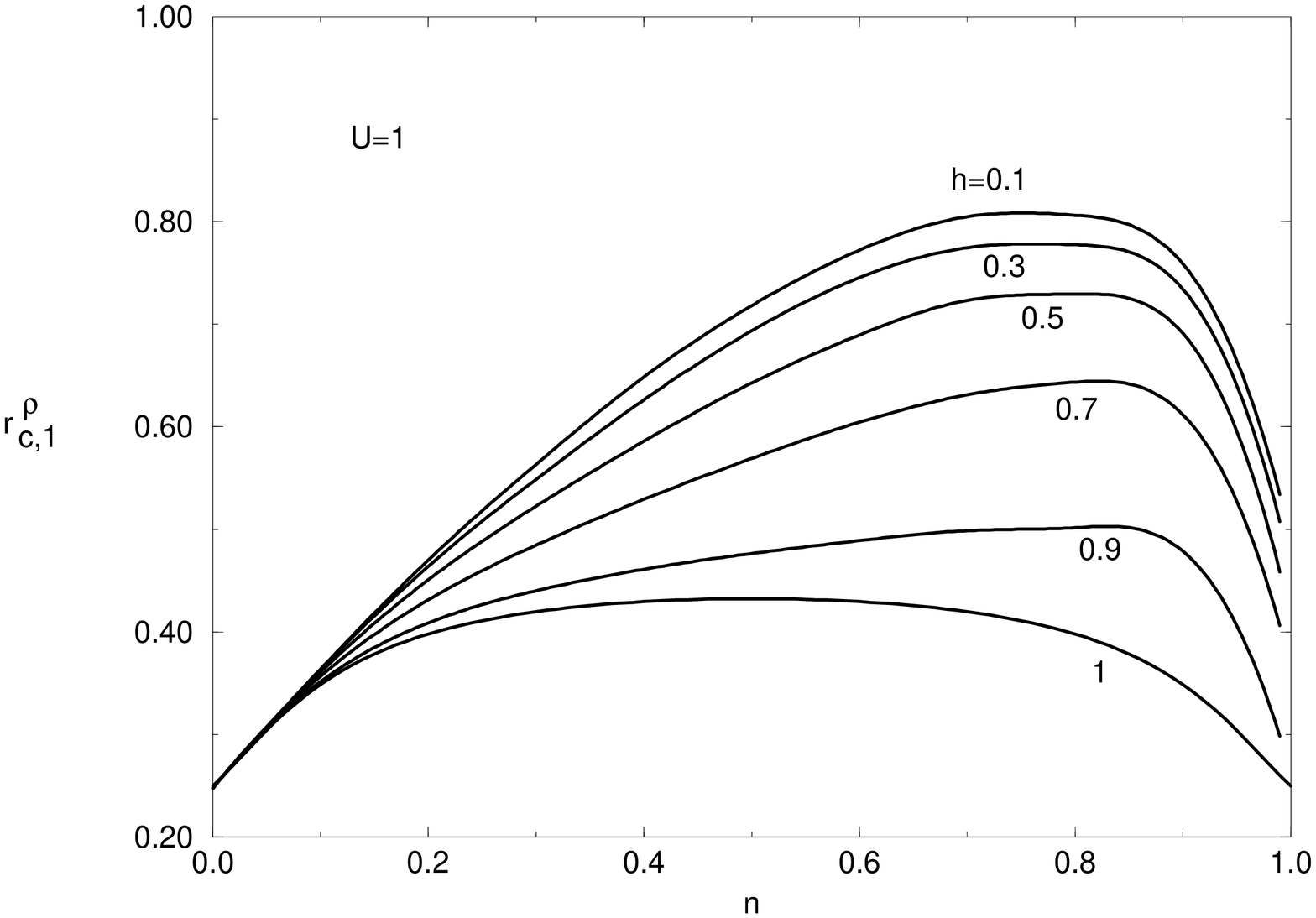,width=3in,angle=-90}} \centerline{(a)}
\end{minipage} \hfill %
\begin{minipage}[b]{.32\linewidth}
\hbox{%
\psfig{figure=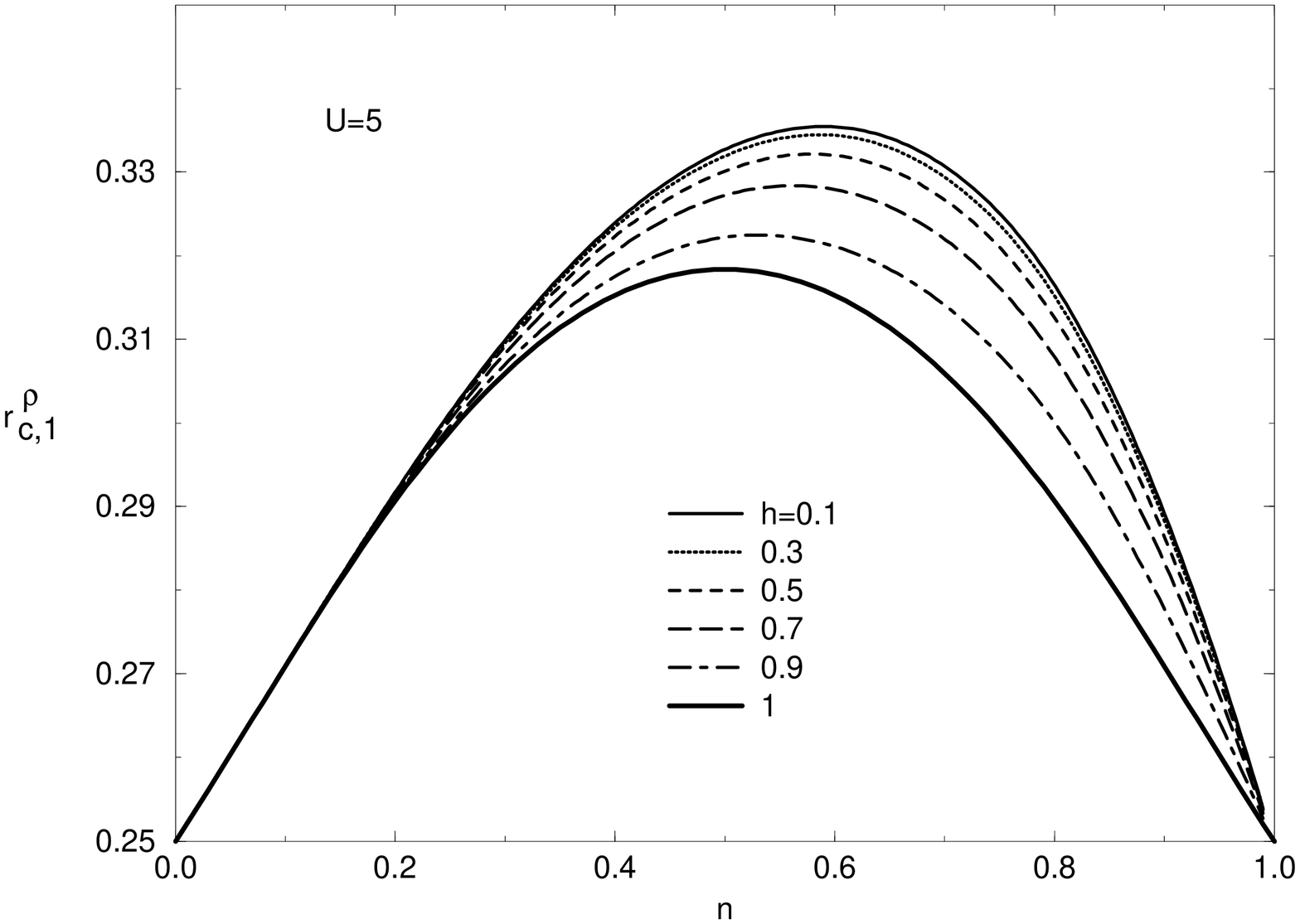,width=3in,angle=-90}}\centerline{(b)}
\end{minipage} %
\end{center}
\caption{The ratio $m_{c,1}^{\rho}/m_{c,1}^{\ast}$ as function of
the electronic density $n$ and for values of the magnetic field
$h=0.1$, $h=0.3$, $h=0.5$, $h=0.7$, and $h=0.9$. The onsite
Coulomb interaction is (a) $U=1$ and (b) $U=5$.}
\label{5fig:fig2}
\end{figure}
\begin{figure}[htbp]
\begin{center}
\leavevmode
\begin{minipage}[b]{.32\linewidth}
\hbox{%
\psfig{figure=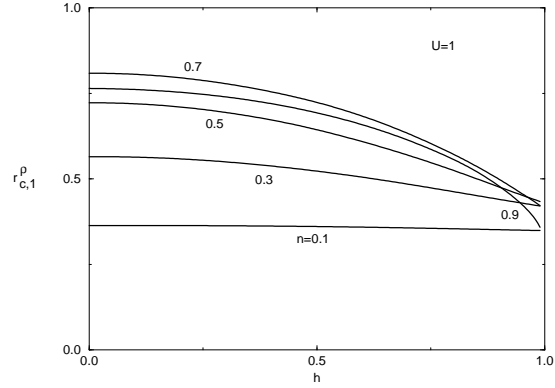,width=3in,angle=-90}} \centerline{(a)}
\end{minipage} \hfill %
\begin{minipage}[b]{.32\linewidth}
\hbox{%
\psfig{figure=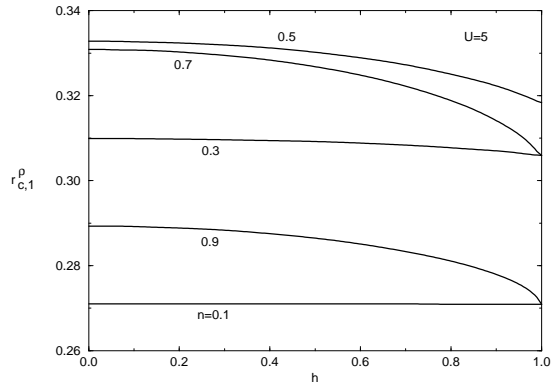,width=3in,angle=-90}}\centerline{(b)}
\end{minipage} %
\end{center}
\caption{The ratio $m_{c,1}^{\rho}/m_{c,1}^{\ast}$ as function of 
the the magnetic field $h$ and for values of the electronic density 
$n=0.1$, $n=0.3$, $n=0.5$, $n=0.7$, and $n=0.9$. The onsite
Coulomb interaction is (a) $U=1$ and (b) $U=5$.}
\label{5fig:fig3}
\end{figure}

\begin{figure}[htbp]
\begin{center}
\leavevmode
\begin{minipage}[b]{.48\linewidth}
\hbox{%
\psfig{figure=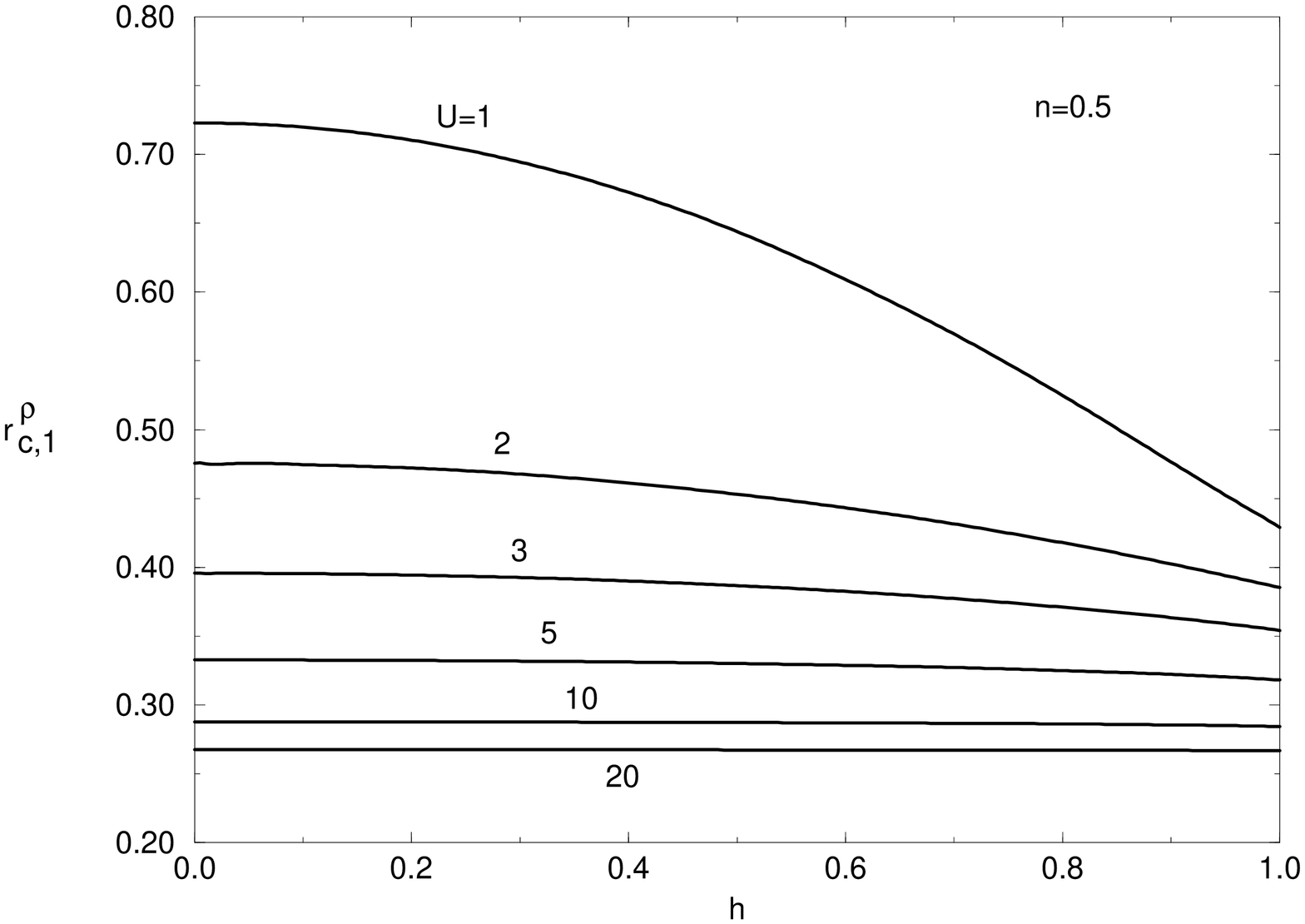,width=3in,angle=-90}} \centerline{(a)}
\end{minipage} \hfill %
\begin{minipage}[b]{.48\linewidth}
\hbox{%
\psfig{figure=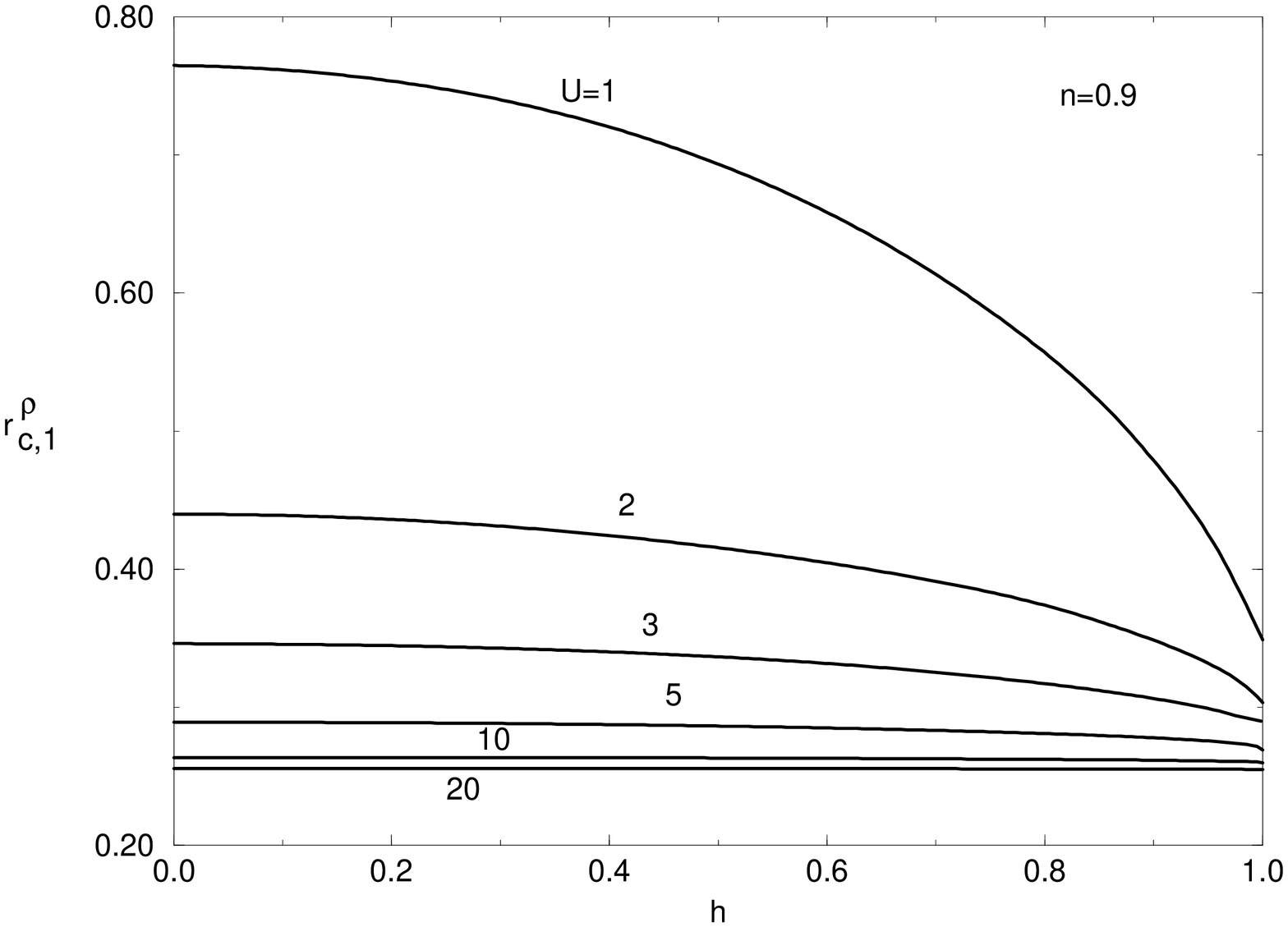,width=3.0in,angle=-90}}\centerline{(b)}
\end{minipage} %
\end{center}
\caption{The ratio $m_{c,1}^{\rho}/m_{c,1}^{\ast}$ as function of the 
the magnetic field $h$ and for values of the onsite Coulomb 
interaction $U=1$, $U=2$, $U=3$, $U=5$, $U=10$, and $U=20$. The 
electronic density is (a) $n=0.5$ and (b) $n=0.9$.}
\label{5fig:fig4}
\end{figure}

\begin{figure}[htbp]
\begin{center}
\leavevmode
\hbox{%
\psfig{figure=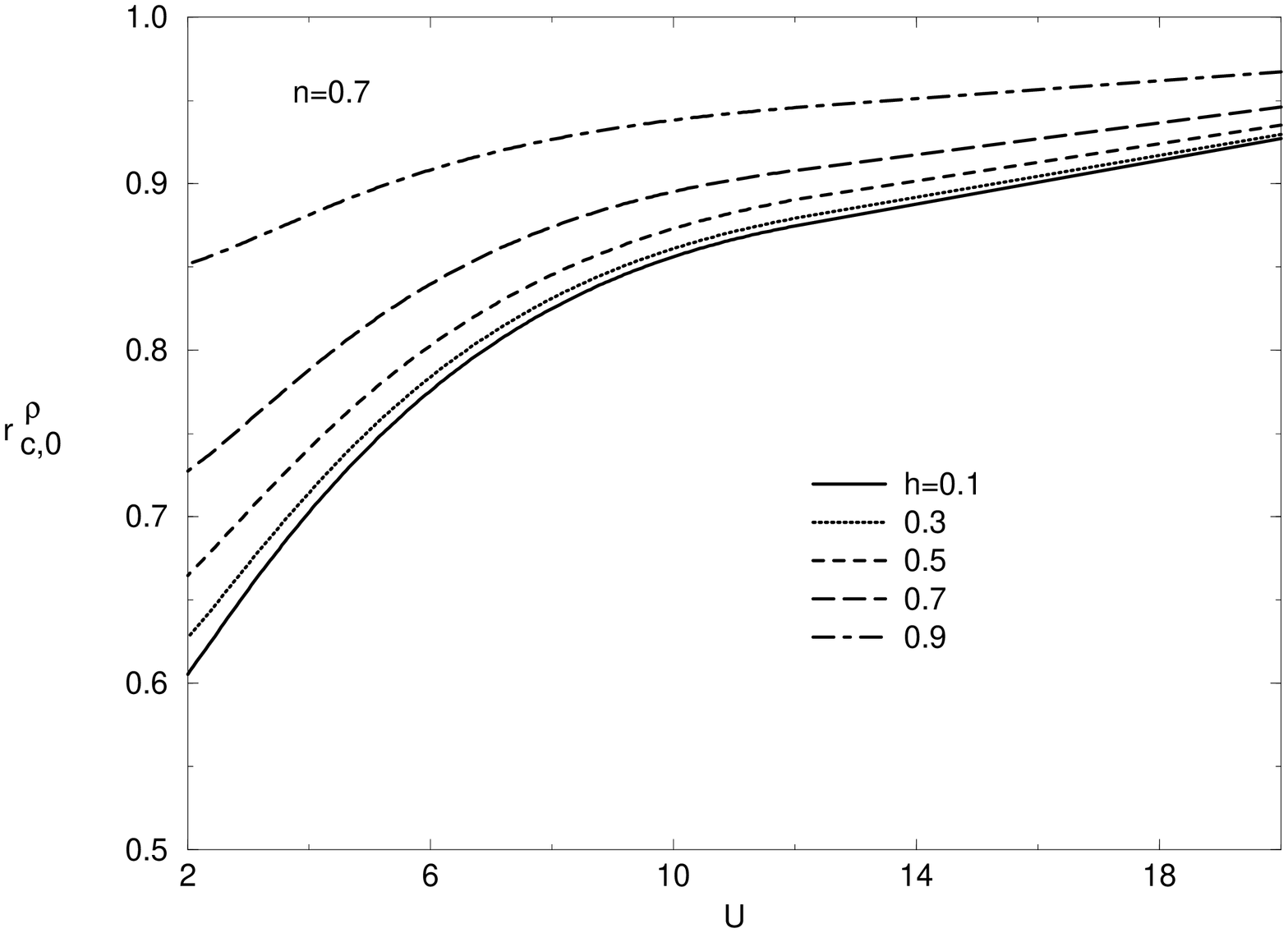,width=4.0in,angle=-90}}
\end{center}
\caption{The ratio $m_{c,0}^{\rho}/m_{c,0}^{\ast}$ as function of 
$U$, for electronic density $n=0.7$, and for values of the 
magnetic field $h=0.1$, $h=0.3$, $h=0.5$, $h=0.7$, and 
$h=0.9$. For other electronic densities, the plots follow the same 
trends as for $n=0.7$.}
\label{5fig:fig5}
\end{figure}

\begin{figure}[htbp]
\begin{center}
\leavevmode
\begin{minipage}[b]{.32\linewidth}
\hbox{%
\psfig{figure=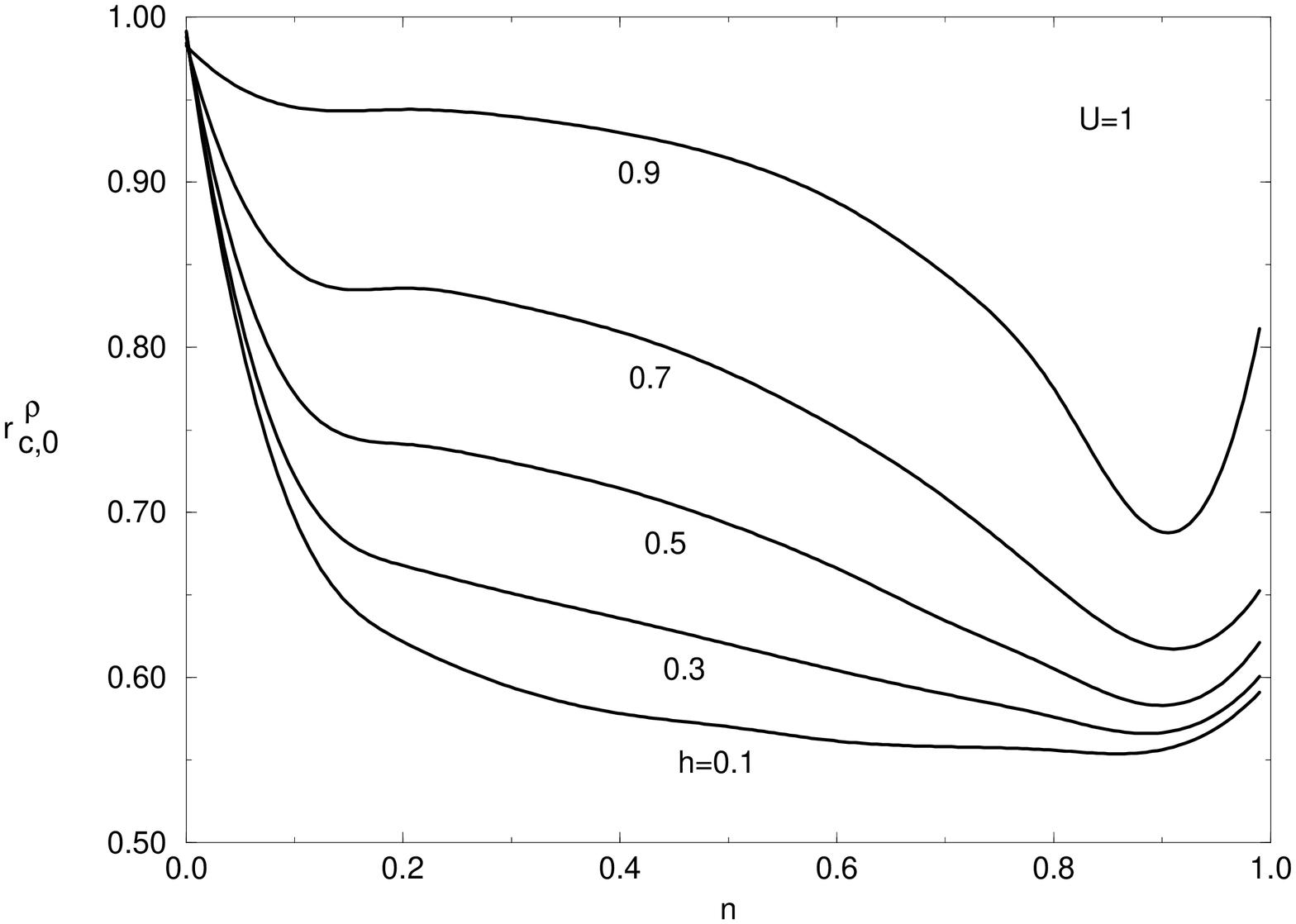,width=3in,angle=-90}} \centerline{(a)}
\end{minipage} \hfill %
\begin{minipage}[b]{.32\linewidth}
\hbox{%
\psfig{figure=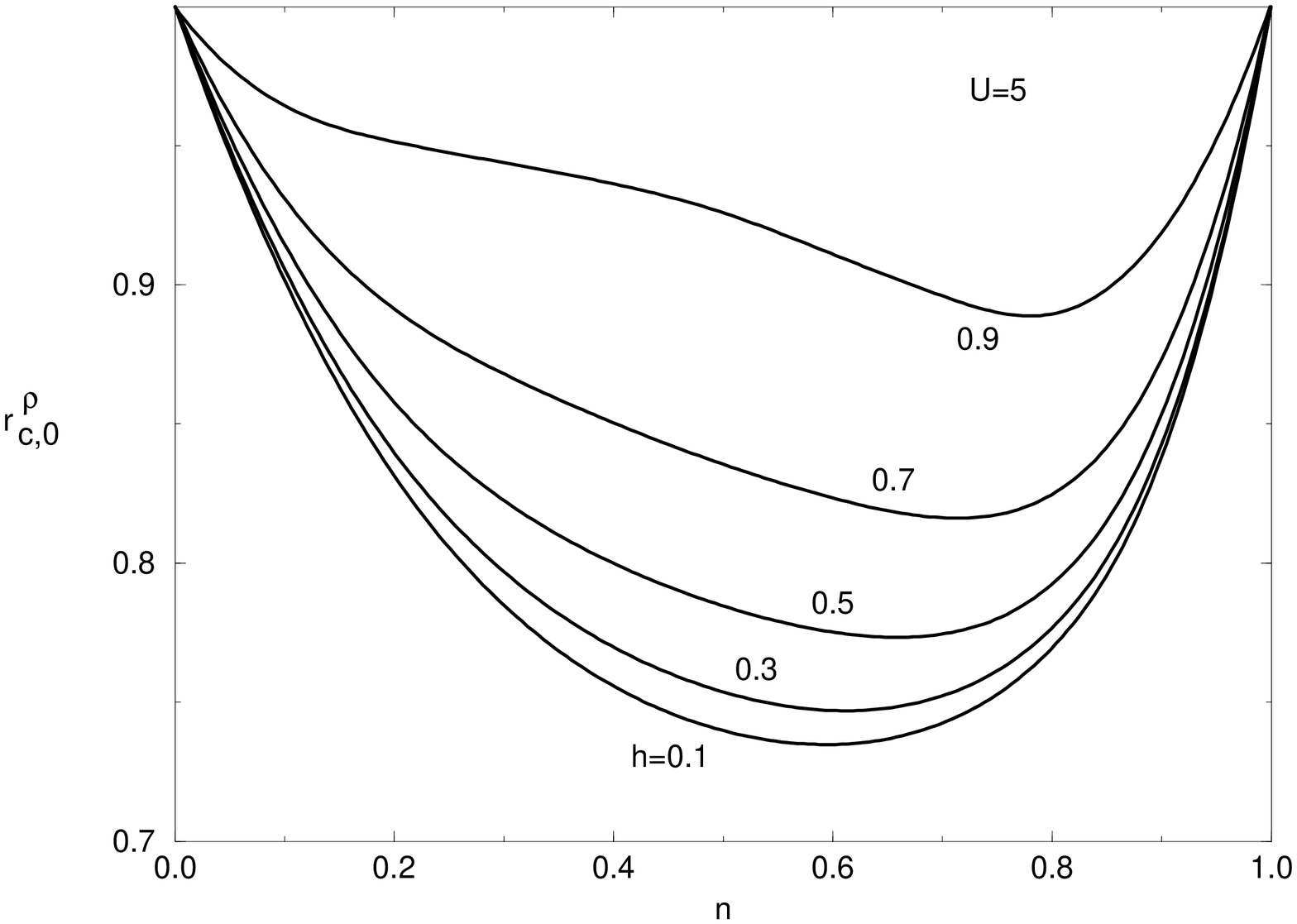,width=3in,angle=-90}}\centerline{(b)}
\end{minipage} %
\end{center}
\caption{The ratio $m_{c,0}^{\rho}/m_{c,0}^{\ast}$ as 
function of the electronic density $n$ and for 
values of the magnetic field
$h=0.1$, $h=0.3$, $h=0.5$, $h=0.7$, and $h=0.9$. The onsite
Coulomb interaction is (a) $U=1$ and (b) $U=5$.}
\label{5fig:fig6}
\end{figure}

\begin{figure}[htbp]
\begin{center}
\leavevmode
\begin{minipage}[b]{.32\linewidth}
\hbox{%
\psfig{figure=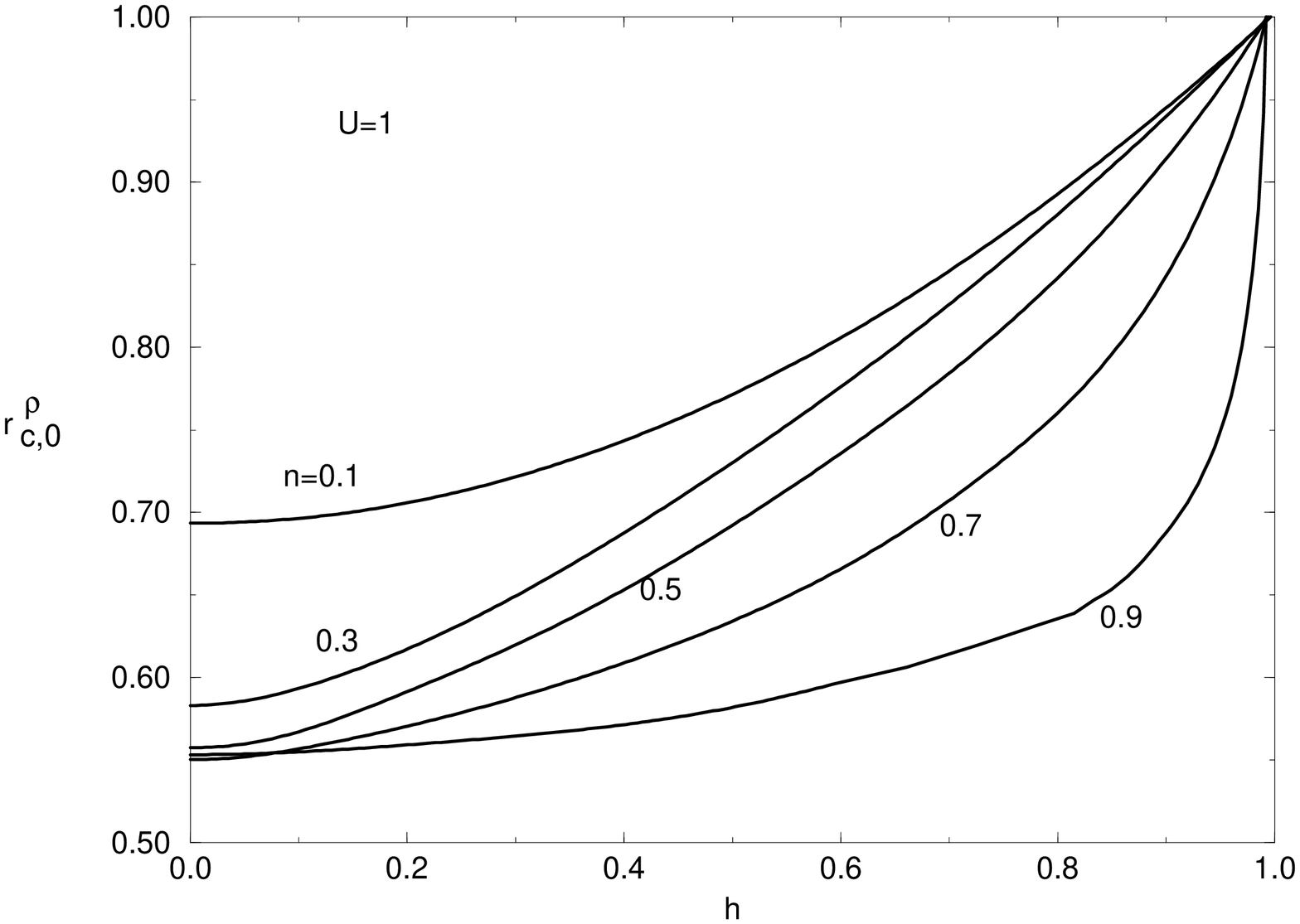,width=3in,angle=-90}} \centerline{(a)}
\end{minipage} \hfill %
\begin{minipage}[b]{.32\linewidth}
\hbox{%
\psfig{figure=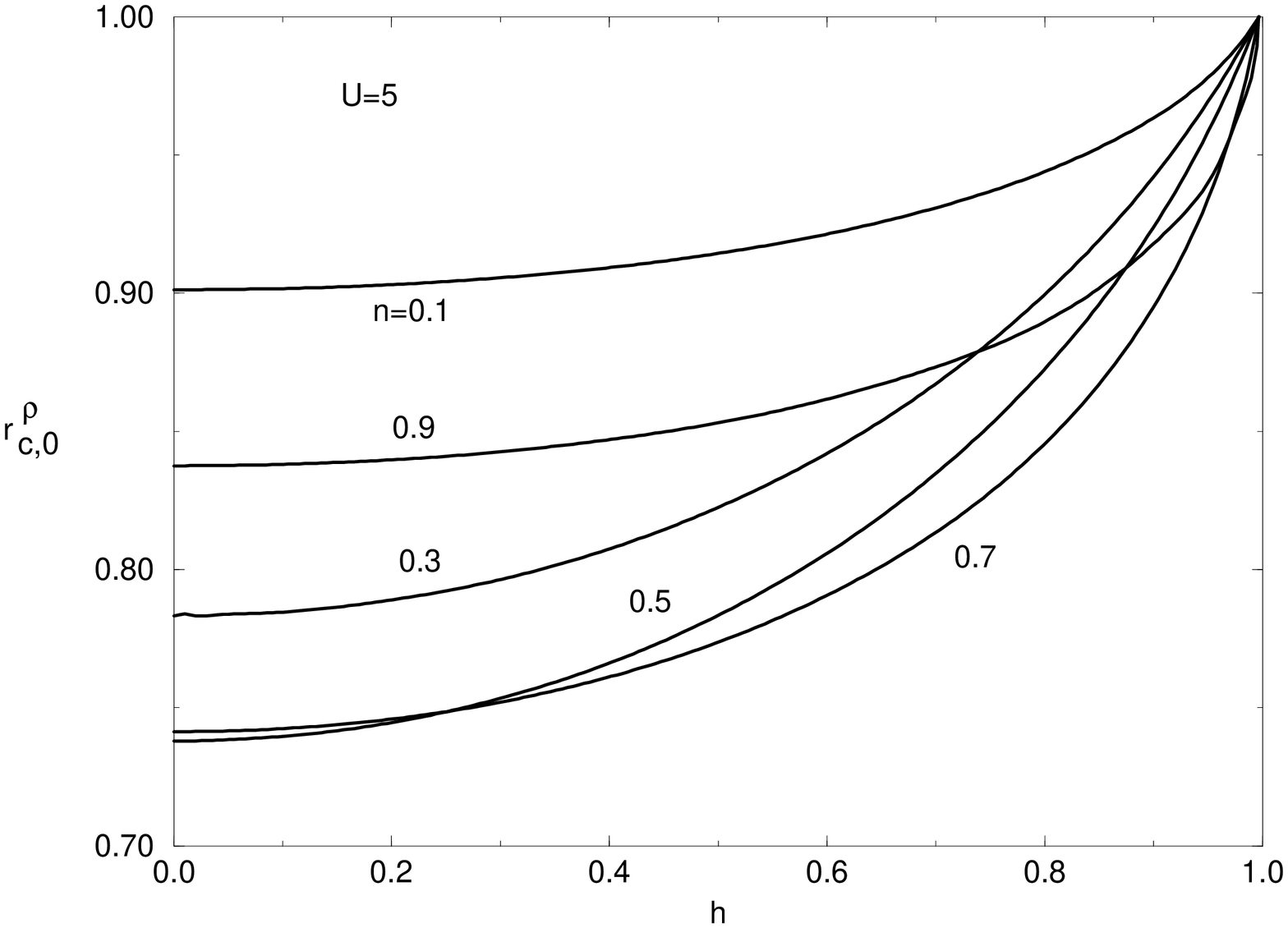,width=3in,angle=-90}}\centerline{(b)}
\end{minipage} %
\end{center}
\caption{The ratio $m_{c,0}^{\rho}/m_{c,0}^{\ast}$ as function 
of the magnetic field $h$ and for values of the electronic 
density $n=0.1$, $n=0.3$, $n=0.5$, $n=0.7$, and $n=0.9$. The onsite
Coulomb interaction is (a) $U=1$ and (b) $U=5$.}
\label{5fig:fig7}
\end{figure}

\begin{figure}[htbp]
\begin{center}
\leavevmode
\begin{minipage}[b]{.48\linewidth}
\hbox{%
\psfig{figure=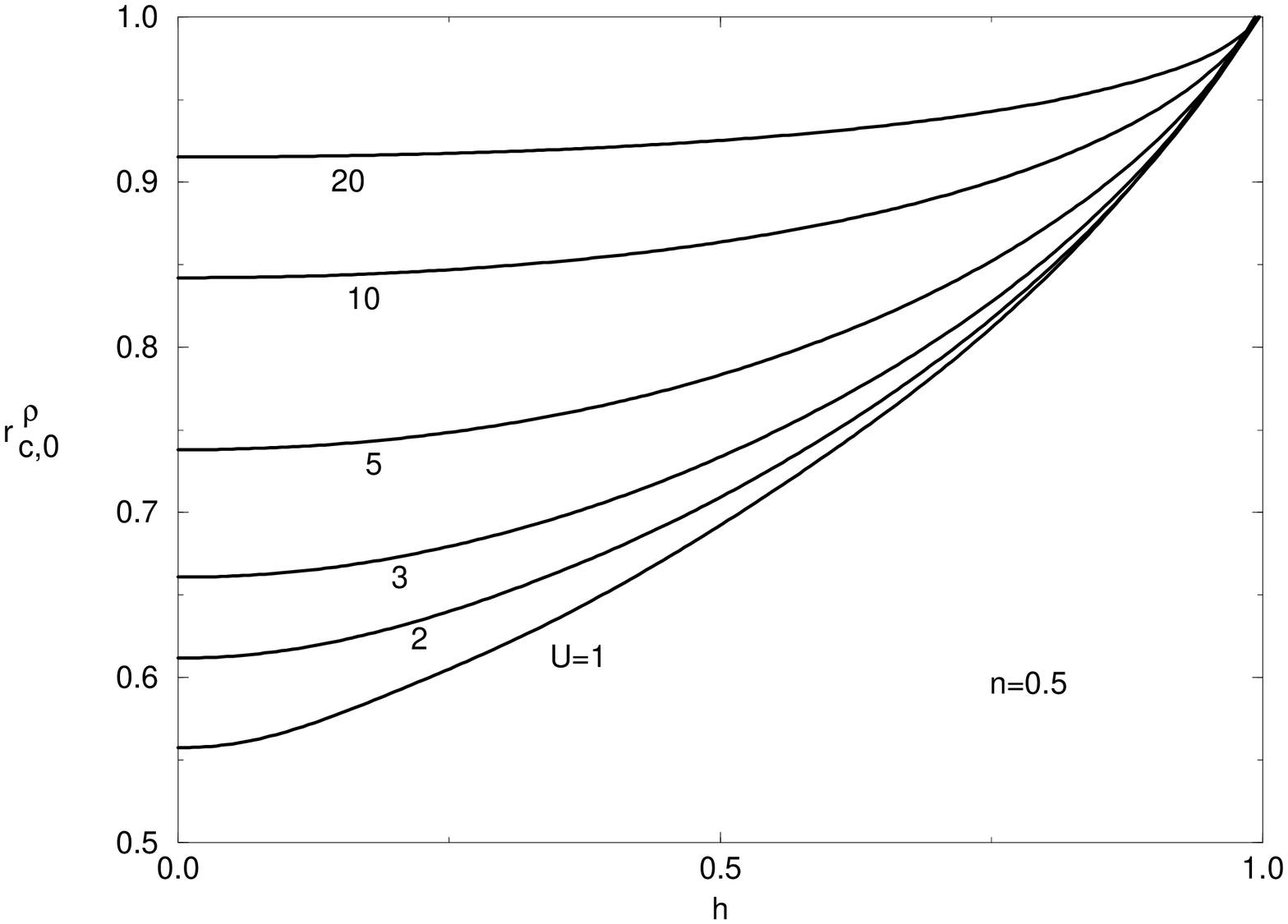,width=3.0in,angle=-90}} \centerline{(a)}
\end{minipage} \hfill %
\begin{minipage}[b]{.48\linewidth}
\hbox{%
\psfig{figure=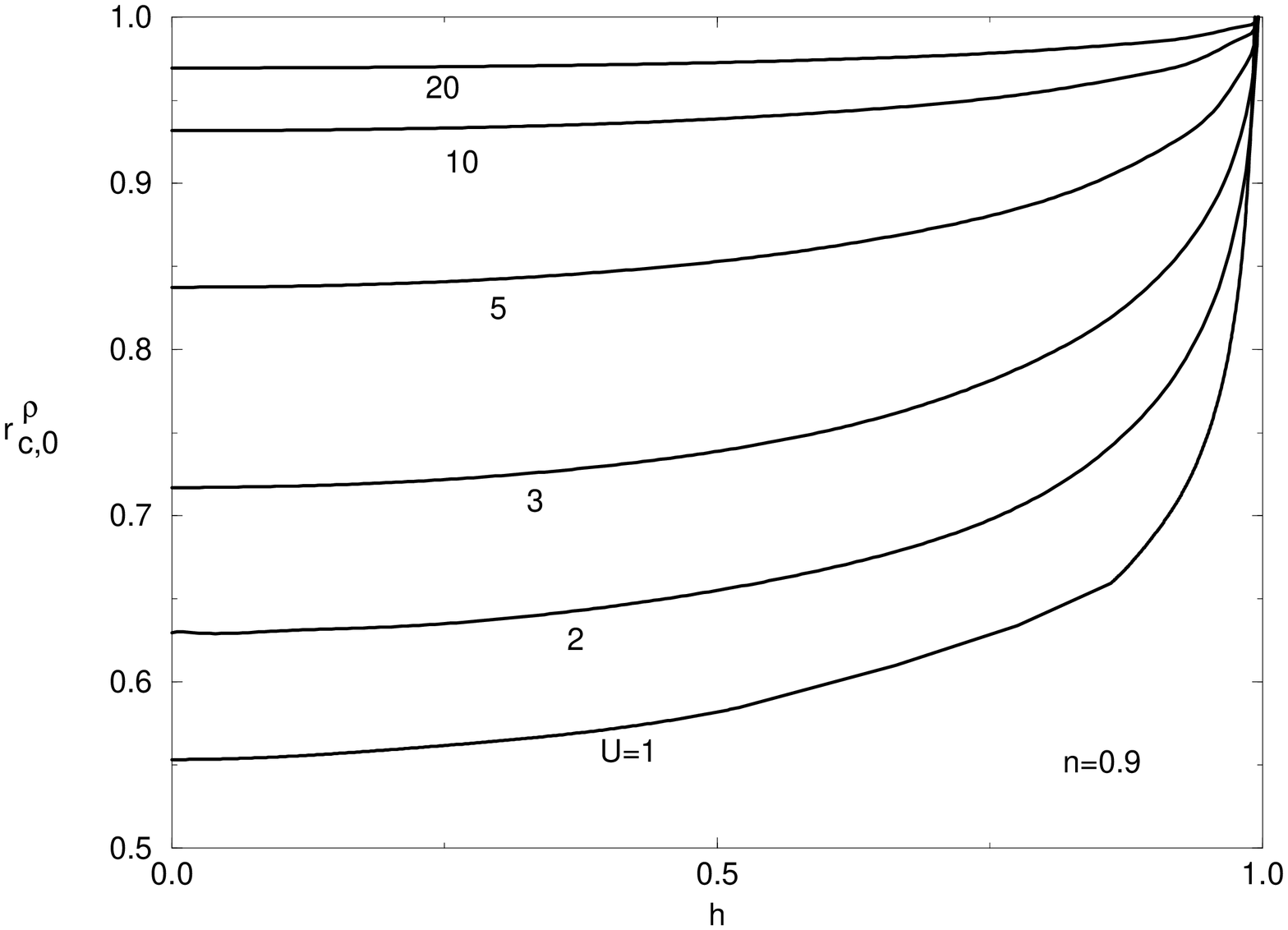,width=3.0in,angle=-90}}\centerline{(b)}
\end{minipage} %
\end{center}
\caption{The ratio $m_{c,0}^{\rho}/m_{c,0}^{\ast}$ as function 
of the the magnetic field $h$ and for values of the 
onsite Coulomb interaction $U=1$, $U=2$, $U=3$, $U=5$, $U=10$, and 
$U=20$. The electronic density is (a) $n=0.5$ and (b) $n=0.9$.}
\label{5fig:fig8}
\end{figure}

\begin{figure}[htbp]
\begin{center}
\leavevmode
\hbox{%
\psfig{figure=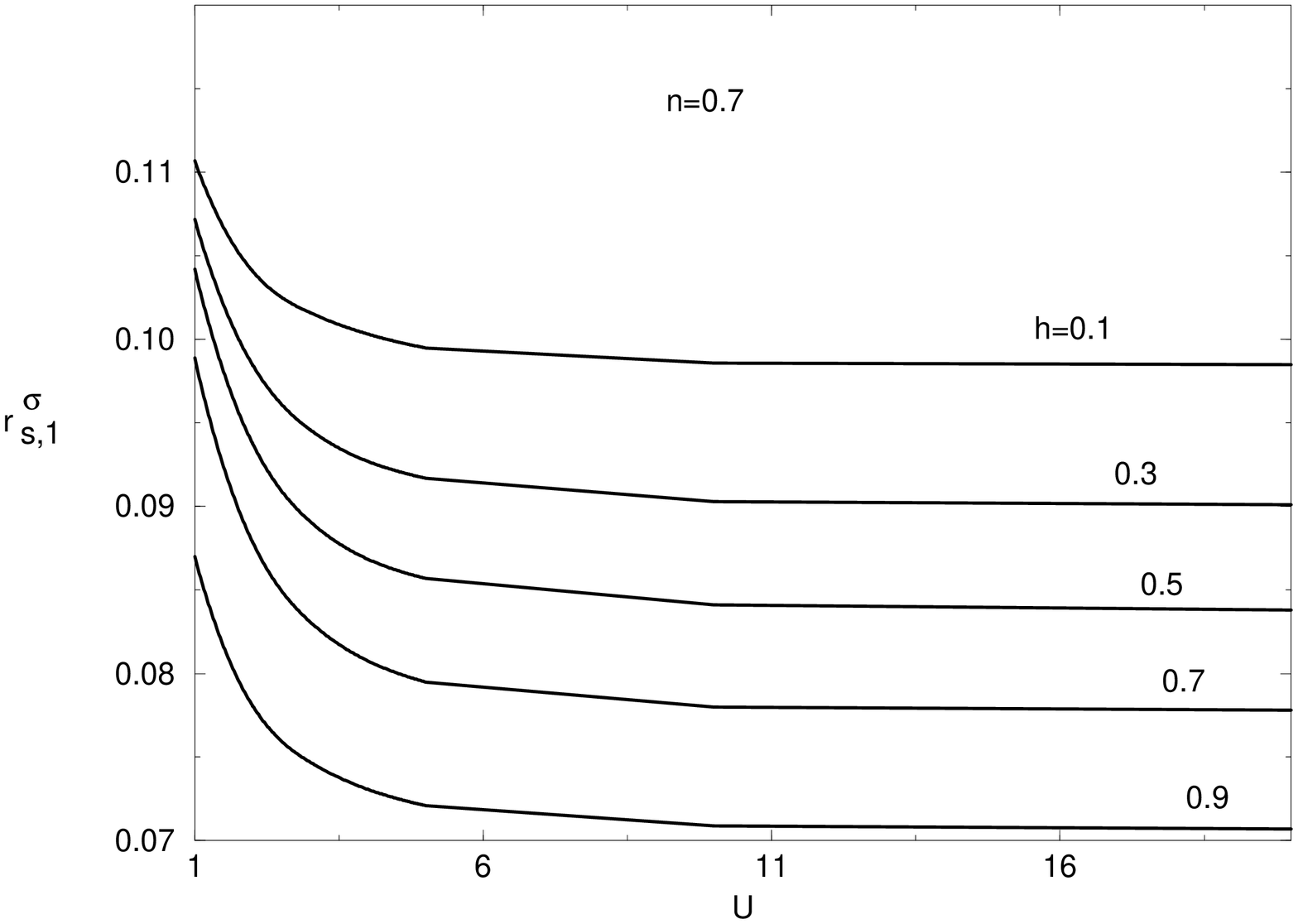,width=4.0in,angle=-90}}
\end{center}
\caption{The ratio $m_{s,1}^{\sigma_z}/m_{s,1}^{\ast}$ as function 
of $U$, for electronic density $n=0.7$, and for values 
of the magnetic field $h=0.1$, $h=0.3$, $h=0.5$, $h=0.7$, 
and $h=0.9$. For other electronic densities, the plots follow the 
same trends as for $n=0.7$.}
\label{5fig:fig9}
\end{figure}

\begin{figure}[htbp]
\begin{center}
\leavevmode
\begin{minipage}[b]{.32\linewidth}
\hbox{%
\psfig{figure=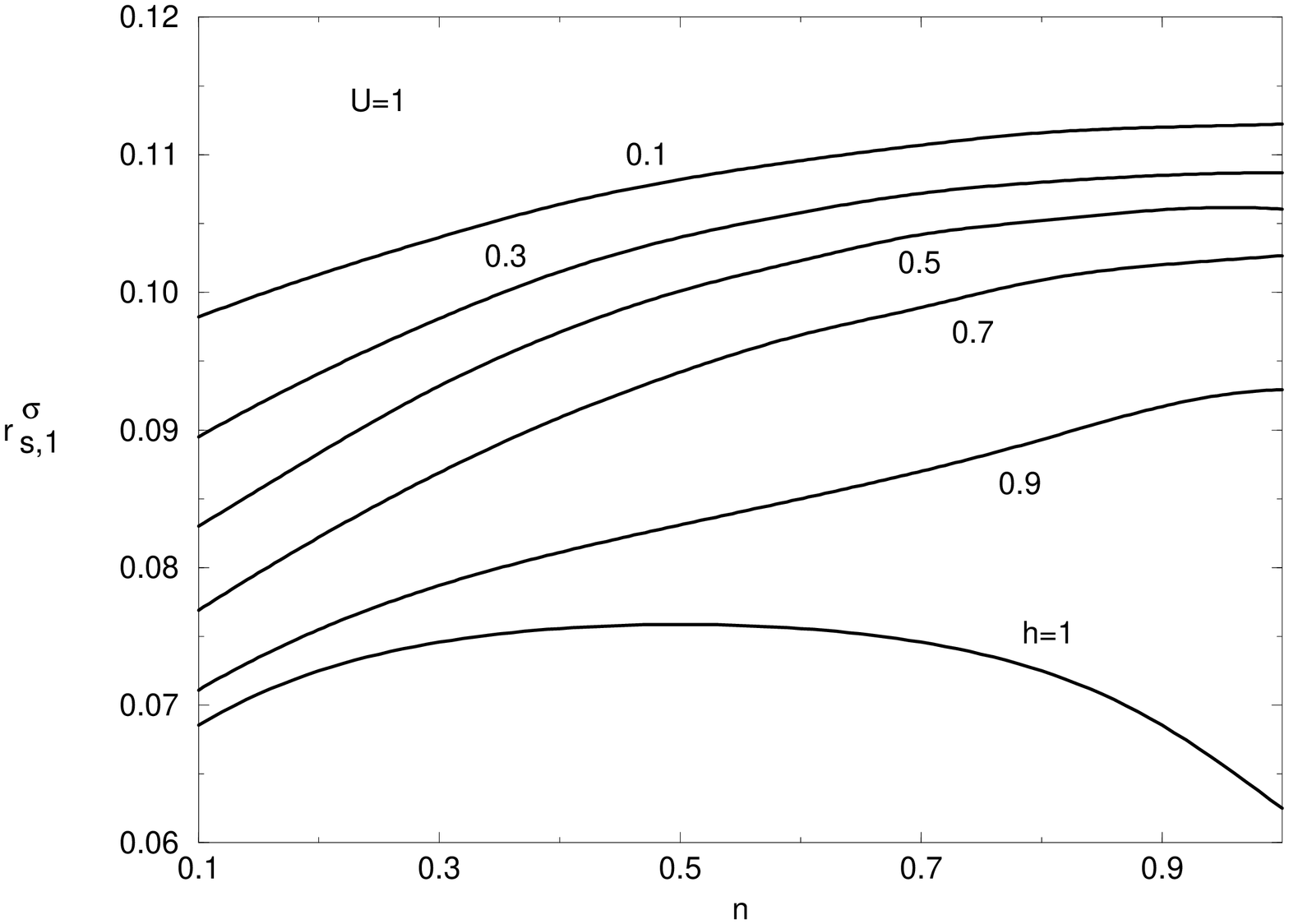,width=3in,angle=-90}} \centerline{(a)}
\end{minipage} \hfill %
\begin{minipage}[b]{.32\linewidth}
\hbox{%
\psfig{figure=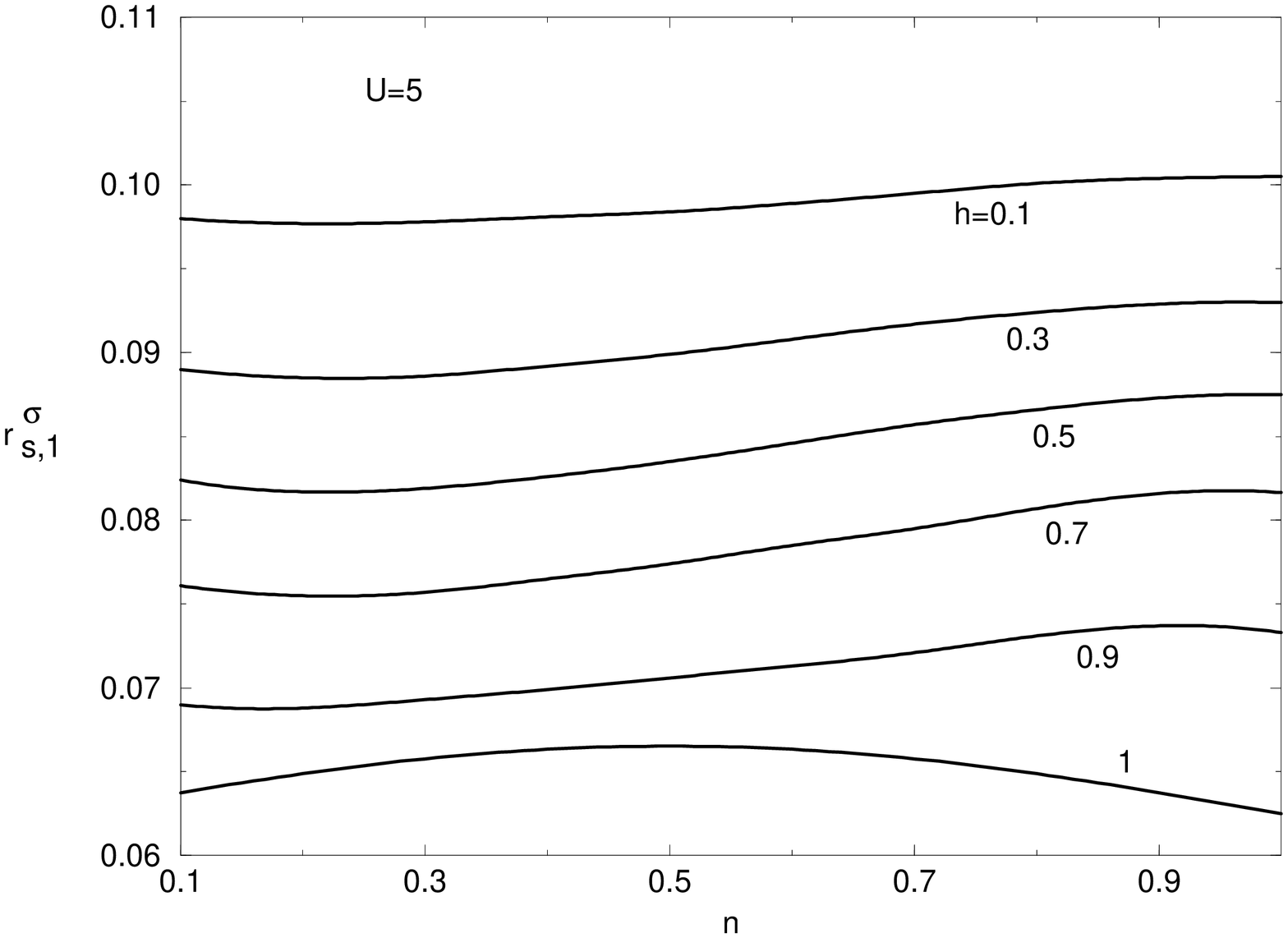,width=3in,angle=-90}}\centerline{(b)}
\end{minipage} 
\end{center}
\caption{The ratio $m_{s,1}^{\sigma_z}/m_{s,1}^{\ast}$ as 
function of the electronic density $n$ and for values of the 
magnetic field $h=0.1$, $h=0.3$, $h=0.5$, $h=0.7$, and 
$h=0.9$. The onsite Coulomb interaction is (a) $U=1$ and (b) $U=5$.}
\label{5fig:fig10}
\end{figure}

\begin{figure}[htbp]
\begin{center}
\leavevmode
\begin{minipage}[b]{.32\linewidth}
\hbox{%
\psfig{figure=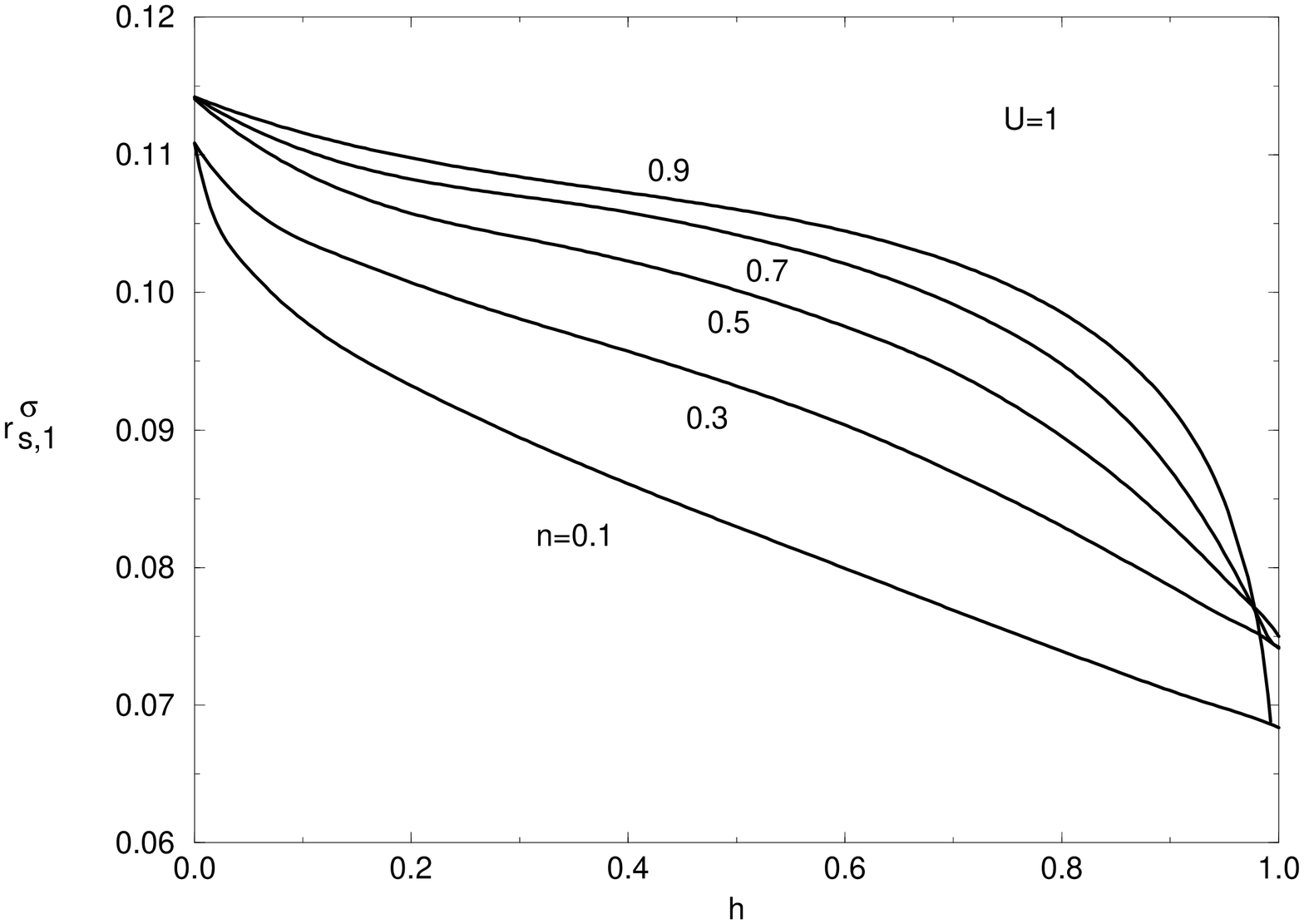,width=3in,angle=-90}} \centerline{(a)}
\end{minipage} \hfill %
\begin{minipage}[b]{.32\linewidth}
\hbox{%
\psfig{figure=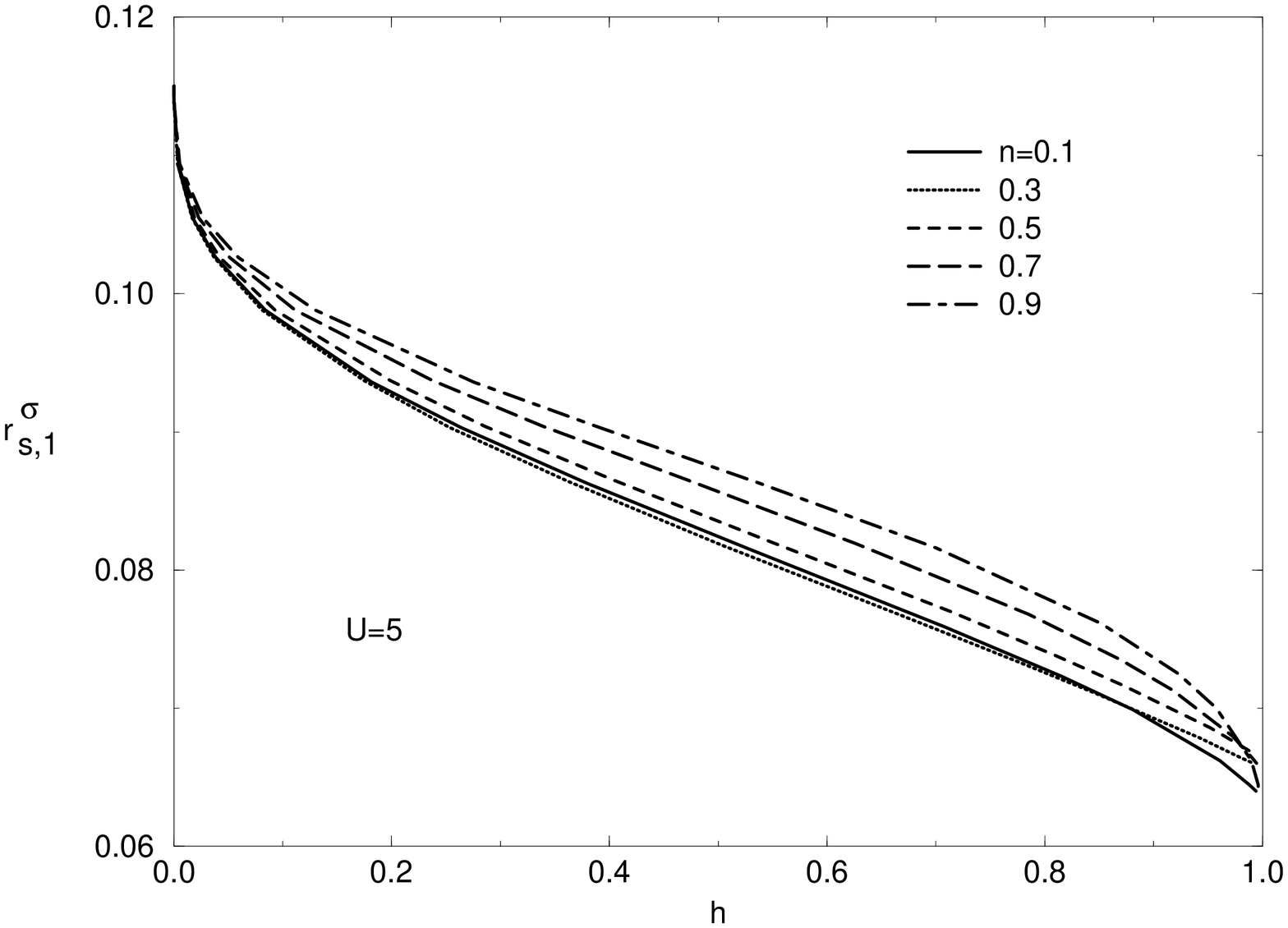,width=3in,angle=-90}}\centerline{(b)}
\end{minipage} %
\end{center}
\caption{The ratio $m_{s,1}^{\sigma_z}/m_{s,1}^{\ast}$ as function 
of the magnetic field $h$ and for values of the electronic 
density $n=0.1$, $n=0.3$, $n=0.5$, $n=0.7$, and $n=0.9$. The onsite
Coulomb interaction is (a) $U=1$ and (b) $U=5$.}
\label{5fig:fig11}
\end{figure}

\begin{figure}[htbp]
\begin{center}
\leavevmode
\begin{minipage}[b]{.32\linewidth}
\hbox{%
\psfig{figure=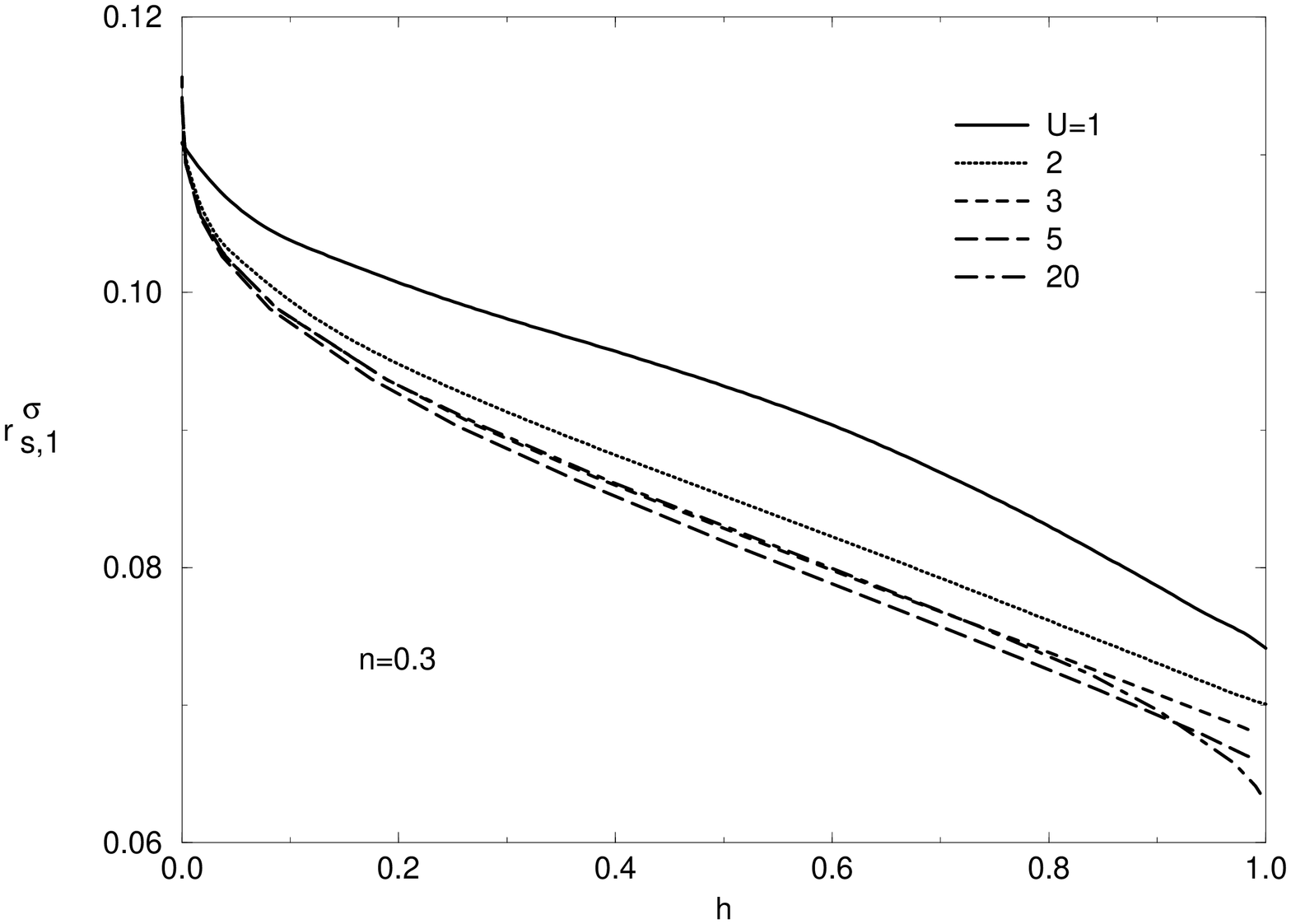,width=3in,angle=-90}} \centerline{(a)}
\end{minipage} \hfill %
\begin{minipage}[b]{.32\linewidth}
\hbox{%
\psfig{figure=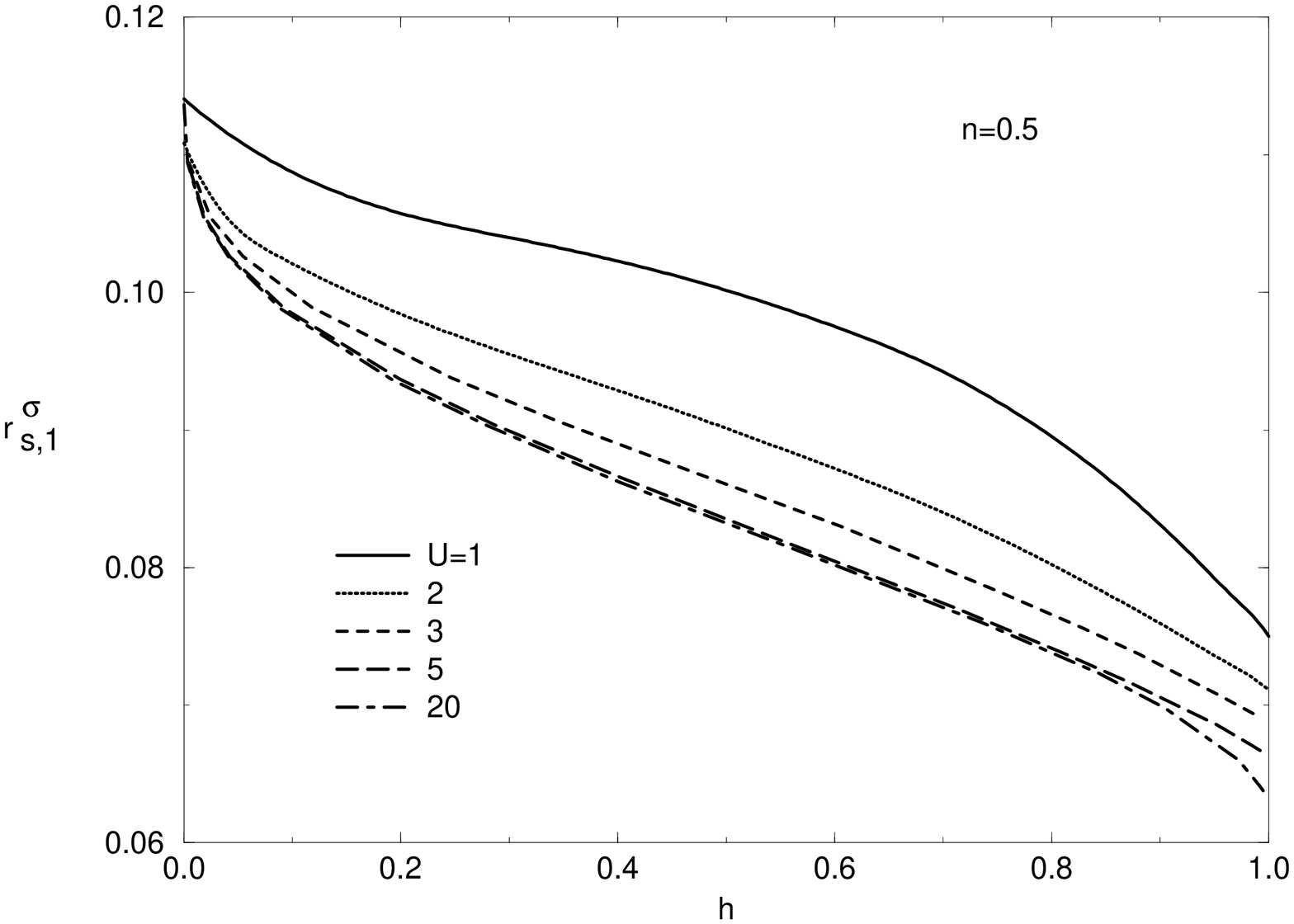,width=3in,angle=-90}}\centerline{(b)}
\end{minipage} %
\end{center}
\caption{The ratio $m_{s,1}^{\sigma_z}/m_{s,1}^{\ast}$ as function 
of the the magnetic field $h$ and for values of the onsite 
Coulomb interaction $U=1$, $U=2$, $U=3$, $U=5$, $U=10$, and $U=20$. 
The electronic density is (a) $n=0.3$ and (b) $n=0.5$.}
\label{5fig:fig12}
\end{figure}

As was referred previously, it can be shown from the results of Chapter
\ref{harmonic}
and other studies \cite{5CarmeloNuno97,5GCFT1,5GCFT2,5conductivity} that the
creation of one $\alpha ,\gamma$ pseudoparticle from
the ground state is a finite-energy excitation which,
to leading order, involves a number $2\gamma $ of electrons.
Therefore, and since the current operators are
of two-electron character and couple to charge and spin
according to the values of the constants
(\ref{5coupling}) and (\ref{5coupling2}), at finite energies 
the $c,1$ and $s,1$ heavy pseudoparticles play the major role in 
charge and spin transport, respectively. On the other hand,
the $\alpha,\gamma >1$ heavy pseudoparticles have a minor role in charge and spin transport.
It follows that in the present section we restrict our study to the
ratios (\ref{5ratio}) involving $\gamma =1 $ heavy pseudoparticles.
We consider the ratios $m_{c,1}^{\rho}/m^{\ast}_{c,1}$ 
and $m_{s,1}^{\sigma_z}/m^{\ast}_{s,1}$. (Note that
$m_{c,1}^{\sigma_z}/m^{\ast}_{c,1}=m_{s,1}^{\rho}/
m^{\ast}_{s,1}=\infty$.) We also consider the case of the 
$c,0$ charge-mass ratio which is closely related to the charge 
stiffness studied in detail by other authors \cite{5Carm4a,5Carm4b}. 
In Figs.\,\ref{5fig:fig1}-\ref{5fig:fig12} these ratios 
are plotted as functions of the onsite repulsion 
$U$ in units of $t$, electronic density $n$, and magnetic 
field $h=H/H_c$. Note that the ratios of the figures are 
smaller than one. 
Moreover, the $\alpha ,1$ mass ratios never achieve the value 
$1$ whereas the $\alpha ,0$ mass ratios tend to one in some 
limits because of the generalized adiabatic principle \cite{5Carm4a,5Carm4b}.


Combined analysis of Figs. \ref{5fig:fig1} and \ref{5fig:fig2} reveals that the 
charge-mass ratio for the $c,1$ pseudoparticle is, for large $U$, 
fairly independent both of the band filling $n$ and magnetic field 
$h$. It is a decreasing function of $U$ and as a function of the
density, $n$, goes through a maximum for a density which is
a decreasing function of $U$. Moreover, 
figures \ref{5fig:fig3} and \ref{5fig:fig4}
show that this ratio is a decreasing function of the magnetic
field. 

In contrast, Figs. \ref{5fig:fig5} - \ref{5fig:fig8} reveal that the 
charge-mass ratio for the $c,0$ pseudoparticle is an increasing 
function of $U$ and of $h$ and as a function of the density, $n$, 
goes through a minimum for a density which is a decreasing 
function of $U$. Note that from Fig. \ref{5fig:fig5} the evolution of 
the $c,0$ pseudoparticles to free spinless fermions as $U$ 
increases is clear. This is signaled by the ratio going to one
as $U\rightarrow\infty$. This behavior follows from the
generalized adiabatic principle \cite{5Carm4a,5Carm4b} and
agrees with the well known decoupling of the BA wave 
function in free spinless 
fermions (in the low-energy sector \cite{5Ricardo97}) and 
localized antiferromagnetic spins \cite{5Ogata90}. 
Figures \ref{5fig:fig7} and \ref{5fig:fig8} also reveal that
in the fully-polarized ferromagnetic limit, $h\rightarrow 1$, 
the ratio goes to one. This mass-ratio behavior also
follows from the generalized adiabatic principle \cite{5Carm4a,5Carm4b}
and confirms that in that limit the onsite Coulomb interations 
play no role in charge transport (they are froozen by the Pauli 
principle) .
 
Note that in the large-$U$ Figs. \ref{5fig:fig2} - (c) and \ref{5fig:fig6} - 
(c) the ratios $m_{c,1}^\rho/m^{\ast}_{c,1}$ and 
$m_{c,0}^\rho/m^{\ast}_{c,0}$, respectively, are almost symmetric 
around the density $n=0.5$. This implies that for large $U$ 
the charge transport properties show similarities in the
cases of vanishing densities and of densities 
closed to one. 

Figure \ref{5fig:fig9} shows that the spin-mass ratio of the $s,1$
pseudoparticles is a decreasing function of $U$ but that it
depends little on $U$ for $U>6$. For large $U$ this ratio
almost does not depend on the density $n$, as revealed by
Fig. \ref{5fig:fig10} - (c). Figures \ref{5fig:fig10} show that, in general,
it is an increasing function of $n$ but that for $h\rightarrow
1$ it has a maximum for an intermediate density. In figures
\ref{5fig:fig11} and \ref{5fig:fig12} this spin-mass ratio is plotted as a 
function of $h$. It is a decreasing function of $h$. 

The transport masses are very sensitive to the effects of 
electronic correlations, as for instance to the metal-insulator 
transition which occurs at zero temperature 
when $n\rightarrow 1$ \cite{5Lieb}. As a direct result of
this transition, $m_{c,0}^{\rho}\rightarrow\infty$ as
$n\rightarrow 1$, as was shown and discussed 
in the works
\cite{5Carm4a,5Carm4b}. Moreover, the zero-temperature charge and
spin stiffnesses, $D^{\zeta}$, 
\cite{5Carm4a,5Carm4b,5Shastry,5Stafford,5Fye,5Millis,5Scalapino}, defined
as
 
\begin{equation}
	D^{\zeta}= \left. \frac 1{2}
	\frac{d^2(E_0/N_a)}{d(\phi/N_a)^2}\right\vert_{\phi=0} \, ,
	\label{5drude}
\end{equation}
where $\phi $ is defined for charge $\zeta =\rho$ and
spin $\zeta =\sigma_z$ in Eq. (\ref{5phis}),
are such that $2\pi D^{\zeta}=\sum_{\alpha}q_{F\alpha,0}
/m^{\zeta}_{\alpha,0}$ \cite{5Carm4a,5Carm4b}. For charge,
$m^\rho_{s,0}=\infty$, and the latter expression reads  
$2\pi D^{\rho}=q_{Fc,0}/m^{\rho}_{c,0}$ and is such that
$D^{\rho}\rightarrow 0$ as $n\rightarrow 1$, satisfying Kohn 
criterion \cite{5Kohn}. These quantities can be computed 
within the pseudoparticle formalism by direct evaluation 
of Eq.\,(\ref{5drude}). They can also be obtained by combining a 
pseudoparticle Boltzmann transport description with linear response 
theory \cite{5Carm4a,5Carm4b}. In order to confirm the validity and
correctness of our formalism, in Appendix \ref{5stiffness} we have 
recovered the charge and spin stiffness expressions $(135)$ - $(137)$ of 
the studies \cite{5Carm4a,5Carm4b} by direct use of Eq. (\ref{5drude}). 

Equations (\ref{5coupling}) and (\ref{5coupling2}) show 
that the Hubbard-chain charge carriers are the $c,\gamma$ 
pseudoparticles. In contrast to the zero-temperature limit 
where the $c,0$ pseudoparticles fully determine the charge
stiffness, we expect that the $c,\gamma$ heavy pseudoparticles 
play an important role in the charge-transport properties at finite 
temperatures \cite{5Castella1,5Castella2,5Zotos}. 

In Appendix \ref{5zeroH} we present simpler equations to define 
the pseudoparticle bands and phase shifts in the limit of zero 
magnetic field. These results show that for $H\rightarrow 0$ and
$\gamma >0$ the bands $\epsilon_{s,\gamma}^0(q)$ collapse to a point. 
This is because both the bandwidth [see Eq. (\ref{5esgh0})] 
and the momentum pseudo-Brillouin zone width (in the ground state)

\begin{eqnarray}
	q_{c ,0} &=& \pi \, ; \hspace{1cm } q_{c,\gamma } = \pi - 2k_F
	\, \hspace{0.5cm} \gamma >0 \, ,\nonumber \\
	q_{s ,0} &=& k_{F\uparrow} \, ; \hspace{1cm } q_{s,\gamma } = 
	k_{F\uparrow } - k_{F\downarrow } \, \hspace{0.5cm} \gamma >0           \,,
\label{5gsbz}
\end{eqnarray} 
go to zero as $H\rightarrow 0$. This behavior is
also present in the Heisenberg chain and, therefore, in that 
model the triplet and singlet exitations are degenerated at 
zero magnetic field \cite{5Faddeev81}. This also holds true for 
the Hubbard chain at $H=0$ and in the limit $U\gg 4t$, where the BA wave 
function factorizes in a spinless-fermion Slater determinant 
and in the BA wave function for the 1D antiferromagnetic 
Heisenberg chain. On the other hand, in the limit $n\rightarrow 1$ 
the bands $\epsilon_{c,\gamma}^0(q)$ (for $\gamma >0$) collapse to a 
point also because
both the bandwidth and the momentum pseudo-Brillouin zone width
[see Eq. (\ref{5gsbz})] go to zero in that limit. 

\section{Kinetic equations for the pseudoparticles}
\label{5kinetic}
\pagestyle{myheadings}
\markboth{6  Zero-temperature transport}{6.5 Kinetic equations for ...}

In the previous sections the quantum-liquid physics for energies 
just above the $\omega_0$ values, Eq. (\ref{5gap}), was described 
in terms of homogeneous pseudoparticle distributions.  The 
pseudoparticles experience only zero-momentum forward-scattering 
interactions at all energy scales. This property is 
absent in Fermi-liquid theory where it holds true only
at low excitation energy when the quasiparticles are well
defined quantum objects \cite{5Pine,5Baym,5Landau1,5Landau2}. This unconventional 
character of integrable models \cite{5Shastry861,5Shastry862} allows us to 
extend the use of the kinetic equations to energy scales just above 
the $\omega_0$ energy values, Eq. (\ref{5gap}), and not only to 
low energies \cite{5Carm4a,5Carm4b}. The results presented
in this section are a generalization of the kinetic-equation
low-energy studies \cite{5Carm4a,5Carm4b}. 

In the final Hilbert subspace of energy $\omega $ relative to 
the initial ground state the Hubbard model can be mapped onto a 
continuum field theory of small energy $(\omega -\omega_0)$
\cite{5GCFT1,5GCFT2}. The time coordinate $t$ of such theory is the 
Fourier transform of the small energy $(\omega-\omega_0)$ which 
corresponds to a finite energy $\omega $ in the original Hubbard model.
The validity of this approach is confirmed by the fact that it fully
reproduces the rigorous results of Section \ref{5baphi}. 

Let us consider excitations described by space and time dependent
pseudoparticle distribution functions, $N_{\alpha,\gamma}(q,x,t)$,
given by

\begin{equation}
N_{\alpha,\gamma}(q,x,t)=N_{\alpha,\gamma}^0(q)+
\delta N_{\alpha,\gamma}(q,x,t)\, ,
\label{5nqxt}
\end{equation}
where $N_{\alpha,\gamma}^0(q)$ is the ground-state distribution.
It follows from the PPT introduced Chapter \ref{harmonic} \cite{5CarmeloNuno97} 
and discussed in the Chapter \ref{harmonic} that the
single-pseudoparticle local energy is given, to first order in the 
deviations $\delta N_{\alpha,\gamma}(q,x,t)$, by

\begin{equation}
\check {\varepsilon}_{\alpha,\gamma} (q,x,t)=
\epsilon_{\alpha,\gamma} (q)+ \frac 1{2 \pi}
\sum_{\alpha',\gamma'}
\int_{-q_{\alpha',\gamma'}}^{q_{\alpha',\gamma'}}dq' 
\delta N_{\alpha',\gamma'}(q',x,t)
f_{\alpha,\gamma;\alpha',\gamma'}(q,q')\, .
\label{5classicale}
\end{equation}

Let ${\cal A}^{\zeta}$ represent the total charge, $\zeta=\rho$, 
or spin, $\zeta=\sigma_z$. It follows from the relations 
(\ref{5nbaixo}) and (\ref{5nc}) involving the pseudoparticle and 
electron numbers that ${\cal A}^{\zeta}$ depends linearly on the 
pseudoparticle deviation numbers. Thus, in the case of 
inhomogeneous excitations described by Eq.\,(\ref{5nqxt}) the 
corresponding expectation value at point $x$ and time $t$, 
$\langle {\cal A}^{\zeta} (x,t)\rangle$, can be written as

\begin{equation}
\langle {\cal A}^{\zeta}(x,t) \rangle=
\langle {\cal A}^{\zeta} \rangle_0+
\frac {N_a}{2 \pi}
\sum_{\alpha',\gamma'}
\int_{-q_{\alpha',\gamma'}}^{q_{\alpha',\gamma'}}dq' 
\delta N_{\alpha',\gamma'}(q',x,t)
{\cal C}_{\alpha',\gamma'}^\zeta \times a^\zeta\, ,
\label{5conserved}
\end{equation}
where $a^\rho=-e$ and $a^{\sigma_z}=1/2$. 

In this``semi-classical'' approach the response to a scalar field, 
$V^{\zeta}(x,t)$, is proportional to the conserved quantity 
${\cal A}^{\zeta}$. As for low energy \cite{5Carm4a,5Carm4b}, in the presence 
of the inhomogeneous potential the force 
${\cal F}^{\zeta}(x,t)_{\alpha,\gamma}$ that acts upon 
the $\alpha,\gamma$ pseudoparticle is given by 
${\cal F}^{\zeta}_{\alpha,\gamma}(x,t)=
-[\partial V^{\zeta}(x,t)/ \partial x]
{\cal C}_{\alpha,\gamma}^\zeta \times a^\zeta$. It
follows that the deviations $\delta N_{\alpha,\gamma}(q,x,t)$ are 
determined by the solution of a system of kinetic equations 
(one equation for each occupied $\alpha,\gamma $
branch) which reads

\begin{eqnarray}
0&=&\frac{\partial N_{\alpha,\gamma}(q,x,t)}{\partial t}+
\frac{\partial N_{\alpha,\gamma}(q,x,t)}{\partial x}
\frac{\partial \check {\varepsilon}_{\alpha,\gamma} (q,x,t)}
{\partial q}-\frac{\partial N_{\alpha,\gamma}(q,x,t)}{\partial q}
\frac{\partial \check {\varepsilon}_{\alpha,\gamma} (q,x,t)}
{\partial x} \nonumber\\
&-&\frac{\partial N_{\alpha,\gamma}(q,x,t)}{\partial q}
\frac {\partial V^{\zeta}(x,t)}{\partial x}
{\cal C}_{\alpha,\gamma}^\zeta \times a^\zeta \, .
\label{5kin}
\end{eqnarray}

Introducing Eq.\,(\ref{5nqxt}) in Eq.\,(\ref{5kin}), expanding
to first order in the deviations $\delta N_{\alpha,\gamma}(q,x,t)$, and
using Eq.\,(\ref{5classicale}) we obtain the following set of 
linearized kinetic equations

\begin{eqnarray}
0&=&\frac{\partial \delta N_{\alpha,\gamma}(q,x,t)}{\partial t}
+v_{\alpha,\gamma}(q) 
\frac{\partial \delta N_{\alpha,\gamma}(q,x,t)}{\partial x}
\nonumber\\
&-&\frac{\partial \delta N_{\alpha,\gamma}(q,x,t)}{\partial q}
\left \{\frac {\partial V^{\zeta}(x,t)}{\partial x}
{\cal C}_{\alpha,\gamma}^\zeta \times a^\zeta 
\right . \nonumber\\
&+&\left . \sum_{\alpha',\gamma'} \frac 1{2\pi}
\int_{-q_{\alpha',\gamma'}}^{q_{\alpha',\gamma'}} dq'
\frac{\partial \delta N_{\alpha',\gamma'}(q',x,t)}{\partial x}
f_{\alpha,\gamma;\alpha',\gamma'}(q,q')
\right \}\, .
\label{5linear}
\end{eqnarray}

The conservation law for $\langle {\cal A}^{\zeta}(x,t) \rangle$ 
leads in one dimension to

\begin{equation}
\frac{\partial \langle {\cal A}^{\zeta}(x,t) \rangle}
{\partial t}+\frac{\langle {\cal J}^{\zeta}(x,t) \rangle}
{\partial x}=0\, ,
\label{5conservation}
\end{equation}
where $\langle {\cal A}^{\zeta}(x,t) \rangle$ is given by 
Eq.\,(\ref{5conserved}) and $\langle {\cal J}^{\zeta}(x,t) 
\rangle$ is the associate current. Multiplying Eq.\,(\ref{5linear}) 
by ${\cal C}_{\alpha,\gamma}^\zeta \times a^\zeta$, summing 
over $\alpha$ and $\gamma$, and integrating over $q$ we find for
$V^{\zeta}(x,t)=0$ and by comparing
the result with Eq.\,(\ref{5conservation}) that the current 
spectrum $j^\zeta_{\alpha,\gamma}(q)$ is given by $a^\zeta$ times 
expression\,(\ref{5jq}). (This expression has been derived from the 
solution of the BA equations with $a^\zeta=1$.) 

This agreement confirms the validity of the above 
low-$(\omega-\omega_0)$ continuum-field theory.
The unusual spectral properties associated with 
the zero-momentum forward-scattering character of the pseudoparticle
interactions follow from the integrability of the Hubbard
chain \cite{5CarmeloNuno97,5Shastry861,5Shastry862}. 

\section{On the conductivity of the Hubbard chain}
\pagestyle{myheadings}
\markboth{6  Zero-temperature transport}{6.6 On the Conductivity of ...}

Solvable 
one-dimensional many-electron models such as the Hubbard chain are 
often used as an approximation for the study of the properties of 
quasi-one-dimensional conductors \cite{5Jacobsen1,5Jacobsen2,5Donovan,5Danilo,5Mori}. Although 
the model has been 
diagonalized long ago \cite{5Takahashi,5Lieb}, the involved form of 
the Bethe-ansatz (BA) wave function has prevented the
direct calculation of 
dynamic response functions, including the charge-charge and 
spin-spin response functions and their associate conductivity spectra. 

Information on low-energy expressions for correlation functions
can be obtained by combining BA with conformal-field theory
\cite{5Frahm90,5Frahm91}. On the other hand, several approaches using 
perturbation theory \cite{5Maldague}, 
bosonization \cite{5Schulz90,5Schulz95,5Gimarchi}, the pseudoparticle formalism 
\cite{5Carm4a,5Carm4b}, scaling methods \cite{5Stafford}, and spin-wave theory 
\cite{5Horsch} have been used to investigate the low-energy transport 
properties of the model away from half filling and at the 
metal -- insulator transition \cite{5Lieb}. Some information 
on the transport properties at finite 
energies has also been obtained by numerical methods 
\cite{5Fye,5Loh,5NunoZFPB}.

To study the optical conductivity of the Hubbard chain we start from its
Lehmann representation, whose real part, ${\cal R}e\, \sigma(\omega)$, is
given by \cite{5Shastry,5Pine}

\begin{eqnarray}
	{\cal R}e\, \sigma(\omega)&=& \frac{2\pi e^2}{\hbar}
	\left[\,D \delta (\hbar \omega)+\frac 1{N_a}\sum_{m \not= 0}
	\vert \langle m \vert \hat{j}^{\rho} \vert 0 \rangle \vert ^2
	\delta ((E_m-E_0)^2-\hbar ^2 \omega ^2)
	\right ]\nonumber\\
	 &\equiv& \frac{2\pi e^2}{\hbar}
	\left[\,D \delta (\hbar \omega)+\sigma_{\rm{reg}} (\omega)
	\right]\,,
\label{5cond}
\end{eqnarray}
with the charge stiffness $D$ given by

\begin{equation}
	D=\frac 1{N_a}\left [ -\frac 1{2} \langle 0 \vert \hat{T} 
	\vert 0 \rangle - \sum_{m \not= 0}
	\frac {\vert \langle m \vert \hat{j}^{\rho} \vert 0 \rangle \vert ^2}
	{E_m-E_0}\right ]\,
\label{5dness}
\end{equation}
and $\hat{T}=-2t\sum_{k,\sigma}\cos(k)c^{\dag}_{k,\sigma}c_{k,\sigma}$ 
being the model kinetic energy (all other 
quantities have been previously defined in the Chapter \ref{hm}).

Following Kohn \cite{5Kohn} and our previous discussion, the value of charge stiffness $D$ is a criterium for the metal-insulator transition. For such transition to occur, $D$ must change from non-zero to zero. According to Eq.\,(\ref{5dness}), this is possible only if the two terms in this equation cancel indentically. For correlated systems with continuos spatial symmetry, the total current operator ${\hat{j}}^{\rho}=\sum_i {\hat{p}}_i/m=\hat{P}/m$
is a constant of motion ($m$ is the particle mass, ${\hat{p}}_i$ is the momentum
of the particle $i$, and $\hat{P}$ is the total momentum). It follows that $[\hat{H},{\hat{j}}^{\rho}]=0$, which implies that ${\hat{j}}^{\rho}\vert m \rangle = {\hat{P}}_m /m \vert m \rangle$ and

\begin{equation}
	{\cal R}e\, \sigma(\omega)= \frac{2\pi e^2}{\hbar} D \delta(\omega)\,,
	\hspace{1cm}  \sigma_{\rm{reg}} (\omega)=0\,, \hspace{1cm} 
	D=-\frac 1{2N_a} \langle 0 \vert \hat{T} 
	\vert 0 \rangle \,.
\label{5conti}
\end{equation} 
Equation\,(\ref{5conti}) then implies that for systems with continuos spatial symmetry a metal-insulator transition can not occur and the optical-conductivity is reduced to the $\omega=0$ absorption peak.
That is, the conductivity for a correlated system with continous spatial symmetry behaves like a independent-electron system.

The situation is different for lattice systems. In this case it may happen that
${\hat{j}}^{\rho}$ is no longer a constant of motion. Then 
$\langle 0 \vert {\hat{j}}^{\rho}\vert m \rangle \not= 0$
for a set of Hamiltonian eigenstates $\vert m \rangle$ and the optical conductivity shows a non-zero regular spectrum 
($\sigma_{\rm{reg}} (\omega)\not= 0$) and a metal-insulator transition may occur.
Metal-insulator transitions can be induced, for instance, by the Peierls
mechanism due to the dimerization of the lattice (the original band is splitted into two with the formation of a gap, due to the electron-phonon interaction), by Anderson localization, due to electron scattering with 
random distributed impurities, or by the Mott mechanism, associated with charge localization produced by electronic correlations. The latter mechanism occurs in the one-dimensional Hubbard model at half-filling.

In the pseudoparticle description of the Hubbard model, the metal-insulator transition is associated with the existance of a gap $\omega_0$
between the $c,0$ and the $c,\gamma$ bands. 
As we have remarked before, at half-filling and zero temperature
the charge stiffness is zero and the real part of the optical conductivity,
$Re \sigma (\omega)$, has a finite energy gap for all values of $U/t$. 
For this band filling the $c,0$-pseudoparticle band is full and charge excitations can only occur between
the low-energy $\alpha,0$ band and the $c,\gamma$ heavy pseudoparticle bands. 
The form of the regular part of the optical conductivity is determined by this type of
excitations. 

As we have
discussed in Section \ref{5rev}, to populate, for example, the
lowest $c,\gamma$ energy-band (the $c,1$ band) with one $c,1$ heavy
pseudoparticle one has to annihilate
two $c,0$ and one $s,0$ pseudoparticles. The energy involved in this process
is exactly $\epsilon^0_{c,1}(0)-\epsilon^0_{c,0}(2k_{F})-
\epsilon^0_{c,0}(-2k_{F})-\epsilon^0_{s,0}(k_{F\downarrow}) =2\mu(n=1)$,
where $\mu(n=1)$ stands for the chemical potential in the limit of
half-filling
and $\epsilon^0_{\alpha,\gamma}(q)$ is the pseudoparticle energy
dispersion.
In addition, this excitation process is characterized by
a topological excitation of the $s$ pseudoparticles of momentum
$-k_{F\downarrow}$ \cite{5Nuno3}, ensuring that the overall excitation
has zero momentum, as required for the transitions induced by the
current operator (Chapter \ref{finitesize}). In general,
the process of creating one $c,\gamma$ pseudoparticle has an minimum
energy cost of $2 \gamma \mu(1)$ and a maximum cost of $2 \gamma \mu(1)+
8 \gamma t$. Thus, in principle, these transitions should show up in
$\sigma_{\rm{reg}}$  as a set of optical bands. 
\begin{figure}[htbp]
\begin{center}
\leavevmode
\hbox{%
\psfig{figure=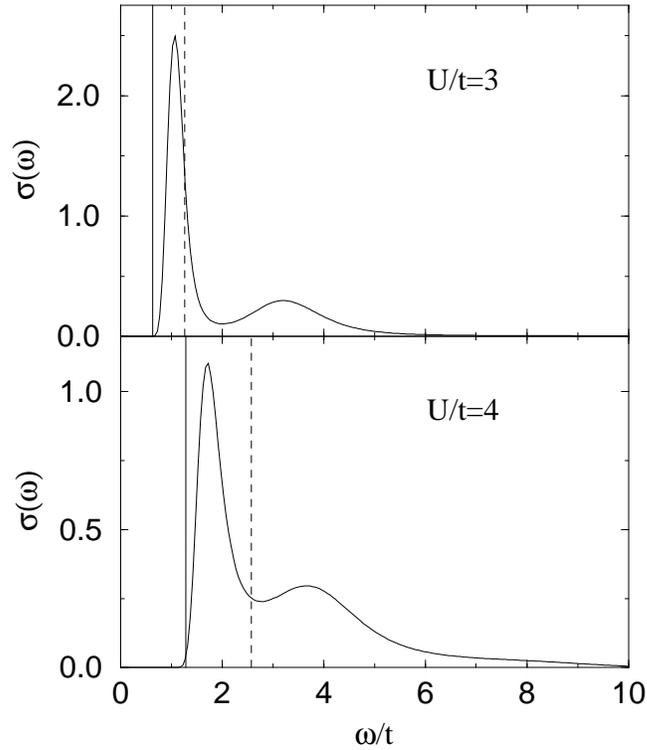,width=4.0in,angle=0}}
\end{center}
\caption{Optical conductivity results for $32$-site Hubbard chains
at half filling for values of $U/t=3$ (top) and $4$ (bottom), computed
using maximum-entropy analytic continuation of quantum Monte Carlo data.
The solid and dashed vertical lines indicate the theoretical predictions
for the lower edges of the first two bands.}
\label{sigmafig}
\end{figure}
As remarked in Section \ref{5rev}, it is possible to 
show that the creation of one $\alpha,\gamma$
pseudoparticle from the ground state involves, to
leading order, a number $2\gamma$ of electrons. Since the
currents are two-electron operators, it follows that
the creation of one $\alpha,1$ pseudoparticle from
the ground state is the most important contribution to the
transport of charge at finite energies. Moreover, for $U \gg 4t$ the
creation of one $c,1$ pseudoparticle is equivalent to the creation
of a single doubly occupied site. Thus, for these values of $U$ only a single
optical band should appear in $\sigma_{\rm{reg}}$. This is an exact result which can be confirmed by numerical studies 
in this regime \cite{5Maldague,5Fye,5Loh}.

In order to investigate the presence of more than
one optical band for small and intermediate values of $U$, recent numerical studies 
have been performed \cite{5conductivity}, 
by use of a quantum Monte Carlo technique for systems
with 32 sites. This corresponds to a considerably larger system than  in
previous exact diagonalization studies \cite{5Fye,5Loh}. The current-current
correlation function has been  
computed in imaginary time and continued to real
frequency using the so called maximum-entropy method \cite{5Maxent}. The
resolution of this method is limited but, nevertheless, gives a
semi-quantitative confirmation of a multi-band optical conductivity. In
Fig.\,\ref{sigmafig}, the results are shown for $U/t=3$ and $4$ where two peaks are
clearly resolved. 

As indicated in the figure, the lower edges of the spectra
agree with the predicted gaps (Chapter \ref{harmonic}), 
and the second peak appears above
the predicted lower edge of the second band, as computed from the
pseudoparticle theory. This theory also allows the evaluation of the 
critical exponents associated with the conductivity edges \cite{5GCFT1,5GCFT2,5conductivity}.
Note that the weight of the first band
is concentrated towards the lower end of the allowed band of width $8t$.
The second peak seen in these results is probably dominated by the second
optical band, but likely contains contributions also from the tail of the
first band, as well as from higher bands that cannot be resolved due to the
limitations of the method.

\begin{appendix}
\input{apjno}
\input{apsma}
\input{apstif}
\input{apzero}
\end{appendix}


%% file: apjno.tex
\chapter{Normal-ordered solution of the BA equations with a flux
$\phi$}
\pagestyle{myheadings}
\markboth{Normal-ordered solution of ...}{Appendix A of Chapter 6}

\label{5normalOrder}

In this appendix we derive the normal-ordered BA equations required
for the evaluation of Eqs.\,(\ref{5deltaj}) and (\ref{5jq}). Writing 
$W^{\phi}(q)$ from Eq.\,(\ref{5exp2}) as

\begin{equation}
W^{\phi}(q)=\frac{dW(q)}{dq}L^{\phi}(q)\, ,
\end{equation}
where $L$ equals $L_{c,0}$, $L_{c,\gamma}$, or $L_{s,\gamma}$ 
when $W$ equals $K$, $R_{s,\gamma}$, or $R_{c,\gamma}$, respectively, 
we find that $W^{1,\phi}(q)$ obeys the following equality

\begin{equation}
W^{1,\phi}(q)=\frac {dW^0}{dq}L^{1,\phi}(q)+
\frac {dW^1}{dq}L^{0,\phi}(q) \, .
\end{equation}
Introducing the above equation in Eq.\,(\ref{5jmean}) 
and writing the distributions functions $N_{\alpha,\gamma}(q)$ 
as $N_{\alpha,\gamma}^0(q)+\delta N_{\alpha,\gamma}(q)$, 
we can expand $J\equiv\langle m\vert \, 
\hat{j}^\zeta\vert m\rangle$ in terms of the pseudomomentum
deviations as

\begin{equation}
J = J^0 + J^1 + J^2 ... \, ,
\label{5Jexp}
\end{equation}
where the first term, $J^1$, of the current normal-ordered 
expansion (\ref{5Jexp}) can after some algebra be written as

\begin{eqnarray}
J^1&=& J^1_0
-2t\sum_{j=\pm 1} {L^{0,\phi}_{c,0}(jq_{Fc})L_{c,0}^1(jq_{Fc})
\over 2\pi\rho_{c,0}(Q)} \sin (Q)
+\nonumber\\
&+&\sum_{\gamma >0} \theta(N_{c,\gamma})Re\, 4t \sum_{j=\pm 1}
\frac{u^2[jr_{c,\gamma} -i\gamma]}
{\sqrt{1-u^2[jr_{c,\gamma} -i\gamma]^2}} 
{L_{c,\gamma}^{0,\phi}(jq_{Fc,\gamma}) 
L^1_{c,\gamma}(jq_{Fc,\gamma})\over 2\pi\rho_{c,\gamma}(r_{c,\gamma})}
\nonumber\\
&-&2t\int_{-q_{Fc}}^{q_{Fc}} dq 
\frac{dK^{(0)}(q)}{dq}\sin(K^{(0)}(q))
{\cal L}^{1,\phi}_{c,0}(q) \, ,  
\label{5jnorm}
\end{eqnarray}
where the functions $2\pi\rho_{c,0}(k)$ and $2\pi\rho_{\alpha,\gamma}(r)$
were defined in Ref. \cite{5CarmeloNuno97} and

\begin{eqnarray}
&&J^1_0=-2t\int_{-q_{c}}^{q_{c}} dq \delta N_c(q)
\frac{dK^{(0)}(q)}{dq}\sin(K^{(0)}(q)) L^{0,\phi}_{c,0}(q) \nonumber\\
&+& \sum_{\gamma >0} Re\, 4t\int_{-q_{c,\gamma}}
^{q_{c,\gamma}} dq \delta N_{c,\gamma}(q)
\frac{u^2[R^{(0)}_{c,\gamma}(q) -i\gamma]}
{\sqrt{1-u^2[R^{(0)}_{c,\gamma}(q) -i\gamma]^2}}
\frac{dR^{(0)}_{c,\gamma}(q)}{dq}
L^{0,\phi}_{c,\gamma}(q) \, .
\end{eqnarray}
The function ${\cal L}^{1,\phi}(q)$ is defined as

\begin{equation}
{\cal L}^{1,\phi}(q)=L^{1,\phi}(q)-W^1(q)
\frac{L^{0,\phi}(q)}{dq}\,.
\end{equation} 

In order to obtain the integral equations for $L^{0,\phi}(q)$ and 
${\cal L}^{1,\phi}(q)$ (with ${\cal L}={\cal L}_{c,0}, 
{\cal L}_{c,\gamma}$, and ${\cal L}_{s,\gamma}$), we start from 
the continuum limit of Eqs. (\ref{5tak1}), (\ref{5tak2}), and 
(\ref{5tak3}) which reads

\begin{eqnarray}
K(q) &=& q
+\phi_\uparrow/N_a-\sum_{\gamma'}\frac 1{2\pi}
\int_{-q_{s,\gamma'}}^{q_{s,\gamma'}}
dq\,'N_{s,\gamma'}(q\,')\, 2\tan^{-1}\left(\frac{\sin(K(q))/u-
R_{s,\gamma'}(q\,')}
{(\gamma'+1)} \right)\nonumber\\
&-&\sum_{\gamma' >0} \frac 1{2\pi}
\int_{-q_{c,\gamma'}}^{q_{c,\gamma'}}
dq\,'N_{c,\gamma'}(q\,')\, 2\tan^{-1}\left(\frac{\sin(K(q))/u-
R_{c,\gamma'}(q\,')}
{\gamma'}\right)\, ,
\label{5int1}
\end{eqnarray}

\begin{eqnarray}
&&2 Re \, \sin^{-1}[(R_{c,\gamma}(q)-i\gamma)u] = q
+ \gamma (\phi_\uparrow+\phi_\downarrow)/N_a-\nonumber\\
&-&\frac 1{2\pi} \int_{-q_c}^{q_c}
dq\,'N_c(q\,')\, 2\tan^{-1}\left(\frac{\sin(K(q\,'))/u-R_{c,\gamma}(q)}
{\gamma}\right)\nonumber\\
&+&\sum_{\gamma'>0} \frac 1{2\pi}
\int_{-q_{c,\gamma'}}^{q_{c,\gamma'}}
dq\,'N_{c,\gamma'}(q\,')\Theta_{\gamma,\gamma'}
(R_{c,\gamma}(q)-R_{c,\gamma'}(q\,')\, ,
\label{5int2}
\end{eqnarray}
and

\begin{eqnarray}
q&=& (\gamma+1)(\phi_\uparrow-\phi_\downarrow)/N_a+
\frac 1{2\pi} \int_{-q_c}^{q_c} dq\,'N_c(q\,')\,
2\tan^{-1}\left(\frac{R_{s,\gamma}(q)-\sin(K(q\,'))/u}
{(\gamma+1)}\right)\nonumber\\
&-&
\sum_{\gamma'}\frac
1{2\pi}\int_{-q_{s,\gamma}}^{q_{s,\gamma}}
dq\,'N_{s,\gamma'}(q\,')\Theta_{\gamma+1,\gamma'+1}
(R_{s,\gamma} (q)-R_{s,\gamma'}(q\,'))\, .
\label{5int3}
\end{eqnarray}

It is convenient to write the function 
$\Theta^{[1]}_{\gamma,\gamma'}(x)$,
defined by Eq. (\ref{derivativeteta}) of Chapter
\ref{PseudoPT}, as follows

\begin{equation}
\Theta^{[1]}_{\gamma,\gamma'}(x) =
\sum_{l}{2b_{l}^{\gamma ,\gamma'}\over 1 +
[x/l]^2} \, .
\label{5T1}
\end{equation} 
We emphasize that comparision term by term of
Eq. (\ref{derivativeteta}) of Chapter
\ref{PseudoPT} with expression 
(\ref{5T1}) above fully defines the coefficients
$b_{l}^{\gamma ,\gamma'}$ and the corresponding 
set of integer numbers $l$.  

Following equation (\ref{5phis}), we have that
$\phi_\uparrow=\phi_\downarrow$ for a charge-probe
current and $\phi_\uparrow=-\phi_\downarrow$ for a spin probe. 
With the above equations written in terms of
$\phi_\uparrow$ and $\phi_\downarrow$, Eq.\,(\ref{5jnorm})
provides both the charge and spin currents. In what follows,
we introduce in the functions $L^{\phi}(q)$ the index $\zeta=
\rho,\sigma_z$
to label the equations for either the charge or the spin current,
respectively. We start by expanding 
Eqs.\,(\ref{5int1}), (\ref{5int2}), and (\ref{5int3}) up
to first order in $\phi$. This procedure reveals that the functions 
$L^{\phi,\zeta}(q)$ obey the following integral equations

\begin{eqnarray}
L^{\phi,\zeta}_{c,0} (q) &=& {\cal C}^\zeta_{c,0}
+\sum_{\gamma'}\frac 1{(\gamma'+1)\pi }
\int_{-q_{s,\gamma'}}^{q_{s,\gamma'}}
dq\,'{N_{s,\gamma'}(q\,')\over {1 + 
[\frac {\sin(K(q))/u-R_{s,\gamma'}(q\,')}{\gamma' +1}]^2}}
\frac{dR_{s,\gamma'}(q\,')}{dq\,'}L_{s,\gamma'}^{\phi,\zeta}(q\,')
\nonumber\\
&+&\sum_{\gamma' >0}\frac 1{\pi \gamma'}
\int_{-q_{c,\gamma'}}^{q_{c,\gamma'}}
dq\,'{N_{c,\gamma'}(q\,')\over {1 +  
[\frac {\sin(K(q))/u-R_{c,\gamma'}(q\,')}{ \gamma'}]^2}}
\frac{dR_{c,\gamma'}(q\,')}{dq\,'}
L_{c,\gamma'}^{\phi(q'),\zeta}(q\,') \, ,
\label{5lint1}
\end{eqnarray}

\begin{eqnarray}
L^{\phi,\zeta}_{c,\gamma} (q) &=& {\cal C}^\zeta_{c,\gamma}+
\frac 1{\pi u \gamma}\int_{-q_{c}}^{q_{c}}
dq\,'{N_c(q\,')\over {1 + 
[\frac{\sin(K(q\,'))/u-R_{c,\gamma}(q)}{ \gamma}]^2}}
\frac{dK(q\,')}{dq\,'}\cos(K(q\,'))L_{c,0}^{\phi,\zeta}(q\,')\nonumber\\
&+&\sum_{\gamma'>0}\sum_l\frac 1{\pi l}
\int_{-q_{c,\gamma'}}^{q_{c,\gamma'}}
dq\,'{N_{c,\gamma'}(q\,') b^{\gamma ,\gamma'}_l\over {1 +
[\frac{R_{c,\gamma}(q)-R_{c,\gamma'}(q\,')}{l}]^2}}
\frac{dR_{c,\gamma'}(q\,')}{dq\,'}L_{c,\gamma'}^{\phi,\zeta}(q')\, ,
\label{5lint2}
\end{eqnarray}
and

\begin{eqnarray}
L^{\phi}_{s,\gamma} (q) &=& {\cal C}^\zeta_{s,\gamma}+\nonumber\\
&+&\frac 1{u(\gamma+1) \pi}
\int_{-q_{c}}^{q_{c}}dq\,'{N_c(q\,')\over {1 + 
[\frac {\sin(K(q\,'))/u-R_{s,\gamma}(q)}{
\gamma+1}]^2}} \frac{dK(q\,')}{dq\,'}\cos(K(q\,'))
L_{c,0}^{\phi,\zeta}(q\,')\nonumber\\
&-&\sum_{\gamma'}\sum_l\frac 1{\pi l}
\int_{-q_{s,\gamma'}}^{q_{s,\gamma'}}
dq\,'{N_{s,\gamma'}(q\,') b^{\gamma +1,\gamma' +1}_l\over {1 +
[\frac{R_{s,\gamma}(q)-R_{s,\gamma'}(q\,')}{l}]^2}}
\frac{dR_{s,\gamma'}(q\,')}{dq\,'}L_{s,\gamma'}^{\phi,\zeta}(q') \, , 
\label{5lint3}
\end{eqnarray}
where the coupling constants ${\cal C}^{\zeta}_{\alpha,\gamma}$
are defined by Eqs. (\ref{5coupling}) and (\ref{5coupling2}).
We again write the distributions functions $N_{\alpha,\gamma}(q)$ of
Eqs.\,(\ref{5lint1}), (\ref{5lint2}), and (\ref{5lint3}) as 
$N_{\alpha,\gamma}^0(q)+\delta N_{\alpha,\gamma}(q)$. This allows 
us to obtain integral equations
for $L^{0,\phi,\zeta}(q)$ and ${\cal L}^{1,\phi}(q)$
(we remark that the functions ${\cal L}^{1,\phi}(q)$
are the same both for $\zeta=\rho,\sigma_z$). It is then
straighforward to find the integral equations obeyed 
by $L^{0,\phi,\zeta}(q)$ and show that $L^{0,\phi,\zeta}(q)$  
can be simply expressed in terms of linear combinations of 
phase shifts. The final result is

\begin{equation}
L_{\alpha,\gamma}^{0,\phi,\zeta}(q) =
{\cal C}^\zeta_{\alpha,\gamma}+
\sum_{\alpha',\gamma'}\sum_{j=\pm 1}j\theta(N_{\alpha',\gamma'}) 
{\cal C}^\zeta_{\alpha',\gamma'}
\Phi_{\alpha,\gamma;\alpha',\gamma'}(q,jq_{F\alpha',\gamma'})\, .
\label{5l0zeta}
\end{equation}
The integral equations obeyed by ${\cal L}^{1,\phi}(q)$ are related
to the integral equations obeyed by $\tilde{{\cal L}}^{1,\phi}(r)$, 
where $r$ equals $\sin(K^{(0)}(q))/u$, $R^{(0)}_{c,\gamma}(q)$, and 
$R^{(0)}_{s,\gamma }(q)$ for ${\cal L}={\cal L}_{c,0}$, 
${\cal L}_{c,\gamma}$, and ${\cal L}_{s,\gamma}$, respectively. 
The functions $\tilde{{\cal L}}^{1,\phi}(r)$ obey the 
following integral equations

\begin{equation}
\tilde{{\cal L}}_{c,0}^{1,\phi}(r)=
\tilde{{\cal L}}_{c,0}^{1,\phi,0}(r)+
\frac 1{\pi} \int_{-r_{s,0}}^{r_{s,0}} dr'
\frac{ \tilde{{\cal L}}_{s,0}^{1,\phi}(r')}
{1+(r-r')^2}\, ,        
\label{5qtilde}
\end{equation}

\begin{equation}
\tilde{{\cal L}}^{1,\phi}_{c,\gamma}(r)=
\tilde{{\cal L}}^{1,\phi,0}_{c,\gamma}(r)-
\frac 1{\pi\gamma u} 
\int_{-r_{c}}^{r_{c}} dr'
\frac{ \tilde{{\cal L}}_{c,0}^{1,\phi}(r')}
{1+(\frac{r-r'}{\gamma})^2} \, ,        
\label{5qgtilde}
\end{equation}
and

\begin{eqnarray}
\tilde{{\cal L}}^{1,\phi}_{s,\gamma} (r)&=&
\tilde{{\cal L}}^{1,\phi,0}_{s,\gamma}(r)-
\frac 1{\pi(\gamma+1)u} \int_{-r_{c}}^{r_{c}} dr'
\frac{ \tilde{{\cal L}}_{c,0}^{1,\phi}(r')}
{1+(\frac{r-r'}{\gamma})^2} \nonumber\\
&-&\sum_l \frac 1{\pi l} \int_{-r_{s,0}}^{r_{s,0}} 
dr'{b^{\gamma +1,1}_l\tilde{{\cal L}}_{s,0}^{1,\phi}(r')\over {1 +
[\frac{r-r'}{l}]^2}} \, ,       
\label{5pgtilde}
\end{eqnarray}
where the free terms $\tilde{{\cal L}}_{c,0}^{1,\phi,0}(r)$,
$\tilde{{\cal L}}^{1,\phi,0}_{c,\gamma}(r)$, and 
$\tilde{{\cal L}}^{1,\phi,0}_{s,\gamma}(r)$ are, respectively, 
given by

\begin{eqnarray}
\tilde{{\cal L}}_{c,0}^{1,\phi,0}(r)&=&
\sum_{\gamma'}\frac1{\pi \gamma'} \int_{-q_{c,\gamma'}}^{q_{c,\gamma'}}
dq' \delta N_{c,\gamma'}(q'){L_{c,\gamma'}^{0,\phi}(q')\over 
{1+[\frac{r-R^{(0)}_{c,\gamma'}(q')}{\gamma'}]^2}}
\frac{dR^{(0)}_{c,\gamma'}(q')}{dq'}\nonumber\\ 
&+&\sum_{\gamma'}\frac1{\pi (\gamma'+1)} 
\int_{-q_{s,\gamma'}}^{q_{s,\gamma'}}
dq' \delta N_{s,\gamma'}(q')\frac{L_{s,\gamma'}^{0,\phi}(q')}
{(1+[\frac{r-R^{(0)}_{s,\gamma' }(q')}{\gamma'+1}]^2)}
\frac{dR^{(0)}_{s,\gamma' }(q')}{dq'}\nonumber\\
&+& \sum_{\gamma'} \theta (N_{c,\gamma'})\frac 1{\gamma' \pi}
\sum_{j=\pm 1}{jL^1_{c,\gamma'}(jq_{Fc,\gamma'})\over
{2\pi\rho_{c,\gamma'}(r_{c,\gamma'})}}
\frac {L_{c,\gamma'}^{0,\phi}(jq_{Fc,\gamma'})}
{(1+[\frac{r-jr_{c,\gamma'}}{\gamma'}]^2)}\nonumber\\
&+&\sum_{\gamma'} \theta (N_{s,\gamma'})\frac 1{(\gamma'+1) \pi}
\sum_{j=\pm 1}{jL^1_{s,\gamma'}(jq_{Fs,\gamma'})\over {
2\pi\rho_{s,\gamma'}(r_{s,\gamma'})}}
\frac {L_{s,\gamma'}^{0,\phi}(jq_{Fs,\gamma'})}
{(1+[\frac{r-jr_{s,\gamma' }}{\gamma'+1}]^2)}\, ,
\end{eqnarray}

\begin{eqnarray}
\tilde{{\cal L}}^{1,\phi,0}_{c,\gamma}(r)&=&
-\frac1{\pi u \gamma} \int_{-q_c}^{q_c}
dq' \delta N_c(q')
\frac{L_{c,0}^{0,\phi}(q')}
{1+[\frac{\sin(K^{(0)}(q'))/u-r}{\gamma}]^2}\cos(K^{(0)}(q'))
\frac{dK^{(0)}(q')}{dq'}\nonumber\\ 
&-&\sum_{\gamma'} \sum_l \frac1{\pi l} 
\int_{-q_{c,\gamma'}}^{q_{c,\gamma'}}
dq' \delta N_{c,\gamma'}(q'){b^{\gamma ,\gamma'}_l
L_{c,\gamma'}^{0,\phi}(q')\over {1 +
[\frac{r-R^{(0)}_{c,\gamma'}(q')}{l}]^2}}
\frac{dR^{(0)}_{c,\gamma'}(q')}{dq'}
\nonumber\\
&-& \frac1{\gamma u \pi}
\sum_{j=\pm 1}{jL^1_{c,0}(jq_{Fc})\over
{2\pi\rho_{c,0}(Q)}}
\frac{L_{c,0}^{0,\phi}(jq_{Fc})}
{(1+[\frac{r-jr_{c,0}}{\gamma}]^2)}\cos (Q)\nonumber\\
&-&\sum_{\gamma'} \theta (N_{c,\gamma'})\sum_l \frac1{\pi l} 
\sum_{j=\pm 1}{j b^{\gamma ,\gamma'}_lL^1_{c,\gamma'}(jq_{Fc,\gamma'})\over
{2\pi\rho_{c,\gamma'}(r_{c,\gamma'})}}
{L_{c,\gamma'}^{0,\phi}(jq_{Fc,\gamma'})
\over {(1 + [\frac{r-jr_{c,\gamma'}}{l}]^2)}} \, ,
\end{eqnarray}
and

\begin{eqnarray}
\tilde{{\cal L}}^{1,\phi,0}_{s,\gamma}(r)&=&
\frac1{\pi u (\gamma +1)} \int_{-q_c}^{q_c}
dq' \delta N_c(q')
\frac{L_{c,0}^{0,\phi}(q')}
{1+[\frac{\sin(K^{(0)}(q'))/u-r}{\gamma +1}]^2}\cos(K^{(0)}(q'))
\frac{dK^{(0)}(q')}{dq'}\nonumber\\
&-&\sum_{\gamma'} \sum_l \frac1{\pi l}
\int_{-q_{s,\gamma'}}^{q_{s,\gamma'}}
dq' \delta N_{s,\gamma'}(q'){b^{\gamma +1 ,\gamma' +1}_l
L_{s,\gamma'}^{0,\phi}(q')\over {1 +
[\frac{r-R^{(0)}_{s,\gamma'}(q')}{l}]^2}}
\frac{dR^{(0)}_{s,\gamma'}(q')}{dq'}
\nonumber\\
&+& \frac1{(\gamma +1) u \pi}
\sum_{j=\pm 1}{jL^1_{c,0}(jq_{Fc})\over
{2\pi\rho_{c,0}(Q)}}
\frac{L_{c,0}^{0,\phi}(jq_{Fc})}
{(1+[\frac{r-jr_{c,0}}{\gamma +1}]^2)}\cos (Q)\nonumber\\
&-&\sum_{\gamma'} \theta (N_{s,\gamma'})\sum_l \frac1{\pi l}
\sum_{j=\pm 1}{j b^{\gamma +1 ,\gamma' +1}_l
L^1_{s,\gamma'}(jq_{Fs,\gamma'})\over
{2\pi\rho_{s,\gamma'}(r_{s,\gamma'})}}
{L_{s,\gamma'}^{0,\phi}(jq_{Fs,\gamma'})
\over {(1 + [\frac{r-jr_{s,\gamma'}}{l}]^2)}} \, .
\end{eqnarray}
Introducing the functions ${\cal L}^{1,\phi}(q)$ obtained 
from Eqs.\,(\ref{5qtilde}), (\ref{5qgtilde}), and (\ref{5pgtilde}) in 
Eq.\,(\ref{5jnorm}) and
keeping terms only up to second order in the density of 
heavy pseudoparticles, we obtain Eq.\,(\ref{5jmean}) with 
$j_{\alpha,\gamma}^\zeta(q)$ given by

\begin{equation}
j_{\alpha,\gamma}^\zeta(q)=v_{\alpha,\gamma}(q)
L^{0,\phi,\zeta}_{\alpha,\gamma}(q)+\sum_{\alpha',\gamma'}
\sum_{j=\pm 1}j \theta (N_{\alpha',\gamma'})
v_{\alpha',\gamma'}
L^{0,\phi,\zeta}_{\alpha',\gamma'}(jq_{F\alpha',\gamma'})
\Phi_{\alpha',\gamma';\alpha,\gamma}(jq_{F\alpha',\gamma'},q)\, .
\label{5jq2}     
\label{5jnorm2}
\end{equation}
Inserting  Eq.\,(\ref{5l0zeta}) in Eq.\,(\ref{5jq2}) we obtain
Eq.\,(\ref{5jq}).

%% file: apsma.tex
\chapter{Static masses for the heavy pseudoparticles}
\label{5static}
\pagestyle{myheadings}
\markboth{Static masses for ...}{Appendix B of Chapter 6}

The $\alpha,\gamma$ pseudoparticle static mass, $m^{\ast}_{\alpha,\gamma}$, is defined in 
Ref. \cite{5CarmeloNuno97} as

\begin{equation}
\frac 1 {m^{\ast}_{\alpha,\gamma}}= \left .
\frac{2t\,d\eta_{\alpha,\gamma}(r)/dr}
{(2\pi \rho_{\alpha,\gamma}(r))^2}\right \vert_{r=r^0}
-\left . \frac{2t\eta_{\alpha,\gamma}(r)
(2\pi\,d\rho_{\alpha,\gamma}(r)/dr)}
{(2\pi \rho_{\alpha,\gamma}(r))^3}\right \vert_{r=r^0} \, ,
\label{5staticdef}
\end{equation}
where the functions $2t\eta_{\alpha,\gamma}(r)$ and 
$2\pi\rho_{\alpha,\gamma}(r)$ are defined in Ref.\, 
\cite{5CarmeloNuno97} and $r^0$ stands for
$W^0_{\alpha,\gamma}(q_{F\alpha,\gamma})$ which
represents $Q$, $r_{c,\gamma}$, and $r_{s,\gamma}$.

After some straightforward algebra, the general expressions 
(\ref{5staticdef}) lead to the following simple expressions 
for $1/m^{\ast}_{\alpha,\gamma}$

\begin{equation}
\frac 1 {m^{\ast}_{c,\gamma}}=\frac 
{-4tu^2/(1+u^2\gamma^2)^{3/2}+\Lambda^{\eta}_{c,\gamma}}
{(2u/\sqrt{1+u^2\gamma^2}-\Lambda^{\rho}_{c,\gamma})^2}\,,
\hspace{1cm}\gamma>0 \,,
\label{51overmc}
\end{equation}
and

\begin{equation}
\frac 1 {m^{\ast}_{s,\gamma}}=\frac 
{\Lambda^{\eta}_{c,\gamma+1}-
\Lambda^{\eta}_{s,\gamma}-\Lambda^{\eta}_{s,\gamma+2}}
{(\Lambda^{\rho}_{c,\gamma+1}-\Lambda^{\rho}_{s,\gamma}-
\Lambda^{\rho}_{s,\gamma+2})^2}\,,\hspace{1cm}
\gamma>0 \,.
\label{51overms}
\end{equation}
In Eqs.\,(\ref{51overmc}) and (\ref{51overms}) the functions
$\Lambda^{\eta}_{\alpha,x}$ and
$\Lambda^{\rho}_{\alpha,x}$ read

\begin{equation}
\Lambda^{\eta}_{\alpha,x}=2\int_{-q_{F\alpha,0}}
^{q_{F\alpha,0}} \frac {dq}{\pi x^3}
\frac{v_{\alpha,0}(q)R^{(0)}_{\alpha,0}(q)}
{[1+(R^{(0)}_{\alpha,0}(q)/x)^2]^2} \, ,
\end{equation}
and 

\begin{equation}
\Lambda^{\rho}_{\alpha,x}=\int_{-q_{F\alpha,0}}
^{q_{F\alpha,0}} \frac {dq}{\pi x}
\frac{1}
{1+[R^0_{\alpha,0}(q)/x]^2} \, ,
\end{equation}
with

\begin{equation}
R^{(0)}_{c,0}(q)=\frac {\sin (K^{(0)}(q))}{u} \, .
\end{equation}

In the limit of fully polarized ferromagnetism, these expressions 
lead to the following closed-form expressions for the static masses

\begin{equation}
\frac 1{m_{c,\gamma}^{\ast}}=\frac {t\pi}
{8[\eta_{1,\gamma}]^2}\left ( -\frac {\pi
+2[\eta_{1,\gamma}]}{\sqrt {1+[u\gamma]^2}}
-\frac {u\gamma\sin(2n\pi)}{[u\gamma]^2+\sin^2(n\pi)}
\right)
\label{5mcast}
\end{equation} 
and
\begin{equation}
\frac 1{m_{s,\gamma}^{\ast}}=\frac {t\pi}
{[\eta_{2,\gamma+1}]}\left ( \frac {1}{\sqrt {1+[u(\gamma+1)]^2}}
-\frac{u(\gamma+1)}{2[\eta_{2,\gamma+1}]}
\frac {\sin(2n\pi)}{[u(\gamma+1)]^2+\sin^2(n\pi)}
\right)\, ,
\label{5msast}
\end{equation}
where

\begin{equation}
\eta_{1,x}=\tan^{-1}[\cot (n\pi)\frac {x u}
{\sqrt{1+u^2 x^2}}]\, ,
\end{equation}
and $\eta_{2,x}=\pi/2-\eta_{1,x}$. 
We remark that the static masses of the $c,\gamma$ pseudoparticles 
are, in general, negative. The static masses of the $\alpha,0$ 
pseudoparticles have been studied in Ref. \cite{5Carm3}.

%% file: apstif.tex
\chapter{Charge and spin stiffnesses at zero temperature}
\label{5stiffness}
\pagestyle{myheadings}
\markboth{Charge and spin ...}{Appendix C of Chapter 6}

In this Appendix we show that the direct use of Eq. (\ref{5drude})
leads to the stiffness expressions $(135)$ - $(137)$ of
Ref. \cite{5Carm4a,5Carm4b}.

The calculation of the charge and spin stiffnesses (\ref{5drude})
requires the expansion of Eq.\,(\ref{5energy}) and of 
Eqs.\,(\ref{5int1}), (\ref{5int2}), and (\ref{5int3}) up 
to second order in $\phi$. As in the case of the charge and spin 
current, both he charge and spin stiffnesses can be computed from 
Eq.\,(\ref{5drude}), and we obtain one or the other depending on the 
coupling constants we choose in Eqs.\, (\ref{5int1}), (\ref{5int2}), 
and (\ref{5int3}). Expanding the ground-state 
energy up to second order in $\phi$, we obtain

\begin{equation}
\left . \frac{d^2(E_0/N_a)}{d(\phi/N_a)^2}
\right \vert_{\phi=0}=
\frac 1{2\pi} \int_{-q_{Fc}}^{q_{Fc}} dq \left[
2t K^{0,\phi \phi}(q) \sin(K^{(0)}(q)) +
2t [K^{0,\phi}(q)]^2 \cos(K^{(0)}(q))\right]\, ,
\label{5e2fi}
\end{equation}
where the function $K^{0,\phi \phi}(q)$ is the second 
derivative of the rapidity function defined by Eq. \,(\ref{5int1}) 
in order to $\phi/N_a$ at $\phi=0$. The functions $K^{0,\phi}(q)$ 
and $K^{0,\phi \phi}(q)$ can be written as

\begin{equation}
K^{0,\phi}(q)=\frac {dK^{(0)}(q)}{dq} L_{c,0}^{0,\phi}(q)\, ,
\label{5k0phi}
\end{equation}
and

\begin{equation}
K^{0,\phi \phi}(q)=\frac d{dq}
\left(\frac {dK^{(0)}(q)}{dq}[L_{c,0}^{0,\phi}(q)]^2\right)+
2\frac {dK^{(0)}(q)}{dq}L^{0,\phi \phi}_{c,0,\ast}\, ,
\label{5k02phi}
\end{equation}
respectively. The use of Eqs.\,(\ref{5k0phi}) and (\ref{5k02phi}) in 
Eq.\,(\ref{5e2fi}) leads then to

\begin{eqnarray}
\left . \frac{d^2(E_0/N_a)}{d(\phi/N_a)^2}
\right \vert_{\phi=0}&=&
\frac1{2\pi}\int_{-q_{Fc}}^{q_{Fc}} dq
2t\sin(K^{(0)}(q))             
2\frac{dK^{(0)}(q)}{dq}L^{0,\phi\phi}_{c,0,\ast}(q)
\nonumber\\
&+&\frac 1{2\pi}\sum_{j=\pm 1} {2t\sin(Q)
[L_{c,0}^{0,\phi}(jq_{Fc})]^2\over
{2\pi\rho_{c,0}(Q)}} \, ,
\label{5e2phi}
\end{eqnarray}
where the function $L_{c,0}^{0,\phi}(jq_{Fc})$ is defined in Appendix 
\ref{5normalOrder}. The function $L^{0,\phi\phi}_{c,0,\ast}(q)$ 
obeys the following integral equation 

\begin{eqnarray}
L^{0,\phi \phi}_{c,0,\ast} (q) &=& \frac 1{2\pi} 
\sum_{j=\pm 1} \frac {j[L_{s,0}^{0,\phi}(jq_{Fs,0})]^2}
{2\pi\rho_{s,0}(r_{s,0})(1+[\sin(K^{(0)}(q))/u-jr_{s,0}]^2)}\nonumber\\
&+&\frac 1{\pi}\int_{-q_{Fs,0}}^{q_{Fs,0}}dq' \frac 
{dR^{(0)}_{s,0}(q')}{dq'}
\frac {L^{0,\phi \phi}_{s,0,\ast} (q')}
{1+[\sin(K^{(0)}(q))/u-R^0_{s,0}(q')]^2} \, ,
\label{5Qast}
\end{eqnarray}
which was obtained by performing the type of expansions developed 
in Appendix \ref{5normalOrder}.
Moreover, $L^{0,\phi \phi}_{s,0,\ast} (q)$ is given by

\begin{eqnarray}
L^{0,\phi \phi}_{s,0,\ast} (q)&=&\frac 1{2 u \pi} \sum_{j=\pm 1}
{j[\cos(Q) L_{c,0}^{0,\phi}(jq_{Fc})]^2\over 
{2\pi\rho_{c,0}(Q)(1+[R^{(0)}_{s,0}(q)-jr_{c,0}]^2)}}\nonumber\\
&-&\frac 1{4\pi} \sum_{j=\pm 1}{j[L_{s,0}^{0,\phi}(jq_{Fs,0})]^2
\over {2\pi\rho_{s,0}(r_{s,0})
(1+[(R^{(0)}_{s,0}(q)-jr_{s,0})/2]^2)}}\nonumber\\
&-&\frac 1{2\pi}\int_{-q_{Fs,0}}^{q_{Fs,0}}dq' 
\frac {dR^{(0)}_{s,0}(q')}{dq'}
\frac {L^{0,\phi \phi}_{s,0,\ast} (q')}
{1+[(R^{(0)}_{s,0}(q)-R^{(0)}_{s,0}(q'))/2]^2}\nonumber\\
&+&\frac 1{\pi u}\int_{-q_{Fc}}^{q_{Fc}}dq' \frac {dK^{(0)}(q')}{dq'}
\frac {\cos (K^{(0)}(q'))L^{0,\phi \phi}_{c,0,\ast} (q')}
{1+(\sin(K^{(0)}(q'))/u-R^{(0)}_{s,0}(q))^2} \, .
\label{5Past}
\end{eqnarray}
Introducing Eqs.\,(\ref{5Qast}) and (\ref{5Past}) in Eq.\,(\ref{5e2phi}) 
we obtain, after some algebra, the following expression
for the (charge and spin) stiffness $D^\zeta$

\begin{equation}
4\pi D^\zeta = \sum_{j=\pm 1}v_{c,0}
[L_{c,0}^{0,\phi,\zeta}(jq_{Fc,0})]^2
+\sum_{j=\pm 1}v_{s,0}[L_{s,0}^{0,\phi,\zeta}(jq_{Fs,0})]^2\, , 
\label{5dzeta}
\end{equation}
where the functions $L^{0,\phi,\zeta}(jq_{F\alpha,0})$ are 
defined by Eq.\,(\ref{5l0zeta}). After some simple
algebra, expression\,(\ref{5dzeta})
can be shown to be the same as expressions $(135)$ - $(137)$
of Ref. \cite{5Carm4a,5Carm4b}.

%% file: apzero.tex
\chapter{The zero magnetic-field solution}
\label{5zeroH}
\pagestyle{myheadings}
\markboth{The zero magnetic-field ...}{Appendix D of Chapter 6}

For the case of zero magnetic field it is 
possible to cast the equations for the energy bands, phase shifts, 
and rapidities in a simpler form. After some algebra, the 
$\epsilon_{s,\gamma}^0(q)$ band (with $\gamma=0,1,2,\ldots,\infty$) 
and the $\epsilon_{c,\gamma}^0(q)$ band with 
($\gamma=1,2,\ldots,\infty$) can at zero magnetic field
be rewritten as

\begin{equation}
\epsilon_{s,\gamma}^0(q) =
-\delta_{\gamma ,0} \left[2t\int_0^{\infty}d \omega
{\cos(\omega R^{(0)}_{s,0}(q))\over \omega\cosh (\omega)}
\Upsilon_1(\omega)\right] \, ,
\label{5esgh0}
\end{equation}

and
\begin{eqnarray}
\epsilon_{c,\gamma}^0(q)&=&
Re\,4t\sqrt{1-u^2(R_{c,\gamma}^{(0)}(q)+i\gamma)^2}-\nonumber\\
&-&4t\int_0^{\infty}d \omega
\frac{e^{-\gamma \omega}}{\omega}\cos(\omega R^{(0)}_{c,\gamma}(q))
\Upsilon_1(\omega) \, ,
\label{5ecgh0}
\end{eqnarray}
where $\Upsilon_1(\omega)$ obeys the integral equation

\begin{eqnarray}
\Upsilon_1(\omega)=\Upsilon_1^0(\omega)+\int_{-\infty}^{\infty}
d \omega' \Upsilon_1(\omega') \Gamma(\omega',\omega) \, .
\label{5up1}
\end{eqnarray}
Here the free term and the kernel read

\begin{equation}
\Upsilon_1^0(\omega)=\frac 1 {2 \pi} \int_{-Q}^{Q}dk 
\sin(k)\sin(\omega \sin(k)/u)\, ,
\label{5up01}
\end{equation}
and

\begin{equation}
\Gamma(\omega',\omega)=\frac {\sin((\omega-\omega')r_{c,0})}
{\pi(\omega-\omega')(1+e^{2\vert \omega' \vert})}\, ,
\label{5gama1}
\end{equation}
respectively. The kernel (\ref{5gama1}) was already obtained
in Ref. \cite{5Carm88} (see Eq. (A5) of that reference). 
Equation \,(\ref{5esgh0}), together with the fact
that in the limit of zero magnetic field the width of the
$s,\gamma >0$ momentum pseudo-Brillouin zone vanishes
[see Eq. (\ref{5gsbz})], shows that the $s,\gamma $ bands collapse for 
$\gamma>0$ and all values of $U$ and $n$ to the point 
zero. 

In this limit it is also possible to cast the integral
equations for the phase shifts, whose expressions are given in Ref.\,
\cite{5CarmeloNuno97}, in the following alternative form

\begin{equation} 
\bar {\Phi}_{c,0;c,0}(r,r')=-B(r-r')+\int_{-r_{c,0}}^{r_{c,0}}dr''
\bar {\Phi}_{c,0;c,0}(r'',r')A(r-r'') \, ,
\label{5c0c0}
\end{equation}

\begin{equation}
\bar {\Phi}_{c,0;c,\gamma'}(r,r')=-\frac 1{\pi}
\tan^{-1}(\frac {r-r'}{\gamma'})+\int_{-r_{c,0}}^{r_{c,0}}dr''
\bar {\Phi}_{c,0;c,\gamma'}(r'',r') A(r-r'') \, ,
\label{5c0cg}
\end{equation}

\begin{eqnarray}
\bar {\Phi}_{c,0;s,\gamma'}(r,r')&=&-\delta_{0,\gamma'}
\frac 1{2\pi}\tan^{-1}[\sinh(\pi /2(r-r'))]
+\nonumber\\
&+&\int_{-r_{c,0}}^{r_{c,0}}\frac {dr''}{\gamma'+1}
\bar {\Phi}_{c,0;s,\gamma'}(r'',r') A(r-r'') \, ,
\label{5c0sg}
\end{eqnarray}

\begin{equation}
\bar {\Phi}_{c,\gamma;c,0}(r,r')=\frac 1{\pi}
\tan^{-1}(\frac {r-r'}{\gamma})-\int_{-r_{c,0}}^{r_{c,0}}
\frac {dr''}{\pi \gamma}
\frac {\bar {\Phi}_{c,0;c,0}(r'',r')}
{1+(\frac {r-r''}{\gamma})^2} \, ,
\label{5cgc0}
\end{equation}

\begin{equation}
\bar {\Phi}_{c,\gamma;c,\gamma'}(r,r')=
\frac 1{2\pi}
\Theta_{\gamma,\gamma'}(r-r')
-\int_{-r_{c,0}}^{r_{c,0}}
\frac {dr''}{\pi \gamma}
\frac {\bar {\Phi}_{c,0;c,\gamma'}(r'',r')}
{1+(\frac {r-r''}{\gamma})^2} \, ,
\label{5cgcg}
\end{equation}

\begin{equation}
\bar {\Phi}_{c,\gamma;s,\gamma'}(r,r')=
-\int_{-r_{c,0}}^{r_{c,0}}
\frac {dr''}{\pi \gamma}
\frac {\bar {\Phi}_{c,0;s,\gamma'}(r'',r')}
{1+(\frac {r-r''}{\gamma})^2} \, ,
\label{5cgsg}
\end{equation}

\begin{equation}
\bar {\Phi}_{s,\gamma;c,0}(r,r')=0 \, ,
\label{5sgc0}
\end{equation}

\begin{equation}
\bar {\Phi}_{s,\gamma;c,\gamma'}(r,r')=
\delta_{0,\gamma}\frac 1{4}
\int_{-r_{c,0}}^{r_{c,0}}dx''
\frac {\bar {\Phi}_{c,0;c,\gamma'}(r'',r')}
{\cosh(\pi /2(r''-r))} \, ,
\label{5sgcg}
\end{equation}

\begin{eqnarray}
\bar {\Phi}_{s,0;s,\gamma'}(r,r')&=& 
\frac 1{\pi} \int_0^\infty d\omega {\sin(\omega[r-r'])
\over \omega(1+e^{2\omega})}
[e^{-\gamma' \omega}+(1-\delta_{0,\gamma'})
e^{-(\gamma'-2)\omega}]\nonumber\\
&+&\frac 1{4}
\int_{-r_{c,0}}^{r_{c,0}}
\frac {dr''}{\gamma'+1}
\frac {\bar {\Phi}_{c,0;s,\gamma'}(r'',r')}
{\cosh(\pi /2(r-r''))} \, ,
\label{5s0sg}
\end{eqnarray}
and

\begin{eqnarray}
&&\bar {\Phi}_{s,\gamma >0;s,\gamma'}(r,r')=
\frac 1{2\pi}
\Theta_{\gamma+1,\gamma'+1}(r-r')
-\nonumber\\
&-& \frac 1{\pi} \int_0^\infty d\omega\frac 
{\sin[\omega(r-r')]}
{\omega(1+e^{2\omega})}e^{-(\gamma'+\gamma)\omega}
[2-\delta_{0,\gamma'}+e^{-2 \omega}+(1-\delta_{0,\gamma'})
e^{2\omega}] \, .
\label{5sgsg}
\end{eqnarray}
The functions $A(r)$ and $B(r)$ are defined as

\begin{equation}
A(r)=\frac 1{\pi} \int_0^{\infty} d\omega \frac {\cos(r\omega)}
{1+e^{2|\omega|}}\, ,
\end{equation}
and

\begin{equation}
B(r)=\frac 1{\pi} \int_0^{\infty} d\omega \frac {\sin(r\omega)}
{\omega (1+e^{2|\omega|)}}\, ,
\end{equation}
respectively.

At $n=1$ we have that $\Upsilon_1(x)=\Upsilon^0_1(x)
=J_1(x/u)$, where $J_1(x/u)$ is the Bessel function 
of order one, and the bands (\ref{5esgh0}) and (\ref{5ecgh0}) are 
obtained in closed form. 

%% file: conclusi.tex
\pagestyle{headings}
\setcounter{chapter}{6}
\chapter{Conclusions}
\label{concl}
\pagestyle{myheadings}
\markboth{7 Conclusions}{}

In this Thesis we have focused our attention on the one-dimensional Hubbard model. As it became clear from Chapters \ref{introdu} and \ref{hm}, the Hubbard Hamiltonian, as introduced by Hubbard, Gutzwiller, and Kanamori 
\cite{6Hubbard63,6Gutzwiller63,6Kanamori63}, is the simplest model for the description of interacting electrons in a lattice. 

Introduced firstly to study itinerant ferromagnetism in transition metals, the Hubbard model revealed later to have an antiferromagnetic ground state
\cite{6Lieb95}. With the
discovery of a new class of superconducting materials, by Berdnoz and M\"{u}ller in 1986 \cite{6Muller93} 
and the proposal, by Anderson \cite{6Anderson97}, that the 2D Hubbard model would retain the essential physics of these new materials, the model gained renewed interest. It soon became clear that, for the values of the physical parameters imposed by the experimental results, the usual theoretical tools would not suffice to deal with this model (and its ``relatives")
\cite{6Assa94} and that, accordingly, 
the search for non-perturbative methods should be pursued. The claim by Anderson that the 2D Hubbard model should present many similarities with its one-dimensional  counterpart
\cite{6Anderson97}, for which an exact solution is available, together with better crystals of quasi one-dimensional materials, and more refined experimental methods (e.g. angle resolved photoemission measurements)
\cite{6Wurzburg}, renewed the interest in the one-dimensional Hubbard model. 

At present, the theoretical goal associated with the study of the Hubbard model is to investigate the role  of electron-electron interactions in relation with the thermodynamic and transport properties of a non-perturbative electron liquid. Based on the exact Bethe-ansatz solution of the Hubbard chain and on its symmetries, we presented 
an algebraic representation for all the eigenstates of the model (Chapter \ref{harmonic}). Our representation is much simpler than the starting Bethe-ansatz wave function. The latter is a complicated sum of permutations written in terms of the original electronic operators, while the former representation is a simple Slater determinant of pseudoholes and heavy pseudoparticles (Chapter \ref{harmonic}). These latter operators are the quantum operators that 
diagonalize the Hubbard Hamiltoninan. In this representation it is then simpler to select and characterize the relevant eigenstates which contribute to a given physical property of the model. For example, the low-temperature thermodynamics for electronic fillings less than one 
and finite magnetization is controlled by the eigenstates with no heavy-pseudoparticles
occupancy; on the other hand, the optical conductivity is controlled, at half filling, by  eigenstates with one $c,1$ heavy pseudoparticle  and two $c,0$ pseudoholes. This kind of characterization is by no means possible for the usual Bethe-ansatz wave function, and it is essencial for a simple calculation of correlation functions using conformal-field theory 
methods
\cite{6Frahm90,6Frahm91,6GCFT1,6GCFT2}.

The pseudoparticle perturbation theory
(PPT) described in Chapter \ref{PseudoPT}, combined with
the results of Chapter \ref{harmonic} and with new results on conformal theory methods \cite{6GCFT1,6GCFT2}, allow the calculation of correlation functions at low $(\omega-\omega_0)$ frequency values (note that $\omega_0$ can be very large). This possibility goes much beyond both bosonization studies \cite{6Giamarchi92}, and the spinon and holon conformal-field theory of Framh and Korepin \cite{6Frahm90,6Frahm91}. The possibility opened up by our pseudoparticle theory, namely that of giving finite energy results for the spectral functions of the model, allows the comparison of finite energy spectral functions in real materials with the results obtained for the Hubbard model. This is a topic that deserves further research.

The use of the algebraic representation introduced in Chapter \ref{harmonic} permits the computation of the energy
eigenvalues for all the eigenstates of the model. This can be accomplished, at least in principle, by means of the  PPT
developed in Chapter \ref{PseudoPT}. This PPT is an expansion
of the Hamiltoninan in the density of excited pseudoparticles, relatively to a suitable reference state. The most convenient choice for the reference state is, for most of the physical properties, the ground state or the generalized ground-state, both introduced in Chapter \ref{harmonic}. 

The normal-ordered PPT Hamiltonian describes the physics associated with the Hilbert subspace spanned by the reference state and all Hamiltonian eigenstates differing from it by the occupancies of a small density of pseudoparticles. Combined with a generalized conformal-field theory, the PPT provides the leading-order expressions of correlation functions for low-energies $(\omega-\omega_0)$. These expressions refer either to the low-energy functions $(\omega_0=0)$
\cite{6Frahm90,6Frahm91,6CarmeloPedro97} or to energy values closed to the energy gaps, $\omega_0$, of Hamiltonian eigenstates with heavy-pseudoparticle occupancy \cite{6GCFT1,6GCFT2}.

Very important for the description of excitations that  change the number of electrons $N_{\uparrow}$ and/or $N_{\downarrow}$ (change of canonical ensemble) and/or the number of pseudoparticles (change of sub-canonical ensemble) is the study of the collective pseudo momentum shifts of value $\pm \pi/N_a$. These momentum shifts are generated by the topological momentum shift operators introduced in Chapter \ref{finitesize}. These are important in two ways: (i) they contribute to the  momentum of state transitions connecting different canonical ensembles and sub-canonical ensembles
\cite{6Nuno96,6Neto96}; (ii) in transitions between ground states, the final pseudoparticle momenta are such that the final ground
state has the lowest possible energy \cite{6Ricardo92}. 
The importance of these types of excitations had not been recognized before in the literature in its full relevance.

The problem of charge and spin transport in the
one-dimensional many-body problem, in general, and in one-dimensional integrable models, in particular, is a difficult and subtle issue \cite{6Castella1,6Castella2,6Zotos,6Kawakami97,6Andrei97}. By using the Bethe-ansatz equations with twisted boundary conditions,  we have obtained a general formula for the
mean value of the charge and spin current operators
(Chapter \ref{transport}). However, the functional character of this expression does not seem amenable to practical calculations. Combining this general functional with the PPT, and performing equivalent calculations to those presented in Chapter \ref{PseudoPT}, we were able to compute the mean value of the charge and spin currents for a given subset of eigenstates of the model. The obtained normal-ordered functional provides the current for states that differ from the reference state by a small density of excited pseudoparticles.  Our current functional provided the values for the
transport mass of the charge and spin carriers. These masses provide important information on the role of electronic
correlations in the transport properties. Our functional representation can also provide information on the
transport of charge and spin at finite temperatures. Althougt the PPT does not cover the full range of temperatures $T$ from $0$ to $\infty$, it provides
some information in some regimes. For example, away from half filling, the regime where PPT can be used is $T\ll v_{Fc,0}q_{Fc,0}$, while at half filling it is $T\ll \Delta$, where $\Delta$ is the smallest of the energy gaps $\omega_0$
derived in Chapter \ref{PseudoPT}. Studies on finite-temperature transport by use of the PPT is a subject of current research. 

The study of the Bethe-ansatz equations, together with the PPT, gave the coupling constants of the pseudoparticles to the external charge and spin probes. These coupling constants provide selection rules for the ground state transitions. Conformal-field theory provides asymptotic expressions for correlation functions, including their $x,t$ (space and time) and $k,\omega$ (momentum and frequency) dependences. However, it does not provide the multiplicative constants determined by the weights of the ground-state transitions, which control these expressions and dependences. Often, due to symmetry and 
selection rules,
these weights and corresponding constants vanish. Therefore, the above couplings provide relevant information on such rules, which goes beyond conformal-field theory.

Both the coupling constants and the mean value of the charge and spin current operators can also be derived from a semi-classical transport description of the quantum liquid. Comparing this approach to the Bethe-ansatz exact solution (PPT theory), one confirms that the kinetic approach gives exact results for small
$(\omega-\omega_0)$, while for a Fermi liquid the same kinetic approach gives exact results for small $\omega$ only. This difference can be traced back to the existence of only forward scattering among the pseudoparticles at all energy scales, and it is peculiar of integrable one-dimensional models.

The formalism presented here is applicable to 
other one-dimensional integrable systems
\cite{6Shlottmann97}. Examples of these are the super-symmetric $t-J$ and the lattice spinless fermion models. Both these models present interesting transport properties, since the charge current operator does not commute with the Hamiltonian. Besides the charge stiffness, there is a finite-energy absorption that can be studied by combinig the methods presented in Chapters \ref{harmonic}, \ref{PseudoPT}, \ref{finitesize}, and
\ref{transport} with the thechniques of conformal-field theory
\cite{6Frahm90,6Frahm91,6GCFT1,6GCFT2}. Also the study of finite-temperature transport can be fulfilled
by developing the apropriate PPT for these models. 
Following the results presented here,
calculations for the lattice spinless fermion model are currently being developed \cite{6Nuno98}.